\documentclass[12pt,a4paper,twoside,openright]{book}
\usepackage{times}
\usepackage[british]{babel}
\usepackage[left=2.5cm,right=2.5cm,top=3cm,bottom=2.5cm,head=2cm,foot=1.25cm]
{geometry}
\usepackage{nopageno}

\usepackage{fancyhdr}
\usepackage[usenames,dvipsnames,svgnames,table]{xcolor}
\definecolor{MYBLUE}{rgb}{0,0,0.6}

\usepackage{minitoc}
\usepackage{cite}
\usepackage[linktocpage=true]{hyperref}
\hypersetup{linkbordercolor=red,citebordercolor=blue,urlbordercolor=cyan}

\usepackage[intlimits]{amsmath}
\usepackage{amsfonts}
\usepackage{mathtools}
\usepackage{amssymb}
\usepackage{slashed}
\usepackage{graphicx}
\usepackage{graphics}
\usepackage{float}
\usepackage{lscape}

\def\cA{{\cal A}}

\def\cD{{\cal D}}

\def\cF{{\cal F}}

\def\cH{{\cal H}}

\def\cJ{{\cal J}}

\def\cO{{\cal O}}
\def\cP{{\cal P}}
\def\cR{{\cal R}}

\def\cT{{\cal T}}
\def\cV{{\cal V}}

\newcommand{\p}{\partial}

\newcommand{\Dslash}{\ensuremath{D\hspace{-1.5ex} /}}
\newcommand{\DslashIndex}{\ensuremath{D\hspace{-1ex} /}}
\newcommand{\Tr}{\ensuremath{\operatorname{Tr}}}

\newcommand{\Det}{\ensuremath{\operatorname{Det}}}
\newcommand{\be}{\begin{equation}}
\newcommand{\ee}{\end{equation}}
\newcommand{\ba}{\begin{eqnarray}}
\newcommand{\ea}{\end{eqnarray}}
\newcommand{\bi}{\begin{itemize}}
\newcommand{\ei}{\end{itemize}}
\newcommand{\xmark}{{\text{\sffamily X}}}
\newcommand{\gb}{\bar{g}}
\newcommand{\Db}{\bar{D}}
\newcommand{\Rb}{\bar{R}}
\newcommand{\dq}{\frac{d^dq}{(2\pi)^d}}

\def\thebibliography#1{\chapter*{\textcolor{MYBLUE}{Bibliography}}\list
{[\arabic{enumi}]}{\settowidth\labelwidth{[#1]}\leftmargin\labelwidth
\advance\leftmargin\labelsep
\usecounter{enumi}}
\def\newblock{\hskip .11em plus .33em minus .07em}
\sloppy\clubpenalty4000\widowpenalty4000}

\usepackage{array,xcolor}
\newcolumntype{L}{>{\raggedleft}p{0.18\textwidth}}
\newcolumntype{R}{p{0.7\textwidth}}
\newcommand\VRule{\color{gray}\vrule width 1.0pt}

\begin{document}

\dominitoc

\begin{titlepage}
\title{\LARGE \bf QUANTUM GRAVITY \\ FROM FUNDAMENTAL QUESTIONS \\ TO PHENOMENOLOGICAL APPLICATIONS}
\author{~\\~\\~\\~\\~\\\bf Nat\'{a}lia Alkofer}
\date{}
\end{titlepage}
\maketitle


\thispagestyle{empty}

~\vspace{140mm}

This thesis has been financially supported by the Netherlands Organization for Scientific Research (NWO) within the Foundation for Fundamental Research on Matter (FOM) grant 13PR3137.

~\vspace{20mm}

ISBN 978-94-92896-57-5

~\vfill

Printed by GROSCHBERGER DRUCK GmbH


\newpage
\thispagestyle{empty}

\begin{center}
{\Large \bf Quantum Gravity \\ from Fundamental Questions \\ to Phenomenological Applications}

\vspace{10mm}

Proefschrift

\vspace{5mm}

ter verkrijging van de graad van doctor

aan de Radboud Universiteit Nijmegen

op gezag van de rector magnificus prof. dr. J.H.J.M. van Krieken,

volgens besluit van het college van decanen

in het openbaar te verdedigen op

\vspace{5mm}

woensdag 19 september 2018

\vspace{5mm}

om 16:30 uur precies

\vspace{5mm}

door

\vspace{5mm}

{\large \bf {Nat\'{a}lia Alkofer}}

\vspace{5mm}

geboren op 6 mei 1987

\vspace{5mm}

te Niter\'{o}i, Brazili\"e
\end{center}


\newpage
\thispagestyle{empty}

~\vspace{150mm}

Promotor: Prof. dr. Renate Loll

Copromotor: dr. Frank Saueressig

\vspace{5mm}

Manuscriptcommissie:

\vspace{2mm}

Prof. dr. Jan Ambj\o rn (K\o benhavns Universitet, Denemarken)

Prof. dr. Alfio Bonanno (Osservatorio Astrofisico di Catania, Italië)

Prof. dr. Ronald H. P. Kleiss

Prof. dr. Jan M. Pawlowski (Universit\"at Heidelberg, Duitsland)

Prof. dr. Andreas Wipf (Friedrich-Schiller-Universit\"at Jena, Duitsland)


\newpage
\thispagestyle{empty}

\begin{center}
{\Large \bf Quantum Gravity \\ from Fundamental Questions \\ to Phenomenological Applications}

\vspace{10mm}

Doctoral thesis

\vspace{5mm}

to obtain the degree of doctor

from Radboud University Nijmegen

on the authority of the Rector Magnificus prof. dr. J.H.J.M. van Krieken,

according to the decision of the Council of Deans

to be defended in public on

\vspace{5mm}

Wednesday September 19, 2018

\vspace{5mm}

at 16:30 hours

\vspace{5mm}

by

\vspace{5mm}

{\large \bf {Nat\'{a}lia Alkofer}}

\vspace{5mm}

born on May 6, 1987

\vspace{5mm}

in Niter\'{o}i, Brazil
\end{center}


\newpage
\thispagestyle{empty}

~\vspace{150mm}

Supervisor: Prof. dr. Renate Loll

Co-supervisor: dr. Frank Saueressig

\vspace{5mm}

Manuscript Committee:

\vspace{2mm}

Prof. dr. Jan Ambj\o rn (University of Copenhagen, Denmark)

Prof. dr. Alfio Bonanno (INAF, Astronomical Observatory of Catania, Italy)

Prof. dr. Ronald H. P. Kleiss

Prof. dr. Jan M. Pawlowski (University of Heidelberg, Germany)

Prof. dr. Andreas Wipf (University of Jena, Germany)


\newpage
\chapter*{Samenvatting}
\thispagestyle{empty}

Ons hedendaagse theoretische begrip van de fundamentele eigenschappen van de natuur is opgesplitst in twee domeinen. Algemene Relativiteitstheorie beschrijft zwaartekracht en de structuur van ruimtetijd. Het is belangrijk voor astrofysica en kosmologie, en daarom is haar belangrijkste toepassing op zeer grote lengteschalen. Bovendien zijn vele van haar experimenteel bevestigde voorspellingen totaal onverwacht, zoals bijvoorbeeld zwarte gaten.

Evenzo beschrijft het Standaardmodel van deeltjesfysica nauwkeurig de fysica op kleine, atomaire en subatomaire lengteschalen. De pogingen om een kwantumtheorie voor gravitatie te formuleren proberen deze twee fundamenten van de theoretische natuurkunde te unificeren in \'e\'en enkele beschrijving van de natuur.

Hoewel Algemene Relativiteitstheorie en de kwantumveldentheorie\"en die het Standaardmodel vormen, voortkomen uit een zeer verschillende aanpak om vrij verschillende verschijnselen te verklaren, onthult nader onderzoek onverwachte analogie\"en. Deze hebben in het verleden verscheidene, in principe ongerelateerde benaderingen voor kwantumgravitatie voortgebracht. Hierbij is het verre van triviaal dat verschillende modellen voor kwantumgravitatie een niet-manifoldachtige structuur van ruimtetijd voorspellen, die een dimensionele reductie op korte lengteschaal vertonen. Deze observatie verdient een dieper begrip.

Bovendien dienen zwarte gaten als een rijke proeftuin voor idee\"en voor kwantumgravitatie. Het begrijpen van eigenschappen als het bestaan of niet bestaan van een singulariteit, of het lot van het verdampen van zwarte gaten door Hawkingstraling, zijn slechts twee van de vele uitdagingen in deze context.

Het onderzoeken van kwantumgravitatie vereist kennis van zowel Algemene Relativiteitstheorie en kwantumveldentheorie. Daarom begint dit proefschrift, na een algemene inleiding in de behandelde onderwerpen, met een kort overzicht van de veldentheoretische aspecten van zwaartekracht die relevant zijn voor de projecten die hier behandeld worden, dat wil zeggen, (i) een studie van interagerende vaste punten voor $f(R)$-zwaartekracht gekoppeld aan materie, (ii) een berekening van de spectrale dimensie gebaseerd op een spectrale actie, (iii) een studie naar het Unruh-effect in verscheidene modellen voor kwantumzwaartekracht, en (iv) de renormalisatiegroep-gemodificeerde metriek voor Schwarzschild-zwarte gaten.

Onder de beschikbare veldentheoretische instrumenten is er \'e\'en essentieel voor het eerste en hoofdproject, namelijk de functionele renormalisatiegroep voor de effectieve gemiddelde actie. Daarom wordt de versie van de functionele renormalisatiegroep die geschikt is om toe te passen op zwaartekracht in wat meer detail besproken. De belangrijkste toepassing van dit formalisme is het asymptotische veiligheidsscenario.
Simpelweg zegt dit dat het gedrag van zwaartekracht op de kortste lengteschalen, en daarom de hoogste energieschalen, gedomineerd worden door een vast punt van de renormalisatiegroep: een verdere toename van de energie verandert de waarde van de koppelingen slechts met een snel afnemende hoeveelheid, waardoor deze uiteindelijk een constante waarde bereiken. Als deze waarden ongelijk aan nul zijn, spreekt men van een interagerend of niet-Gau\ss isch vast punt van de renormalisatiegroep.
In het geval van zwaartekracht is het belangrijk op te merken dat het asymptotische veiligheidsscenario voorspelt dat de dimensieloze constante van Newton een eindige waarde benadert, en dat de theorie interagerend is op de hoogste energieschalen.

En ander belangrijk resultaat geboekt v\'o\'or aanvang van de projecten in dit proefschrift is de observatie dat een dergelijk vast punt een dimensionele reductie omvat. Deze eigenschap wordt gekwantificeerd door een concept dat ontwikkeld is in de tak van de wiskunde die fractale geometrie\"en bestudeert, namelijk de spectrale dimensie. Het achterliggende idee hierbij is dat de terugkeerwaarschijnlijkheid van een willekeurig pad op een dergelijke gegeneraliseerde ruimte, afhangt op een grootheid die ge\"interpreteerd zou kunnen worden als een dimensie van zo'n ruimte. Aangezien deze gerelateerd is aan de propagator van een willekeurig bewegend testdeeltje, wordt deze spectrale dimensie genoemd.

\thispagestyle{empty}

Uitgerust met deze instrumenten is een studie naar de niet-Gau\ss ische vaste punten van de renormalisatiegroep uitgevoerd, in de context van $f(R)$-zwaartekracht gekoppeld aan een willekeurig aantal scalaire, Dirac- en vectorvelden.
Deze opzet geeft heldere schattingen van welke zwaartekracht-materiesystemen een niet-Gau\ss isch vast punt voorspellen dat geschikt is om een theorie asymptotisch veilig the houden.
De analyse gebruikt een exponenti\"ele splitsing van de metriek en behoudt een zeven-parameterfamilie van \emph{coarse-graining} operatoren. Wanneer deze parameters op nul worden gezet, vertonen zwaartekracht gekoppeld aan Standaard\-model-materie en vele uitbreidingen voorbij het Standaardmodel een niet-Gau\ss isch vast punt waarvan de eigenschappen sterk overeenkomen met die van zuivere zwaartekracht.
Omgekeerd vertonen geen van de fenomenologisch interessante zwaartekracht-materiesystemen een stabiel vast punt als zogenoemde spectraal aangepaste regulatoren worden toegepast.
De gepresenteerde analyse geeft uitsluitsel over conflicterende resultaten uit eerdere literatuur door aan te tonen dat de verkregen interagerende vaste punten in de twee methodes tot verschillende klassen behoren, en dat slechts \'e\'en van hen stabiel is onder toevoeging van hogere-orde operatoren in de actie.
Verder tonen de resultaten aan dat, net als in het geval van zuivere zwaartekracht, ook onder toevoeging van scalar-, fermion- en vectorvelden globale oplossingen bestaan voor discrete verzamelingen van parameterwaarden.
We bestuderen het asymptotische, grote-kromminggedrag van de vaste puntfuncties, en geven voorbeelden van globale oplossingen.

In het volgende project berekenen we de spectrale dimensie voor klassieke spectrale acties zoals die voorkomen in bijna-commutatieve meetkunde. We analyseren de propagatie van  \linebreak[4] spin-0-,  
spin-1- and spin-2-velden, en tonen aan dat een niet-triviale spectrale dimensie al op klassiek niveau opduikt. Voor lange diffusietijden vinden we de verwachte waarde vier. Voor korte diffusietijden wordt de spectrale dimensie volledig gedomineerd door hoge-energie-eigenschappen van de spectrale actie, en dit geeft een spectrale dimensie van nul voor alle bosons onder beschouwing. Dit resultaat kwantificeert eerdere beweringen dat hoge-energiebosons niet pro\-pageren.

Het Unruh-effect is de theoretische voorspelling dat een versnelde detector deeltjes in het vacu\"um detecteert. Het spectrum van deze deeltjes is hierbij thermaal, met een effectieve temperatuur strikt proportioneel met de eigenversnelling van de waarnemer.
In \'e\'en van de projecten van dit proefschrift, namelijk de analyse van het Unruh-effect voor scalardeeltjes in verscheidene modellen voor kwantumgravitatie, onderzoeken we of een verandering in spectrale dimensie observabele consequenties heeft. We tonen aan dat Lorentz-invariante correcties aan de tweepuntsfunctie een karakteristieke vingerafdruk geven in de ge\"induceerde detectiefrequentie van de versnelde detector, terwijl de Unruh-temperatuur onveranderd blijft. In het algemeen vertonen modellen voor kwantumgravitatie met een dynamische dimensionele reductie een onderdrukte detectiefrequentie bij hoge energie\"en, terwijl de frequentie versterkt wordt bij Kaluza-Kleintheorie\"en met compacte extra dimensies.

Dit gedrag wordt gekwantificeerd door de introductie van het nieuwe begrip ``Unruh-di\-mensie'' als de effectieve ruimtetijddimensie gezien door het Unruh-effect. Deze dimensie blijkt gerelateerd, maar niet identiek, aan de spectrale dimensie die gebruikt wordt om ruimtetijd in kwantumgravitatie te karakteriseren. We bespreken de fysische oorsprong van deze effecten en hun relevantie voor de verdamping van zwarte gaten.

De cyclus van projecten die we hier presenteren wordt afgesloten door een nauwkeuriger blik op de eigenschappen van zwarte gaten in het asymptotische veiligheidsscenario. In het bijzonder geven we opheldering over de observatie dat renormalisatiegroep-gemodificeerde Schwarzschild-zwarte gaten een prototypisch voorbeeld zijn van een Hayward-geometrie. De laatste wordt genoemd als een model voor niet-singuliere zwarte gaten in kwantumgravitationele fenomenologie.
Verder bespreken we kort de rol van de cosmologische constante in het proces van renormalisatiegroepmodificatie.
We benadrukken dat deze niet-singuliere zwarte gaten vele eigenschappen delen met zogenoemde Planck-sterren: hun effectieve geometrie omvat op logische wijze de een-luscorrecties gevonden met behulp van effectieve veldentheorie, hun Kretschmann-scalar is begrensd, en de singulariteit van het zwarte gat is vervangen door een regulier De Sitterdomein.

\thispagestyle{empty}


\newpage
\thispagestyle{empty}


\newpage
\chapter*{Summary}
\thispagestyle{empty}

The current theoretical understanding of the fundamental properties of 
nature is split into two domains. General relativity 
describes gravity and the structure of spacetime. 
It is important for astrophysics and cosmology, and therefore its main 
application is at very large length scales. Furthermore, many of its
observationally confirmed predictions, as, {\it e.g.}, black holes, came totally 
unexpected. 
Similarly, the standard model of particle physics, describes accurately 
the physics at small, atomic and subatomic, length scales. The attempts to 
formulate a quantum theory for gravity 
seek to unify these two pillars of theoretical physics into one single 
description of nature. 
Although general relativity and the quantum field theories constituting 
the standard model originate from the very distinct efforts to explain 
quite different phenomena, a closer look reveals
unanticipated analogies which in the past have initiated several, 
in principle, disconnected approaches to quantum gravity. Hereby the fact 
that various quantum gravity models arrive at a non-manifold like
structure of spacetime exhibiting dimensional reduction at short
length scales is far from trivial and deserves a deeper understanding. 
Furthermore, black holes
serve as a rich testing ground for quantum gravity ideas.
Understanding features like the existence or non-existence 
of a singularity or the fate of black hole evaporation by Hawking 
radiation, are only two of the many challenges posed in this context.  

Investigating quantum gravity requires a comprehension of both,
general relativity and quantum field theory. Therefore this thesis 
starts, after a general introduction to the treated topics, with a 
brief review of the field theoretical aspects of gravity which are relevant
for the projects described here, {\it i.e.}, (i) a study of interacting
fixed points for $f(R)$ gravity coupled to matter, (ii) a calculation 
of the spectral dimension based on a spectral action, (iii) an 
investigation of the Unruh effect in different quantum gravity models,
and (iv) the metric for renormalisation group improved Schwarzschild 
black holes. 
  
Amongst the available field-theoretical 
tools one is central to the first and major project, namely the 
functional renormalisation group for the effective average action.
Therefore the version of the functional renormalisation group
suitable to be applied to
gravity is introduced in some more detail. The most important application
 of this formalism is the asymptotic safety scenario.
In short, it states that the behaviour of gravity at the shortest length
and thus highest energy scales is dominated by a renormalisation
group fixed point:
a further increase of the energy scale does change the values of the 
couplings only by a quickly decreasing amount, and they eventually reach
constant values. If these values are non-vanishing one speaks about 
an interacting or non-Gau\ss ian renormalisation group fixed point.
In the context of gravity it is important to note that the asymptotic
safety scenario predicts that the dimensionless Newton constant 
approaches a non-vanishing 
value, and the theory is interacting at highest energy scales.

A further important result obtained previous to the start of this
thesis' projects is the observation that such a fixed point entails
a dimensional reduction. This property may be quantified by a
concept developed in the branch of mathematics studying fractal 
geometries,  the spectral dimension. The underlying idea is hereby
that the return probability of a random walk on such a generalised 
space depends on a quantity which might be interpreted as a 
dimension on such a space. As it is related with the propagator
of the randomly walking test particle it is called spectral dimension.  

Equipped with these tools  a study of the non-Gau\ss ian
renormalisation group fixed points arising within the framework
of $f(R)$-gravity minimally coupled to an arbitrary number of
scalar, Dirac, and vector fields has been performed.
Based on this setting comprehensive estimates are presented which
gravity-matter  systems give rise to  non-Gau\ss ian renormalisation group
fixed points  suitable for rendering the theory asymptotically safe. 
The analysis employs an exponential split of the metric
and retains a seven-parameter family of coarse-graining
operators. For vanishing parameters, gravity coupled to  the matter 
content of the standard model 
of particle physics and many beyond the standard model extensions exhibit 
non-Gau\ss ian renormalisation group fixed points
whose properties are strikingly similar to the case of pure gravity. 
Conversely, none of the phenomenologically interesting gravity-matter systems
exhibits a stable fixed point if so-called spectrally adjusted regulators 
are employed.  The presented analysis resolves conflicts on  previous
results in the literature by demonstrating that the obtained interacting
fixed points in the two settings belong to different
classes, and only one of them is stable under the inclusion of higher-order operators
in the action.
Furthermore, it is demonstrated that, as in the pure gravity case, also with
scalar, fermion and vector fields added, global quadratic solutions exist for
discrete sets of parameter values.
The asymptotic, large-curvature behaviour of the fixed point functions is 
analysed, and  examples for global solutions are provided. 

\thispagestyle{empty}

In the next project the generalised spectral dimension is computed for
classical spectral actions obtained within the framework of
almost-commutative geometry. Analysing the propagation of spin-0, spin-1, and
spin-2 fields, it is demonstrated that a non-trivial spectral dimension
arises already at the classical level. For long diffusion times
the expected value four is obtained. For short diffusion times 
the spectral dimension
is completely dominated by the high-momentum properties of the
spectral action, yielding a vanishing spectral dimension for all
considered bosons. This result quantifies
earlier claims that high-energy bosons do not propagate.

The Unruh effect is the theoretical prediction that an accelerated
detector measures particles in the vacuum. The spectrum of these particles
is hereby thermal with an effective temperature strictly proportional
to the proper acceleration of the observer.
In one of this thesis' projects, namely the analysis of the Unruh effect
for scalar particles in different quantum gravity models, it is 
investigated whether  a change in the spectral dimension has  observable 
consequences.  It is shown 
that, while the Unruh temperature is unchanged, Lorentz-invariant
corrections to the two-point function leave a characteristic signature
in the induced detection rate of the accelerated detector. Generically,
quantum gravity models exhibiting dynamical dimensional reduction show  
a suppression of the Unruh rate at high energy while the rate is enhanced
in Kaluza-Klein theories with compact extra dimensions.

This behaviour is
quantified by introducing the novel concept of a ``Unruh dimension''
as the effective spacetime dimension seen by the  Unruh effect.
It turns out that this dimension  is related, though not identical,
to the spectral dimension used to characterise spacetime in quantum gravity.
The physical origin of these effects and their relevance for black hole
evaporation is discussed.

The cycle of the projects presented here is concluded by a 
closer look into the properties of black holes within the asymptotic safety
scenario. In particular, the observation is elucidated that  
correspondingly renormalisation group improved Schwarzschild black holes 
constitute a prototypical example of a Hayward geometry. The latter has 
been advocated as a model for 
non-singular black holes within quantum gravity phenomenology.
Furthermore, the role of the cosmological constant in the
renormalisation group improvement process is briefly discussed.
It is emphasised that these  non-singular black holes share many features
of a so-called Planck star: their effective geometry naturally  incorporates
the one-loop corrections found in the effective field theory framework,
their Kretschmann scalar is bounded, and the black hole singularity is
replaced by a regular de Sitter patch.

\thispagestyle{empty}


\newpage
\clearpage{\thispagestyle{empty}\cleardoublepage}


\pagenumbering{roman}
\setcounter{page}{1}
\tableofcontents


\newpage
\thispagestyle{empty}


\listoffigures


\newpage
\thispagestyle{empty}


\listoftables


\newpage
\thispagestyle{empty}
~
\newpage


\pagenumbering{arabic}
\setcounter{page}{1}
\chapter{{\color{MYBLUE}{Introduction}}}
~\vspace{-10mm}

\rightline{\footnotesize{\it{``No question about quantum gravity is more 
difficult}}}
\rightline{\footnotesize{\it{than the question, ``what is the question?"."}}}
\rightline{\footnotesize{\it{(J.\ Wheeler)}}}
\minitoc

\section{A Remark on the Renormalisability of Gravity}

This decade has seen the experimental verification of two long-standing 
theoretical predictions: first, on July 4th, 2012, CERN announced the 
discovery of the Higgs boson \cite{Aad:2012tfa,Chatrchyan:2012xdj}, 
and second, on September 14th, 2015, a  gravitational wave was for the first 
time identified by the LIGO detector \cite{Abbott:2016blz}. 

Whereas the first of these experimental results is based on the standard model (SM) 
of particle physics\footnote{The SM was developed through the 
work of many scientists, see, {\it e.g.}, \cite{Griffiths:2008zz} for a 
didactical introduction.}
which is a collection of quantum gauge field theories,
the second rests on Einstein's version of classical gravity
called general relativity (GR), see, {\it e.g.}, \cite{Rindler:2006km}. 
These two theories constitute the main pillars of contemporary theoretical physics.

Although these two discoveries each mark a triumph of theoretical and 
experimental physics, taken together they also bring to mind one of the most 
challenging open questions of theoretical physics: how can one combine these two 
extraordinarily successful theories into a unified one describing quantum gravity? 

At a first glance GR and quantum gauge field theories seem
to be quite different. Whereas GR relates gravitational forces to
geometric properties of spacetime the SM connects the electromagnetic, 
the weak and the strong force to the exchange of vector particles, the
photon, the $W^{\pm} / Z$, and the gluons, respectively. Looking deeper, one 
realises puzzling similarities, however.

Gravity is also a gauge theory if one denotes by this expression a field theory whose 
Euler-Lagrange equations are covariant under an infinite-dimensional group 
of transformations. Furthermore, as noted by Fierz and Pauli already in 1939
\cite{Pauli:1939xp,Fierz:1939ix}, the relativistic field theory of a spin-two field 
in Minkowski space leads to the linearised form of Einstein's equation. Therefore
in the weak-field limit the gravitational force can be visualised as the exchange of
a graviton, in very much the same fashion a perturbative treatment of the 
interactions in the SM relates them to a vector boson exchange. The
difference in the spin of the exchanged particle then also nicely provides, at 
least for the tree-level scattering amplitude, an explanation
why gravity is only attractive and the other three forces in nature  possess 
attractive and repulsive channels each.
Therefore an understanding of GR as field theory is legitimate, 
and actually the second chapter of this thesis is devoted to elucidate this 
relation in more detail.

The investigation of gauge theories, on the other hand, revealed that they 
possess a geometrical interpretation, too. Besides the possibility of formulating
them in terms of fibre bundles, the geometry of spacetime and the gauge groups
play an essential role in identifying topologically non-trivial field configurations
as, {\it e.g.},  instantons (see, {\it e.g.}, \cite{Coleman:1978ae}) 
with non-trivial 
homotopy properties or Kran - van Baal calorons \cite{Kraan:1998pm} with 
non-trivial holonomy as well. Further geometrical properties are uncovered 
by the use of loop variables like Wilson, Polyakov and/or 't Hooft loops in
lattice gauge field theories, see, {\it e.g.}, \cite{Gattringer:2010zz}. 

It is interesting to note that in gravity and gauge theories the very first
challenge to be met is the removal of unphysical degrees of freedom. A consistent
procedure has been developed by DeWitt \cite{DeWitt:1967ub} for gravity and by 
Faddeev and Popov \cite{Faddeev:1967fc} for gauge theories at approximately the
same time. Both introduced auxiliary fields which nowadays are known as 
Faddeev-Popov ghosts. Fixing the gauge in either gravity or gauge theories 
but then introducing the ghost field as a parameter in the gauge transformation
constitutes the so-called BRST transformations \cite{Becchi:1974md,Tyutin:1975qk}
which turn out to be a symmetry of the gauge-fixed action of gravity and gauge theories.
As the Noether charge derived from the BRST symmetry is a nilpotent operator 
a BRST cohomology can be defined. The related classification of states into BRST 
singlets and quartets then may provide a mechanism to identify the physical state 
space and to avoid unitarity violations by cancelation of BRST quartets in the 
$S$-matrix, see, {\it e.g.}, \cite{Nakanishi:1990qm} and references 
therein.\footnote{For the possibility of non-perturbative
BRST quartets in QCD see
\cite{Alkofer:2011pe,Alkofer:2011uh,Alkofer:2011wq,Alkofer:2013ih}.}

The gauge theory of strong interactions, quantum chromodynamics (QCD), provides a
beautiful example of renormalisability. QCD is asymptotically  free, a phrase which 
describes the fact that the interaction gets weaker when the momentum scale is 
increased. As a consequence, the processes in a scattering experiment at
very high energies can be treated within perturbation theory. In a perturbative
treatment the infinities arising in a quantum field theoretical calculation 
of QCD amplitudes
can be consistently removed by introducing counterterms for the so-called
primitively divergent Green's functions of the theory. Therefore one can 
provide to a given order in perturbation theory a consistent calculation 
scheme which provides finite numbers and thus predictions for observables. 
The comparison of these with experimental results have led to an impressive
verification of QCD. Asymptotic freedom also implies that QCD in itself is a quantum
field theory which is valid up to arbitrarily high energies. This is a 
desirable property, however, the question whether every quantum field theory (QFT) 
should fulfil it cannot be answered yet.

What about renormalisability of gravity?
As a matter of fact it is not complicated to treat an appropriately defined 
deviation of the spacetime metric $g_{\mu\nu}$ from the Minkowski metric 
$\eta_{\mu\nu}$ as a quantum field in Minkowski spacetime. Starting from
the Einstein-Hilbert (EH) action and performing a perturbative expansion in the
Newton coupling one obtains non-renormalisable expressions in the sense that 
novel infinities appear in every further order of the perturbative expansion. 
Clearly, cutting the perturbative expansion  at some low order 
an effective field theory arises \cite{Donoghue:1994dn} which is expected
to be able to describe quantum corrections in the limit of small gravitational fields
accurately. Nevertheless, taken the 
theory at face value it has no predictive power as infinitely many 
constants would be required in principle. The situation is similar 
to the strong interaction where chiral perturbation theory (see, {\it e.g.},
\cite{Scherer:2002tk} and references therein) provides meaningful results
at low energies although without any doubt QCD is the exact theory.

It is worthwhile to have a closer look at a perturbative treatment of gravity and why it
can make predictions only at low energies when the corresponding  approach is 
understood as an effective field theory. Newton's constant possesses (in units where
$\hbar = c=1)$ the dimension of (mass)$^{-2}$. The 
definition of the Planck mass $m_{Pl}$ is given by $G_N=1/m_{Pl}^2$. 
As gravity is very weak, which is expressed as Newton's constant being small, 
the expectation is that the gravitational effective field theory is  applicable as long as 
one deals with momenta very much below the Planck mass. However, every method 
to obtain results from a QFT for interacting fields relies on some 
approximations\footnote{There are a few exceptions of exactly solvable QFTs in two spacetime 
dimensions.}, and, in principle, one would need
to show that the errors induced by the method are small. But in QFT this is 
impossible.\footnote{This remark also applies to perturbation 
theory. The perturbation expansion leads to results in terms of 
asymptotic series, and there is no
a priori possibility to determine until which order residual terms are small.}
What really is done is the justification of the employed approximation 
{\it a posteriori} by comparing to experimental results. Up to now there 
is no experimental result for quantum gravity, and correspondingly there is no 
possibility of confirming the validity of the used approximation.

Of course, a meaningful theory should give finite answers. But one should note
that this is a statement on the exact solution and allows only for definite 
conclusions in the unrealistic situation that the theory can be solved exactly.
Perturbative infinities may or may not have something to do with the consistency
of the theory. Or phrased otherwise, perturbative renormalisability is neither a 
necessary nor a sufficient condition for the consistency of a theory. 

The example of chiral perturbation theory being the low-energy effective field
theory of QCD might give rise to the anticipation that the ``true'' theory of 
gravity is renormalisable, and GR arises from taking the 
appropriate low-energy limit. However, a fundamental theory is not necessarily
renormalisable, see, {\it e.g.}, \cite{Huang:1998qk} for a didactical
presentation of Wilson's interpretation
of renormalisability. Following the ideas of Wilson, the renormalisable theories are effective
low-energy theories because renormalisability is nothing else than the statement 
that one can calculate low-energy quantities without knowing the true high-energy
behaviour. As long as only a few ``coarse-grained'' quantities originating from 
the underlying theory influence the calculation of low-energy observables, one can
substitute the unknown quantities and remove thereby the infinities by counterterms
such that physically motivated normalisation conditions are met. Note that hereby
experimental input is necessary to fix numerical values within the normalisation
conditions. It has to be emphasised that this view implies that the exact solution 
to an ultimate theory should not contain any infinities any more. Nevertheless, 
finding an indisputably renormalisable version of gravity (or more precisely, 
of gravity coupled to the matter known to us) would be an enormous step forward 
in our understanding of nature.
 
\section{Asymptotic Safety}
\label{AsySafe}

The fact that GR is not renormalisable within perturbation theory 
\cite{Heisenberg:1938xla,tHooft:1974toh,Deser:1974cz,Goroff:1985th}
can have two quite different reasons: GR or modifications thereof
is in a quite drastic manner
not the correct theory of gravity.\footnote{The addition of terms which are quadratic
in the curvature can render the such amended theory perturbatively renormalisable
\cite{Stelle:1976gc}, however, the theory contains then propagating ghosts and violates
unitarity.} On the other hand, there is the possibility that the failure of renormalisation 
is an artefact of applying perturbation theory.  However,
it might very well be that, although GR (or some variation of it) is 
renormalisable beyond perturbation theory, nevertheless nature has chosen in 
addition some to us yet unknown construction principles for space and time at the
shortest possible length scales and time intervals.

Already within the SM, for which substantial experimental input 
is available, non-pertur\-bative approaches come with outstanding challenges. 
Obviously, for gravity the problem of applying methods beyond perturbation theory
is even harder, both, on the conceptual as on the technical level. Similarly to
treatments in particle physics one may roughly categorise many of the related 
efforts in  three classes. First, alike to lattice QFT one attempts to calculate 
the basic properties resulting from  the underlying theory
on  basis of the Euclidean version of the functional integral by employing Monte-Carlo 
({\it i.e.}, statistical) estimates. 
One successful effort in this direction is
causal dynamical triangulation (CDT) \cite{Ambjorn:2012jv}. Second, the Hamiltonian 
of the theory is used as starting point for all further investigations. This line of 
research led eventually to loop quantum gravity 
\cite{Rovelli:1997yv,Ashtekar:2004eh,Gambini:2011zz,Rovelli:2014ssa}.
Third, one can make use of functional methods, {\it e.g.}, the so-called functional 
renormalisation group (RG) equation \cite{Wetterich:1992yh,Morris:1993qb}
has been employed since the seminal work  \cite{Reuter:1996cp}
in many studies of the so-called asymptotic safety  scenario \cite{Weinberg}, 
see, {\it e.g.}, \cite{Niedermaier:2006wt,Reuter:2012id,Percacci:2017fkn}  
for recent comprehensive descriptions.

The basic idea behind asymptotic safety is easiest explained for a theory with 
only one coupling constant. The RG equation implies that 
this coupling constant will depend on the renormalisation scale. To be consistent 
through this thesis I will call this scale immediately $k$.\footnote{In most 
textbooks on QFT and within perturbation theory the canonical
name for this scale is~$\mu$.} It proves useful to further introduce the dimensionless
number $t:= \ln (k/k_0)$ where $k_0$ is some reference scale. (In almost all 
applications it is not necessary to specify it.) As the coupling $g(k)$ depends
on $k$, its derivative with respect to $t$ will be in general a function of the 
scale $k$ and of the coupling,
\be
\frac {dg}{dt} =: \beta (g,k) \, ,
\ee 
and thus defines the so-called beta function.\footnote{In case the coupling is not 
dimensionless one usually multiplies it with the appropriate power of the scale 
$k$ and defines the beta function from the thus dimensionless version of the
coupling.} Its behaviour as a function of the 
coupling close to its zeros characterises typically 
the behaviour of the theory either in the 
IR or in the ultraviolet (UV). In QCD one has for example to leading order 
in the coupling
\be
\beta (\alpha_S) \propto - \alpha_S^2
\ee
such that the proportionality constant is positive. As now the beta function has a zero and is
negative for physical values of the coupling one infers that 
increasing the scale decreases the coupling, and for asymptotically large scales
the interaction will be switched off. This is known as asymptotic freedom. On the other
hand, one might have a zero of the beta function at vanishing coupling but the 
beta function is positive. The most prominent example is quantum electrodynamics
(QED), its beta function starts
out also quadratic but is positive. This implies a divergence in the coupling 
known as Landau pole \cite{Landau:1959fi}.\footnote{As a consequence of his 
discovery Landau declined QFT as an useful tool.} To the best
of our knowledge this Landau pole in QED can only be avoided if the coupling
is strictly zero everywhere (see, {\it e.g.}, \cite{Gockeler:1997dn,Nagy:2012ef}),
a scenario which is called RG triviality, and is the motivation for many
investigations in beyond-the-standard-model (BSM) approaches because an 
imbedding of QED (or any other appearing U(1) gauge theory) into a grand unified
theory (GUT) as a non-Abelian gauge theory is avoiding the Landau pole
problem.\footnote{At this point it is interesting to note that this problem
might also have a solution within the asymptotic safety scenario of gravity
\cite{Harst:2011zx,Eichhorn:2017lry}.}  
Calculating the beta function for  gravity from an expansion in Newton's 
coupling one notices for small couplings a situation similar to the one in QED,
the $\beta$-function for small values of the coupling is positive. 
Of course, this is a direct reflection 
of the perturbative non-renormalisability of gravity.

As gravity is a highly non-linear theory it is far from clear that the 
$\beta$-function keeps on rising and Newton's coupling runs inevitably into a Landau
pole. Alternatively, the beta function may acquire a second zero if expressed in 
the dimensionless coupling $g_k = k^2 G_N(k)$, respectively, in $d$ spacetime dimensions
$g_k = k^{d-2} G_N(k)$, at some finite value of $g=:g^\star$. Assuming that at finite 
values of the scale $k$ one has $g(k)< g^\star$ then a further increase of the scale 
drives the coupling $g$ towards $g^\star$: one has found an UV fixed point (FP) of 
the coupling, and the theory is UV complete.
Such a scenario provides the easiest example of Weinberg's notion of asymptotic 
safety \cite{Weinberg}. And, as a matter of fact, convincing evidence for the 
asymptotic safety  scenario has been gathered: since the seminal work  
\cite{Reuter:1996cp} many studies of this scenario in increasingly 
sophisticated approximations have been performed up to today thereby verifying and
further clarifying this concept.

At this point a few remarks are in order. The first one is about notation: a FP
at vanishing coupling is called Gau\ss ian fixed point (GFP). Correspondingly one
at a non-vanishing and finite coupling (or several non-vanishing couplings) is referred to
as non-Gau\ss ian fixed point (NGFP). Second, a theory
might contain more than one coupling constant. As a matter of fact, a general 
{\it ansatz} consistent with the required symmetries may contain infinitely
many couplings. In chapter \ref{Chap4} such an action, namely the 
so-called $f(R)$-gravity action, will be discussed. In order that the asymptotic safety
scenario stays predictive it is additionally required that only finitely many 
couplings are relevant, because then only this finite number of couplings 
has to be adjusted to phenomenological values in order to perform a quantitative
calculation. Technically, this condition refers to the statement that the RG flow
remains within the space of all couplings on a finite-dimensional hypersurface.
In the vicinity of NGFP the so-called relevant directions and the UV 
critical hypersurface can be determined from a linear approximation to the RG flow,
for the details of such an analysis see section \ref{AsympSafe}.
Third, a NGFP is always non-perturbative irrespectively of the value of $g^\star$
(or the $g_i^\star$ in the multidimensional case). This is immediately evident 
from the fact that the precise numerical value depends in general on the employed 
RG scheme.\footnote{There is the possibility to keep FP values parametrically
small, see, {\it e.g.}, \cite{Litim:2014uca}, which allows definite conclusions
on these FPs also from the perturbative expansion.} 
Fourth, the running of Newton's coupling and the cosmological constant might 
have impact even on astrophysical and cosmological scales, see, {\it e.g.}, 
\cite{Reuter:2004nx,Bonanno:2017pkg,Bonanno:2018gck} and on black hole spacetimes, 
see, {\it e.g.},  \cite{Koch:2014cqa,Bonanno:2016ako} for a recent synopsis of 
related investigations. A possible 
resolution of the black hole singularity \cite{Saueressig:2015xua} will be
discussed in chapter \ref{Chap7}. And last but not least, as will 
be now discussed in the next subsection, the existence of a NGFP comes with 
certain implications for the anomalous dimensions of the fields. The latter
might be interpreted from the point of view that the theory undergoes a certain 
type of dimensional reduction.

\section{Dimensional Reduction}

As stated above, and as will become evident in the next chapter of this thesis,
at the NGFP Newton's constant behaves like $G_N(k) \propto k^{2-d}$,
and the cosmological constant like $\Lambda(k) \propto k^2$.  
This is an exact consequence of asymptotic safety which exclusively relies on the 
scale-independence at the UV FP. In \cite{Lauscher:2005qz} this
fact was related to an effective fractal spacetime structure in asymptotically safe
gravity. This connection has been further elucidated 
\cite{Litim:2006dx,Reuter:2011ah,Rechenberger:2012pm,Calcagni:2013vsa}
and can be considered as firmly established.
Different than on a manifold, on fractal spaces different definitions of the 
dimensions do not coincide. Two widely-used concepts besides the topological
dimension, the Hausdorff and the
spectral dimension are defined and briefly described in appendix \ref{MathSpecDim},
for some relations for the spectral dimension see also section \ref{SpecDimUV}.
Postponing the technical details to chapter \ref{SpecDimSpecAct}
and the appendix \ref{MathSpecDim}, I only want to mention at this point
that strong deviations from the tree-level behaviour as, {\it e.g.}, large 
anomalous dimensions and the related non-canonical momentum dependence of 
propagators, lead to a sizeable discrepancy between the topological and the spectral 
dimension.

Not only asymptotic safety but in virtually all approaches to quantum gravity 
and quantum gravity inspired models dimensional flows are a common feature, see,
{\it e.g.}, \cite{Carlip:2009kf,Carlip:2012md} for a synopsis of relevant results.
The most prominent example of a dimensional flow occurs in Kaluza-Klein theories 
where the ``effective'' dimensionality of spacetime increases for length scales smaller than 
the compactification radius.\footnote{A Kaluza-Klein model with a ``large'' 
extra dimension has been studied within functional RG in \cite{Master,ADS}.}
The opposite behaviour, on the other hand, is seen as a dynamical dimensional reduction 
with the spectral dimension of spacetime decreasing at short distances. 
Besides in asymptotic safety such a characteristic feature has been seen 
within CDT where a random walk sees a two-dimensional spacetime at short distances while 
long  walks exhibit a four-dimensional behaviour \cite{Ambjorn:2005db},
within loop quantum gravity 
\cite{Modesto:2008jz,Caravelli:2009gk,Magliaro:2009if,Calcagni:2013dna,
Calcagni:2014cza,Ronco:2016rtp}, 
Ho\v{r}ava-Lifshitz gravity \cite{Sotiriou:2011mu,Sotiriou:2011aa}, 
causal set theory \cite{Eichhorn:2013ova,Carlip:2015mra,Belenchia:2015aia},  
$\kappa$-Minkowski space \cite{Benedetti:2008gu,Anjana:2015ios,Anjana:2015msa},  
non-local gravity theories \cite{Modesto:2011kw,Modesto:2015ozb}, 
minimal length models \cite{Padmanabhan:2015vma}, and based on the 
Hagedorn temperature seen by a gas of strings \cite{Atick:1988si}.
Dimensional reduction based on spectral actions\footnote{See 
section \ref{sec:SpecAct} for a brief description.}
\cite{Kurkov:2013kfa,Alkofer:2014raa} will be discussed in chapter \ref{SpecDimSpecAct}.

It is important to note that one should not understand this dimensional reduction 
as a choice of directions and thus a breaking of Lorentz invariance. Based on the 
fact that in most models of quantum gravity spacetime is not a
manifold at short distances, and furthermore on the 
observation that the spectral dimension is the Hausdorff-dimension of the
theory's momentum space \cite{Amelino-Camelia:2013gna,Amelino-Camelia:2013cfa},
a better visualisation of this kind of dimensional reduction is the following:
whereas for large distance scales and therefore in the IR the density of the 
spectrum coincides with the one of the spectrum of a free theory 
in the $d$-dimensional spacetime manifold, the number of states grows less than
na\"ively expected in the UV. The spectral density compares then much
better with the one of a free theory in a smaller number of dimensions. And this closes
then also the circle with respect to renormalisability: it is evident that a weaker
spectral growth will lead to weaker UV divergencies thus rendering the 
theory renormalisable. If the spectral dimension is small enough the theory may 
become even finite.

\newpage
\thispagestyle{empty}


\chapter{{\color{MYBLUE}{On Field Theoretical Aspects of Gravity}}}
\label{Chap1}
~\vspace{-10mm}

\rightline{\footnotesize{\it{``Phantasie ist wichtiger als Wissen.}}}
\rightline{\footnotesize{\it{Wissen ist begrenzt, Phantasie aber umfa\ss t die ganze Welt."}}}
\rightline{\footnotesize{\it{(A.\ Einstein)}}}
\minitoc

\section[The Einstein-Hilbert Action and Modified Theories of Gravity]
{The Einstein-Hilbert Action and Modified Theories of \\ Gravity}

\subsection{General Relativity}

The vacuum Einstein field equations can be derived from the Hilbert action
\be
\label{EHaction} 
S_{EH} = \frac 1 {16\pi G_N} \int d^d x \sqrt{|g|} ( -R + 2 \Lambda  )
\ee
by variation with respect to the metric $g_{\mu\nu}$. Hereby $G_N=6.674 \times 10^{-11} \, 
{\mathrm{m^3/s^2kg}}$ is Newton's constant and $g$ is a shorthand for $\det (g)$.
Using the form with the absolute value below the square root covers the cases with 
Lorentzian and Euclidean signature of the metric. Furthermore, $R$ denotes the scalar 
curvature, {\it i.e.}, the trace of the Ricci tensor, and $\Lambda$ denotes the
cosmological constant. Note that $d=2$ constitutes a special case because then the
term proportional to $R$ in the action is a topological invariant. 

\subsection{Modifications and Additions to the Einstein-Hilbert Action}
\label{ModGrav}

The construction principle of GR, namely covariance of the equations
with respect to general coordinate transformations, allows to add to the action 
an infinite number of diffeomorphism-invariant
terms which are then generally of higher order in the number
of derivatives acting on the metric. The most general action build from
fourth-order terms is
\be
\label{4action} 
S_{(4)} = \frac 1 {16\pi G_N} \int d^d x \sqrt{|g|} 
\left( \alpha R^2 + \beta R_{\mu\nu}R^{\mu\nu}
+\gamma R_{\mu\nu\sigma \rho}R^{\mu\nu\sigma \rho}+
\delta \, \, D^2 R \right)
\ee
where $\alpha , \beta , \gamma$ and $\delta$ are some constants. The last term is a total
derivative, and is thus typically ignored. Furthermore, the other three terms can be 
rewritten such that 
\be
E=R_{\mu\nu\sigma \rho}R^{\mu\nu\sigma \rho} - 4 R_{\mu\nu}R^{\mu\nu} +R^2
\ee
can be split off,
\be
\label{4actionE} 
S_{(4)} = \frac 1 {16\pi G_N} \int d^d x \sqrt{|g|} 
\left( a_1 R_{\mu\nu}R^{\mu\nu} + a_2 R^2 + a_3 E \right)
\ee
with $a_1=4\beta +\gamma$, $a_2=\alpha-\gamma$ and $a_3=\gamma$.  In four dimensions
and under the usual
assumptions of neglecting boundary contributions only the first two terms contribute
to the field equations. 

Of course, further extensions with 6th-, 8th- and higher order derivatives are possible
without violating diffeomorphism invariance. An extension of a qualitatively different
type is given by $f(R)$-gravity with the action containing a function of the curvature
scalar,
\be
\label{fRaction} 
S_{f(R)} = \int d^d x \sqrt{|g|}  \, f(R) \, .
\ee
Within such modifications of the gravity action one can distinguish between the case
where either $f(R)$ is a polynomial or a more general function. Note that the EH 
action falls into the class of a polynomial $f(R)$-gravity with $f$ being a polynomial of order one. A 
typical example for a non-polynomial $f(R)$-gravity is 
$f(R) = R^2 \ln R + \ldots $. As a matter of fact, such a term can appear
as a part of the fixed function at the NGFP in the asymptotic safety scenario 
\cite{Dietz:2013sba,Gonzalez-Martin:2017gza},
see section \ref{GlobalFF} below.

\subsection{Background Metric, Tensor Decomposition, and Gauge Fixing }
\label{secBack}

Throughout this thesis I will use an approach in which the gravitational degrees of freedom
are carried by the metric. As a symmetric rank-two tensor it transforms 
accordingly under coordinate transformations, resp., diffeomorphisms. This immediately
implies that the metric contains besides physical components also unphysical ones.
Technically this is reflected by the fact that the second variation of the action,
the Hessian, 
possesses zero modes which are proportional to infinitesimal coordinate transformations. 
Therefore, inverting the Hessian is not possible, phrased otherwise the calculation
of the (tree-level) propagator cannot be done. This situation is alike in 
electrodynamics for which the gauge zero modes prevent the na\"ive determination of 
the tree-level photon propagator and a gauge-fixing term is needed to
make that propagator well-defined. Formally, the solution for gravity is similar, 
{\it i.e.}, one adds a gauge-fixing term, however, the technical details are more 
intricate. 

As mentioned in the introduction, in perturbation theory one usually expands around
the Minkowski metric,
\be
\label{LinSplitM}
g_{\mu\nu}= \eta_{\mu\nu} + \sqrt{16 \pi G_{\rm N}} \,  h_{\mu\nu}  \, ,
\ee
and assumes that all components of $h_{\mu\nu}$ are small. The rescaling with the 
term $\sqrt{16 \pi G_{\rm N}}$ is done to obtain the standard canonical 
dimension of a boson propagator for the propagator of the fluctuating part of the 
metric. In case a Wick rotation is performed 
already at the level of the action, the metric is correspondingly expanded 
around the one in Euclidean space, 
\be
\label{LinSplitE}
g_{\mu\nu}= \delta_{\mu\nu} + \sqrt{16 \pi G_{\rm N}} \,  h_{\mu\nu}  \; . 
\ee

Non-perturbative techniques like the functional RG typically
employ the background field formalism. In the case of gravity 
one splits the metric into a background\footnote{Throughout 
this thesis quantities constructed from the background metric will be 
denoted with a bar. Indices are lowered and raised with the background metric, {\it e.g.},
$h_{\mu\nu} = \bar g_{\mu\rho} h^\rho_{~\nu}$.}  $\bar g_{\mu \nu}$
and a fluctuation $h_{\mu \nu}$, the latter  not necessarily being small 
\cite{Reuter:1996cp,Niedermaier:2006wt}:
\be
\label{LinSplit} 
g_{\mu \nu}  = \bar g_{\mu \nu} + h_{\mu \nu} \,\,.
\ee
Hereby, the background can be also the one of flat space. A convenient 
non-trivial choice, employed in  chapter \ref{Chap4}, is to take
a $d$-dimensional sphere. However, the linear split of the metric, 
(\ref{LinSplit}) has not been used. The reason is the following: as the 
size of the fluctuating metric is not bounded\footnote{In the functional
integral one sums over all metric fluctuations.} a decomposition like (\ref{LinSplit}) 
 effectively takes metrics with different signatures into account. 
Of course, there are good reasons to avoid that the functional integral
also sums over spaces with different signatures. A choice that 
ensures that $g_{\mu\nu}$ and $\bar g_{\mu\nu}$ have the same signature, 
even if the fluctuations are large, is the exponential split \cite{Percacci:2015wwa}
\be
\label{ExpSplit}
g_{\mu\nu}= \bar g_{\mu\rho} ( e^{\gb^{-1}h} )^\rho{}_\nu \, \, . 
\ee 
Obviously, expanding the right hand side of (\ref{ExpSplit}) up to linear
order and neglecting higher orders the linear split (\ref{LinSplit}) is obtained.
The difference becomes visible beyond the linear order, one has
\ba
g_{\mu\nu} &=&  \bar g_{\mu\nu} + h_{\mu\nu} + \tfrac 1 2 h_{\mu \rho} h^\rho_{~\nu}
+\ldots \, , \\
g^{\mu\nu} &=&  \bar g^{\mu\nu} - h^{\mu\nu} + \tfrac 1 2 h^{\mu \rho} h_\rho^{~\nu}
+\ldots \, .
\ea
Different than in the linear split, also the covariant metric, and not only its inverse,
is non-linear in the fluctuating field $h^\rho_{~\nu}$.

Before the concrete realisation of gauge-fixing can be introduced a somewhat 
technical but important issue needs to be discussed. In the context of 
gravitational waves in the linearised theory as well as in the beginning 
of this section the existence of non-physical modes in the fluctuating part
of the metric $h_{\mu \nu}$ was already mentioned. In addition, this tensor possesses
different irreducible components if it comes to the transformation properties 
under the Lorentz transformations. As the propagating modes have spin 2 one 
can conclude the spin 1 and spin 0 components of the metric have to be unphysical.
There are several ways to achieve a decomposition of the fluctuation 
fields into their irreducible spin components. In the following I will use
the well-established York decomposition \cite{York:1973ia}
\be
h_{\mu\nu} = h^{TT}_{\mu\nu} + \bar D_\mu\xi_\nu + \bar D_\nu\xi_\mu +
(\bar D_\mu \bar D_\nu  -\frac{1}{d} \bar g_{\mu\nu} \bar D^2 )\sigma +
\frac{1}{d} \bar g_{\mu\nu} h \,,
\label{york}
\ee
where the superscript $TT$ denotes the traceless and transverse part, $\bar D_\mu$ 
is the background covariant derivative (containing the Levi-Civita connection), 
and $\xi_\mu$ is a transverse vector field:
\be
\bar D^\mu h^{TT}_{\mu\nu} =0 , \quad \bar g^{\mu\nu} h^{TT}_{\mu\nu} =0 , \quad 
\bar D^\mu \xi_\mu =0 \, .
\ee
Hereby, $h^{TT}_{\mu\nu}$ is the spin 2 degree of freedom, $\xi_\mu$ describes
the spin 1 part,
and $\sigma$ and $h =\bar g^{\mu\nu} h_{\mu\nu}$ have spin 0. The latter two scalar
fields are not gauge independent but
\be
s = h - \bar D^2 \sigma 
\ee
is. 

For further use below it is already noted here that on a sphere as background
and in the York decomposition the two lowest eigenmodes of $\sigma$ and the lowest 
vector mode $\xi_\mu$ of $-\bar D^2$ do not change the right hand side of \eqref{york}. 
These modes must then be removed by hand in order to make the decompositions into 
irreducible spin components bijective. 

The York decomposition viewed as a change of variable $h_{\mu\nu} \to (h^{TT}_{\mu\nu},
\xi_\mu,\sigma,h)$ has a non-vanishing Jacobian determinant. On a background
manifold which is 
of the Einstein type, {\it i.e.}, for which $\Rb_{\mu\nu} \propto \gb_{\mu\nu}$, one can
calculate this Jacobian to be \cite{Percacci:2017fkn}
\be
\label{YorkJacobian}
\cJ = \Det_{(1)}\left(-\bar D^2 - \tfrac{\Rb}{d}\right)^{1/2} 
\Det_{(0)}\left(\left(-\bar D^2\right) \left(-\bar D^2 - \tfrac{\Rb}{d-1} \right)
\right)^{1/2}
\ee
where the subscript denotes to which spin the determinant has to be attributed. 
The necessity of a non-vanishing Jacobian is also evident from the fact that the 
fields $\xi_\mu$  and $\sigma$  have non-standard dimensions. The following change 
of variables, 
\be
\label{HattedFields}
\hat \xi_\mu = \sqrt{-\bar D^2 - \tfrac{\Rb}{d}}\, \xi_\mu \,\, , \qquad 
\hat \sigma = \sqrt{ \left(-\bar D^2\right) \left(-\bar D^2 - \tfrac{\Rb}{d-1} \right)}
\sigma \, 
\ee
has a Jacobian which exactly cancels the one of the York decomposition, and the 
``hatted'' fields possess standard dimensions. It should be emphasised that a 
non-local transformation like \eqref{HattedFields} is not allowed for physical fields,
it would change the spectrum of the theory. On the other hand, $\xi_\mu$ and $\sigma$ 
are unphysical auxiliary fields which act like ``book-keeping devices'', and therefore
such a transformation is only required to be bijective.

As in gauge theories the gauge procedure is most efficiently performed by adding 
a gauge-fixing term $S_{\rm gf}$ to the action.  A standard choice is\footnote{ When
added to the EH action the form 
$S_{\rm gf} = \frac{1}{32 \pi G_N \alpha} \int \ldots $
is usually chosen. The form given in \eqref{gf0} is related to this 
one by an appropriate rescaling of the gauge parameter $\alpha$. }
\be
\label{gf0}
S_{\rm gf} = \frac{1}{2\alpha}\int d^d x \sqrt{\bar g}\,\bar g^{\mu\nu} F_\mu F_\nu \, , 
\qquad F_\mu = \bar D_\rho h^\rho{}_\mu-\frac{\beta +1}{d} \bar D_\mu h\, .
\ee
Hereby, setting $\alpha \to 0^+$ is called Landau gauge, it enforces the gauge condition 
strictly. Choosing the second gauge parameter as $\beta=1$ one arrives at the so-called
Landau - de Donder gauge. In case of a flat background it enforces the de Donder 
condition $\partial _\rho h^\rho{}_\mu-\frac{2}{d} \partial_\mu h =0$.  
Employing the limit $\beta \rightarrow - \infty$ which imposes strongly
the gauge condition $h=const.$ offers a further simplification and is therefore sometimes
referred to as ``physical gauge''. This is then also the gauge
chosen for the investigation described in chapter \ref{Chap4}.

One should, however, point out that the attribute ``physical gauge'' is by no means 
unique. In \cite{Wetterich:2016vxu} it has been argued that a gauge-fixing term 
which is strictly quadratic in the fluctuating field is the most appropriate one for 
certain studies within the functional RG (see the next chapter for a brief introduction
to the functional RG). The author refers to his term 
as ``physical gauge-fixing'', see, {\it e.g.}, \cite{Wetterich:2016ewc}.

Before rewriting the gauge-fixing condition $F_\mu$ with the fields of the York 
decomposition a remark on the different possible ways of introducing a Laplace operator 
is in order, see appendix \ref{App.Laplace} for a brief review. The simplest choice for
general tensor fields is  based on the covariant derivative containing the Levi-Civita
connection (for fermions this has to replaced by the spin connection, see the
following section \ref{MatterGrav}), and the background Laplacian is then defined as
\be
\label{DefLaplace}
\Delta = - \bar D^2 \, .
\ee
Unfortunately, this version of the Laplacian does not possess the most convenient 
properties. Especially, one wants that the corresponding differential operator
preserves the rank and the symmetries of the tensor it is acting on. On the other hand, 
for the purpose  of this thesis, a general discussion of this topic related to the 
definition of the 
Laplace-Beltrami operator and the Lichnerowicz Laplacians ({\it cf.} appendix  
\ref{App.Laplace}) would only divert the presentation unnecessarily. This is 
because on manifolds of the Einstein type and especially on maximally symmetric spaces
like spheres the relation of Lichnerowicz Laplacians to the Laplacian \eqref{DefLaplace}
is relatively simple. For example, on all Einstein manifolds the operator
\be
\Delta_{(1)} = \Delta + \frac {\bar R} d
\ee
has the desired properties for vectors, and on maximally symmetric spaces
\be
\Delta_{(2)} = \Delta + \frac {2\bar R} {d-1}
\ee
for rank-2 tensors. Restricting in the following to maximally symmetric spaces 
and employing the first of the two definitions above one obtains
\be
F_\mu = - \left( \Delta_{(1)} - \frac {2 \bar R} d \right) \xi_\mu 
- \bar D_\mu \left( \frac {d-1}d \left( \Delta - \frac {\bar R}{d-1} \right) \sigma
+ \frac \beta d h \right) \, .
\ee
Note that on Einstein manifolds and thus also on maximally symmetric spaces
the differential operator in the first term becomes 
$\Delta - \frac { \bar R} d$. Later on we will see that such structures propagate
to the flow equation, see, {\it e.g.}, the denominator of the last term in 
\eqref{gravflow}.

The Faddeev-Popov ghost action is derived by exponentiating the functional determinant
implied by the gauge-fixing. Following the standard procedure one introduces
Grassmannian-valued vector fields which will be denoted as $C_\mu$ and 
$\bar C_\mu$. The Faddeev-Popov operator for the employed gauge-fixing 
is a non-minimal operator, the ghost action reads
\be
S_{gh} = - \int d^dx \, \sqrt{\bar g} \, \, \bar C^\mu\left( \,
\delta_\mu^\nu \, \bar D^2 +\left(1 - 2 \, \tfrac{\beta+1}{d}\right) 
\bar D_\mu \bar D^\nu+ {\bar R}_\mu ^{~\nu}  \right) \, C_\nu \, .
\ee 
Decomposing the ghosts into their transversal and longitudinal parts, 
$C_\nu = C_\nu^T + \bar D_\nu  \, \frac 1 {\sqrt{\Delta}} \, \hat c^{L}$, one obtains
\be
S_{gh} =  \int d^dx \, \sqrt{\bar g} \, \,\left(  \bar C^{T\mu} \left( \,
\Delta_{(1)} - \frac {2 \bar R} d \right) \, C^T_\mu 
+ 2 \frac{d-1-\beta } d \, \, \bar {\hat c}^{L} \left( \Delta - \frac {\bar R}{d-1-\beta }
\right) \hat c^{L} \right)  \, . 
\ee 
Hereby, the longitudinal part has been introduced such that this change of variables
possesses unit Jacobian.

With $S_{gh}$ being the ghost action the functional integral for the generating 
functional reads
\ba
Z[j^{\mu\nu}, \bar \eta^ \mu, \eta^\mu ; \bar g_{\mu \nu} ]= &&
\int {\cal D} h_{\mu \nu} {\cal D} C_\mu {\cal D} \bar C_\mu
 \exp \bigl( -S_{\rm grav} [h _{\mu \nu}; \bar g_{\mu \nu}] - S_{gf} -
S_{gh} [h _{\mu \nu},C_\mu,\bar C_\mu ; \bar g_{\mu \nu}  ]
\nonumber \\
&& \hspace{35mm} + \int d^dx \, \sqrt{\bar g} \, \, (j^{\mu\nu}h_{\mu \nu}  
+ \eta^\mu \bar C_\mu + \bar \eta^\mu C_\mu ) \bigr) 
 \,\,.
\ea
In a next step one introduces a source coupled to the fluctuating part of the metric and
performs  the Legendre transform of the $\ln Z$ to obtain the effective action
$\Gamma$ ({\it cf.} appendix \ref{App.FRGderiv}),
\be
\Gamma [\bar h_{\mu \nu} , \bar C_\mu , C_\mu  ; \bar g_{\mu \nu} ] = 
-\ln Z[j^{\mu\nu}, \bar \eta^ \mu, \eta^\mu ; \bar g_{\mu \nu}] 
+ \int d^dx \, \sqrt{\bar g} \, \, (j^{\mu\nu}\bar h_{\mu \nu}  
+ \eta^\mu \bar C_\mu + \bar \eta^\mu C_\mu ) \,\, ,
\ee
where $\bar h_{\mu \nu} := \langle h_{\mu \nu}  \rangle$ denotes the expectation value 
of the fluctuating part of the metric. 
Usually the effective action $\Gamma$ is considered to be a functional of 
the expectation value of the total metric $g_{\mu \nu}$
and $\bar g_{\mu \nu}$ instead of $\bar h_{\mu \nu}$ and $\bar g_{\mu \nu}$.
It can be shown \cite{Reuter:1993kw,Litim:1998nf,Reuter:1996cp,Freire:2000bq,Litim:2002hj}
that the effective action in the usual
formalism, {\it i.e.}, the generating functional of the one-particle irreducible Green
functions,
$\Gamma [g_{\mu \nu} ]$, is obtained by setting $g_{\mu \nu} = \bar g_{\mu \nu}$ or,
equivalently, $\bar h _{\mu \nu} =0$.

\bigskip

\subsection{Coupling Matter to Gravity}
\label{MatterGrav}

In order to couple matter fields minimally to gravity the respective matter Lagrangian densities
in Minkowski space are modified by substituting (i) the integration $\int d^dx$ by the
(invariant) integration $\int d^d x \sqrt{|g|}$, (ii) the partial derivative $\partial_\mu$
by the covariant derivative $D_\mu$, and (iii) a four-vector product by contraction as,
{\it e.g.}, in $A_\mu A^\mu$ by $g^{\mu\nu} A_\mu A_\nu$ where $g^{\mu\nu}$ is the inverse
of the metric $g_{\mu\nu}$. 
As all research projects presented in this thesis focus on the 
UV behaviour of gravity-matter systems the mass of the matter fields will
be neglected if not stated explicitly otherwise. In addition, only the minimal coupling
of matter to gravity will be taken into account.

Amongst all matter fields the easiest to couple are the scalar fields because for them
the covariant derivative and the partial derivative coincide. Therefore to couple 
$N_S$ massless scalar fields  
minimally one only needs to take into account the factor $\sqrt{|g|}$ and the metric 
for lowering and raising the indices
\be
\label{Sscalar}
S^{\rm scalar}  =  \frac{N_S}{2} \int d^d x \, \sqrt{|g|} \, g^{\mu\nu}  \,
\left(  \partial_\mu \phi \right) \left( \partial_\nu \phi \right) \, .
\ee 
As matter self-interactions are neglected it actually makes no difference whether
Abelian or non-Abelian gauge fields are considered, and as masses are also not taken
into account whether the vector fields belong to a gauge theory in the Coulomb, 
confining or Higgs phase. Of course, gauge-fixing is required to remove the
gauge zero modes, and linear covariant gauges provide the easiest possibility.
However, as for QFT in curved spacetimes
(see also the next section) even for Abelian gauge fields the Faddeev-Popov 
operator does not decouple and one needs to take into account ghosts. Therefore 
$N_V$ massless vector fields contribute with
\be
\label{Svector}
S^{\rm vector} =  N_V  \int d^d x \, \sqrt{|g|} \, \left( 
\frac 1 4 F^{\mu\nu} F_{\mu \nu}  + \frac 1 {2 \lambda} ( D^\mu A_\mu)^2
+ \bar c \, (- D^2) \, c \right) \, 
\ee
to the total action. Hereby, the field strength tensor is given by
$F_{\mu \nu} = D_\mu A_\nu - D_\nu A_\mu$. 
The parameter $\lambda$ is the usual one for gauge-fixing
to linear covariant gauge (as in QED). From the expression \eqref{Svector} 
it is obvious that the Faddeev-Popov gauge ghosts decouple from the gauge 
field $A_\mu$ (as in QED) but couple to the Levi-Civita connection and thus
to the metric. 

For fermions the covariant derivative contains the spin connection,
see, {\it e.g.}, \cite{Birrell:1982ix}
for an introduction to Dirac fermions in curved spacetimes using either 
coordinate dependent Dirac matrices or vielbeins. In appendix
\ref{App.Dirac} it is described how to construct this connection explicitly 
on spheres employing the vielbein formalism, see \eqref{eqB5}. 
Denoting by small Greek letters the Dirac matrices 
which fulfil the general Clifford algebra
$\{\gamma_\mu(x),\gamma_\nu(x)\}=2g_{\mu\nu}(x)$ and with capital Greek letters
the ones which obey the flat-space Clifford algebra
$\{\Gamma_a,\Gamma_b\}=2\eta_{ab}$, resp., 
$\{\Gamma_a,\Gamma_b\}=2\delta_{ab}$,
the spinor covariant derivative is defined via 
\be
\label{CovDerivSpin}
D_\mu \psi = e_\mu^{~a}\partial_a \psi + \frac 1 2 
\omega_{\mu ab }\Sigma^{ab} \psi \, , 
\quad
\Sigma^{ab} = \frac 1 4 [\Gamma^a,\Gamma^b] \, ,
\ee
where $\omega_{\mu  ab}$ denotes the vielbein connection which can be expressed
via the vielbeins $e_\mu^{~a}$ and the Levi-Civita connection 
$\Gamma_{\mu\rho\sigma}$ as
\be
\label{SpinConn}
{\omega_\mu^{~a}}_b = e_\nu^{~a}  \partial_\mu e^\nu_{~b} 
+ e_\nu^{~a}  {\Gamma_\mu^{~\nu}}_\rho e^\rho_{~b} \, .
\ee 
Decomposing a product of two Dirac matrices, $\Gamma^a \Gamma^b$, into 
the anticommutator and the commutator,
one can straightforwardly show that \cite{Birrell:1982ix}
\be
\label{Dslash}
 - \Dslash^{~2} = - D^\mu D_\mu + \frac R 4 = \Delta + \frac R 4 
\ee
on all manifolds where the spin connection can be defined. On such manifolds
with Euclidean signature the eigenvalues of $\Dslash $ are purely imaginary 
and come in pairs of opposite sign, {\it i.e.}, complex-conjugated to each 
other. In appendix \ref{App.Dirac} explicit expressions are given for the
case of a sphere as underlying manifold. Of course, \eqref{Dslash} allows to determine the eigenvalues and eigenfunctions 
of the Laplacian in case they are known for the operator $\Dslash$. 
With the covariant derivative as defined above the action for $N_D$ 
minimally coupled and massless fermions is
\be
\label{Sfermion}
S^{\rm fermion} =   \, N_D \,  \int d^d x \,  \sqrt{|g|} \, 
\bar \psi \, i \, \Dslash \,\, \psi \, .
\ee
The actions \eqref{Sscalar}, \eqref{Svector} and \eqref{Sfermion} will be 
used in chapter \ref{Chap4} as ansatz for the effective action coupling 
matter fields to gravity.

Variation of the classical action for matter fields,
$S^{\rm matter} = S^{\rm scalar} + S^{\rm fermion} + S^{\rm vector}$,
with the three terms on the right hand side as given above, 
yields these fields' energy-momentum tensor, more precisely this
symmetric and covariantly conserved tensor is given by
\be
\label{Tmunu}
T^{\rm matter}_{\mu\nu} = - \frac 2 {\sqrt{|g|}}
 \frac {\delta S^{\rm matter}}{\delta g^{\mu\nu}} \, .
\ee
By construction one has $D^\mu T^{\rm matter}_{\mu\nu} = 0$.

\section{Some Basic Aspects of Quantum Field Theory in Curved Spacetimes}
\label{QFTinCST}

As for this topic there are several textbooks available, see, {\it e.g.},
\cite{Birrell:1982ix,Mukhanov:2007zz,Parker:2009uva},
I will focus here on the aspects which are relevant for this thesis,
especially for the Unruh effect treated in chapter \ref{Chap6}.

The classical action for matter fields has been given in the last section.
In most applications of QFT in curved spacetimes (respectively,
in curved spaces obtained after a Wick rotation) the metric and thus the 
curvature is treated as classical background. Typical examples for such spacetimes
are models for the expanding universe and black hole geometries. 
On spacetimes which admit a foliation and Cauchy hypersurfaces the canonical
formalism of Minkowski space QFT can be straightforwardly 
generalised by applying corresponding changes to the definition of the
momentum field and the canonical equal-time commutation and anti-commutation 
relations. 

In QFT, in flat space one introduces a 
decomposition of the field according to positive and negative frequency
solutions. In the standard treatment the latter are chosen as plane waves, see,
{\it e.g.}, \cite{Ryder:1985wq,Peskin:1995ev}. The coefficients of this
decomposition are then elevated to creation and annihilation operators 
with the corresponding algebra. This then defines the field operator 
which fulfils due to this construction the canonical equal-time
commutation relation for bosons and anti-commutation relations for fermions.
In case of an underlying curved spacetime the decomposition of the field therefore
will depend on the solutions to the classical field equations in that given
background. 

In curved spacetimes a special connection between the statistics and the dynamics 
of the fields manifests itself: 
based on the fact that the algebra of creation and annihilation operators  
is built on commutators for bosons  (anti-commutators for fermions)
it turns out that the curved spacetime dynamics leads for bosons (fermions) in 
that spacetime but freely moving otherwise to a Bose-Einstein 
(Fermi-Dirac) distribution \cite{Birrell:1982ix,Mukhanov:2007zz,Parker:2009uva}.  
Although {\it a priori} such a relation is not present in Minkowski space
the requirement of a continuous limit from curved spacetimes to Minkowski
space makes such a connection necessary, {\it cf.} the corresponding discussion 
in \cite{Parker:2009uva}.

A relatively simple example, elaborated in \cite{Parker:2009uva}, 
for this statement can be provided by considering a Robertson-Walker  metric 
\be
ds^2 = dt^2 - a^2(t) d\vec x^{~2}
\ee
with an artificially tuned scale factor such that 
\be
a(t) = \begin{cases} a_i \quad {\rm for} \quad t\le t_i \\ 
a_f  \quad {\rm for} \quad t \ge t_f \end{cases} ,
\ee 
or phrased otherwise, for early and late times one has a static Minkowski space.
At some early time $t<t_i$ the field operator possesses the decomposition
\be
\phi(x) = \sum _{\vec k} \left( A_{\vec k} f^{(i)}_{\vec k} (x) 
+ A^\dagger_{\vec k} f^{(i) \star}_{\vec k} (x) \right)
\ee
with $f^{(i)}_{\vec k} (x) $ being a positive-frequency solution at these
early times,
\be
f^{(i)}_{\vec k} (x) \propto \frac 1 {\sqrt{V\, a_i^3}} \frac 1 {\sqrt{2\omega_{i\, k}}}
e^{i(\vec k \cdot \vec x - \omega_{i\, k} t)} \, ,
\ee
where $V$ is the quantisation volume,\footnote{Typically chosen as a cube 
of length $L$ and endowed with periodic boundary conditions} and 
$\omega_{i\, k} = \sqrt{\vec k^{~2} / a_i^2 + m^2}$,
~ $m$ being the field's mass. The corresponding decomposition at late times $t>t_f$
can be written as 
\be
\phi(x) = \sum _{\vec k} \left( a_{\vec k} f^{(f)}_{\vec k} (x) 
+ a^\dagger_{\vec k}  f^{(f)\star}_{\vec k} (x) \right)
\ee
and analogously
\be
f^{(f)}_{\vec k} (x) \propto \frac 1 {\sqrt{V\, a_f^3}} \frac 1 {\sqrt{2\omega_{f\, k}}}
e^{i(\vec k \cdot \vec x - \omega_{f\, k} t)} \, ,
\qquad \omega_{f\, k} = \sqrt{\vec k^{~2} / a_f^2 + m^2} \, .
\ee
Both sets of creation and annihilation operators can be related by 
Bogoliubov transformation, 
\be
a_{\vec k} = \alpha_{\vec k} A_{\vec k} + \beta^\star_{\vec k} 
A^\dagger_{-\vec k}
\ee
as well as the hermitian conjugate relation for $a^\dagger_{\vec k}$.
Requiring the standard algebra\footnote{One should nevertheless bear in mind
that these two sets of operators lead to an unitary inequivalent representation
for the canonical (anti-)commutation relation, see, {\it e.g.}, \cite{Miransky:1994vk}.}
\be
[A_{\vec k},A^\dagger_{\vec k'}]_{\mp} = [a_{\vec k},a^\dagger_{\vec k'}]_{\mp}
= \delta_{\vec k \, \vec k'}
\ee
leads to the condition $|\alpha_{\vec k}|^2 \mp |\beta_{\vec k}|^2 =1$ where
the upper (lower) sign refers to bosons (fermions). 

Assuming that at early times no particles were present the state 
is the corresponding vacuum annihilated by $A_{\vec k}$ for all $\vec k$,
\be
A_{\vec k} |0_i\rangle =0  \, .
\ee
A straightforward calculation now reveals that the late time vacuum defined
by
\be
a_{\vec k} |0_f\rangle = 0
\ee
contains particles with a spectrum according to the Bose-Einstein (Fermi-Dirac)
distribution independent of the details of time dependence of the scale factor 
$a(t)$. Furthermore, as the metric is translationally invariant and isotropic
the number of particles with opposite momenta is equal. This then leads to 
the interpretation of the physical effect: particles have been created in pairs,
or more precisely in particle-antiparticle pairs. The number density of created
particles is hereby (see, {\it e.g.}, \cite{Parker:2009uva} for a didactical 
presentation of the derivation)
\be
\langle n \rangle _{t>t_f} = \sum_{\vec k} |\beta_k|^2 \, .
\ee
To calculate the Bogoliubov coefficient $\beta_k$ one then needs, of course, 
the explicit time dependence of $a(t)$.


It has to be emphasised that at any intermediate time $t$ with $t_i<t<t_f$ the 
number of created particles is not a well-defined concept because it is
not measurable by the principles of quantum theory. Choosing a short measuring time
the number of detected particles will be determined by Heisenberg's uncertainty 
principle because the energy (and thus particle number) fluctuations are 
very large. On longer time scales the change of $a(t)$ is sizeable, the 
related Bogoliubov coefficients are changed drastically and thus a well-defined
particle number is prevented.

The basic lessons from this simple example can be taken over to the Unruh
effect \cite{Unruh:1976db} (see, {\it e.g.}, \cite{Crispino:2007eb} for a recent 
review) which is employed in chapter \ref{Chap6} to find an answer to the 
question whether the dimensional reduction seen in many models of quantum 
gravity is an observable effect. The Unruh effect states that 
an uniformly accelerating observer or detector sees a thermal spectrum of 
particles in 
the vacuum of an initial observer. In appendix \ref{App.UnruhCoord} it is
shown that the comoving frames of such an accelerating observer are best
described in Rindler coordinates. One can now take the Rindler spacetime 
as an underlying manifold in itself, and thus formulate the whole setup
as a quantum field theoretical problem in curved spacetime. As can be 
shown with the help of a Bogoliubov transformation
\cite{Mukhanov:2007zz}, a thermal spectrum results. The Unruh temperature 
which is strictly proportional to the proper acceleration $a$, $T=a/2\pi$,
is thereby a strict consequence of the problem's geometry. (This is 
also verified with an independent argument in chapter \ref{Chap6}.) 
The energy of the created particles which then excite the Unruh-DeWitt 
detector is provided by the engine which accelerates the detector.

\section{Almost-Commutative Geometry and Spectral Actions}
\label{sec:SpecAct}

Although noncommutative geometry, see, {\it e.g.}, 
\cite{Connes:1994yd,Marcolli,vanSuijlekom:2015iaa}, 
 is a branch of mathematics it has 
applications in elementary particle physics and gravity. In this 
section I will shortly introduce some related ideas and review 
the concepts which will be used later in the thesis. 
As already stated in the introduction, both, the SM of 
particle physics being a collection of gauge theories, and gravity
can be formulated to a large part in a geometrical way.  
Noncommutative geometry as a generalisation of the usual geometry
may then serve as an {\it ansatz} for formulating BSM theories 
which are rooted in mathematical principles.

Building on the well-established correspondence 
(see, {\it e.g.}, \cite{Connes:1980ji}) that for a compact Hausdorff space
$M$ the commutative algebra of continuous functions on this space,
$C(M)$, contains the same information as the space itself, 
the underlying idea of noncommutative 
geometry is to generalise this relation to noncommutative algebras. 
This approach provides then concepts to apply these noncommutative 
algebras via the definition of noncommutative spaces within physics. 
One of the most important notions is the one of a spectral triple 
describing a noncommutative manifold. Such a triple $(\cA,\cH,\cD)$
consists of an algebra $\cA$ that is represented as bounded operators 
on a Hilbert space $\cH$ on which also a (Euclidean)  Dirac operator 
$\cD$  acts. Hereby the latter can take a very general form,
$\cD=\Dslash + E$ with $E$ being an endomorphism.\footnote{An 
explicit example of such a Dirac operator will be used in chapter
\ref{SpecDimSpecAct}.} 
Leaving it at such an abstract level many types 
of noncommutative manifolds will be allowed.

In almost-commutative geometry \cite{Connes:1994yd} (see
\cite{Connes:1992ex,Jureit:2007qm,Sakellariadou:2013ve,vandenDungen:2012ky} 
for reviews) 
the possible spaces are restricted such that they contain Riemannian 
spin manifolds $M$.
These are spaces that locally look like the Euclidean space $\mathbb R^d$
and a Riemannian metric $g_{\mu\nu}$ exists as well as spinors are admitted. 
The noncommutative
part in the total space $M \times F$ is a finite space $F$. With respect
to physics it is related to the internal degrees of freedom. Mathematically
its defining structure is a finite-dimensional, in general noncommutative,
algebra $\cA_F$. This algebra possesses usually a representation based
on $N \times N$ matrices. To complete the so-called finite spectral triple one
introduces a finite-dimensional left module $\cH_F$ ({\it i.e.}, the
$N \times N$ matrices representing the algebra act on $\cH_F$), and
a hermitian $N\times N$ matrix, $D_F : \cH_F \to  \cH_F$.
This is summarised as
\be
F := \left( \cA_F, \cH_F, D_F \right) \, .
\ee
The algebra related to the Riemannian
spin manifolds $M$ is chosen to be the one of smooth complex functions 
on $M$ denoted by $C^\infty(M,\mathbb C)$. The Hilbert space compatible with 
this algebra is $L^2(M,S)$ which consists of smooth spinor-valued 
functions.\footnote{It is not guaranteed that for a manifold $M$ 
the Hilbert space $L^2(M,S)$ exists.}
The number of components of this spinor depends on the dimensions $d$ of
the manifold, it is $2^{\lfloor d/2 \rfloor}$ where $\lfloor \ldots \rfloor$ 
is the floor function, {\it cf.} also the construction of spinors on spheres
as presented in appendix \ref{App.Dirac}. 
The mani\-fold $M$ possesses a Levi-Civita connection (the unique connection 
on $M$ that is compatible with the metric), and if $L^2(M,S)$ exists also
a spin connection. The Dirac operator $\cD$ is then  constructed from the 
covariant derivative (\ref{CovDerivSpin}) tensored with non-commutative 
matrices.  This is the decisive step in the construction, hence the approach is 
named as almost-commutative geometry. 
In addition, the Dirac operator may contain endomorphisms depending on further
(matter) fields. 
To summarise this construction, a decisive building block of
an almost-commutative geometry is the so-called canonical spectral triple
serving as definition of the Riemannian  spin manifolds $M$,
\be
M:=(C^\infty(M,\mathbb C), L^2(M,S), \cD) 
\ee
which then constitutes the commutative part of the almost-commutative 
geometry. 

The spectral action principle 
related to the almost-commutative geometry uses, first,  
as much as possible of the structure present in the spectral triple
as a guiding principle to include BSM physics, offers, second, 
an approach for unifying gravity and elementary particle physics, and
contains, third, the SM in the low energy limit 
\cite{Chamseddine:1996rw,Chamseddine:1996zu}. 
These conditions can be met by spectral action of the form
\cite{Chamseddine:1996rw,Chamseddine:1996zu}
\be
S_{\chi,\Lambda}=\Tr \bigl( \chi(\cD^2/\Lambda^2) \bigr) \, .
\label{SpAct1}
\ee
Hereby, $\chi$ is a positive function and  $\Lambda$ an appropriate 
UV cutoff. As usual, the trace is defined by the sum over eigenvalues of $\cD$.
In chapter \ref{SpecDimSpecAct} such traces will be performed with the
help of heat kernel techniques. For those spectral actions which contain
the SM as low-energy limit the cutoff $\Lambda$ is identified as the scale 
related to a GUT which appears in an intermediate step.
Note that  depending on the choice for the almost-commutative manifold 
the related spectral actions result in different types of particle physics models. 
It has been shown that suitably tuned choices of the manifold lead to 
a low-energy limit of the spectral action identical or containing 
the SM minimally  coupled to gravity 
\cite{Connes:2006qv,Chamseddine:2006ep,Chamseddine:2007ia,Chamseddine:2012sw,Stephan:2013rna}.
As mentioned it is possible in such a framework to include physics 
BSM
\cite{Chamseddine:2010ud,Devastato:2013oqa,Chamseddine:2013kza,Chamseddine:2013rta} 
as, {\it e.g.}, supersymmetry \cite{Ishihara:2013asa,Beenakker:2014yla,Beenakker:2014zla,Beenakker:2014ama}.
The renormalisation of spectral actions has been investigated in
\cite{vanSuijlekom:2011uu,vanSuijlekom:2011kc,vanSuijlekom:2012xb,Estrada:2012te,Suijlekom:2014ata}.
The phenomenology of the resulting low-energy effective actions has  been studied, 
{\it e.g.}, in \cite{Nelson:2010ru,Lambiase:2013dai,Chamseddine:2014nxa}. 
Some detailed discussion of the UV cutoff $\Lambda$ is provided 
in \cite{DAndrea:2013rix}.
Further generalisations have also been investigated, {\it e.g.}, an extension 
to non-commutative spaces built from non-associative algebras has been 
studied in \cite{Farnsworth:2013nza}. Hereby, the simplest non-associative algebra,
the octonions, lead to a spectral action describing gravity coupled to a $G_2$ gauge 
theory with eight fermions, one singlet and a multiplet in the seven-dimensional
fundamental representation.

\section{Spectral Dimension in the Ultraviolet}
\label{SpecDimUV}

As mentioned in the introduction the spacetimes of many quantum gravity models 
are not manifolds but more general structures, {\it e.g.}, in the asymptotic safety 
scenario of gravity one obtains an effective fractal spacetime structure 
\cite{Lauscher:2005qz}, see also 
\cite{Litim:2006dx,Reuter:2011ah,Rechenberger:2012pm,Calcagni:2013vsa}.
On fractal spaces different definitions of a dimension do not coincide, and
two related widely-used concepts, the Hausdorff and the
spectral dimension are defined and briefly described in appendix \ref{MathSpecDim}.

To determine the spectral dimension the quantum spacetime is equipped with an 
artificial diffusion process for a test particle. In a next step the  return probability 
$\cP (T)$ of the particle as a function of the diffusion time $T$ is calculated. The 
definition of the spectral dimension is  obtained in the limit of vanishing diffusion time.
From \eqref{Ppower} one obtains
\be
\label{def:spectraldimension}
d_s = -2 \, \lim_{T \rightarrow 0}  \frac{d \ln \cP (T) }{d \ln T} \, . 
\ee
As explained in appendix  \ref{MathSpecDim} on a manifold the spectral dimension 
agrees with the topological dimension $d$. 

The properties of the spacetime within a given quantum gravity model will in 
general depend on the length scales probed by the diffusing particle. This can 
then be captured by the generalised spectral dimension $D_s(T)$ in which 
simply the limit  $T\to 0$ is omitted. In many approaches to quantum gravity
$D_s(T)$ interpolates between $D_s = 4$ on macroscopic scales and 
$D_s = 2$ at short trans-Planckian distances \cite{Carlip:2009kf,Carlip:2012md}.
Based on this observation multi-scale geometries as a phenomenological model 
of quantum gravity inspired spacetimes have been investigated 
\cite{Calcagni:2012rm}. 

For the study of the spectral dimensions derived from spectral actions 
presented in chapter~\ref{SpecDimSpecAct}  it is important to note that 
the spectral dimension bears a close relation to the two-point correlation 
function $\widetilde G$ of the diffusing particle. The corresponding 
formalism will be discussed in detail in section \ref{sec:specdim}. 
For a massless scalar  particle propagating on a four-dimensional Euclidean space one has 
$\widetilde{G} = p^{-2}$, which results in a scale-independent spectral dimension 
$D_s = 4$. Non-trivial $D_s$-profiles are created if the two-point correlation 
function acquires an anomalous dimension. Based on this close connection, 
the interpretation of the spectral dimension as the Hausdorff dimension of 
the momentum space has been advocated in \cite{Amelino-Camelia:2013gna}. 
Notably, a non-trivial spectral dimension does 
not necessarily involve the breaking of Lorentz invariance, since 
$\widetilde{G}(p^2)$ may be a function of the momentum four-vector squared 
and thus a Lorentz invariant quantity.
However, this function can in principle have more general forms than 
those allowed in a local QFT.
One relevant example is a two-point function arising in a non-local field theory, 
defined as a theory
whose equations of motion have an infinite number of derivatives.
This form is ubiquitous in causal set studies \cite{Aslanbeigi:2014zva}.

The fictitious nature of the diffusion process underlying the spectral dimension 
then raises the crucial question whether the flow of the spectral dimension 
can be seen in a physical observable quantity. The main goal of chapter 
\ref{Chap6}  is to explicitly demonstrate that this is indeed the case. 
The non-trivial momentum profiles of the propagators leave an imprint in 
the Unruh effect  felt by an accelerated detector. More precisely, the effective 
dimension of spacetime seen by the Unruh detector is determined 
to a large extent by the spectral dimension.
\newpage
\thispagestyle{empty}


\chapter{{\color{MYBLUE}{Functional Renormalisation Group}}}
~\vspace{-10mm}

\rightline{\footnotesize{\it{``Es gibt nur eine Landstra\ss e der Wissenschaft, 
und nur diejenigen haben Aussicht ihren}}}
\rightline{\footnotesize{\it{hellen Gipfel zu erreichen, die die Erm\"udung 
beim Erklettern ihrer steilen Pfade nicht scheuen."}}}
\rightline{\footnotesize{\it{(K.\ Marx)}}}
\minitoc

In this chapter the use of non-perturbative RG methods for gravity will be shortly 
reviewed. The functional RG is well suited for such an investigation of quantum effects 
in  gravity for mainly two reasons: first, the investigation of gravity and
gravity-matter systems needs to cover scales which differ by many orders of magnitude.
Thus for determining properties of those systems
a continuum method is preferred, and especially the RG is suitable for such
studies. Second, as discussed in chapter \ref{Chap1}, gravity is perturbatively 
non-renormalisable and might be renormalisable non-perturbatively. Therefore
a non-perturbative method is required. 

It has to be noted that since the seminal work \cite{Reuter:1996cp}
has been published, increasingly sophisticated approximations for applying 
the functional RG to gravity have been studied. 
This starts with investigations based on the EH action 
\cite{Souma:1999at,Falkenberg:1996bq,Reuter:2001ag,Lauscher:2001ya,Litim:2003vp,
  Bonanno:2004sy,Eichhorn:2009ah,Manrique:2009uh,
  Eichhorn:2010tb,Groh:2010ta,Manrique:2010am,Christiansen:2012rx,Codello:2013fpa,
  Christiansen:2014raa,Becker:2014qya,Falls:2014zba,
  Falls:2015qga,Christiansen:2015rva,
  Gies:2015tca,Benedetti:2015zsw,
  Pagani:2016dof,Denz:2016qks,Falls:2017cze,
  Knorr:2017fus}, 
continues with studying extensions by higher-derivative and higher-order curvature terms 
\cite{Lauscher:2002sq,Reuter:2002kd,
  Codello:2006in,Codello:2007bd,Machado:2007ea,Codello:2008vh,    
 Niedermaier:2009zz,Benedetti:2009rx,Benedetti:2009gn,Benedetti:2009iq,
 Benedetti:2010nr,Rechenberger:2012pm,
  Ohta:2013uca,Falls:2013bv,Benedetti:2013jk,
  Falls:2014tra, Eichhorn:2015bna,Ohta:2015efa,
  Falls:2016wsa,Falls:2016msz,Christiansen:2016sjn,
  Gonzalez-Martin:2017gza,Becker:2017tcx}, 
and the construction of fixed functions including an infinite number of 
coupling constants 
\cite{Reuter:2008qx,Benedetti:2012dx,Demmel:2012ub,Dietz:2012ic,Bridle:2013sra,
 Dietz:2013sba,Demmel:2014sga,Demmel:2014hla,Demmel:2015oqa,Ohta:2015fcu,Labus:2016lkh,
 Dietz:2016gzg,Christiansen:2017bsy,Knorr:2017mhu,Falls:2017lst}, 
as well as including the notorious Goroff-Sagnotti two-loop counterterm 
\cite{Gies:2016con},
or studies based on the foliation structure 
\cite{Manrique:2011jc,Rechenberger:2012dt,Biemans:2016rvp,Biemans:2017zca,
Houthoff:2017oam}. 

The main motivation, however, for the project presented in the next chapter is 
the fact that despite many related studies of gravity-matter systems
\cite{Dou:1997fg,Percacci:2002ie,Narain:2009fy,Daum:2010bc,
  Folkerts:2011jz,Harst:2011zx,Eichhorn:2011pc,
  Eichhorn:2012va,Dona:2012am,Henz:2013oxa,Dona:2013qba,
  Percacci:2015wwa,Labus:2015ska,Oda:2015sma,Meibohm:2015twa,Dona:2015tnf,
  Meibohm:2016mkp,Eichhorn:2016esv,Henz:2016aoh,Eichhorn:2016vvy,
  Christiansen:2017gtg,Eichhorn:2017eht,Christiansen:2017qca,
  Eichhorn:2017ylw,Eichhorn:2017lry,Eichhorn:2017egq,Christiansen:2017cxa}, 
the current understanding is far from satisfactory. In particular, a systematic 
study of the predictive power of the gravity-matter FPs along the lines of 
$f(R)$-gravity, which played a pivotal role in the case of pure gravity, was still 
mostly missing. The goal of the next chapters is to contribute substantially to
such an analysis.\footnote{For some related work in this direction also 
see \cite{Schroder:2015xva}.}

Before going into more details it should be noted that the functional RG has found
widespread use in almost all areas of physics. Besides gravity this includes especially
the areas of elementary particle physics (and hereby the SM including QCD as well as BSM
physics), of ultracold atom gases, of condensed-matter physics, and of statistical 
physics in general. The interested 
reader is referred to one of the many well written review articles and books
in this field, as, {\it e.g.}, 
\cite{Berges:2000ew,
 Pawlowski:2005xe,Gies:2006wv,Delamotte:2007pf,Kopietz:2010zz,Polonyi:2012ff}.

It is important to note that the vast majority of investigations based on  functional RG 
equations (FRGEs) has been performed with a metric with Euclidean signature, and
this will be also assumed in this and the following chapter.

\section{The Wetterich Equation for the Effective Average Action}

The main tool employed in the next chapter is a suitable projection of the Wetterich 
equation \cite{Wetterich:1992yh,Morris:1993qb} adapted to the case of gravity 
\cite{Reuter:1996cp}.  The major steps of a derivation for generic fields
can be found in appendix \ref{App.FRGderiv}, it is given by 
\be
\label{FRGE}
\partial_t \Gamma_k = \frac 1 2 \, {\mathrm {STr}} 
\left( \left(\Gamma_k^{(2)} + \cR_k\right)^{-1} \partial_t \cR_k \right) \, .
\ee
Hereby, $k$ is the RG scale and $t$ is defined as $t = \ln(k/k_0)$ 
with $k_0$ being an arbitrary reference scale. 
$\Gamma_k$ denotes the effective average action (see also below),
and $\Gamma_k^{(2)}$ its second variation with respect to the fluctuating fields. 
In the compact notation of \eqref{FRGE} all Lorentz and spinorial as well
as field type indices are suppressed, it is understood that  $\Gamma_k^{(2)}$
is a matrix in the space of all field components. Correspondingly, 
$\cR_k$ denotes a matrix-valued regulator  (and the notation 
$R_k(z)$ is reserved for real functions being one of the coefficients 
in the ``regulator matrix'').
The symbol ``STr'' denotes the sum of a trace over all continuous and 
discrete bosonic degrees minus the trace over fermionic (or in case of ghosts,
Grassmannian) degrees of freedom. For the latter case one should note that 
the use of a suitable matrix-valued regulator is required. In the fermionic 
channels $\Gamma_k^{(2)}$ is off-diagonal, and together with the same 
structure in the regulator $\cR_k$ one obtains a non-vanishing trace, however, 
with a minus sign and a factor two canceling the 1/2 in front of the 
right hand side of \eqref{FRGE}. Hereby, the matrix $\cR_k$ 
contains the regulators $R_k$ for the different types of fields 
which ensure that the flow is governed  by integrating out quantum fluctuations. 
In this thesis I will use the ``flat'' Litim regulator \cite{Litim:2000ci,Litim:2001up} 
\be
R_ {k} (z) = ({k}^2 - z)\, \Theta \, ({k}^2 - z) \,\,.
\label{LitimReg}
\ee
Its scale derivative is given by
\be
\partial_t R_ {k} (z) = 2 \,{k} ^2 \,\Theta ({k} ^2 - z)  \,\,.
\ee
In the calculations presented in chapter \ref{Chap4} the step function 
$\Theta (k^2 - z)$ will allow for some significant simplifications in the 
expressions appearing as summands in the traces, and, more importantly, it will
turn the traces into finite sums. 

In the following the background formalism for 
$\Gamma_k[g_{\mu \nu} , \bar C_\mu , C_\mu  ; \bar g_{\mu \nu} ]$ 
as discussed at the end of section \ref{secBack} will be used. 
The background metric will be used in the following way in the regulator function: 
one constructs first the background covariant Laplacian $\Delta = -\bar D^2$ 
and uses its eigenmodes and eigenvalues to set the RG scale $k$. 
This implies that the regulator function $R_k$ depends on the background metric. 
The background is then eliminated by identifying it with the physical average metric  
in the final equations after solving the FRGE. The conventional effective action 
$\Gamma [g]$ is then obtained in the limit $k\to 0$ by setting $g_{\mu \nu}=\bar g_{\mu \nu}$,
{\it i.e.}, $\Gamma [g]=\Gamma_{k\to 0} [g ;  g ] $.
This dynamical adjustment of the background metric implements 
in an approximate way the background independence required for a theory of quantum 
gravity.

The employed {\it ansatz} for the effective average action consists of two 
parts, 
\be
\label{eq:EAA}
\Gamma_k = \Gamma_k^{\rm grav} + \Gamma^{\rm matter}_k \, ,
\ee
the gravitational and the matter one. Hereby, $\Gamma_k^{\rm grav}$ 
comprises three components which consist of the gravity action 
but with running couplings, the gauge-fixing (gf) and the ghost (gh)  terms,
\be
\Gamma_k^{\rm grav} =  \Gamma_k^{\rm class} \,\, + 
\Gamma_k^{\rm gf} + \Gamma_k^{\rm gh}  \,  .
\label{GammaGravGeneral}
\ee
As stated above well-studied choices for $\Gamma_k^{\rm class}$ include 
the EH action and its various extensions, see section \ref{ModGrav}.
In chapter \ref{Chap4} $f(R)$-gravity as defined in the action \eqref{fRaction} 
will be investigated.

Following the discussion of coupling matter to gravity in section \ref{MatterGrav}
the matter part of the effective average action, 
\be
\label{GammaMat2}
\Gamma^{\rm matter}  =  \Gamma^{\rm scalar}  +  \Gamma^{\rm fermion}  +  
\Gamma^{\rm vector} \, ,
\ee
describes the coupling of $N_S$ scalar fields $\phi$, $N_D$ Dirac fermions $\psi$, and 
$N_V$ (Abelian) gauge fields $A_\mu$ to gravity. As discussed in section \ref{MatterGrav}
the vector part includes a gauge-fixing term and ghosts $\bar{c}$ and $c$. 
As Abelian gauge invariance is respected by the way the vector fields will be treated
(see, {\it e.g.}, \cite{Percacci:2017fkn} for a corresponding discussion)  without
loss of generality Feynman gauge will be used for simplicity. 

Based on the reasons provided in the beginning of section \ref{MatterGrav} matter field 
renormalisation  and matter self-interactions will be neglected in the following.
As a consequence the effective average action (\ref{GammaMat2}) will not depend
on the RG scale $k$ but, of course, contributes to the flow of the gravitational 
couplings.

\section{The Asymptotic Safety Scenario for Gravity}
\label{AsympSafe}


The asymptotic safety mechanism was first suggested by Weinberg in the context of 
gravity \cite{Weinberg,Weinberg:1996kw}. In section \ref{AsySafe} the underlying 
idea has already been explained: if the high-energy behaviour of a theory 
is controlled by a FP which fulfils certain conditions, the theory 
is free from unphysical UV divergences. 
In this section the focus is on the key ingredient of the asymptotic safety scenario, 
the NGFPs (interacting FPs) of the RG flow. 
Hereby, the predictive power encoded in the properties of the NGFP may be comparable 
to the one known from perturbatively renormalisable QFTs, see 
\cite{Niedermaier:2006wt,Codello:2008vh,Litim:2011cp,Percacci:2011fr,
Reuter:2012id,Nagy:2012ef,Percacci:2017fkn} 
for respective reviews.

First of all, at NGFPs the  dimensionless couplings $g_i$
(where the index is simply meant as counting label in the case of several couplings)
approach fixed values denoted by $g_i^\star$ if the RG ``time'' $t$ is 
increased. As $\partial_t g_i = \beta_i(\{g_i\})$,  by definition, 
 the NGFPS are zeros of the theory's beta functions $\beta_i(\{g_i\})$,
{\it i.e.}, 
\be
\beta_j (\{g_i^\star \}) = 0.
\ee
Obviously, if all couplings are close enough to the values approached at a NGFP 
a linearised treatment of the RG equation in terms of the $g_i-g_i^\star$ will become
accurate. These linearised flows can be encoded in the stability matrix
\be
\label{Bmat} 
{\bf B}_{nm} \equiv \left. \partial_{g_m} \, \beta_{g_n}\right|_{g = g^*} \, .
\ee
Let us assume for the moment that the eigenvalues of this matrix are real. Then it 
is obvious that for a negative eigenvalue $\lambda$ the corresponding eigenmode which is a 
linear superposition of the couplings $g_i$ approaches its FP value 
according to $\exp( \lambda t) = \exp(- |\lambda | t)$. In the direction of eigenmodes
with positive eigenvalues the flow trajectory is exponentially fast repelled. 
Correspondingly, the negative of the eigenvalues $\theta_n=-\lambda_n$ are called
stability coefficients. 

This makes obvious that the asymptotic safety scenario requires that not all 
possible couplings are assumed, the unstable eigendirections ${\mathbf e}_i$ in the 
space of couplings with positive eigenvalue, {\it i.e.} negative stability coefficients,
must be exactly zero. 
The above considerations also explain why 
one defines the stability coefficients as minus the eigenvalues 
of ${\bf B}$.\footnote
{As $t=\ln(k/k_0)$ the FP values are approached
with a power law in the scale $k$ as $k\to \infty$. In analogy to the scaling
exponents in statistical systems the stability coefficients $\theta_n$ 
are sometimes also called scaling exponents.} 

However, in general the stability coefficients are not real but may occur
in complex conjugate pairs. Those with a positive real part are the 
``relevant directions'', and it is this real part which determines 
the approach to FP. The imaginary part leads then to a trigonometric
function of $t$, and overall the RG trajectories are ``spiralling'' into
the FP.

The above described situation is typical for the EH truncation 
with the two couplings $\Lambda_k = \lambda_k \, k^2$ and $G_k = g_k \, k^{2-d}$.
The stability matrix takes then a simple 2$\times$2 form
\be
{\bf B} =
\begin{pmatrix}
\hspace{2mm} \frac {\partial \, \beta_{\lambda}} {\partial \, \lambda} &
\frac {\partial \, \beta_{\lambda}} {\partial \, \mathrm {g}} 
\hspace{2mm} \vspace{5mm} \\
\hspace{2mm} \frac {\partial \, \beta_{\mathrm {g}}} {\partial \, \lambda} &
\frac {\partial \, \beta_{\mathrm {g}}} {\partial \, \mathrm {g}} \hspace{2mm}
\hspace{2mm}
\end{pmatrix}
\hspace{1cm} \mathrm {evaluated ~ at} \hspace{1cm} ( \, g^{\ast} \, , \, \lambda^{\ast} \, ) \,\, .
\ee
In EH truncation with de Donder gauge and a Litim regulator one 
has a NGFP with  
$g^{\ast}=0.707$, $\lambda^{\ast}=0.193$ and two stability coefficients as 
complex conjugate pair, $\theta_{0,1} = 1.475 \pm 3.043$ \cite{Fischer:2006fz}. 

\begin{figure}[ht]
\includegraphics[width=\textwidth]{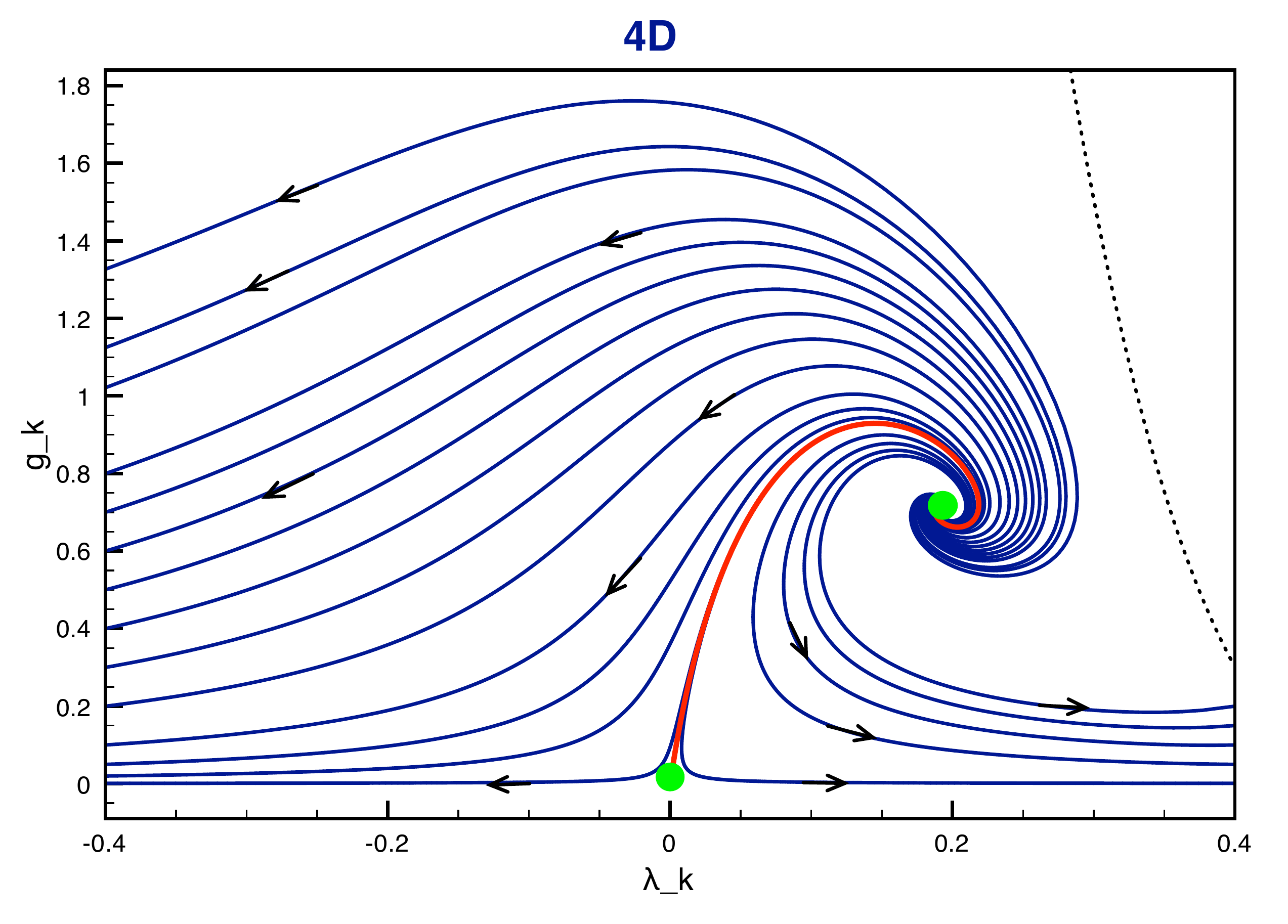}
\vspace{-7mm}
\caption[The phase diagram in EH truncation for d=4.]
{\label{fig:3.5}  Shown is the phase diagram for  the running gravitational coupling 
$g_k$ and the cosmological constant $\lambda_k$ in four dimensions 
for the Wetterich equation in the Einstein-Hilbert truncation.
The separatrix  (full red line) connects the NGFP and the GFP marked by green dots. 
Arrows indicate the direction of the RG flow with decreasing $k \to 0$ .
Adapted from \cite{Master}. }
\end{figure}

The flow diagram in figure \ref{fig:3.5} depicts the NGFP at 
positive $g^\star > 0, \lambda^\star > 0$ which acts as the UV completion of 
all RG trajectories with positive Newton's constant. Lowering the RG scale 
$k$ the flow undergoes a crossover towards the 
GFP situated in the origin. In the vicinity of the GFP the 
{ dimensionful} coupling constants $G_k$ and $\Lambda_k$ become independent 
of $k$, so that classical GR is recovered in the IR. 
Depending on whether the RG trajectory ends at the GFP or flows to its left (right) a 
zero or negative (positive) IR value of the cosmological constant is recovered.

\newpage
\thispagestyle{empty}


\chapter{{\color{MYBLUE}{Asymptotically Safe $f(R)$-Gravity Coupled to Matter}}}
\label{Chap4}
~\vspace{-15mm}

\centerline{Most of this chapter is based on the following publication
\cite{Alkofer:2018fxj}:}
\vspace{1mm}
\centerline{N.~Alkofer and F.~Saueressig.}
\centerline{Asymptotically Safe $f(R)$-Gravity Coupled to Matter I: the Polynomial Case.}
\centerline{arXiv:1802.00498 [hep-th].}

\vspace{2mm}

\centerline{The material presented in section \ref{GlobalFF} is unpublished,
a corresponding publication is in preparation \cite{Alkofer:2018baq}.}

\minitoc

\section{Objective and Key Results}
\label{sec:introFR}

In this chapter 
the flow equation for $f(R)$-gravity with an arbitrary number of 
minimally coupled matter fields is obtained and  the resulting FP structure 
will be analysed. Due to the arguments in favour for the exponential split 
\eqref{ExpSplit} of the metric into background and fluctuating part instead of 
the linear one, the exponential split will be used in the following.  The
related  RG equation for 
$f(R)$-gravity, but without  coupling of matter, has been recently investigated 
  \cite{Ohta:2015efa,Ohta:2015fcu}.
In order to facilitate the comparison with these previous studies 
of the pure gravity case the technical implementation (like, {\it e.g.}, gauge
fixing) will be followed. In particular, $d$-dimensional spheres 
will be used as backgrounds such that the operator-valued traces can be computed 
as sums over eigenvalues of the corresponding differential operators. The 
construction incorporates a $7$-parameter family of coarse-graining operators. The free 
(endomorphism) parameters implement relative shifts in the eigenvalue spectra. 
Essentially, they determine which fluctuating modes are integrated out to drive the flow 
at the scale $k^2$. The consequences of using different coarse-graining operators can 
then be studied systematically. In addition, it will be shown by explicit solution
that certain sets of these parameters exist which allow for a global quadratic solution
of the function $f(R)$ at the NGFP. This latter function will be called fixed function 
in analogy to the FPs for a finite number of couplings. 

Besides improving the understanding of the influence of matter on the structure 
of the NGFP found in $f(R)$-gravity \cite{Ohta:2015efa,Ohta:2015fcu}, a 
second objective of this project is to elucidate the 
 observation that different endomorphism parameters 
lead to different effects on the NGFP
\cite{Dona:2012am,Dona:2013qba}.
It is a key result of this project that $f(R)$-gravity coupled to matter gives 
rise to at least {two different families of ``universality'' classes}. While these 
classes are virtually indistinguishable at the level of the EH 
truncation, the inclusion of higher-derivative operators, and hereby most 
important the $R^2$-term, reliably disentangles the two families. The 
``gravity''-type family has a stable extension under the inclusion of 
higher-derivative operators, and the expansion exhibits a rapid convergence 
in terms of the position of the FP and its critical exponents. Moreover, 
the family comes with a low number of free parameters associated with relevant 
operators. Provided that the coarse-graining operator is chosen as the 
Laplacian, it is found that many phenomenologically interesting gravity-matter 
systems actually possess a NGFP belonging to this family 
(see table \ref{Tab.mainresults}). 
In contrast, the ``matter-dominated'' family of FPs turns out to be 
unstable under the inclusion of higher-derivative terms. It is then shown 
explicitly, that changing the coarse-graining operator may have the effect 
of leaving the domain supporting the ``gravity-type'' FPs 
replacing them by a FP from the matter-dominated family, 
see figure \ref{Fig.typeIIannihilation}. 

However, it should be emphasised that this does not provide insights on how 
the coarse-graining operator should be chosen. Nevertheless, this result is
important in two aspects:  it offers a natural explanation for the qualitatively 
different results on gravity-matter systems in the literature: depending on 
the precise setup, the computation may probe a FP belonging to either 
of the two distinguished families. If this FP happens to be of 
``matter-dominated'' type one may then expect that the FP may become 
unstable once the approximation exceeds a certain degree of sophistication. 
Second, the correct choice of the coarse-graining operator and thus of the 
endomorphism parameters in the regulator can be either based on a mathematical
or a physical argument. As no obvious mathematical argument has turned up yet 
(besides the assumption that the used truncations are still too simplistic)
the above mentioned result of different ``universality'' classes  might 
provide the basis for a physical argument how to implement the coarse-graining 
correctly.

Another key result is given by the explicit form of global solutions
for the fixed function at the NGFP which
is found to be amazingly close to a purely quadratic solution, resp., to 
a solution of the type 
$f^\star(R) =  g_0^\star + g_1^\star R + g_2^\star R^2 + g_L^\star R^2 \log R$.
It seems indeed even possible that deviations from either of the two forms
are purely due to the used approximations.

These key features might also persist in computations which resolve background and 
fluctuation vertices of the effective average action. At this point it is,
however, important to note that the fixed functions in a recent similar approach
also based on performing the traces as spectral sums on a sphere but using 
a linear split of the metric and a vertex expansion \cite{Christiansen:2017bsy}
look qualitatively different. In addition, in \cite{Christiansen:2017cxa}
it was argued, also on the basis of a vertex expansion but around flat backgrounds,
that gravity dominates the high-energy behaviour largely independent of the matter 
fields. Nevertheless, also in this investigation a pronounced scheme dependence
visible as a dependence on technicalities such as the chosen gauge, the regularisation
and/or momentum cutoff was found. Nonetheless, the authors of \cite{Christiansen:2017cxa}
did see compelling evidence for the main feature, namely, that asymptotic safety
for gravity-matter systems follows from asymptotic safety of pure gravity. There 
is clearly a tension between this very recent study and the results presented in this
chapter. This will be discussed further in the concluding chapter~\ref{Chap:Concl}.
 
\section{Flow Equation for Gravity-Matter Systems in the $f(R)$-Truncation}
\label{sec:Gamma}

In this section the ansatz for the effective average action (section  \ref{sec:setup}) and  
the operator traces entering into the projected flow equation (section  \ref{sec:RGgrav})
will be presented. The properties of the regularisation schemes employed in this work
 are discussed in section \ref{sec:endomorphism}.

\subsection{Project Outline}
\label{sec:setup}

This study focuses on the RG flow of $f(R)$-gravity supplemented by minimally 
coupled matter fields. 
The basis for this work is an effective average action $\Gamma_k$ consisting 
of two parts, 
\be
\label{Gammaans}
\Gamma_k = \Gamma_k^{\rm grav} + \Gamma^{\rm matter}_k \, ,
\ee
where $\Gamma^{\rm grav}_k$ is the gravitational part of the effective average action and 
$\Gamma^{\rm matter}_k$ encodes the contribution of the matter fields. Based on the
discussion of the tree-level actions given in chapter \ref{Chap1} 
the gravitational part of the action is chosen as
\be
\Gamma_k^{\rm grav} = \int d^dx \, \sqrt{g} \, f_k(R)  \,\, + 
\Gamma_k^{\rm gf} + \Gamma_k^{\rm gh}  \,  ,
\label{GammaGrav} 
\ee
where $f_k(R)$ is an arbitrary, scale-dependent function of the Ricci scalar $R$ 
and the action is supplemented by suitable gauge-fixing and ghost terms. This sector 
is taken to be identical to the one studied in \cite{Ohta:2015fcu}. 
The matter sector, 
$ \Gamma^{\rm matter}  =  \Gamma^{\rm scalar}  
+  \Gamma^{\rm fermion}  +  \Gamma^{\rm vector}$, 
follows also closely the tree-level forms given in section \ref{MatterGrav} and 
contains $N_S$ scalar fields $\phi$, $N_D$ Dirac fermions $\psi$, and $N_V$ 
(Abelian) gauge fields $A_\mu$, including appropriate gauge-fixing to Feynman 
gauge and ghosts $\bar{c}$ and $c$. Their actions are given by
\begin{subequations}  \label{mattersector}
	\begin{align} \label{Gscalar}
	\Gamma^{\rm scalar}  = &  \, \, 
	\frac{N_S}{2} \int d^d x \sqrt{g} g^{\mu\nu} \left(  D_\mu \phi \right) \left( D_\nu \phi \right) \, ,  \\ \label{Gfermion}	
	\Gamma^{\rm fermion} = &  \, \, i \, N_D \,  \int d^d x \sqrt{g} \, \bar \psi  \Dslash \, \psi \, , \\ \label{Gvector}
\Gamma^{\rm vector} = & \, \, N_V  \int d^d x \sqrt{g} \, 
\left( \frac 1 4 F^{\mu\nu} F_{\mu \nu}
 + \frac 1 2 ( D^\mu A_\mu)^2 
+ \bar c \, (- D^2) \, c
\right) \, .  
	\end{align}
\end{subequations}

In the following the field renormalisation for matter fields and therefore contributions
from their anomalous  dimensions will be neglected.
Although it would be not complicated to keep them, {\it e.g.}, in the calculations
presented in chapter \ref{Chap4} I decided otherwise for two reasons.
First, matter self-interactions are likely to provide an important contribution to the
matter anomalous dimensions, and it is, however, far beyond the scope of the present
thesis to include those self-interactions.
Second, in the following I either focus on the impact of matter on the gravitational
couplings and only the ``classical'' (tree-level) matter propagators of certain
models for quantum gravity will be employed.


The evaluation of the FRGE  \eqref{FRGE}  for the ansatz \eqref{Gammaans} can be 
significantly  simplified by choosing a suitable background. Since the used 
{\it ansatz} 
projects the full RG flow onto functions of the Ricci scalar, it is convenient to work 
with $\bar g_{\mu\nu}$ being a one-parameter family of metrics on the maximally 
symmetric $d$-sphere with arbitrary radius $a$. In this case the Riemann tensor 
and the Ricci tensor are determined by scalar curvature $\Rb$,
\be
\label{max:sym}
\bar R _{\mu\rho\nu\sigma} = \frac {\bar R}{d(d-1)}
\left( \bar g_{\mu\nu}\bar g_{\rho\sigma} - \bar g_{\mu\sigma }
\bar g_{\nu\rho} \right) \, ,  \qquad 
\bar R _{\mu \nu} = \frac {\bar R}{d} \, \bar g_{\mu\nu} \, .
\ee
Hereby, the scalar curvature is covariantly constant, $\bar D_\mu \bar R=0$,  
and it is related to the radius $a$ of the sphere,
\be
\bar R =  \frac 1 {a^2} d (d-1) \, . \label{curv}
\ee
The volume $V_d$ of the background is given by
\be
\label{volsphere}
V_d = \frac{2 \, \pi^{(d+1)/2}}{\Gamma((d+1)/2)} \, a^d =  \frac{2 \, 
\pi^{(d+1)/2}}{\Gamma((d+1)/2)} \, \left( \frac{d (d-1)}{\Rb} \right)^{d/2} \, . 
\ee
Defining the Laplacian $\Delta = -\gb^{\mu\nu} \Db_\mu \Db_\nu$ (see appendix 
\ref{App.Laplace} for a brief discussion of possible choices for the Laplacian), 
the complete set of eigenvalues $\lambda_\ell^{(s)}$ together with their 
degeneracies $M_\ell^{(s)}$ for $\Delta$ acting on irreducible spin 
representations have been determined in \cite{Rubin:1984tc,Camporesi:1995fb}. 
This data is collected in table \ref{Tab.ev}. In particular, the spectrum of 
the Laplacian acting on spinor fields is obtained from the eigenvalues 
of $-\bar \Dslash~^2$ \cite{Camporesi:1995fb},
\be
\lambda_l^{-\bar {\DslashIndex}~^2} = \frac{1}{a^2} \left(l + \tfrac{d}{2} \right)^2
\; , \qquad 
M_l^{-\bar \DslashIndex~^2} = 2^{\lfloor d/2+1 \rfloor}
\frac{(\ell+d-1)!}{(d-1)! \, \ell !} \, , \qquad \ell = 0,1,\ldots \, , 
\ee
in combination with the Lichnerowicz formula 
$-\bar \Dslash~^2 = \Delta + \Rb/4$. Here $\lfloor \ldots \rfloor$ denotes 
the floor function.
\begin{table}[t!]
\centering{
        \begin{tabular}{c|ccc}
        spin $s$  & $\lambda_\ell^{(s)}$ & $M_\ell^{(s)}$ & \\[2mm] \hline \hline
\\[-2mm]
        $0$ & \qquad $\frac 1 {a^2} \ell (\ell+d-1)$ \quad \;
            & \qquad $\frac{(\ell+d-2)!}{(d-1)! \, \ell !}(2\ell +d-1)$ \quad  \;
            &  $\ell = 0,1, \ldots$  \\[2mm]
 $\tfrac{1}{2}$ & $\frac 1 {a^2} (\ell^2 + d\ell + \tfrac{d}{4})$
                & $ 2^{\lfloor d/2+1 \rfloor}\frac{(\ell+d-1)!}{(d-1)! \, \ell !}$
                & $\ell = 0,1, \ldots$\\[2mm]
            $1$ & $\frac 1 {a^2} \left(\ell (\ell+d-1) -1 \right)$
                & $\frac{(\ell+d-3)!}{(d-2)! (\ell +1)!}(2\ell +d-1)(\ell+d-1)\ell$
                & $\ell = 1,2,\ldots$ \\[2mm]
            $2$ &  \; \; $\frac 1 {a^2} \left(\ell (\ell+d-1) -2 \right)$  \; \; 
                &  \; \;
                $\tfrac{(d+1)(d-2)(l+d)(l-1)(2l+d-1)(l+d-3)!}{2(d-1)!(l+1)!}$  \; \;
                & $\ell = 2,3,\ldots$ \\[2mm] \hline \hline
        \end{tabular}
}
\caption[Eigenvalues $\lambda_\ell^{(s)}$ and 
their degeneracy $M_\ell^{(s)}$ for the Laplacian 
$\Delta = -\gb^{\mu\nu} \Db_\mu \Db_\nu$ acting on
scalars ($s=0$), Dirac fermions ($s=1/2$), transverse vectors ($s=1$) and 
transverse-traceless matrices ($s=2$).]
{\label{Tab.ev} Eigenvalues $\lambda_\ell^{(s)}$ and 
their degeneracy $M_\ell^{(s)}$ for the Laplacian 
$\Delta = -\gb^{\mu\nu} \Db_\mu \Db_\nu$ acting on
scalars ($s=0$), Dirac fermions ($s=1/2$), transverse vectors ($s=1$) and 
transverse-traceless matrices ($s=2$). The bosonic results are taken 
from \cite{Rubin:1984tc} while the fermionic case has been derived in 
\cite{Camporesi:1995fb}.}
\end{table}

Finally, the computation is simplified by decomposing the fluctuation fields 
into their  irreducible spin components, see \cite{Lauscher:2001ya} for an 
extended discussion. For 
the gravitational fluctuations  this is achieved by the York decomposition, see 
(\ref{york}),
which expresses $h_{\mu\nu}$ in terms of a transverse-traceless tensor $h^{TT}_{\mu\nu}$ 
(spin $s=2$), a transverse vector $\xi_\nu$ (spin $s=1$) and two scalar fields $\sigma,h$
\cite{York:1973ia}. Hereby,   $h^{TT}_{\mu\nu}$  and $\xi_\nu$ are 
subject to  differential constraints, see section \ref{secBack}.
Similarly, a vector field $A_\mu$ is decomposed into a transverse vector 
$A_\mu^T$ and a scalar $a$ according to
\be
\label{Tdec}
A_\mu = A_\mu^T + \bar{D}_\mu \, a \, , \qquad \Db^\mu A_\mu^T = 0 \, .  
\ee  
Notably, not all eigenmodes of the Laplacian contribute to the decompositions 
\eqref{york} and \eqref{Tdec}. A constant mode $a$ drops out of the transverse 
decomposition \eqref{Tdec} while in the York decomposition the two lowest 
eigenmodes of $\sigma$ and the lowest vector mode $\xi_\mu$ of the Laplacian 
do not change the right hand side of \eqref{york}. These zero modes must 
then be removed by hand in order to make the decompositions into irreducible 
spin components bijective. Moreover, the decompositions give rise to 
operator-valued Jacobians. On a spherical background these are given by
\be
\label{fieldredef}
\cJ^{\rm vec} = \Det_{(1)}\left(\Delta - \tfrac{\Rb}{d}\right)^{1/2} \, , \quad
\cJ^{\sigma} = \Det_{(0)}\left(\Delta^2 - \tfrac{\Rb}{d-1} \Delta\right)^{1/2} \, , 
\quad 
\cJ^{a} = \Det_{(0)}\left(\Delta\right)^{1/2} \, . 
\ee

\goodbreak

\subsection{Trace Contributions from the Gravitational and Matter Sector}
\label{sec:RGgrav}

Given the ansatz \eqref{Gammaans} the flow of $\Gamma_k$ will be sourced
 by quantum fluctuations in the gravitational and matter sector
\be
\label{deftraces}
\partial_t \Gamma_k = T^{\rm grav} + T^{\rm matter} \, . 
\ee
The construction of the gravitational sector follows 
\cite{Ohta:2015efa,Ohta:2015fcu}. In this setting, $\Gamma_k^{\rm grav}$
is supplemented by a classical gauge-fixing term
\be
\label{gf1}
\Gamma^{\rm gf}=\frac{1}{2\alpha}\int d^d x \sqrt{\bar g}\,
\bar g^{\mu\nu}F_\mu F_\nu \, , \qquad F_\mu = 
\bar D_\rho h^\rho{}_\mu-\frac{\beta +1}{d} \bar D_\mu h\, .
\ee
Expressing $F_\mu$ in terms of the component fields \eqref{york} one has
\ba
F_\mu &=& -\left( \Delta   -\tfrac{\bar R}{d} \right)\xi_\mu
- \tfrac{1}{d} \bar D_\mu
\left(\left[(d-1)\Delta - \Rb \right]\sigma+ \beta \, h\right).
\label{gf}
\ea
Following \cite{Benedetti:2011ct} this suggests to recast the scalar fields in 
terms of a gauge invariant field $s$ and a gauge dependent degree of freedom $\chi$
\be
\label{scalarsector}
s= h + \Delta \sigma \, , \qquad \chi = 
\frac{[(d-1)\Delta - \Rb] \sigma + \beta h}{(d-1-\beta)\Delta - \Rb}
\ee
where the denominator in $\chi$ is fixed by requiring that the transformation 
has a constant Jacobian. Expressing the gauge-fixing term in terms of 
these fields leads to
\be
\Gamma^{\rm gf} = \frac{1}{2\alpha}\int d^d x \sqrt{\bar g} \, 
\Big\{ \xi^\mu \left[  \Delta - \tfrac{\Rb}{d} \right]^2 \xi_\mu + 
\tfrac{(d-1-\beta)^2}{d^2} \chi \, 
\left[ \Delta \left( \Delta - \tfrac{\Rb}{(d-1-\beta)} \right)^2 \right] \,\chi
\Big\} \, . 
\ee

The ghost action associated with the gauge-fixing \eqref{gf1} is obtained in the 
standard way. Restricting to terms quadratic in the fluctuation fields, it reads
\ba
\Gamma^{\rm ghost}&=&\int d^dx\sqrt{\bar g} \, \, \bar C^\mu\left[ \,
\delta_\mu^\nu \, \bar D^2 
+\left(1-2 \, \tfrac{\beta+1}{d}\right) \bar D_\mu \bar D^\nu+
\tfrac {\bar R} d \, \delta_\mu^\nu\right]C_\nu .
\ea
Decomposing the ghosts into their transversal and longitudinal part, 
$C_\nu = C_\nu^T + \bar D_\nu  \, C^{L}$, 
one obtains
\ba
\Gamma^{\rm ghost} = - \int d^dx\sqrt{\bar g} \, \,  
\left\{   \bar C^{T \mu} \left[ \Delta  - \tfrac {\bar R}{d}  
\right]C^T_\mu + 2 \, \tfrac{d-1-\beta}d \, \bar C^{L} \left[  \Delta - 
\tfrac {\bar R}{d-1-\beta} \right] \, \Delta \, C^{L} \right\} .
\ea

The gravitational sector is completed by the expansion of 
$\Gamma^{\rm grav}_k[g] = \Gamma^{\rm grav}_k[\gb] + \cO(h) + 
\Gamma_k^{\rm quad}[h;\gb]+\ldots$. For the exponential split 
\eqref{ExpSplit} the terms quadratic in the fluctuation fields are given by 
\cite{Ohta:2015fcu} 
\ba
\Gamma_k^{\rm quad} &=& \int d^dx \sqrt{\gb} \,  
\Big\{ -\tfrac{1}{4} \, f'(\bar R) \, h_{\mu\nu}^{TT} \, 
\Big[\Delta +\tfrac{2}{d(d-1)}\bar R \Big]\,  h^{TT\, \mu\nu}
\nonumber \\ && 
+ \tfrac{d-1}{4d} \, s \, \Bigg[ \tfrac{2(d-1)}{d} \, f''(\bar R ) \, 
\Big(\Delta -\tfrac{\bar R}{d-1}\Big)
+ \tfrac{d-2}{d}f'(\bar R )\Big] \Big[ \Delta-\tfrac{\bar R}{d-1} \Big] \, s
\nonumber \\  && 
+ h \, \Big[ \tfrac{1}{8} f(\bar R )-\tfrac{1}{4d} \bar  R \, 
f'(\bar R) \Big] \, h \Big\} \, .
\label{HessianGrav}
\ea
Note that all terms containing the spin-1 component $\xi_\mu$ canceled out. 

At this stage, it is useful to collect the determinants arising from the various spin 
sectors. The transverse vector sector receives contributions from the Jacobian in the 
transverse-traceless decomposition \eqref{fieldredef}, from the transverse ghosts, and 
from $\xi^\mu$ in the gauge-fixing term. All one-loop determinants have the same form, 
such that they combine according to
\be
\label{vecdet}
\Det_{(1)}\left(\Delta - \tfrac{\Rb}{d}\right)^{1/2} \, 
\Det_{(1)}\left(\Delta - \tfrac{\Rb}{d}\right) \, 
\Det_{(1)}\left(\Delta - \tfrac{\Rb}{d}\right)^{-1} = 
\Det_{(1)}\left(\Delta - \tfrac{\Rb}{d}\right)^{1/2} \, . 
\ee
In the scalar sector, one combines the contributions from $\chi$, 
the longitudinal ghost $C^L$, and the scalar determinants from the 
field decompositions in the transverse-traceless and ghost decomposition
\be
\begin{split}
& \Det_{(0)}\left(\Delta\right)^{-1/2} \,
\Det_{(0)}\left(\Delta - \tfrac{\Rb}{d-1-\beta} \right)^{-1} \cdot 
 \Det_{(0)}\left(\Delta\right) \, 
  \Det_{(0)}\left(\Delta - \tfrac{\Rb}{d-1-\beta} \right) \cdot \\ &
  \Det_{(0)}\left(\Delta\right)^{1/2} \,
   \Det_{(0)}
  \left(\Delta - \tfrac{\Rb}{d-1}\right)^{1/2} \cdot \Det_{(0)}\left(\Delta\right)^{-1} = 
  \Det_{(0)}\left(\Delta - \tfrac{\Rb}{d-1}\right)^{1/2} \, ,
  \end{split}
\ee
where the $\cdot$ is used to separate the contributions from the various sectors. Note 
that the remaining scalar determinant can be absorbed by the field redefinition 
$s \rightarrow \tilde s = \left[\Delta - \tfrac{\Rb}{d-1} \right]^{1/2} s$, which 
simplifies the contribution of the scalar sector in \eqref{HessianGrav}. The cancellation 
of the scalar determinants is actually \emph{independent} of the choice of the gauge-fixing parameter $\beta$. It solely relies on the field redefinition \eqref{scalarsector} 
used to disentangle the gauge invariant and gauge dependent field contributions.

The structure of the Hessians is further simplified by adopting ``physical gauge'' 
$\beta \rightarrow - \infty$, $\alpha \rightarrow 0$. From \eqref{scalarsector} 
one finds that the limit $\beta \rightarrow - \infty$ aligns $\chi$ and $h$ 
such that $\chi \propto h$. Subsequently evoking the Landau limit 
$\alpha \rightarrow 0$ then ensures that the $h^2$ term appearing in 
$\Gamma_k^{\rm quad}$ does not contribute to the flow equation. In this way the 
contributions of the fields in the gravitational sector is maximally decoupled: 
$\Gamma_k^{\rm quad}$ gives the contributions for $h^{\rm TT}_{\mu\nu}$ and $s$, 
while the gauge-fixing term determines the quantum fluctuations of the transverse 
vector $\xi_\mu$ and scalar $\chi$.

The final ingredient in writing down the projected flow equation \eqref{FRGE} is the 
regulator $\cR_k(\Box)$. Since one of the main objectives of this work is to 
understand the role of different coarse-graining operators, the 
operators
\be
\Box_{S,D,V,T}^{G,M} \equiv \Delta - \alpha_{S,D,V,T}^{G,M} \bar R \, 
\ee 
are introduced which, besides the Laplacian, also contain an endomorphism parameter 
$\alpha_{S,D,V,T}^{G,M}$. Here the superscript indicates if the operator 
belongs to the gravitational (G) or matter sector (M) while the subscript 
gives the spin of the corresponding fields. In case of ambiguities, 
additional numbers to the spin index are added. The regulator 
$\cR_k(\Box)$ is then fixed through the replacement rule
\be
\label{TypeIrep}
\Box \mapsto P_k(\Box) \equiv \Box + R_k(\Box) \, ,
\ee
where it is understood that the endomorphism parameters contained in the coarse 
graining operators $\Box$ may differ for different fields.

Based on \eqref{HessianGrav} and \eqref{vecdet} one has now all 
ingredients for writing down the gravitational contribution to the flow 
of $f_k(R)$. The gravitational sector gives rise to three contributions 
associated with the transverse-traceless fluctuations $h_{\mu\nu}^{TT}$, 
the gauge invariant scalar $s$ and the vector determinant \eqref{vecdet}:
\be
T^{\rm grav} = T^{\rm TT} + T^{\rm ghost} + T^{\rm sinv} \, . 
\ee
The explicit expressions for the traces are given by
\begin{subequations}\label{gravitytraces}
	\begin{align}
	T^{\rm TT} = &  \, \frac{1}{2}\Tr_{(2)}  \left[ 
	\left(f'(\bar R)( P_k^T
		+\alpha_T^G \bar R+\tfrac{2}{d(d-1)}\bar R)\right)^{-1}
	\partial_t \left( f'_k(\bar R)  R_k^T \right) \right] \, , \\
T^{\rm sinv} = & \, 
\frac 1 2 \Tr^{\prime\prime}_{(0)} \left[ \left(f''_k(\bar R) (P_k^S
+\alpha_S^G\bar R-\tfrac{1}{d-1}\bar R) +
\tfrac{d-2}{2(d-1)}f'_k(\bar R)\right)^{-1}  
\partial_t \left( f''_k(\bar R) R_k^S \right)  \right] \, ,
\\
T^{\rm ghost} = - & \,   \frac 1 2 \Tr^{\prime}_{(1)} \left[ 
\left(P_k^V +\alpha_V^G \bar R -\tfrac{1}{d}\bar R\right)^{-1} \partial_t R_k^V
\right]  \, . 
	\end{align}
\end{subequations}
Here the number of primes on the traces indicate the number of modes 
which have to be discarded. The subscript on the traces, on the other hand, 
specifies the spin of the fields. By construction, the result agrees with 
\cite{Ohta:2015fcu}.


The contribution of the minimally coupled matter fields \eqref{mattersector} 
to the gravitational flow can be constructed along the same lines as in the 
gravitational sector: one first decomposes the vector field into its 
transverse and longitudinal parts according to \eqref{Tdec}, computes 
the Hessians $\Gamma^{(2)}$, and determines the regulator function 
according to the prescription \eqref{TypeIrep}. The resulting contribution 
is given by
\be
T^{\rm matter} = T^{\rm scalar} + T^{\rm Dirac}  + T^{\rm vector} 
\ee
where
\begin{subequations}\label{mattertraces}
	\begin{align}
	T^{\rm scalar} = &  \, \frac{N_S}{2} \, \Tr_{(0)}  
\left[ (P_k^S+\alpha_S^M \bar R)^{-1} \, \partial_t R_k^S \right] \, , \\
	T^{\rm Dirac} = & \, - \frac{N_D}{2}  \Tr_{(1/2)} 
\left[ (P_k^D+\alpha_D^M \bar R + \tfrac{1}{4} \Rb)^{-1} \, \partial_t R_k^D \right] 
\, , \\ 
	T^{\rm vector} = & \, \frac{N_V}{2}  \Tr_{(1)} 
\left[ (P_k^{V_1}+\alpha_{V_1}^M \bar R + 
\tfrac{1}{d} \Rb)^{-1} \, \partial_t R_k^{V_1} \right] +
	\frac{N_V}{2}  \Tr_{(0)}^\prime \left[ (P_k^{V_2}+
\alpha_{V_2}^M \bar R)^{-1} \, \partial_t R_k^{V_2} \right] 
\nonumber \\  
& \, - N_V  \Tr_{(0)}^\prime \left[ (P_k^{V_2}+\alpha_{V_2}^M \bar R)^{-1} 
\, \partial_t R_k^{V_2} \right].
    \end{align}
\end{subequations}
The three traces in $T^{\rm vector}$ capture the contribution from the 
transverse vector field, the longitudinal modes, and ghost fields, respectively. 
Again the number of primes indicates that the corresponding number of lowest 
eigenmodes should be removed from the trace. In addition, 
each sector contains its own endomorphism parameter $\alpha$. Following 
\cite{Ohta:2015efa,Ohta:2015fcu},  a ``mode-by-mode'' 
cancellation between the matter and ghost modes has been implemented, 
so that the corresponding 
traces come with the same number of primes and endomorphism parameter. 
The full, projected flow equation is then obtained by substituting 
\eqref{gravitytraces} and \eqref{mattertraces} into \eqref{deftraces}. 

\subsection{Constraining the Coarse-Graining Operator}
\label{sec:endomorphism}

Notably, the values of the endomorphism parameters $\alpha$ may not be 
chosen arbitrarily. On physical grounds one requires that
\begin{enumerate}
	\item[1.] For any fluctuation contributing to the operator traces in the 
flow equation the argument of the regulator 
$\cR_k(\Box)$, $\Box = \Delta - \alpha \Rb$, should be positive-semidefinite.
\end{enumerate}
and
\begin{enumerate}
	\item[2.] The denominators appearing in the trace-arguments should be 
free of poles on the support of $\Box$. In other words, the ``mass-type'' 
terms provided by the background curvature should not correspond to a 
negative squared-mass.
\end{enumerate}
At first sight the second condition may seem somewhat less compelling since these types 
of singularities are removed when the flow equation is expanded in powers of the 
background curvature. Taking into account that the approximate solutions of the flow 
equation arising from this procedure should ultimately have an extension to solutions of 
the full flow equations, constraining the coarse-graining operator to those which do not 
give rise to such extra singularities is a sensible requirement.

Practically, the first condition translates into the requirement that $\alpha \Rb$ must 
be smaller than the lowest eigenvalue contributing to a given trace. Taking into account 
the omitted lowest eigenmodes (indicated by the primes in  \eqref{gravitytraces}) the 
resulting constraints in the gravitational sector are
\be
\label{eqc1g}
\alpha_T^G \le \frac 2{d-1} \, , \qquad 
\alpha_S^G \le \frac{2(d+1)}{d(d-1)} \, , \qquad 
\alpha_V^G \le \frac{2d+1}{d(d-1)} 
 \, .
\ee 
Analogously, the endomorphism parameters in the matter sector should satisfy
\be
\label{eqc1m}
\alpha_S^M \le 0 \, , \qquad 
\alpha^M_D \le \frac{1}{4(d-1)} \, , \qquad
\alpha^M_{V_1} \le \frac{1}{d} \, , \qquad
\alpha^M_{V_2} \le \frac{1}{d-1} \, . 
\ee
Here different bounds for fields with the same spin arise due to a 
different number of fluctuation modes excluded from the traces.

The second condition is evaluated by replacing $P_k \rightarrow k^2$ and subsequently 
writing the propagators in terms of the dimensionless curvature $r \equiv \Rb k^{-2}$. 
The denominators then take the form $(1+(c_d+\alpha)r)$ where the constants $c_d$ depend 
on the trace under consideration and can be read off from \eqref{gravitytraces} and 
\eqref{mattertraces}. For fixed background curvature $\Rb$ and $k \in [0,\infty[$ 
the dimensionless curvature takes values on the entire positive real axis 
$r \in [0,\infty[$. The absence of poles results in the condition 
$\alpha \ge - c_d$. In the gravitational sector this entails
\be
\label{eqc2g}
\alpha_T^G \ge -\frac 2{d(d-1)} \, , \qquad 
\alpha_S^G \ge \frac{1}{d-1} \, , \qquad 
\alpha_V^G \ge \frac{1}{d} 
\, ,
\ee
while for the matter fields the bounds are
\be
\label{eqc2m}
\alpha_S^M \ge 0 \, , \qquad 
\alpha^M_D \ge - \frac{1}{4} \, , \qquad
\alpha^M_{V_1} \ge - \frac{1}{d} \, , \qquad
\alpha^M_{V_2} \ge 0 \, . 
\ee
The bound on $\alpha_S^G$ reported in \eqref{eqc2g} may be less stringent 
though, since the quoted value does not take into account possible contributions 
from the function $f_k(\Rb)$ which can only be computed at the level of solutions. 
Notably, both sets of conditions \eqref{eqc1g}, \eqref{eqc1m} and 
\eqref{eqc2g},\eqref{eqc2m} can be met simultaneously. This requires 
non-zero endomorphism parameters $\alpha_S^G$ and $\alpha_V^G$ though.

For latter use two widely used choices for the coarse-graining 
operators termed ``type I'' and ``type II'' (see \cite{Codello:2008vh} for a detailed 
discussion) are introduced. In this case the endomorphism parameters are chosen as
\begin{subequations}\label{endomorphism}
	\begin{align}\label{endtypeI}
	& \mbox{type I:}  \quad \alpha^G_T = \alpha^G_S = \alpha^G_V  =  
\alpha^M_D = \alpha^M_{V_1} = \alpha^M_{V_2} = \alpha^M_S = 0 \, , \\ \label{endtypeII}
		& \mbox{type II:} \quad
		\alpha^G_T = - \tfrac{2}{d(d-1)} \, , \; 
		\alpha^G_S = \tfrac{1}{d-1} \, , \; 
		\alpha^G_V = \tfrac{1}{d} \, , \;  \nonumber \\ & \qquad \qquad
		\alpha^M_D = - \tfrac{1}{4} \, , \; 
		\alpha^M_{V_1} = - \tfrac{1}{d} \, , \;
		\alpha^M_{V_2} = \alpha^M_S = 0 \, .
	\end{align}
\end{subequations}
For the type I choice the coarse-graining operator $\Box$ agrees with the 
Laplacian acting on the corresponding spin fields. The type II coarse-graining 
operator is tailored in such a way that it removes the scalar curvature from the 
propagators.\footnote{The use of a type I and type II coarse-graining operator 
should not be confused with a ``change of the regulator function''. 
One way to fix the values of $\alpha$ is provided by the 
principle of equal lowest eigenvalues \cite{Demmel:2014hla}, but 
herein the endomorphisms are treated as free parameters.} By construction it 
satisfies both conditions 1 and 2.

In order to trace the dependence of the RG flow on the choice of coarse 
graining operator, a one-parameter family of coarse-graining operators 
is introduced:
\be
\label{regtypei}
\mbox{type I:} \quad
\alpha^G_T = - \tfrac{2c}{d(d-1)} \, , \; 
\alpha^G_S = \tfrac{c}{d-1} \, , \; 
\alpha^G_V = \tfrac{c}{d} \, , \; 
\alpha^M_D = - \tfrac{c}{4} \, , \; 
\alpha^M_{V_1} = - \tfrac{c}{d} \, , \;
\alpha^M_{V_2} = \alpha^M_S = 0 \, .
\ee
It contains one free parameter $c$ and interpolates 
continuously between a coarse-graining operator of type I for $c=0$ and 
type II for $c=1$. In particular, this construction will be very useful 
in order to understand the FP structure of gravity-matter 
systems in section \ref{sec:fixedpoints}.

\section{Evaluating Operator Traces as Spectral Sums}
\label{sec:spectrum}

The next step consists in explicitly evaluating the traces 
\eqref{gravitytraces} and \eqref{mattertraces} and rewrite them as 
explicit functions of the scalar curvature. The main result is the 
partial differential equation \eqref{pdfgeneral} and its restriction 
to four dimensions \eqref{pdf4d} which governs the scale-dependence 
of $f_k(R)$ in the presence of minimally coupled matter fields.  

Our computation follows the strategy 
\cite{Reuter:2008qx,Benedetti:2012dx,Benedetti:2013jk,Demmel:2014sga,
Ohta:2015efa,Ohta:2015fcu} and performs the traces as sums over 
eigenvalues of the corresponding Laplacians. Furthermore, 
a Litim-type regulator \eqref{LitimReg}  is employed.
For finite $k$, the presence of the step-function in the regulator entails
that only a finite number of eigenvalues contribute to the mode sum. Moreover, the 
propagators are independent of $\Delta$ and can be pulled out of the sums. As a 
consequence the traces reduce to finite sums
over the degeneracies of the eigenvalues, possibly weighted by the corresponding 
eigenvalue. These sums take the form
\be
\label{sumdeg}
S^{(s)}_{d}(N) \equiv \sum_{\ell = \ell_{\rm min}}^{N} \, M_\ell^{(s)} \, , \qquad
\widetilde{S}^{(s)}_{d}(N) \equiv \sum_{\ell = \ell_{\rm min}}^{N} \, 
\lambda_\ell^{(s)} \, M_\ell^{(s)} \, ,
\ee
were $N$ is a (finite) integer determined by the regulator, and the eigenvalues and 
degeneracies are listed in table \ref{Tab.ev}. In the matter sector, all traces have the 
structure $S^{(s)}_d(N)$ while the gravitational sector gives rise to both types of 
contributions. The occurrence of contributions of the form $\widetilde{S}^{(s)}_{d}(N)$ 
can be traced back to the presence of scale-dependent coupling constants in the regulator 
functions which only occur in the gravitational sector and are absent in the matter 
traces. In this section the sums \eqref{sumdeg} will be used 
to explicitly evaluate the right
hand side of the flow equation by summing over the eigenvalues of the differential 
operators. 

Carrying out the sums for scalars ($s=0$), Dirac fermions ($s=1/2$), 
transverse vectors $(s = 1)$, and transverse-traceless tensors ($s=2$) results in
\begin{subequations}\label{mattersums}
	\begin{align}
	S^{(0)}_{d}(N) = & \, \left(2N+d\right) \frac{(N+d-1)!}{d! \, N!} \, ,	\\
	S^{(1/2)}_{d}(N) = & \, 2^{\lfloor d/2+1 \rfloor} \, \frac{(N+d)!}{d! \, N!} 
\, , \\	
	S^{(1)}_{d}(N) = & \, 1 + \frac{d-1}{d!} \, 
\frac{(2N+d) \, (N^2+dN-1)\, (N+d-2)!}{(N+1)!} \, , \\
	S^{(2)}_{d}(N) = & \,\tfrac{(d+2)(d+1)}{2} + 
\tfrac{(d+1)(2N+d)\big((d-2) (N^2+dN) - (d+2)(d-1)\big) (N+d-2)!}
{2 \, d! \, (N+1)!} \, .  
	\end{align}
\end{subequations}
These results may readily be confirmed by applying proof by induction techniques.
 All expressions are polynomials of order $d$ in $N$. The sums weighted by the 
eigenvalues can be performed in the same way. In this case it suffices to consider the 
cases $s=0$ and $s=2$, yielding
\begin{subequations}\label{gravitonsum}
	\begin{align}
	\widetilde{S}^{(0)}_{d}(N) = & \, \frac{ (2N + d) (N + d)!}{a^2 \, (d+2) \, (d-1)! (N-1)!} \, , \\
	\widetilde{S}^{(2)}_{d}(N) = & \,
	- \tfrac{2d(d+1)}{a^2}
	+ \tfrac{(d+1) (2N+d) \big((d-2)(N^4 + 2dN^3 +(d^2-d-5)N^2 -d(d+5)N) 
+ 4 (d-1) (2 + d) \big) (N+d-2)!}{2 a^2 (2 + d) ( d-1)! (N+1)!} \, . 
	\end{align}
\end{subequations}
These expressions are again polynomials in $N$ of order $d+2$. The increased order 
thereby compensates the factor $a^2$ such that both \eqref{mattersums}
and $\eqref{gravitonsum}$ exhibit the same scaling behaviour as $R \rightarrow 0$.

The value $N$ at which the sums are cut off is given by the largest integer 
$N^{(s)}_{\rm max}$ satisfying the inequality 
$\lambda^{(s)}_{N^{(s)}_{\rm max}} - \alpha \Rb \le k^2$. 
Substituting the eigenvalues listed in table \ref{Tab.ev} and solving this 
condition for $N^{(s)}_{\rm max}$ yields
\begin{subequations}\label{Ncutoff}
	\begin{align}
	N^{(0)}_{\rm max} = & \, - \tfrac{d-1}{2} - p^{(0)} + 
\tfrac{1}{2} \sqrt{d^2-2d+1 + 4 d (d-1) \left(\tfrac{1}{r} + \alpha\right) 
+ q^{(0)}} \, , \\
	N^{(1/2)}_{\rm max} = & \, - \tfrac{d}{2} - p^{(1/2)} + \sqrt{ d (d-1) 
 \left( \tfrac{1}{r} + \tfrac{1}{4} +  \alpha\right) + \tfrac{1}{4} \,  
q^{(1/2)} } \, , \\
	N^{(1)}_{\rm max} = & \, - \tfrac{d-1}{2} -  p^{(1)} + 
\tfrac{1}{2} \sqrt{d^2-2d+5 + 4 d (d-1) \left(\tfrac{1}{r} + \alpha\right) + q^{(1)} } 
\, , \\
	N^{(2)}_{\rm max} = & \, - \tfrac{d-1}{2} -  p^{(2)} + 
\tfrac{1}{2} \sqrt{d^2-2d+9 + 4 d (d-1) \left(\tfrac{1}{r} + \alpha\right) + q^{(2)} } 
\, ,
	\end{align}
\end{subequations}
\begin{figure}[t!]
	\includegraphics[width=0.48\textwidth]{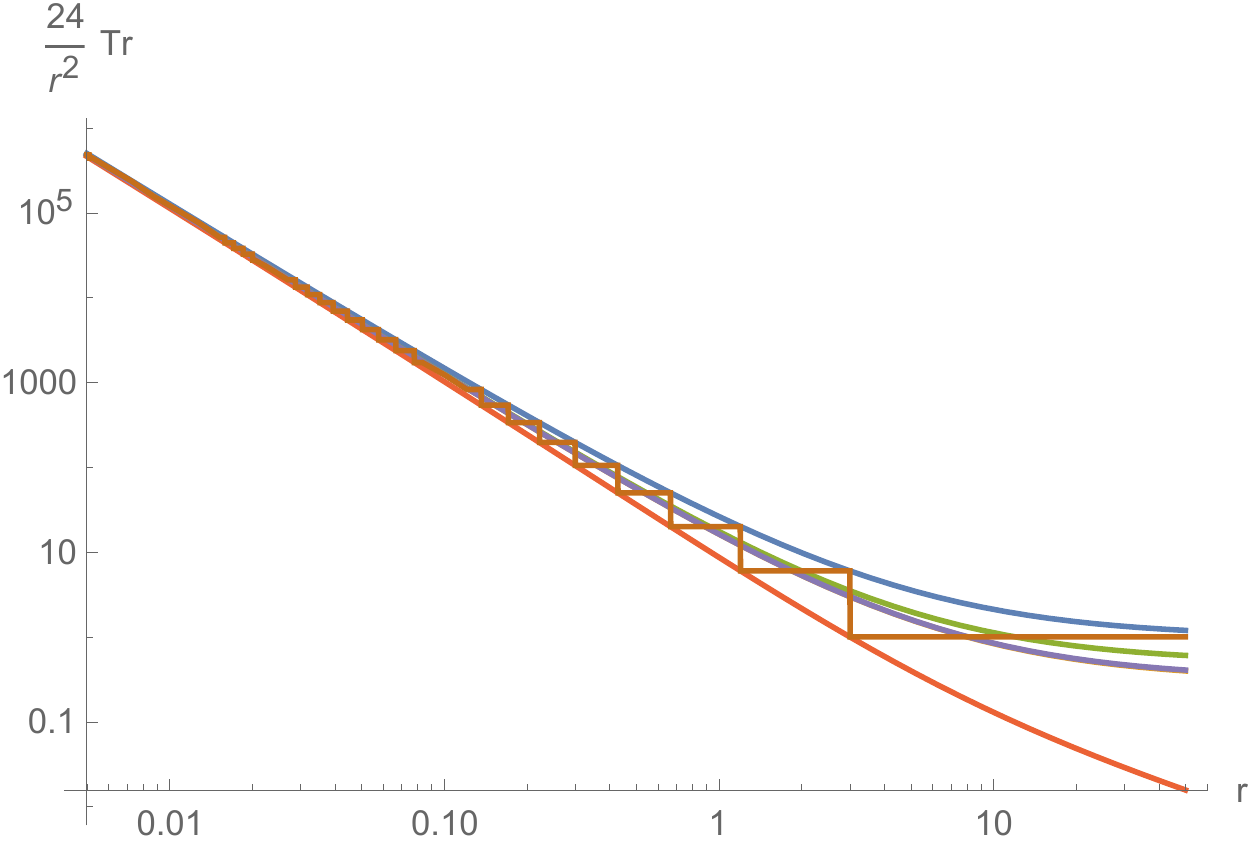}
	\includegraphics[width=0.48\textwidth]{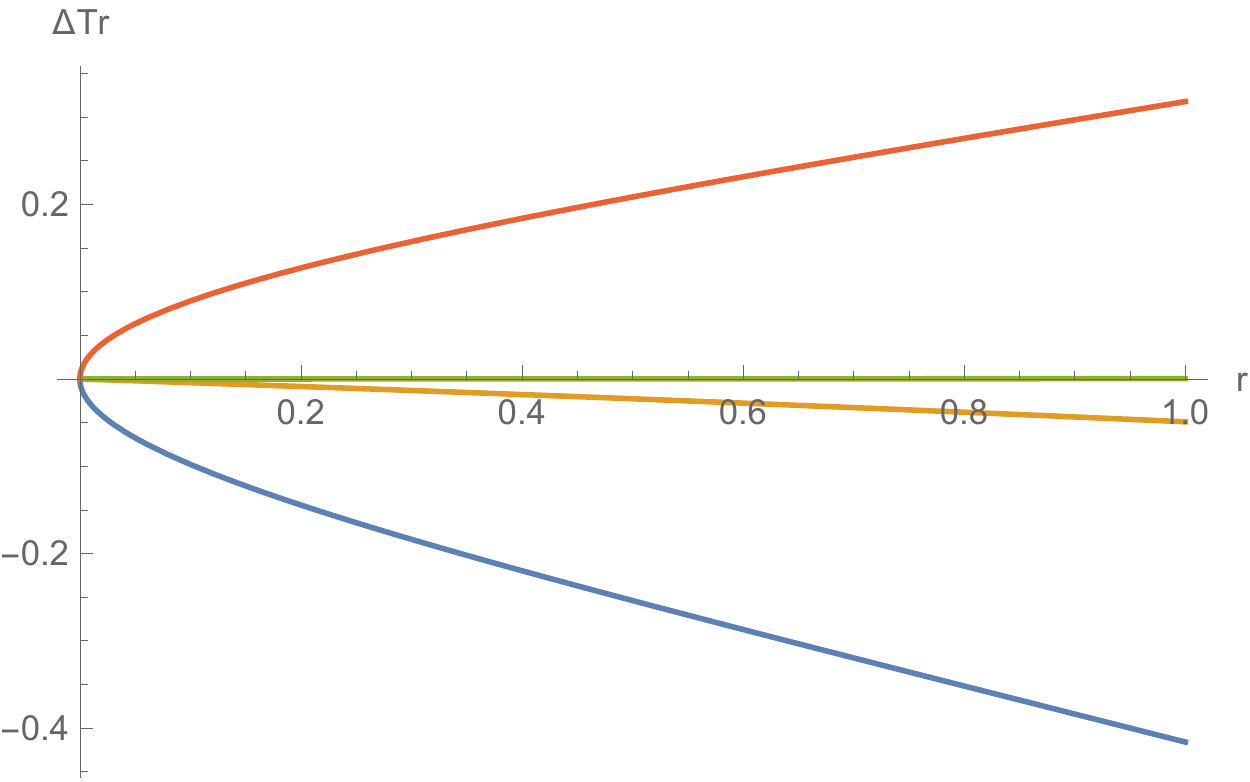}
	\caption[Comparison of smoothing procedures applied to the 
four-dimensional scalar trace.]{Comparison of smoothing procedures applied 
to the four-dimensional scalar trace without endomorphism,
 $\Tr_{(0)}[\theta(k^2-\Delta)]$, as a function of the dimensionless curvature 
$r=\Rb/k^2$. The left panel depicts the staircase behaviour of the sum over 
modes (horizontal lines). The upper (lower) staircase approximations interpolate 
between the upper (lower) points of the discrete result. The early-time expansion 
of the heat kernel, as well as the averaged ($q^{(0)}=0$) and optimised averaged  
($q^{(0)}=-2$) interpolations lie between these extreme curves. 
The right diagram displays the difference $\Delta\Tr$ obtained from 
evaluating the scalar trace with the early-time expansion of the heat kernel (reference) 
and (from bottom to top) the upper staircase, averaged, optimised averaged,  
and the lower staircase interpolation. On the shown interval the relative difference 
between the optimised averaged interpolation and the early-time expansion of the heat 
kernel \eqref{heatbench} is smaller than $6 \times 10^{-4}$. \label{fig.smoothing}}
\end{figure}
with $p^{(s)} = q^{(s)} = 0$ and $r\equiv\Rb/k^2$. The sums then extend up to the integer 
part of these bounds which results in a discontinuous structure in the flow equation. 
This is illustrated in the left panel of figure \ref{fig.smoothing}. Since for any fixed 
dimension $d$ the expressions \eqref{mattersums} and \eqref{gravitonsum} reduce to 
polynomials in $N$, one may substitute the corresponding thresholds \eqref{Ncutoff} and 
treat the resulting expressions as being continuous in the dimensionless curvature $r$. 
For $p^{(s)} = q^{(s)} = 0$ this results in the upper staircase interpolation shown as 
the top curve in the left diagram of  figure \ref{fig.smoothing}. The lower staircase 
curve (connecting the lower points of the discontinuous steps) is obtained from setting 
$p^{(s)} = -1$ and $q^{(s)} = 0$. The  interpolation used in 
\cite{Benedetti:2012dx,Ohta:2015efa,Ohta:2015fcu,Demmel:2015oqa} 
averages the sums \eqref{sumdeg} evaluated at  
$N_{\rm max}^{(s)}$ and $N_{\rm max}^{(s)}-1$ setting $p^{(s)} = q^{(s)} = 0$. 
This procedure removes the non-analytic terms from the sums
once the volume $V_d$ has been factored out. This averaging 
interpolation will be used in the following, see, however, appendix \ref{App.fRA}
for two other interpolation schemes. 
It is convenient to define sums tailored to the 
averaged interpolation, setting
\be
\label{averagedinterpolationsums}
T^{(s)}_d(N) \equiv \tfrac{1}{2} \left( S^{(s)}_d(N) + S^{(s)}_d(N-1) \right) \, , 
\qquad \widetilde{T}_d(N) \equiv \tfrac{1}{2} \left( \widetilde{S}^{(s)}_d(N) + 
\widetilde{S}^{(s)}_d(N-1) \right) \, . 
\ee

At this stage, it is instructive to compare the evaluation of the operator traces in 
terms of spectral sums to the results obtained from the early-time expansion of the heat 
kernel. Applying standard Mellin-transform techniques \cite{Reuter:1996cp,Codello:2008vh} 
one has
\be
\label{trheatkernel}
\Tr_{(s)} \theta(k^2 - \Delta) = \frac{k^d}{(4 \pi)^{d/2}} \, V_d \, f_{(s)}(r;d)
\ee
where
\be
\label{frd}
f_{(s)}(r;d) = \left( \frac{{\rm tr} \, {\bf a}_0^{(s)}}{\Gamma(d/2+1)} +
 \frac{{\rm tr} \, {\bf a}_2^{(s)}}{ \Gamma(d/2)} \, k^{-2} \right) + \mathcal{O}(r^2)
\ee
and the coefficients ${\bf a}_n^{(s)}$ can be found in \cite{Lauscher:2001ya}. 
In particular, for a scalar field in $d=4$ dimensions the early-time expansion 
of \eqref{trheatkernel} gives
\be
\label{heatbench}
f_{(0)}(r;4) = \tfrac{1}{2} + \tfrac{1}{6}r + \tfrac{29}{2160} r^2 \, .  
\ee
Generically, evaluating \eqref{mattersums} for averaging interpolation reproduces 
the leading term in \eqref{frd} while the subleading coefficient multiplying $r$ 
will match for specific values $q^{(s)} \not = 0$, only. This suggests an 
optimised averaged interpolation function where
\be
\label{qopt}
q^{(s)} = -\tfrac{2}{3} (d-1) \, . 
\ee
The results obtained from the various interpolations are then compared in the 
right panel of figure \ref{fig.smoothing}. 
As expected the optimised interpolation leads to an expansion 
which gives the best approximation to the early-time heat kernel.

Based on the mode sums \eqref{mattersums} and \eqref{gravitonsum} together 
with the cutoffs \eqref{Ncutoff} it is rather straightforward to write down 
the explicit form of the traces \eqref{gravitytraces} and \eqref{mattertraces}. 
Introducing the convenient abbreviation
\be
\cV \equiv \frac{d!}{2 \, (4 \pi)^{d/2} \, \Gamma(d/2+1)} 
\left( \frac{r}{d(d-1)} \right)^{d/2}
\ee
and using that the volume of the $d$-sphere may be written as 
$V_d = \tfrac{2}{d!} \Gamma(d/2+1) (4 \pi)^{d/2} a^d$ one sees that 
$V_d \, k^d \, \cV = 1$, which then allows to extract the volume factor from 
the traces rather easily. The matter traces \eqref{mattertraces} then 
evaluate to
\begin{subequations}\label{matterevaluated}
	\begin{align}
	T^{\rm scalar} = & \, V_d \, k^d \,  \cV \, \frac{N_S}{1+ \alpha_S^M r} 
\, T_d^{(0)}(N) \, , \\
	T^{\rm Dirac} = & \, - V_d \, k^d \,  \cV \, \frac{N_D}{1 + 
\left(\alpha^M_D + \tfrac{1}{4}\right) r} \, T_d^{(1/2)}(N) \, , \\
	T^{\rm vector} = & \, V_d \, k^d \,  \cV \, N_V 
\left(\frac{1}{1 + (\alpha^M_{V_1} + \tfrac{1}{d})r} \, 
T^{(1)}_d(N) - \frac{1}{1 + \alpha^M_{V_2}r} \left(T_d^{(0)}(N) - 1 \right) \right) \, . 
	\end{align}
\end{subequations}
Here $N$ represents the cutoff obtained from the corresponding spin representation. 
The last term in $T^{\rm vector}$ originates from removing the lowest scalar 
eigenmode from the trace encoding the contributions of the longitudinal vector 
field. The evaluation of the gravitational traces proceeds along the same lines. 
Denoting derivatives with respect to the RG time $t$ by a dot, one has
\begin{subequations}\label{gravevaluated}
	\begin{align}
	T^{\rm TT} = & \, \tfrac{1}{2} \, \tfrac{V_d \, k^d \, \cV}
{f_k^\prime \left(1 + \left(\alpha^G_T + \tfrac{2}{d(d-1)}\right)r\right)}
	 \left(
	\big( (1+\alpha^G_T r) \dot{f}^\prime_k + 2 f^\prime_k \big) \, T^{(2)}_d(N) 
	- k^{-2} \, \dot{f}_k^\prime \, \widetilde{T}^{(2)}_d(N) 
	\right) 
\, , \\
	T^{\rm ghost} = & \, - \tfrac{V_d \, k^d \, \cV}{1 +
 \left( \alpha_V^G - \tfrac{1}{d} \right) r} \left(T^{(1)}_d(N) - 
\tfrac{1}{2}d(d+1) \right) \, , \\
	T^{\rm sinv} = & \, \frac{1}{2} \, \tfrac{V_d \, k^d \, \cV}
{f_k^{\prime\prime} \left(1 + \big(\alpha^G_S - \tfrac{1}{d-1}\big)r \right) 
+ \tfrac{d-2}{2(d-1) k^2} \, f_k^\prime}
\left(
\big( (1+\alpha^G_S r) \dot{f}^{\prime\prime}_k + 2 f^{\prime\prime}_k \big) \, 
T^{(0)}_d(N) 
- k^{-2} \, \dot{f}_k^{\prime\prime} \, \widetilde{T}^{(0)}_d(N) 
\right) \, . 	
	\end{align}
\end{subequations}
Here $T^{\rm sinv}$ contains the contribution from \emph{all scalar modes}. The two 
lowest eigenmodes are removed by adding
\be
\begin{split}
\Delta 	T^{\rm sinv} = & \,- \frac{1}{2} \, \tfrac{V_d \, k^d \, \cV}{f_k^{\prime\prime} 
\left(1 + \big(\alpha^G_S - \tfrac{1}{d-1}\big)r \right) + \tfrac{d-2}{2(d-1) k^2} \, 
f_k^\prime}
\left(
\big( (1+\alpha^G_S r) \dot{f}^{\prime\prime}_k + 2 f^{\prime\prime}_k \big) \, (d+2)
- \tfrac{d+1}{d-1} \, r \, \dot{f}_k^{\prime\prime} 
\right) \,
\end{split}
\ee
to the flow equation. Based on the explicit results for the traces, the flow 
equation for $f_k(R)$ can be written as
\be
\label{pdfgeneral}
\begin{split}
V_d \, \dot{f}_k = T^{\rm TT} + T^{\rm ghost} + T^{\rm sinv} + \Delta 	T^{\rm sinv} + T^{\rm scalar} + T^{\rm Dirac} + T^{\rm vector} \, . 
\end{split}
\ee
Note that this result is valid for general dimension $d$, retains the dependence on all 
endomorphism parameters and can easily be adapted to any interpolation scheme by 
specifying the corresponding expressions for $N$ according to \eqref{Ncutoff}. The 
partial differential equation \eqref{pdfgeneral} constitutes the main result of this 
section. It generalises the construction \cite{Ohta:2015efa,Ohta:2015fcu} to general 
dimension $d$ and the presence of minimally coupled matter fields.

In order to facilitate the further analysis, the explicit form of 
\eqref{pdfgeneral} in $d=4$ and the averaged interpolation used in 
\cite{Ohta:2015efa,Ohta:2015fcu} will be given. The result is conveniently written 
in terms of the dimensionless quantities
\be
r = \Rb k^{-2} \, , \qquad \varphi_k(r) = k^{-d} f_k(\Rb) \, .
\ee
Following the structure \eqref{pdfgeneral} one has
\be
\label{pdf4d}
\begin{split}
& \, \dot{\varphi} + 4 \varphi - 2 r \varphi^\prime = \cT^{\rm TT} + \cT^{\rm ghost} 
+ \cT^{\rm sinv} + \cT^{\rm scalar} + \cT^{\rm Dirac} + \cT^{\rm vector} \, . 
\end{split}
\ee
Here the $\cT$ constitute the dimensionless counterparts of the traces $T$ 
divided by the factor $V_d k^d$. They are given explicitly with all endomorphism 
parameters kept in \eqref{gravflow} and \eqref{matterflow} in appendix~\ref{App.fRA}.
For type I regulators, {\it i.e.}, all $\alpha=0$ they read:
\begin{subequations}\label{gravflowTypeI}
	\begin{align}\label{gravTTtypeI}
	\cT^{\rm TT} = & \, \frac 1 {(4\pi)^2} 
	\frac{5}{2} \, \frac{1}{1 +\tfrac{1}{6} r} \left(1 - \frac{1}{6} r\right)
\left(1  - \frac{1}{12} r\right) \\ \nonumber & \; \;
		+ \frac 1 {(4\pi)^2} \frac{5}{12} \, 
\frac{\dot{\varphi}^\prime + 2 \varphi^\prime - 2 r \varphi^{\prime\prime}}{\varphi^\prime} 
\left(1- \frac{2}{3} r\right)\left(1  - \frac{1}{6} r\right) \, , \\
 \label{gravsinvI}
	\cT^{\rm sinv} = & \, 
	\frac 1 {(4\pi)^2}  \frac{1}{2 } 
	\frac{ \varphi^{\prime\prime}}{\left(1 - \tfrac{1}{3}r\right) \varphi^{\prime\prime} 
+ \tfrac{1}{3} \varphi^\prime} 
	\left(1 - \frac{1}{2} r\right)
	\left(1 + \frac{11}{12} r\right)
	 \\ \nonumber & \; \;
	 +  \frac 1 {(4\pi)^2} \frac{1}{12} 
	 \frac{\dot{\varphi}^{\prime\prime} - 2 r \varphi^{\prime\prime\prime}}
{\left(1 - \tfrac{1}{3} r\right) \varphi^{\prime\prime} + \tfrac{1}{3} \varphi^\prime} 
	 \left(1 + \frac{3}{2} r\right)
	 \left(1 - \frac{1}{3} r\right)
	 \left(1 - \frac{5}{6} r\right) \, , 
	 \\
	\cT^{\rm ghost} = & \, - \frac 1 {(4\pi)^2} \frac{1}{48} \, 
\frac{1}{1 - \tfrac{1}{4} r} \, \left( 72 + 18 r - 19 r^2  \right) \, , 
	\end{align}
\end{subequations}
together with the matter results
\begin{subequations}\label{matterflowI}
	\begin{align}
\cT^{\rm scalar} = & \, \frac 1 { (4\pi)^2} \frac {N_S} 2 \, 
  \left( 1 +  \tfrac{1}{4} r \right) \left(1 +  \tfrac{1}{6} r\right) \, , \\ 
\label{TdiracI}
\cT^{\rm Dirac} = & \, - \frac{1}{(4\pi)^2} \, 2N_D \, \left(1 + \tfrac{1}{6} r\right) \, , \\
\cT^{\rm vector} = & \, \frac{1}{(4\pi)^2} \frac{N_V}{2 } \, 
\bigg( \frac{3}{1 + \tfrac{1}{4} r} 
\left(1 + \tfrac{1}{6}  r \right) 
\left(1 + \tfrac{1}{12} r\right) 
-  \left( 1 + \tfrac{1}{2} r \right) \left(1  - \tfrac{1}{12} r\right) 
\bigg) \, ,  
	\end{align}
\end{subequations}
which provides a good impression how these quantities look for some general values 
of the endomorphism parameters, see \eqref{gravflow} and  \eqref{matterflow}
in appendix \ref{App.fRA} .
Here primes and dots denote derivatives with respect to $r$ and $t$, respectively, 
and all arguments and subscripts have been suppressed in order to aid the readability of 
the expressions. The result \eqref{gravflow} agrees with the beta functions reported in 
\cite{Ohta:2015fcu} and \eqref{matterflow} constitutes its natural extension to minimally 
coupled matter fields. With the result \eqref{pdf4d} at our disposal, one now has all the 
prerequisites to study the FP structure of gravity-matter systems at the level 
of $f(R)$-gravity. 
 
Let us highlight the main properties of \eqref{pdfgeneral}. Inspecting the gravitational 
sector \eqref{gravflowTypeI}, resp., \eqref{gravflow}, one finds that the function $\varphi_k$ enters the traces in form 
of its first, second, and third derivative with respect to $r$. As a consequence, a 
constant term in $\varphi_k$ does not appear on the right hand side of the flow equation. 
This implies in particular that the propagators of the fluctuation fields do not contain 
contributions from a cosmological constant. This particular feature is owed to the 
interplay of the exponential split (removing the contribution of the cosmological 
constant  from the propagator of the transverse-traceless fluctuations) and the physical 
gauge $\beta \rightarrow -\infty$ (removing the $hh$-term from the gravitational sector 
\eqref{HessianGrav}).

For the specific regulator \eqref{LitimReg}, the evaluation of the spectral sums $S$ 
results in polynomials that are at most quadratic in $r$ while the sums within 
$\widetilde{S}$ terminate at order $r^3$. This feature has already been observed in 
\cite{Codello:2007bd,Machado:2007ea} where it was found that evaluating the flow equation 
of $f(R)$-gravity for a Litim-type regulator required the knowledge of a finite number 
of heat-kernel coefficients only. In this sense, it is expected that the Litim regulator 
leads to similar features when evaluating the operator traces as spectral sums. 

An interesting feature of the averaged interpolation is that the contribution of the 
Dirac fermions is given by \emph{a polynomial of first order} in the dimensionless 
curvature $r$. This particular property can be traced to a highly non-trivial 
cancellation between the propagator and the factors of the spectral sum $T^{(1/2)}_d$. As 
a consequence, the Dirac fields will not contribute to the flow equation at order $r^2$ 
and higher. This particular feature is specific to the averaged interpolation and absent 
in other interpolation schemes (cf.\  \eqref{Tdirac1} and \eqref{Tdirac2} in appendix 
\ref{App.fRA}). Owed to the investigation of the FP properties in terms of the 
matter deformation parameters introduced in \eqref{dgdl} and \eqref{dbetadef} this 
feature will not be essential when studying non-trivial RG fixed 
points in the sequel.

Setting the derivatives with respect to the RG time to zero, \eqref{pdf4d} reduces to a 
third order differential equation for $\varphi_*(r)$. The order of the equation is 
determined by the scalar contribution arising in the gravitational sector. Casting the 
resulting expression into normal form by solving for $\varphi^{\prime\prime\prime}$ one 
finds that the equation possesses four fixed singularities situated at
\be
\label{fixedsing}
r^{\rm sing}_1 = - \frac{1}{\alpha^G_S + \tfrac{3}{2}} \, , \quad
r^{\rm sing}_2 = 0 \, , \quad 
r^{\rm sing}_3 = - \frac{1}{\alpha^G_S -\tfrac{5}{6}} \, , \quad  
r^{\rm sing}_4 = - \frac{1}{\alpha^G_S - \tfrac{1}{3}}  \, \, . 
\ee
Solutions obtained from solving the differential equation (in normal form) numerically 
are typically well-defined on the intervals bounded by these singular loci only. 
Extending a solution across a singularity puts non-trivial conditions on the initial 
conditions of the FP equation. Based on the singularity counting argument 
\cite{Dietz:2012ic}, stating that each first order pole on the interval 
$r \in [0,\infty[$ fixes one free parameter, it is then expected that \eqref{pdf4d} 
admits a discrete set of global fixed functionals. 

\section{Fixed Point Structure of $f(R)$-Gravity Matter 
Systems in the Polynomial Approximation}
\label{sec:fixedpoints}

In this section the FP structure of \eqref{pdf4d} arising within 
polynomial approximations of the function $\varphi_k(r)$ will be discussed. 
The general framework is 
introduced in section \ref{sect.fR41} while the FP structure arising at the level 
of the EH truncation and polynomial approximations up to order $N=14$ are 
investigated in sections \ref{sect.fR42} and \ref{sect.fR43}, respectively. The main focus 
is on matter sectors containing the field content of the SM of particle 
physics and its most commonly studied phenomenologically motivated extensions (cf.\ 
tables \ref{Tab.3} and \ref{Tab.mainresults}).

The key ingredient in realising the asymptotic safety mechanism is a NGFP of the 
theories' RG flow. At the level of the partial differential equation 
\eqref{pdf4d}, such FPs correspond to global, isolated, and $k$ stationary 
solutions $\varphi_*(r)$. The existence and properties of such fixed functionals will be 
considered in section \ref{GlobalFF}. 

\subsection{Polynomial $f(R)$-Truncation: General Framework}
\label{sect.fR41}

First, we follow a different strategy and perform an expansion of 
$\varphi_k(r)$ in powers of $r$, terminating the series at a finite order $r^N$: 
\be
\label{polyexpansion}
\varphi_k(r) = \frac{1}{(4\pi)^2} \sum_{n=0}^N \, g_{n}(k) \, r^n \, , 
\qquad \dot{\varphi}_k(r) = \frac{1}{(4\pi)^2} \sum_{n=0}^N \, \beta_{g_n} \, r^n \, . 
\ee
By construction, the $k$-dependent dimensionless couplings $g_n(k)$ satisfy
$\partial_t g_n \equiv \beta_{g_n}$, $n=0, \ldots, N$.
The explicit expressions for the $\beta_{g_n}$ as a function of the couplings are 
obtained as follows. First, the ansatz \eqref{polyexpansion} is substituted into  
\eqref{pdf4d} which is subsequently expanded in powers of the dimensionless curvature $r$ 
up to order $r^N$. Equating the coefficients of the terms proportional to $r^n$, 
$n=0, \ldots N$, results in $N+1$ equations depending on $\beta_{g_n}$ and $g_n$, 
$n=0, \ldots N$. Solving this system of algebraic equations for $\beta_{g_n}$ determines 
the beta functions as a function of the couplings $g_n$. Since the resulting algebra is 
straightforward but quickly turns lengthy, these manipulations are conveniently done by a 
computer algebra program.

As explained in section \ref{AsympSafe},
the most important property of the beta functions $\beta_{g_n}(g_0, \ldots, g_N)$ are 
their FPs $g^* = \{g_0^*, \ldots, g_N^*\}$ where, by definition, the beta functions vanish.
In addition,  the stability matrix ${\bf B}$  \eqref{Bmat}
 governs the linearised RG flow in the vicinity of the FP. As the 
stability coefficients $\theta_n$ are minus the eigenvalues of ${\bf B}$, eigendirections 
with ${\rm Re}(\theta_n) > 0$ (${\rm Re}(\theta_n) <0$) attract (repel) the flow as 
$k \rightarrow \infty$. Thus stability coefficients with positive real part are linked to 
``relevant directions'' associated with free parameters which have to be determined 
experimentally. Ideally, FPs underlying an asymptotic safety construction should 
come with a low number of free parameters. This implies in particular that the number of 
relevant directions should saturate when the order of the polynomials appearing in 
\eqref{polyexpansion} exceeds a certain threshold in $N$. For pure gravity, this test has 
been implemented in the seminal works \cite{Codello:2007bd,Machado:2007ea} and later on 
extended in \cite{Codello:2008vh,Falls:2013bv,Falls:2014tra,Falls:2017lst}. A systematic 
investigation for gravity-matter systems is still missing though. In the remainder of 
this section, a two-fold search strategy is followed. In section \ref{sect.fR42}, 
first,  matter sectors which give rise to a suitable NGFP at the level of the 
EH action are identified. Based on these initial seeds the stability of these NGFPs under 
the addition of higher-order scalar curvature terms for phenomenologically interesting 
gravity-matter systems is investigated in section \ref{sect.fR43}.
 
The fact that the right hand side of \eqref{pdf4d} is independent of $g_0$,
leads to the 
peculiar feature that $\partial_\lambda \beta_\lambda|_{g = g_*} = -4$ while  
$\partial_\lambda \beta_{g_n} = 0$. This structure ensures that the stability matrix 
always gives rise to a stability coefficient $\theta_0 = 4$, independent of the order 
$N$ of the polynomial expansion.  
 
\subsection{Fixed Point Structure in the Einstein-Hilbert Truncation}
\label{sect.fR42}

First, the FP structure entailed by \eqref{pdf4d} at the level of 
the EH truncation will be investigated. 
In this case the function $\varphi_k(r)$ is approximated 
by a polynomial of order one in $r$,
\be
\label{EHansatz}
\varphi_k(r) = \frac{1}{16 \pi \, g_k} \, \left( 2 \lambda_k - r \right) \, . 
\ee 
The scale-dependent dimensionless cosmological constant $\lambda_k$ and Newton's constant 
$g_k$ are related to their dimensionful counterparts $\Lambda_k$ and $G_k$ by 
$\Lambda_k = \lambda_k \, k^2$ and $G_k = g_k \, k^{-2}$. The beta functions controlling 
the scale-dependence of $g_k$ and $\lambda_k$ in the presence of an arbitrary number of 
minimally coupled matter fields are readily obtained from substituting the ansatz 
\eqref{EHansatz} into the partial differential equation \eqref{pdf4d} and projecting the 
result onto the terms independent of and linear in $r$, respectively. The resulting 
equations take the form
\be
\partial_t \lambda_k = \beta_\lambda(g_k,\lambda_k) \, , \qquad 
\partial_t g_k = \beta_g(g_k,\lambda_k) \, , 
\ee
where
\be
\label{betaEH}
	\beta_\lambda = - \left(2 - \eta_N \right) \lambda +
 \frac{g}{24 \pi} \left( 12 - 5 \eta_N + 6 d_\lambda \right) \, , \qquad
 \beta_g = \left(2+\eta_N\right)g \, . 	
\ee
The anomalous dimension of Newton's constant, $\eta_N \equiv G_k^{-1} \partial_t G_k$ 
can be cast into the standard form \cite{Reuter:1996cp},
\be
\eta_N = \frac{g \, B_1}{1 - g \, B_2} \, ,
\ee
where $B_1$ and $B_2$ are $\lambda$-independent coefficients depending on the choice 
of coarse-graining operator,
\begin{subequations}
	\begin{align}
	\mbox{type I:} \qquad  & B_1 = - \frac{1}{24 \pi} \left(43 - 4 d_g \right) \, ,  
\qquad  B_2 = \frac{25}{72 \pi} \, ,\\
	\mbox{type II:} \qquad & B_1 = - \frac{1}{24 \pi} \left(62 - 4 d_g \right) \, ,  
\qquad  B_2 = \frac{35}{72 \pi} \, , 
	\end{align}
\end{subequations}
and the parameters $d_g$ and $d_\lambda$ summarise the matter content of the model
\be
\label{dgdl}
d_\lambda =  N_S + 2 N_V -4 N_D \, , \qquad
\begin{array}{ll}
	\mbox{type I:} \; & d_g = \tfrac{5}{4} N_S - \tfrac{5}{4} N_V - 2 N_D \\[1.1ex]
	\mbox{type II:} \; & d_g = \tfrac{5}{4} N_S - \tfrac{7}{2} N_V + N_D
\end{array} \, . 
\ee

At this stage the following remarks are in order. The expression for $d_\lambda$ is 
independent of the choice of coarse-graining operator and agrees with the heat-kernel 
based computations \cite{Codello:2008vh}. Essentially, $d_\lambda$ entails that each 
bosonic degree of freedom contributes to the running of the cosmological constant with a 
weight $g/(4\pi)$  while each fermionic degree of freedom contributes with the same 
factor but opposite sign. The results for $d_g$ differ from the ones based on the 
early-time expansion of the heat-kernel \cite{Codello:2008vh} where 
$d_g^{\rm type \, I} = N_S - N_V - N_D$ and $d_g^{\rm type \, II} = N_S - 4 N_V + 2 N_D$. 
This feature just reflects the fact that the evaluation of the spectral sums based on the 
averaged staircase agrees with the early-time expansion of the heat-kernel at leading 
order only. One observes, however, that for both choices of coarse-graining operator all 
fields contribute with their characteristic signature, so that the resulting picture is qualitatively similar.

As a second remarkable feature, the beta functions \eqref{betaEH} do not contain 
denominators of the form $(1-c \lambda)^n$ ($c > 0$) which typically lead to a 
termination of the flow at a finite value $\lambda = 1/c$ \cite{Reuter:2001ag}. As a 
consequence the flow is well-defined for any value of $\lambda$ and gives rise to a 
globally well-defined flow diagram \cite{Gies:2015tca}. Moreover, the mechanism for 
gravitational catalysis \cite{Wetterich:2017ixo} is not realised in the present framework.

Owed to their simple algebraic structure, the FPs of the beta functions 
\eqref{betaEH} can be found analytically. They possess a GFP 
located at $(g^*,\lambda^*) = (0,0)$ whose stability coefficients are given by the 
canonical mass dimension of the dimensionful Newton's constant and cosmological constant. 
In addition the system exhibits \emph{a single NGFP for any given matter sector}. For a 
coarse-graining operator of type I (vanishing endomorphisms) this FP is situated 
at
\be
\label{EH:NGFP}
g^* = \frac{144 \pi}{179 - 12 \, d_g} \, , \qquad 
\lambda^* = \frac{33+9 d_\lambda}{179-12 \, d_g} \, ,
\ee
while its stability coefficients obtained from \eqref{Bmat} are
\be
\label{EH:theta}
\theta_0 = 4 \, , \qquad \theta_1 = \frac{358 - 24 \, d_g}{3 \, (43-4 \, d_g)} \, . 
\ee
The corresponding expressions for the type II coarse-graining operator are 
obtained along the same lines and have a similar structure. 

\begin{figure}[t!]
	\includegraphics[width=0.48\textwidth]{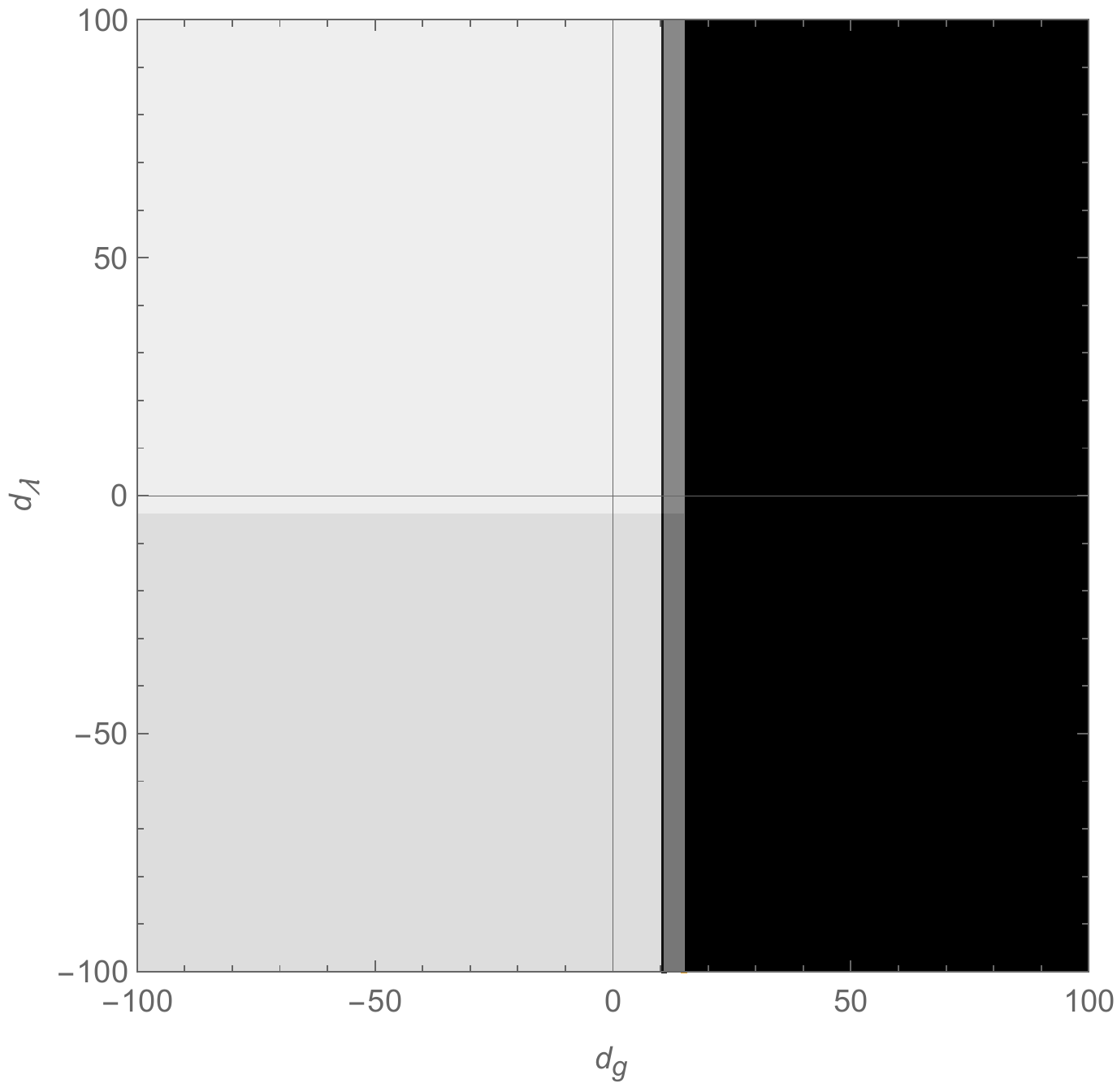}
	\includegraphics[width=0.48\textwidth]{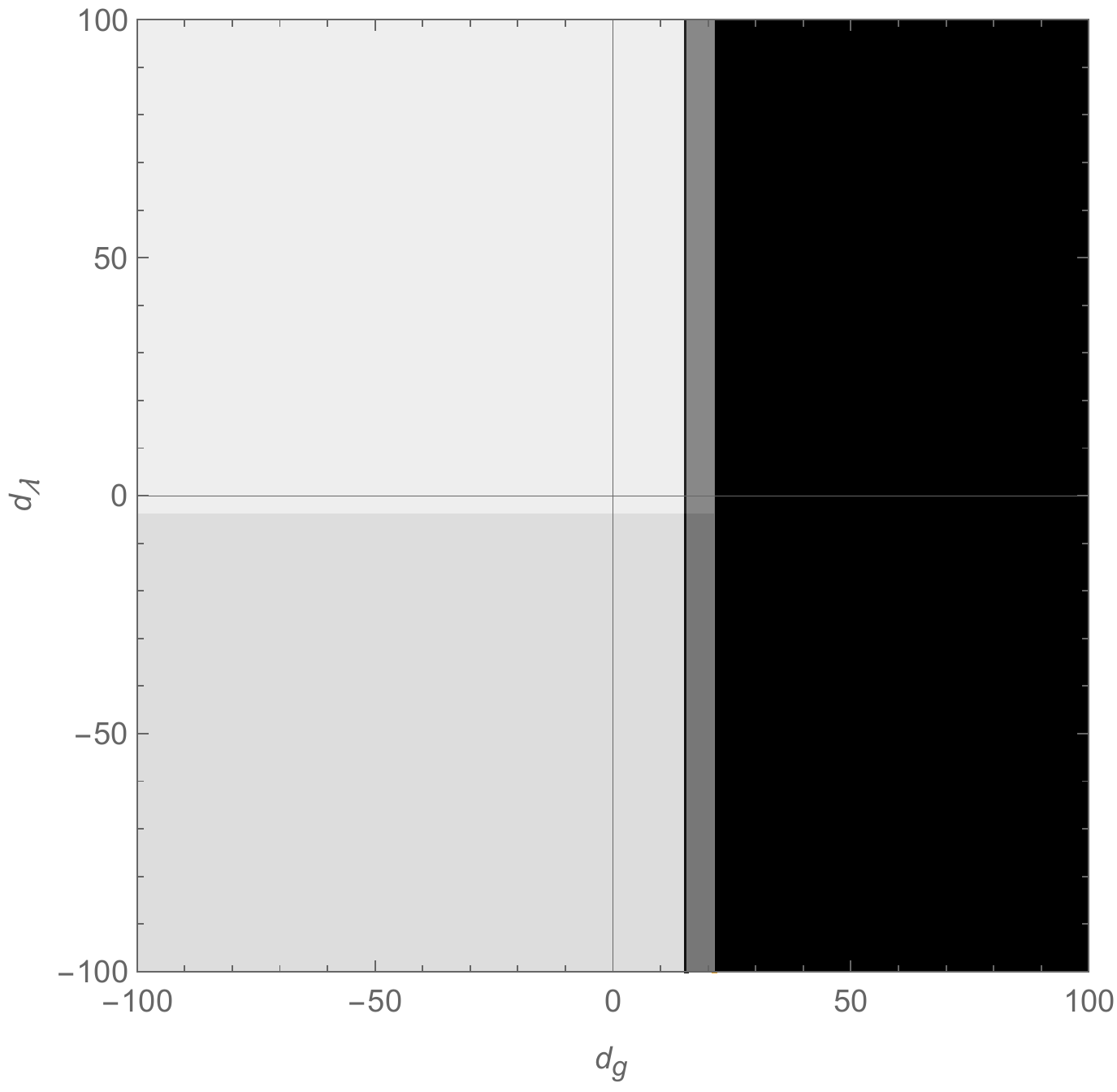}
	\caption[Illustration of the FP structure at the level of the 
EH truncation for a coarse-graining operator of type I and type II 
respectively.]
{Illustration of the FP structure arising from the system \eqref{pdf4d} at the 
level of the EH truncation for a coarse-graining operator of type I (left) 
and type II (right) respectively. The matter content of the model is encoded in the 
parameters $d_g, d_\lambda$ defined in  \eqref{dgdl}. The black region does not support a 
NGFP with $g^* > 0$. In the dark gray region the NGFP is a saddle point with
 $\theta_1 < 0$. The gray and light gray regions support an UV attractive NGFP with 
$g^* \lambda^* < 0$ and $g^* \lambda_* > 0$, respectively. Notably, the qualitative 
FP structure, classified in terms of $d_g, d_\lambda$ is independent of the 
choice of coarse-graining operator. \label{fig.3}}
\end{figure}

The properties of the NGFPs \eqref{EH:NGFP} as a function of the matter content are 
illustrated in figure \ref{fig.3}.
Quite remarkably, the FP structure resulting from the type I and type II coarse 
graining operator is \emph{qualitatively identical} provided that the matter content of 
the model is encoded in the deformation parameters \eqref{dgdl}.
It is determined by three separation lines, $L_1, L_2, L_3$ situated at
\be
	\begin{tabular}{llll}	\mbox{type I:}  & \qquad $L_1: \, d_g = \tfrac{179}{12}$\, ,  & \qquad $L_2: \, d_g = \tfrac{43}{4}$\, ,  & \qquad $L_3: \, d_\lambda = - \tfrac{11}{3}$ \, ,  \\
	\mbox{type II:} &  \qquad $L_1: \, d_g = \tfrac{64}{3}$\, ,  & \qquad $L_2: \, d_g = \tfrac{31}{2}$\, , &  \qquad $L_3: \, d_\lambda = - \tfrac{11}{3}$ \, . 
	\end{tabular}
\ee
For matter sectors located to the left (right) of $L_1$ the NGFP is situated at 
$g^* > 0$ ($g^* < 0$), respectively. If $g^* < 0$, the corresponding FP is 
disconnected from the physically viable low-energy regime and may therefore not be 
suitable for controlling the high-energy behaviour of physically interesting theories. 
Thus this case will be discarded from the further analysis. Matter sectors sitting in the 
region bounded by the lines $L_1$ to the right and $L_2$ to the left support a saddle 
point where $\theta_1 < 0$ while for matter systems to the left of $L_2$ the NGFP is UV-
attractive in both $g$ and $\lambda$. The horizontal line $L_3$ separates the regions 
where the NGFPs come with $\lambda_* g_* < 0$ (lower-left region) and $\lambda_* g_* > 0$ 
(upper-left region), respectively.

Figure \ref{fig.3} makes it also apparent that the systems where scalar matter is coupled 
to gravity possesses an upper bound on the number of scalar fields ($N_S^{\rm max} = 14$ 
for type I and $N_S^{\rm max} = 21$ for type II). If $N_S$ exceeds these bounds the NGFP 
is located in the region $g^* < 0$. While the FP is still present, it is no 
longer suitable for realising a phenomenologically interesting gravity-matter system.

\begin{table}[t!]
\scalebox{0.785}{
        \renewcommand{\arraystretch}{1.4}
                \begin{tabular}{p{3.94cm}||c|c|c||c|c|c|c||c|c|c|c}
                model & \multicolumn{3}{c||}{matter content} & \multicolumn{4}{c||}{type I coarse-graining} & \multicolumn{4}{c}{type II coarse-graining} \\
                        & \, $N_S$ \, & \, $N_D$ \, & \, $N_V$ \, & \, $d_g$ \, & \, $d_\lambda$ \, & \, $g_*\lambda_*$ \, &  \, $\theta_1$ \, & \, $d_g$ \, & \, $d_\lambda$ \, & \, $g_*\lambda_*$  \, &  \, $\theta_1$ \, \\  \hline \hline
                        pure gravity & 0 & 0 & 0 &  0 & 0 & $0.47$ & $2.78$ & $0$ & $0$ & $0.23$ & $2.75$ \\ \hline
                        SM & 4 & ${45}/{2}$ & 12 & $-55$ & $-\,62$ & $-0.34$ & $2.13$ & $-{29}/{2}$ & $-62$ & $-1.28$ & $2.39$  \\ \hline
                        SM, DM & 5 & ${45}/{2}$ & 12 &  $-\,{215}/{4}$ & $-61$ & $-0.34$ & $2.13$ & $-{53}/{4}$ & $-61$ & $-1.36$ & $2.41$ \\ \hline
                        SM, $3\,\nu$ & 4 & 24 & 12 & $-58$ & $-68$ & $-0.34$ & $2.12$ & $-13$ & $-68$ & $-1.54$ & $2.41$ \\ \hline
                        SM, $3\,\nu$, DM, axion & 6 & 24 & 12 & $-{111}/{2}$ & $-66$ & $-0.36$ & $2.13$ & $- {21}/{2}$ & $-66$ & $-1.74$ & $2.45$ \\ \hline
                        MSSM & 49 & ${61}/{2}$ & 12 & $-{59}/{4}$ & $-49$ & $-1.45$ & $2.33$ & ${199}/{4}$ & $-49$ & $-$ & $-$ \\ \hline
                        {SU(5) GUT} & {124} & {24} & {24} &  $77$ & $76$ & $-$ & $-$ & $95$ & $76$ & $-$ & $-$ \\ \hline
                        {SO(10) GUT} & {97} & {24} & {45} & $17$ & $91$ & $-$ & $-$ & $-
{49}/{4}$ & $91$ & $2.37$ & $2.42$ \\ \hline \hline
                \end{tabular}
}
        \caption{\label{Tab.3} FP structure arising from the field content of 
commonly studied matter models. The SM and its extensions by a small 
number of additional matter fields support NGFPs with very similar properties.}
\end{table}

Details for the NGFPs
found for distinguished gravity-matter systems are summarised in table \ref{Tab.3}.
The list covers the cases of pure gravity, gravity coupled to the field content of the 
SM of particle physics, and phenomenologically motivated matter sectors 
arising in frequently studied candidates for BSM physics. The 
latter supplement the field content of the SM by additional scalar fields 
(dark matter (DM) or axion candidates), right-handed neutrinos, supersymmetric partners of the SM 
fields leading to the minimally supersymmetric standard model 
(MSSM)\footnote{Following \cite{Dona:2013qba}, we consider the MSSM 
where the Higgs sector is standard-model like. We also verified that extending the 
Higgs sector to an $SU(2)$ doublet (corresponding to $N_S = 53$ and $N_D = \tfrac{65}{2}$) does not change the qualitative picture.}, or fields required 
in the realisation of GUTs based on the gauge groups SU($5$) or 
SO($10$).  By substituting the matter field content listed in the second to fourth column 
of table \ref{Tab.3} into the maps \eqref{dgdl} and checking the resulting coordinates in 
figure \ref{fig.3} readily shows that many of these models give rise to a NGFP which is UV 
attractive for both Newton's constant and the cosmological constant 
(i.e., $\theta_1 > 0$). The exceptions are the GUT-type models (type I coarse-graining 
operator) and the MSSM and SU($5$) GUT (type II coarse-graining operator) which lead to 
NGFPs with $g^* < 0$ and thus fail the test of asymptotic safety at the level of the 
EH truncation. Table \ref{Tab.3} provides the starting point for 
investigating which of the gravity-matter FPs are stable if 
higher-order scalar curvature terms are included in the ansatz for $\varphi_k(r)$. 

\subsection{Gravity-Matter Fixed Points in the Presence of an $R^2$-Term}
\label{sect.fR44}

Owed to the special property that the beta functions for the dimensionless couplings 
$g_n$, $n \ge 1$ are independent of $g_0$, the polynomial expansion 
\eqref{polyexpansion} to order $N=2$ also gives rise to a two-dimensional subsystem of 
beta-functions which closes on its own. One may then study the FP structure for 
$g_1$ and $g_2$ arising from 
\be
\label{betar2}
\beta_{g_1}(g_1,g_2)|_{g = g^*} = 0 \, , \qquad \beta_{g_2}(g_1,g_2)|_{g = g^*} = 0 \, , 
\ee
for arbitrary matter sectors. Once a FP $(g_1^*,g_2^*)$ is obtained its 
coordinates may be substituted into the beta function $\beta_{g_0}(g_0,g_1,g_2)$. Solving 
$\beta_{g_0}(g_0^*,g_1^*,g_2^*) = 0$ for $g_0^*$ then determines the value of $g_0$ 
uniquely.

The triangular shape of the stability matrix furthermore guarantees that the stability 
coefficients $\theta_1$, $\theta_2$ obtained from the $g_1$-$g_2$ subsystem carry over to 
the full system. As a result the stability coefficients from the $N=2$ expansion are 
$\theta_0 = 4, \theta_1, \theta_2$ where the latter depend on the specific matter content 
and choice of coarse-graining operator. 

Following the strategy of the last subsection, the matter contribution to the 
beta functions \eqref{betar2} is encoded 
by $d_g$, introduced in \eqref{dgdl}, supplemented by
\be
\label{dbetadef}
d_\beta \equiv N_S + 2 N_V \, . 
\ee
Note that this parameter is actually independent of the choice of coarse-graining 
operator. Moreover, it is independent of the number of Dirac fields which is owed to the 
cancellation between numerator and denominator observed in  \eqref{Tdirac}. Since all 
matter fields contribute to $d_\beta$ with a positive sign all matter models are located 
in the upper half-plane $d_\beta \ge 0$ with $d_\beta = 0$ realised by pure gravity and 
gravity coupled to an arbitrary number of Dirac fields. The map 
$(N_S,N_V,N_D) \mapsto (d_\lambda, d_g, d_\beta)$ is actually bijective such that any 
particular matter sector is uniquely characterised by either its field content or its 
coordinates $(d_\lambda, d_g, d_\beta)$. 

\begin{figure}[t!]
\centering
\includegraphics[width=0.48\textwidth]{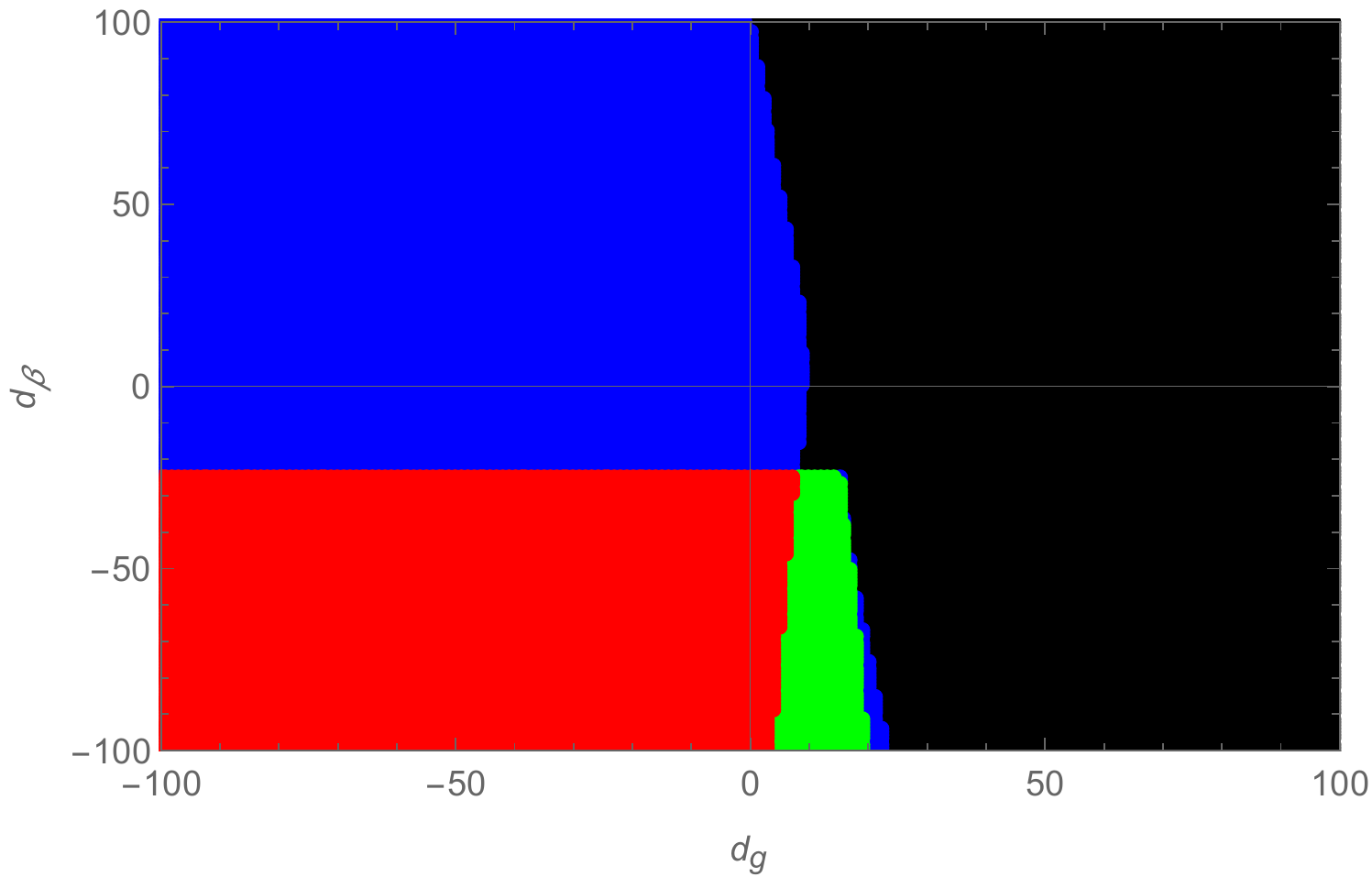} \,
\includegraphics[width=0.48\textwidth]{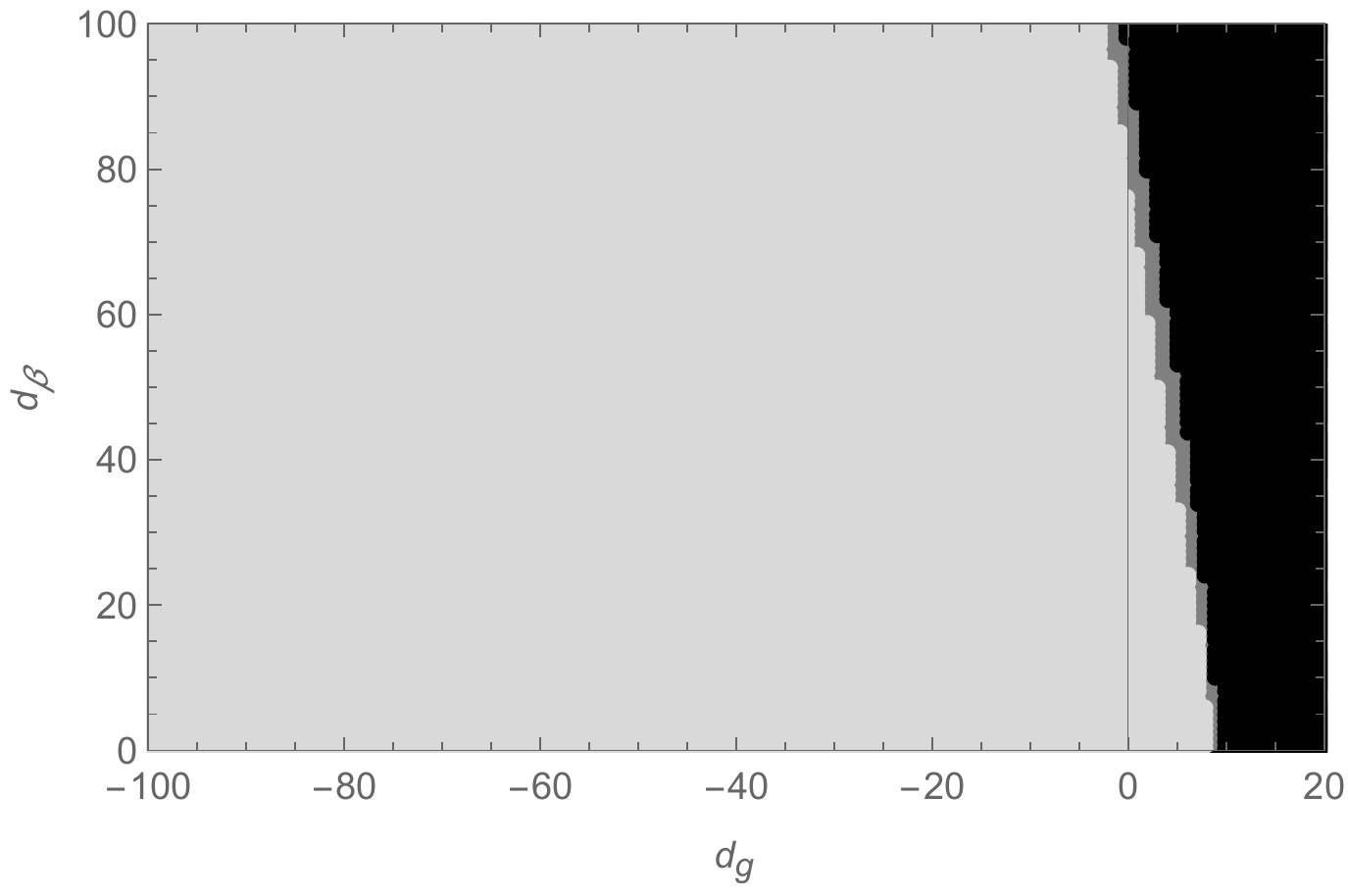} \\
\includegraphics[width=0.48\textwidth]{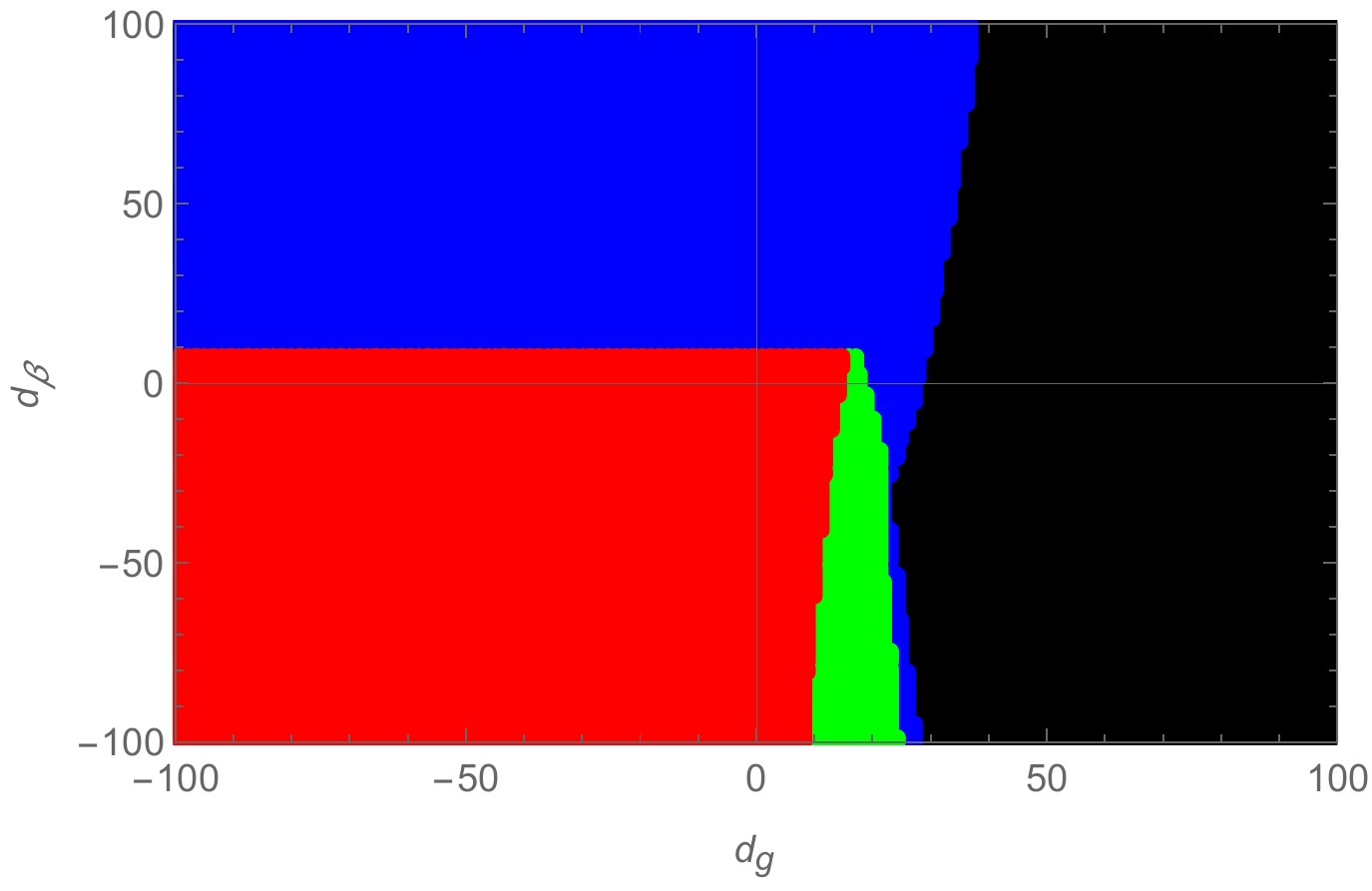} \,
\includegraphics[width=0.48\textwidth]{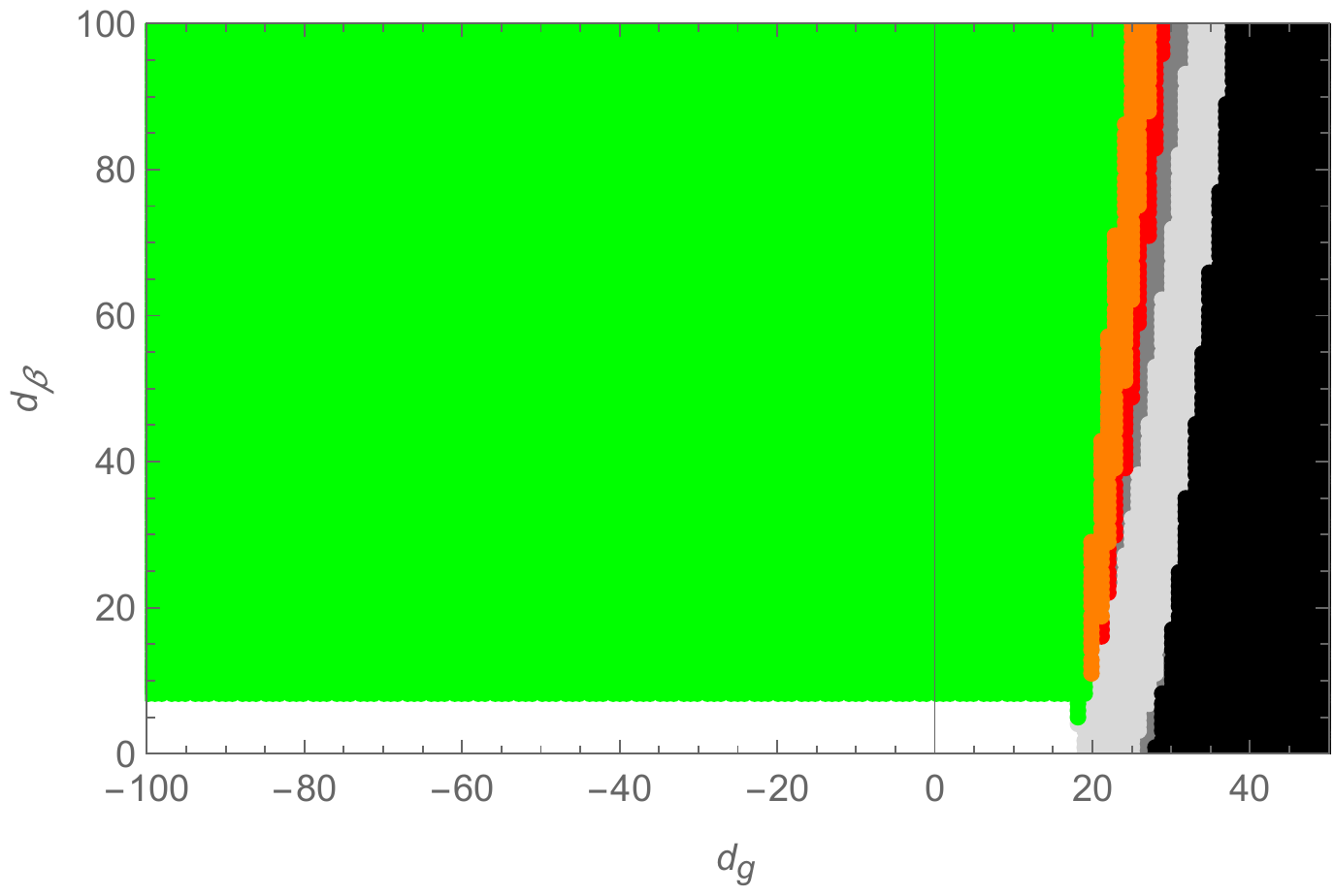}
\caption[NGFPs arising in the polynomial $\varphi_k(r)$ approximation at order 
$N=2$ for a coarse-graining operator of type I and type II.]
{\label{Fig.r2class} 
NGFPs arising in the polynomial $\varphi_k(r)$ approximation at order $N=2$ for a coarse 
graining operator of type I (top line) and type II (bottom line). In the left column the 
colours black, blue, green, and red indicate that the matter sector supports zero, one, 
two, and three NGFPs situated in the region with positive Newton's constant. The right 
column displays the stability properties of the NGFPs with $d_\beta > 0$. For points 
shaded dark grey, light grey, and green the $\theta_1, \theta_2$ subsystem has zero, one, 
and two UV attractive eigendirection with real stability coefficients. In the orange 
region the eigenvalues of the NGFP are complex.}
\end{figure}

Keeping the values of $d_g, d_\beta$ general, the analysis of \eqref{betar2} shows that 
the reduced system can have at most three (five) solutions for a coarse-graining operator 
of type I (type II). Applying the selection criteria that the FP coordinates are 
real and obey $g_1^* < 0$, the number of candidate NGFPs is shown in the left column of 
figure \ref{Fig.r2class}.
Besides the physically interesting region where $d_\beta \ge 0$, the diagrams also show 
the FP structure for $d_\beta < 0$. 
The numerical analysis reveals that there are at most 3 candidate solutions satisfying 
the selection criteria of a real positive Newton's constant. 

The stability properties of the NGFPs arising from matter sectors supporting a single 
candidate NGFP are displayed in the right column of figure \ref{Fig.r2class}. Disregarding 
the boundary region adjacent to the black region where no admissible NGFP is found 
reveals an intricate difference between the two coarse-graining operators. Focusing on a 
generic FP in the upper-left region one has $\theta_2 < 0$ for type I and 
$\theta_2 > 0$ for type II, i.e., the two cases lead to two and three UV relevant 
directions, respectively. The role of the small band of $d_g-d_\beta$-values supporting 
multiple NGFPs (white region in the lower-right diagram) will be clarified below.

For the matter sectors highlighted in table \ref{Tab.3}, the addition of the $r^2$-term 
does not lead to new bounds on the admissible FP structure, i.e., all models 
passing the EH test are situated in the region in the $d_g$-$d_\beta$-plane 
which supports a unique extension of the FP seen for $N=1$ to $N=2$. The sign of 
the new stability coefficient depends on the choice of coarse-graining operator though: 
in the type I case $\theta_2 < 0$ while the type II has $\theta_2 > \theta_1 > 0$ 
indicating that the new direction is UV relevant with a large, positive stability 
coefficient.

\begin{figure}[t!]
	\includegraphics[width=0.45\textwidth]{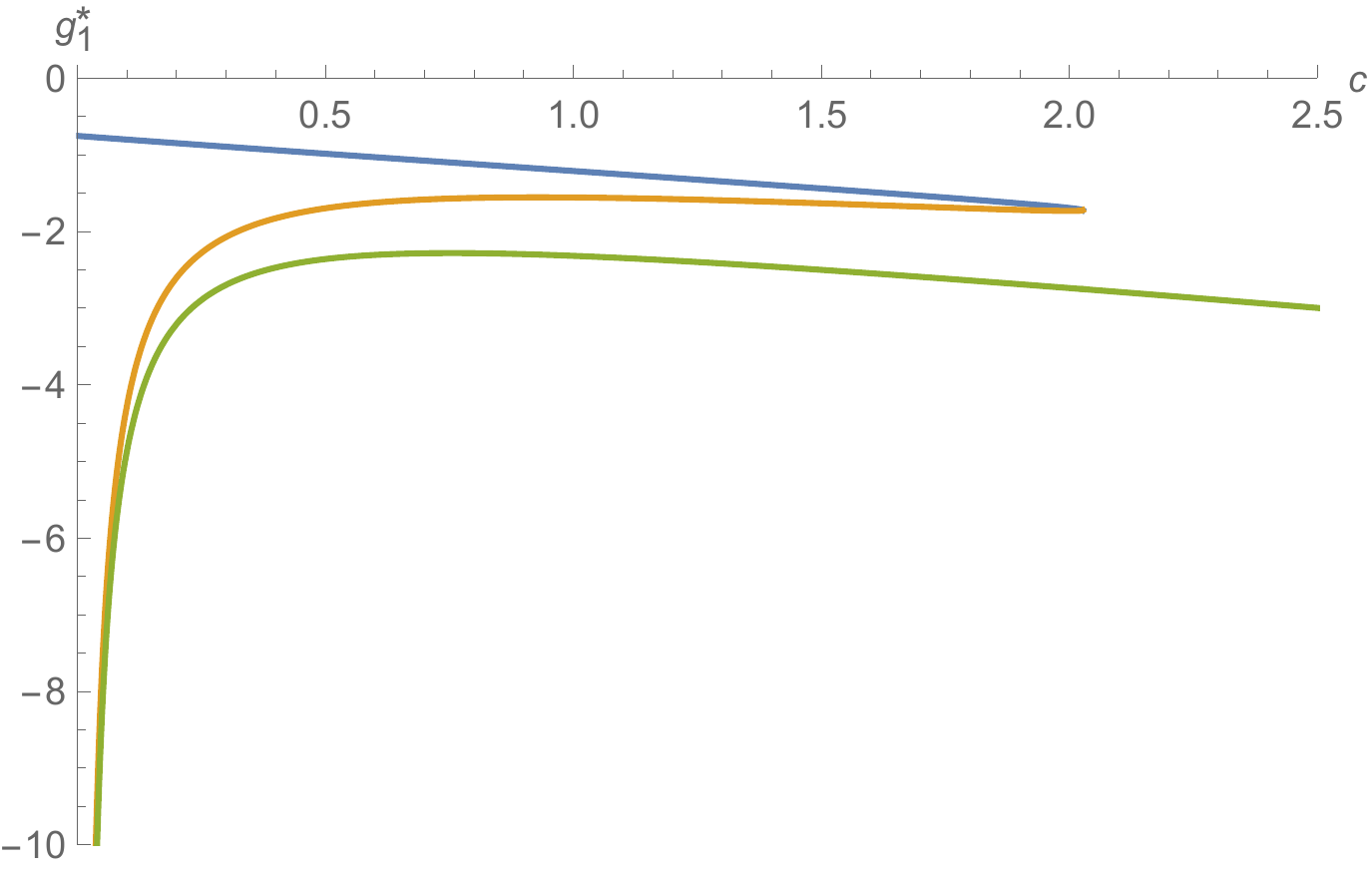} \; \; \;
	\includegraphics[width=0.45\textwidth]{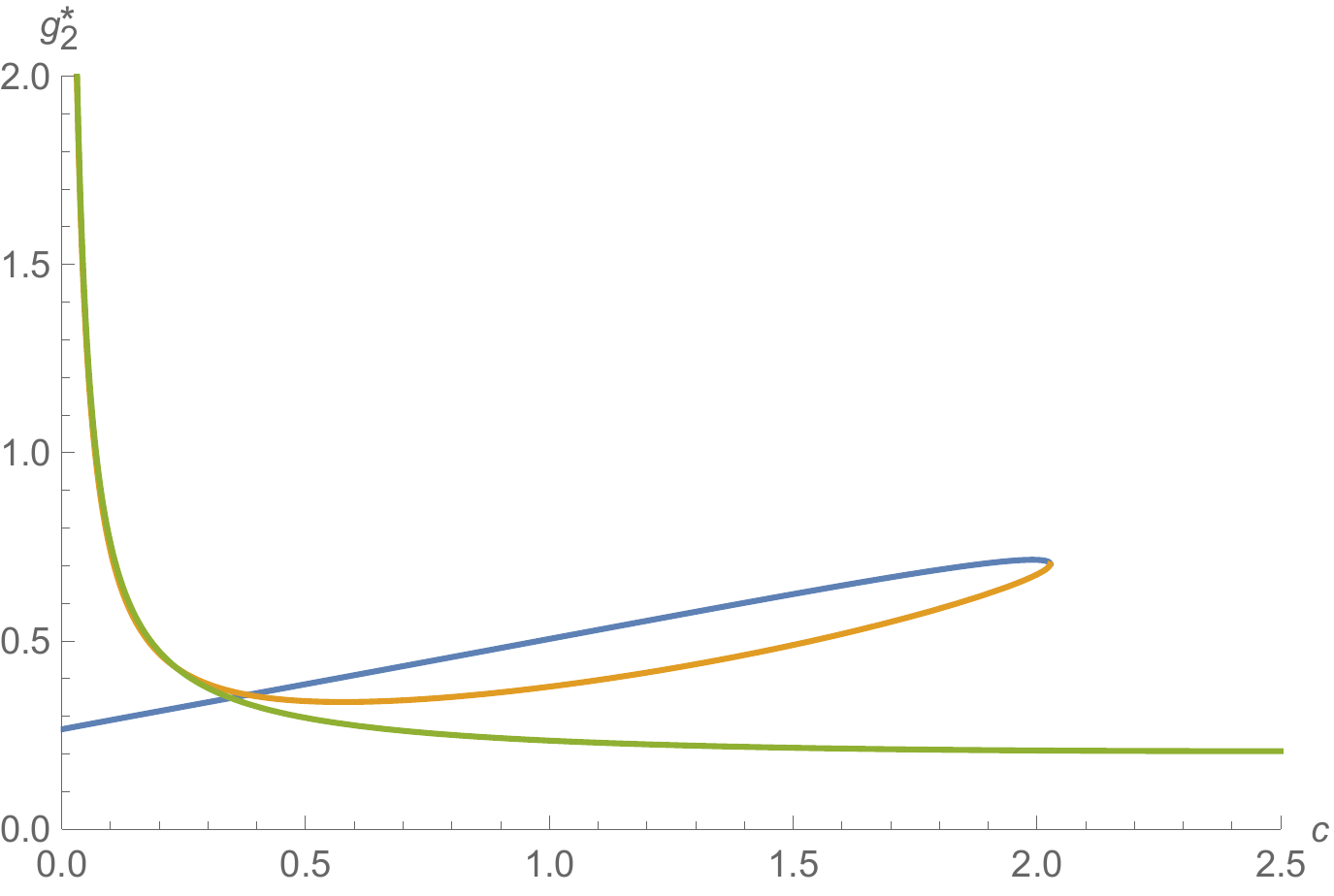} \\[1.4ex]
	\includegraphics[width=0.45\textwidth]{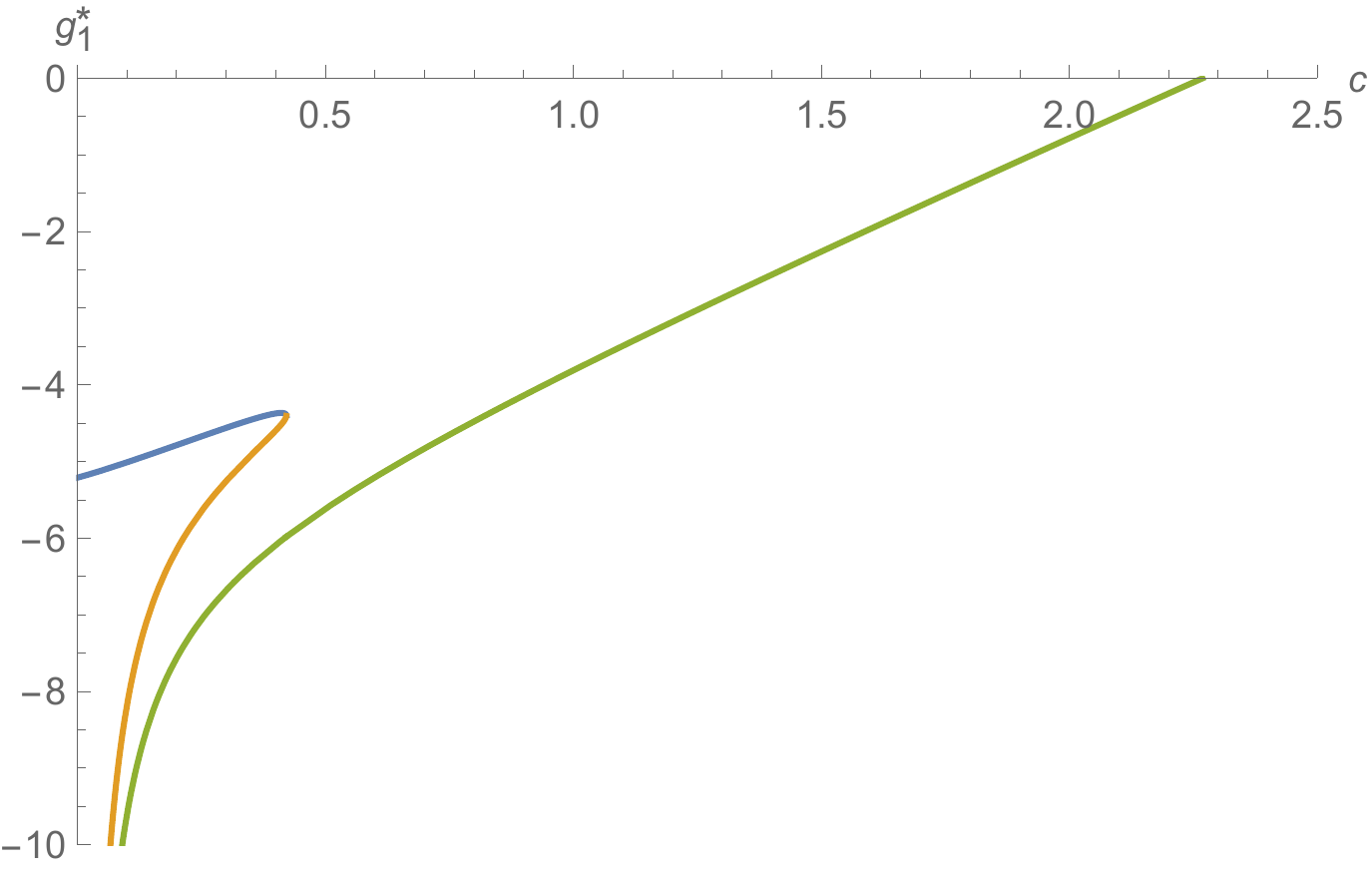}  \; \; \;
	\includegraphics[width=0.45\textwidth]{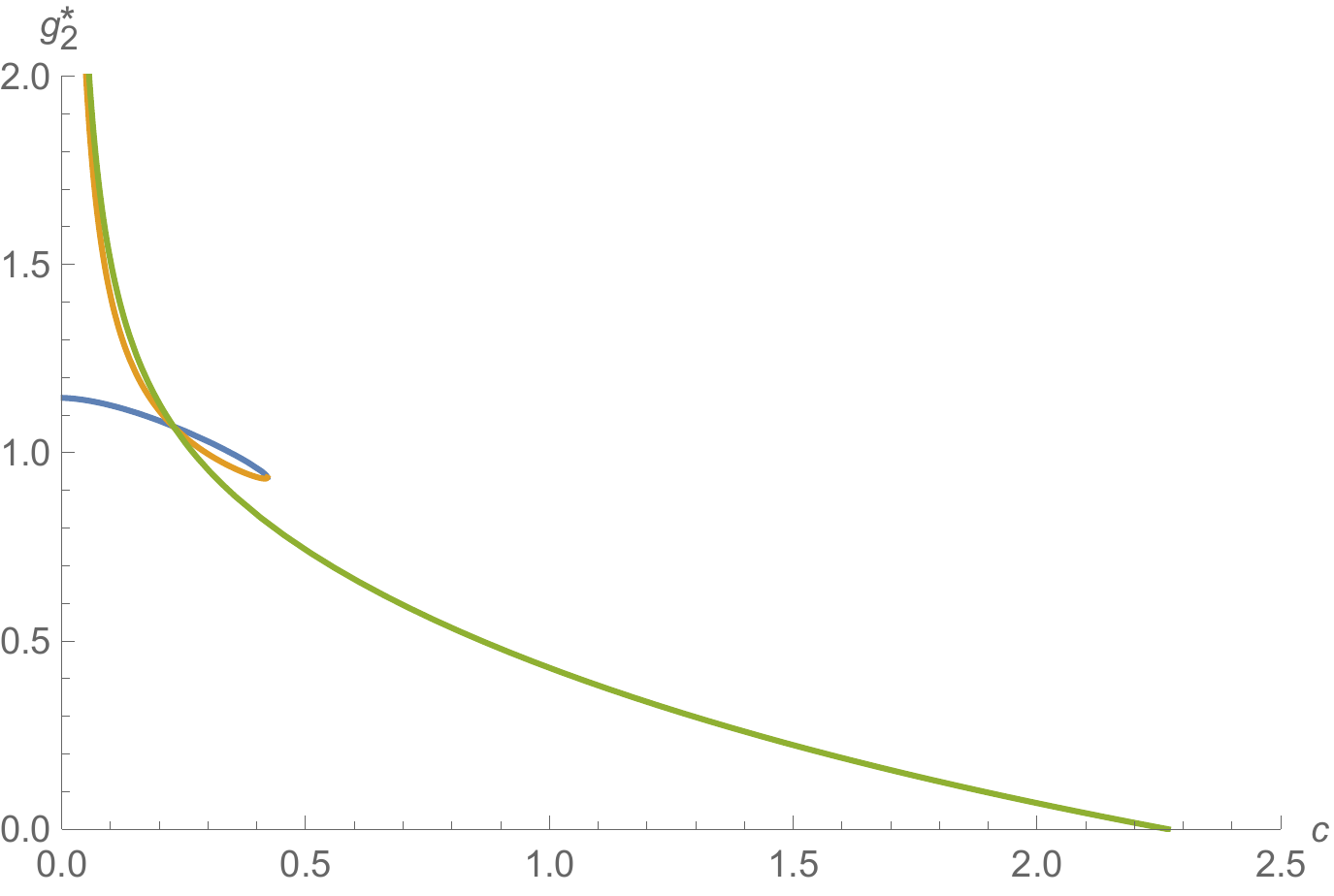}
	\caption[FP structure obtained for $N=2$ as a function 
of the deformation parameter $c$ interpolating between a type I ($c=0$) and 
type II ($c=1$) coarse-graining.]
{\label{Fig.typeIIannihilation} 
FP structure obtained for $N=2$ as a function of the deformation parameter $c$ 
interpolating between a type I ($c=0$) and type II ($c=1$) coarse-graining. The cases of 
pure gravity $(N_S = 0, N_D = 0, N_V = 0)$ and the SM 
$(N_S = 4, N_D = \tfrac{45}{2}, N_V = 12)$ are shown in the top and bottom row, 
respectively. The deformation of the NGFP appearing in the type I analysis is depicted by 
the blue line. For $c > 0$ there are two additional NGFP moving in from infinite. One of 
these FPs annihilates the type I FP at a finite value of $c$. For pure 
gravity this annihilation occurs at $c > 1$ while for the other gravity-matter models 
listed in table \ref{Tab.3} the annihilation is at $c < 0$. As a result the systems 
resulting from the type II coarse-graining again possess a unique NGFP (green line). This 
FP does not admit a convergent extension to higher orders of $N$, however.}
\end{figure}
At this stage, it is natural to inquire about the relation of the NGFPs seen in the type 
I and type II case. For this purpose, we resort to the interpolating coarse-graining 
operators constructed from \eqref{regtypei}. For $N=2$, the subsystem of equations 
determining the position of the NGFPs in the $g_1$-$g_2$-plane is sufficiently simple that 
all of its five roots can be found for general deformation parameter $c$. The 
corresponding implicit expressions allow to trace the position of the NGFP seen for type 
I coarse-graining ($c=0$) as a function of the deformation parameter $c$. 
Figure \ref{Fig.typeIIannihilation} depicts the $c$-dependence of the FP 
structure obtained for two characteristic examples, pure gravity ($d_\beta = 0$) in the 
top row and gravity coupled to the matter content of the SM ($d_\beta = 28$) 
in the bottom row, respectively. The key structure encountered in the analysis is rather 
universal. For $c=0$ the system has a single NGFP which is the one displayed in the top 
line of figure \ref{Fig.r2class}. Once $c$ is increased an additional pair of NGFPs moves 
in from infinity (orange and green lines). At a finite value of $c$ one of these new 
FPs (orange line) annihilates the $c=0$ solution (blue line). For $c$ larger 
than this critical value one is again left with a single NGFP (green line). 

If $d_\beta \le 7$ this annihilation occurs at $c > 1$ while for $d_\beta \ge 8$ the two 
FPs annihilate before the type II coarse-graining operator is reached. Since all 
phenomenologically interesting matter sectors are located at $d_\beta \ge 8$ we see that 
the NGFPs found in the type II computation \emph{are not continuously connected} to their 
type I counterparts. Anticipating results from the next subsection, these 
two disconnected families of NGFPs will be named 
``gravity-type'' (blue line) and ``matter-dominated'' (green line), respectively.  

\subsection{Gravity-Matter Fixed Points for Selected Matter Sectors}
\label{sect.fR43}

The final part of this analysis investigates the stability of the NGFPs characterised in 
the previous subsections under the inclusion of further powers of the dimensionless 
curvature $r$ in the polynomial ansatz \eqref{polyexpansion}. A detailed numerical 
analysis determining the polynomial solution approximating the FP up to $N=14$ 
and its  critical exponents up to $N=9$ 
revealed a strikingly simple structure: {for gravity-type NGFPs the position and 
stability coefficients characterising the FP converge rapidly when $N$ is 
increased. For the matter-dominated NGFPs no such convergence pattern could be 
established.} In order to arrive at this result extending the order of the polynomials 
beyond $N=2$ is crucial.

\begin{table}[h]
\scalebox{0.86}{
        \renewcommand{\arraystretch}{1.4}
                \begin{tabular}{p{3.94cm}|c|c|c||c|c||c|c}
                        model & \multicolumn{3}{c||}{matter content} & \multicolumn{2}{c||}{\; type I coarse-graining \; } & \multicolumn{2}{c}{\; type II coarse-graining \;} \\
                        & \, $N_S$ \, & \, $N_D$ \, & \, $N_V$ \, & \; \; EH \; \; & \; \; $f(R)$  \; \; &  \; \; EH  \; \; &  \; \; $f(R)$  \; \; \\  \hline \hline
                        pure gravity & 0 & 0 & 0 &  $\checkmark$ & $\checkmark$ & $\checkmark$ & $\checkmark$   \\ \hline
                        SM & 4 & ${45}/{2}$ & 12 & $\checkmark$ & $\checkmark$ & $\checkmark$ & $\left(\xmark\right)$  \\ \hline
                        SM, DM & 5 & ${45}/{2}$ & 12 &  $\checkmark$ &  $\checkmark$ & $\checkmark$ & $\left(\xmark\right)$ \\ \hline
                        SM, $3\,\nu$ & 4 & 24 & 12 & $\checkmark$ &  $\checkmark$ & $\checkmark$ & $\left(\xmark\right)$ \\ \hline
                        SM, $3\,\nu$, DM, axion & 6 & 24 & 12 & $\checkmark$ &  $\checkmark$ & $\checkmark$ & $\left(\xmark\right)$ \\ \hline
                        MSSM & 49 & ${61}/{2}$ & 12 & $\checkmark$ & $\checkmark$ & $\xmark$ & $\xmark$  \\ \hline
                        {SU(5) GUT} & {124} & {24} & {24}  & $\xmark$ & $\xmark$ & $\xmark$ & $\xmark$ \\ \hline
                        {SO(10) GUT} & {97} & {24} & {45} & $\xmark$ & $\xmark$ &  $\checkmark$ &  $\left(\xmark\right)$ \\ \hline \hline
                \end{tabular}
}
        \caption[Summary of results on the stability of
NGFPs appearing for the matter content of the SM of particle physics and its 
phenomenologically motivated extensions.]{
\label{Tab.mainresults} Summary of results on the stability of 
NGFPs appearing for the matter content of the SM of particle physics and its 
phenomenologically motivated extensions. Checkmarks $\checkmark$ indicate that the setup 
possesses a suitable NGFP which converges for increasing $N$. The symbol $\xmark$ shows 
that there is no NGFP at the level of the EH ($N=1$) approximation while a 
$(\xmark)$ implies that the NGFP seen at $N=1$ does not exhibit convergence when $N$ is 
increased.}
\end{table}
\begin{table}[p]
\scalebox{0.88}{
	\begin{tabular}{c|ccc|cccc}
		\;\; $N$\;\; & \;  $g_0^*$   \; & \; $g_1^*$ \;  & \; $g_2^*$ \; & \;  $g_3^* \times 10^{-4}$ \; 
		& \; $g_4^* \times 10^{-4}$ \; & \; $g_5^* \times 10^{-4}$ \; & \; $g_6^* \times 10^{-4}$ \; \\ 
		\hline \hline
		$1$ & \; $-7.2917$ \; & \; $-5.8264$ \; & \\
		$2$ & $-6.7744$  & $-5.2122$ & \; $1.1455$ \; \\
		$3$ & $-6.7795$  & $-5.2617$ & $1.1601$ & $50.466$  \\
		$4$ & $-6.7737$  & $-5.2577$ & $1.1550$ & $49.161$ & $-2.7013$ \\
		$5$ & $-6.7742$  & $-5.2598$ & $1.1559$ & $51.122$ & $-2.4926$ & $0.3313$ \\
		$6$ & $-6.7755$  & $-5.2611$ & $1.1571$ & $51.929$ & $-1.9180$ & $0.4268$ & $0.1426$ \\ 
		$7$ & $-6.7764$  & $-5.2632$ & $1.1582$ & $53.712$ & $-1.5336$ & $0.7152$ & $0.1999$ \\ 
		$8$ & $-6.7775$  & $-5.2646$ & $1.1592$ & $54.700$ & $-1.0696$ & $0.8557$ & $0.3065$ \\
		$9$ & $-6.7781$  & $-5.2657$ & $1.1599$ & $55.663$ & $-0.7932$ & $1.0079$ & $0.3586$ \\
		$10$ & $-6.7786$  & $-5.2665$ & $1.1605$ & $56.249$ & $-0.5615$ & $1.0959$ & $0.4091$ \\
		$11$ & $-6.7789$  & $-5.2671$ & $1.1608$ & $56.693$ & $-0.4174$ & $1.1654$ & $0.4382$ \\
		$12$ & $-6.7792$  & $-5.2674$ & $1.1611$ & $56.973$ & $-0.3142$ & $1.2084$ & $0.4602$ \\
		$13$ & $-6.7793$  & $-5.2677$ & $1.1612$ & $56.717$ & $-0.2486$ & $1.2383$ & $0.4737$ \\
		$14$ & $-6.7794$  & $-5.2678$ & $1.1613$ & $55.729$ & $-0.2049$ & $1.2572$ & $0.4830$ \\
		\hline \hline
			\multicolumn{8}{c}{}\\[-2ex]
	\end{tabular}
}

\scalebox{0.88}  {	
\begin{tabular}{c|cc|ccccc}
		%
		\; \; $N$\;\; & $\theta_0$ & $\theta_1$ & $\theta_2$ & $\theta_3$ & $\theta_4$ & $\theta_5$ & $\theta_6$ \\ \hline \hline
		$1$ & \;\;\; \;\;\; $4$ \;\;\; \;\;\; & \;\;\; $2.127$ \;\;\;\\
		$2$ & $4$ & $2.339$ & $-1.671$  \\
		$3$ & $4$ & \; $2.274$ \; & $-1.727$ & $-6.013$ \\
		$4$ & $4$ & $2.279$ & \; $-1.808$ \; & \; $-5.905$ \; & \; $-9.308$ \; \\
		$5$ & $4$ & $2.280$ & $-1.809$ & $-5.928$ & $-9.330$ & $-11.956$ 
		\\
		$6$ & $4$ & $2.279$ & $-1.797$ & $-5.916$ &  $-9.297$ & \;  $-12.146$ \;  & \; $-14.293$ \; \\ 
		$7$ & $4$ & $2.278$ & $-1.791$ & $-5.888$ &  $-9.283$ & $-12.070$ & $-14.628$ \\
		$8$ & $4$ & $2.277$ & $-1.784$ & $-5.874$ &  $-9.248$ & $-12.061$ & $-14.519$ \\
		$9$ & $4$ & $2.276$ & $-1.780$ & $-5.856$ &  $-9.225$ & $-12.018$ & $-14.512$ \\
		\hline \hline
\end{tabular}
}
\caption[FP structure of $f(R)$-gravity coupled to the matter content 
of the SM of particle physics.]
{\label{Tab.FPstandardmodel} 
FP structure of $f(R)$-gravity coupled to the matter content of the standard 
model of particle physics $(N_S=4, N_D=45/2, N_V = 12)$ and a type I cutoff. The fixed 
point exhibits the same stability properties as in the case of pure gravity. Note that 
the polynomial coefficients for the constant, linear and quadratic term are of 
${\mathcal{O}}(1)$ whereas $g_3^*$ is already smaller than 1\% .}
\end{table}
Besides the rapid convergence of the polynomial expansion with regard to the position 
and stability coefficients of the NGFP, the data shows that the FP has the same 
predictive power as the one found in the case of pure gravity: it comes with two relevant 
parameters. These characteristic properties are shared by the FPs found for the 
other matter sectors carrying ticks in table \ref{Tab.mainresults}. Their characteristic 
properties are compiled in appendix~\ref{App.fRB}.

Table \ref{Tab.mainresults} summarises the consequences of this general result for 
phenomenologically interesting gravity-matter models introduced in table \ref{Tab.3}. The 
key insights are the following: for pure gravity where a gravity-type NGFP 
persists for both coarse-graining operators, one consequently has one stable NGFP 
solution in both cases. The characteristics of these NGFPs, including their position and 
stability coefficients, are tabulated in tables \ref{Tab.grav.I} and \ref{Tab.grav.II} of 
appendix~\ref{App.fRB}, respectively. Focusing on the case of type I coarse-graining and 
the  gravity-matter models selected in table \ref{Tab.3}, it is found that all NGFPs seen 
at the level of the EH approximation have a stable extension to polynomial 
$f(R)$-gravity. For gravity supplemented by the matter content of the SM, 
this is strikingly demonstrated in table \ref{Tab.FPstandardmodel}. 
One particular property of the polynomial 
solutions $\varphi(r)$ associated with the gravity-type NGFPs
is peculiar, and also important for the investigation in the next section
\ref{GlobalFF}. Based on the partial 
differential equation \eqref{pdf4d} one finds that the coefficients $g_0^*$, $g_1^*$ and 
$g_2^*$ are of order unity with $g_1^* < 0$ corresponding to a positive Newton's 
coupling. The coefficients $g_n^*, n > 2$ are significantly smaller. E.g., 
$g_3^*/g_2^* \approx 10^{-3}$ and the numerical values of further coefficients 
rapidly approaches zero. Thus the solutions $\varphi(r)$ are essentially second order 
polynomials in the dimensionless curvature $r$.

\begin{figure}[h]
\centering
\includegraphics[width=0.48\textwidth]{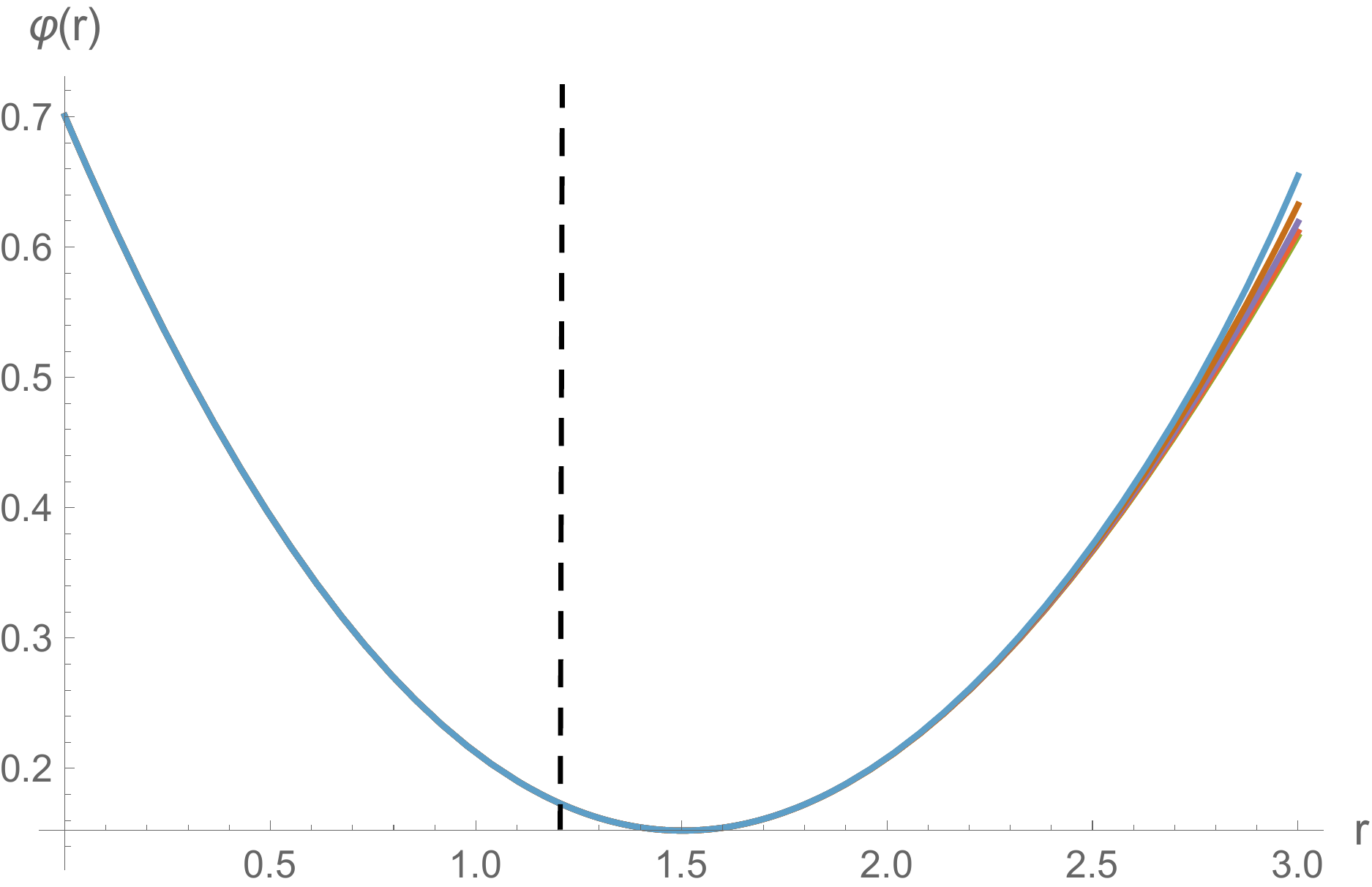} \,
\includegraphics[width=0.48\textwidth]{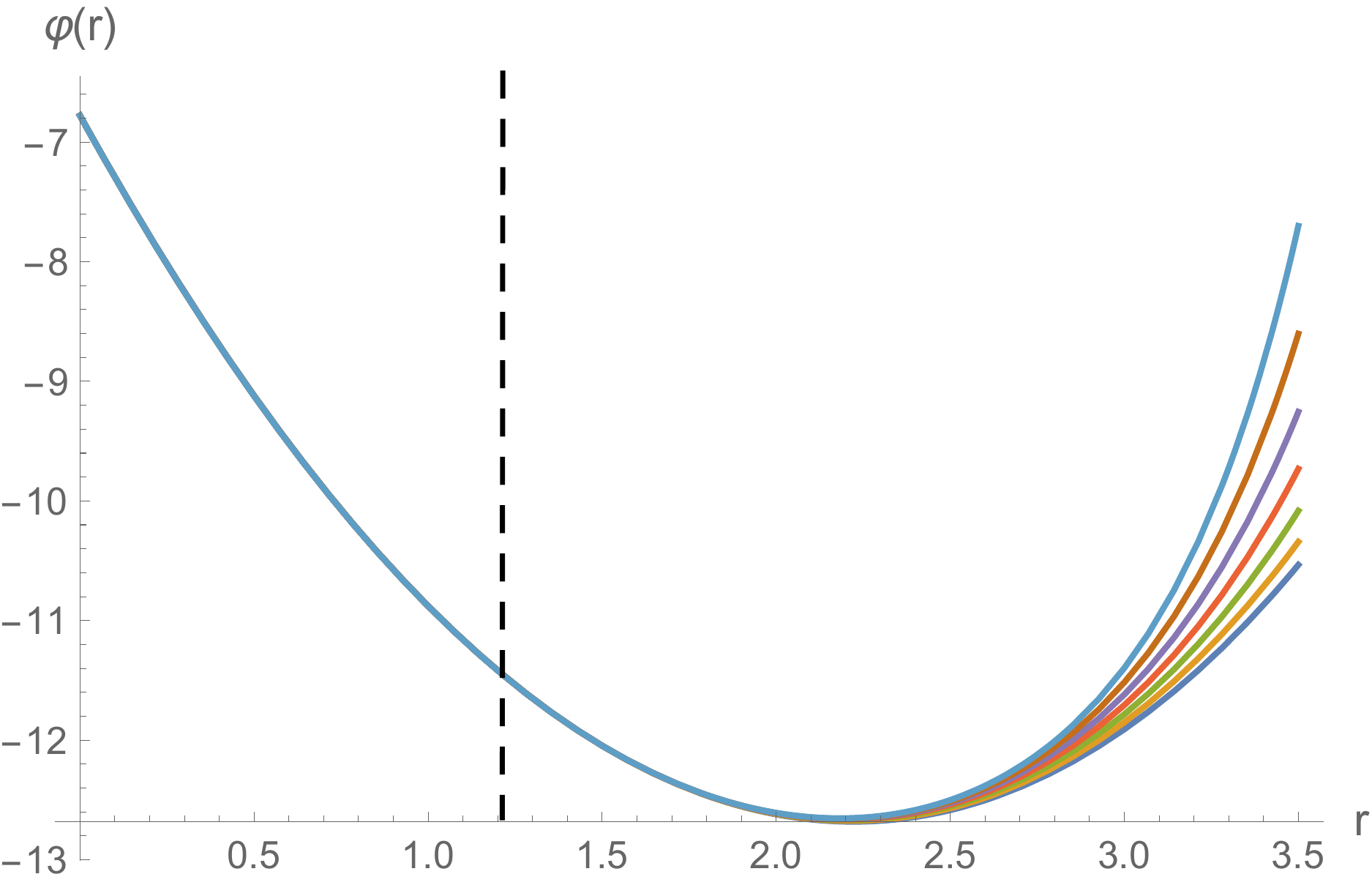}
\caption[Fixed functions  arising from the polynomial expansion of $\varphi(r)$ for a 
type I coarse-graining operator. The cases of pure gravity and gravity coupled to the 
matter content of the SM are shown.]
{\label{Fig.conv} 
Fixed functions  arising from the polynomial expansion of $\varphi(r)$ for a type I 
coarse-graining operator. The cases of pure gravity and gravity coupled to the matter 
content of the SM are shown in the left and right diagram, respectively. From 
bottom to top the curves result from the expansions up to $N=8,9,10,11,12,13,14$. The 
polynomial approximation provides a convergent solution of the FP equation which 
extends up to the moving singularity where $\varphi^\prime(r) = 0$.
}
\end{figure}
The polynomials $\varphi(r)$ for increasing values of $N$ arising for pure gravity and 
gravity coupled to the matter content of the SM are shown in the left and 
right diagram of figure \ref{Fig.conv}, respectively. For small values of $r$ the 
polynomial expansion shows a rapid convergence. Notably, both  solutions exhibit a local 
minimum at $r \approx 1.50$ and $r \approx 2.2$, respectively. Inspecting 
\eqref{gravTTtypeI}, 
one finds that this minimum corresponds to a moving singularity. In order for the fixed 
functional to extend to a global solution the zero of $\varphi^\prime$  must be canceled 
by a corresponding zero in the numerator. For the type I case where $\alpha_T^G=0$ such a 
cancellation occurs automatically at $r=3/2$. The interplay between the moving 
singularity and this cancellation in the case of pure gravity (right diagram of 
figure \ref{Fig.conv}) then leads to a polynomial solution whose convergence properties 
are better than expected on the grounds of the moving singularity.\footnote{Note that the 
radius of convergence displayed in figure \ref{Fig.conv} is independent of the fixed 
singularities given in \eqref{fixedsing}. The construction of the polynomial solution is 
not based on the normal form of the FP equation so that these singular loci are 
irrelevant for determining the convergence structure of the solution.}

\section{Global Fixed Functions for $f(R)$-Gravity Matter Systems}
\label{GlobalFF}

In this section the computation of two types of global fixed functions is presented.
The existence of global quadratic solutions is shown by constructing explicitly two 
example sets of respective parameters $\alpha$.
And second, two examples of a numerical solution are presented. 
Hereby the discussion of the singular points of the flow equation
and its asymptotic behaviour for large scalar curvatures turns
out to be the crucial element.

\subsection{Global Quadratic Solutions}

For pure gravity global fixed functions which are polynomials of quadratic
order in $r$ have been found \cite{Ohta:2015fcu}. As will be discussed in more detail
below, for the purpose of finding such solutions the parameters $\alpha$ are treated 
as free parameters, respectively, they are fixed by the requirement of the existence 
of an exact quadratic solution for the fixed function. The necessary choices for the
coarse-graining parameters are hereby not motivated by physics. The existence
of this special type of solutions,
however, provides a reason why the polynomial approximations plotted above
as well as the numerically obtained solutions described in the next subsection are 
very close to a global purely quadratic fixed function.

For a global quadratic function, the third derivative vanishes and thus
the second summand of \eqref{gravsinv} does not contribute.\footnote{In all other non-trivial solutions
this term (which stems from the conformal mode) is the leading one in the normal form because
in this and only this term a third-order derivative, {\it i.e.}, $\varphi'''(r)$,
appears.} To be specific, the ansatz
\be
\varphi^\star (r) = \frac 1 {(4\pi)^2} \,\, (g_0^\star + g_1^\star r + g_2^\star r^2)
\label{quadAnsatz}
\ee
is chosen, and the coarse-graining parameters $\alpha$ are determined such that
\eqref{quadAnsatz} becomes exact. 
In case $\varphi^\star(r)$ is a polynomial,  the differential equation 
\eqref{pdf4d} determining it can be rewritten as
\be
\label{PolyRat}
\frac {{\cal P}_{num}(r) }{{\cal P}_{den}(r) } = 0 
\ee
by bringing all terms in the equation to a common denominator.
{\it I.e.}, \eqref{pdf4d} can be formulated as the requirement that the ratio of 
two polynomials vanish. This can be solved
in two steps: first, solve for ${{\cal P}_{num}(r) }=0 $, and second, keep only those
solutions where all roots of ${{\cal P}_{den}(r) }$ ({\it i.e.}, the potential singularities
of this equation) coincide with roots of the numerator.

In the case of a quadratic  fixed function, $ {{\cal P}_{num}(r) }$ is a
fifth-order polynomial,\footnote{\label{FootCanc}
The l.h.s.\ of the flow equation (\ref{pdf4d}) is for the ansatz (\ref{quadAnsatz}) not
a polynomial of order $N=2$ as na\"ively expected because the term proportional to $r^2$ cancels:
$4 \varphi(r) -2 r \varphi '(r) = 4g_0+2r g_1 $.}
 and its six coefficients can be determined by a discrete set of values
for $g_0^\star$, $g_1^\star$, $g_2^\star$, $\alpha_T^G$, $\alpha_V^G$ and $\alpha_S^G$.
For pure gravity five different
solutions for a globally quadratic fixed function have been identified \cite{Ohta:2015fcu}.
Quite surprisingly, in all five solutions found in this reference
the potential singularities given by the zeros of the denominator are canceled by the numerator.
On the other hand, for two of these five solutions the eigenperturbations lead to a differential
equation with four instead of three fixed singularities, and therefore such eigenperturbations 
will likely not
exist globally. For another of these five solutions $\alpha_T = (11 + \sqrt{265})/54 \approx 0.505 > 2/3$,
{\it i.e.}, the inequality for a positive argument of the regulator function is violated. This leaves
two solutions, and the corresponding values for the parameters are given in the respective
first lines of tables \ref{tabQ1} and \ref{tabQ5}. The exact values of these parameters are, respectively,
{\begin{align}
\label{Q1}
\alpha_S^G=\tfrac{5\sqrt{265}-73}{216} & , & \alpha_V^G= \tfrac{67-2 \sqrt{265}}{108} & , & 
\alpha_T^G=\tfrac{11-\sqrt{265}}{54}  & , & \\
g_0^\star=\tfrac{49+\sqrt{265}}{96} & , & 
g_1^\star =- \tfrac{4141+121\sqrt{265}}{5184} & , &g_2^\star= \tfrac{67795 + 3583 \sqrt{265}}{279936},
\nonumber
\end{align}}
or
\be
\alpha_S^G=-\tfrac 3 {47} \, ,\,\,\, \alpha_V^G = - \tfrac {83}{564}  \, , \,\,\,  
\alpha_T^G=- \tfrac {53}{94}\, , \,\,\,  g_0^\star = \tfrac {89}{72} \, , \,\,\,
g_1^\star = - \tfrac {101}{94}  \, ,\,\,\, g_2^\star = \tfrac {1414}{6627} \,  . \label{Q5}
\ee
As this will be important below the value for the minimum of the fixed functions
is given \footnote{Note that there is a typo (sign error) in (4.3) of \cite{Ohta:2015fcu}.}
\be
r_{min} = - \frac {g_1^\star}{2g_2^\star} =
\begin{cases}
\frac 3 {20} ( 25 - \sqrt{265}) \approx 1.3082 \, &\\
\frac{141}{56} \approx 2.5179 \,\,\,\,\, .& \\
\end{cases}
\ee
Recalling that the right hand side of \eqref{pdf4d} is independent of $g_0$ and 
depends only on ratios of the fixed function and its derivatives, 
for a  global quadratic solution one can rewrite the equation
for  the fixed function such that the parameters $g_i^\star$ appear only in the ratio
$r_{min} = - {g_1^\star}/{2g_2^\star}$  on the left hand side
since
\be
 \frac{ \varphi^\prime -  r \varphi^{\prime\prime}} {\varphi^\prime}  =
 \frac{r_{min}}{r_{min}-r}
\ee
and
\be
\frac{ \varphi^{\prime\prime}}{\left(1+ \left(\alpha^G_S - \tfrac{1}{3}\right)r\right)
\varphi^{\prime\prime} + \tfrac{1}{3} \varphi^\prime} =
 \frac 1{1 + \alpha_S^G r - r_{min}/3} \,\, .
\ee

For the solution (\ref{Q1}) one of the zeros of the second summand of \eqref{gravTT}
occurs exactly at $r_{min}$ and thus the
potential singularity is canceled. The singularity in the scalar term occurs at negative values of
$r$ and is thus of no concern. For the solution (\ref{Q5}) the potential pole due to the scalar term
appears also exactly at $r_{min}$, and the same is true for the first term in
\eqref{gravTT} and the term \eqref{gravghost}.
With these values of endomorphism parameters, for the pure gravity case, the terms on the left hand side  conspire to yield
\be
\cT^{TT}+\cT^{sinv}+\cT^{ghost}=
\frac 1 {(4\pi)^2} \left(\frac {89}{18} - \frac {101}{47} r \right)
\ee
which, of course, solves then the equation for the fixed function for the parameters $g_0^\star$ and
$g_1^\star$ given in (\ref{Q5}).

The usefulness of the above considerations becomes immediately clear when adding fermions, {\it i.e.},
when adding
\be
\cT^{dirac}=
\frac {-2 N_D} {(4\pi)^2} \left(  1  + (\alpha_D^M + \frac 1 { 6})  r \right)  \,
\ee
respectively,
\be
\frac 1 {(4\pi)^2}
\begin{cases}
- 2 N_D + \frac 1 6 N_D r & {\mathrm {type \, II~\, reg.}} \\
-2 N_D - \frac 1 3 N_D r & {\mathrm {type \, I~\, reg.} } \,\,\,\,\, .\\
\end{cases}
\ee
A global quadratic solution can be now easily obtained by keeping the ratio
$  {g_1^\star}/{g_2^\star}$
and thus $ r_{min} $ fixed.  One simply maintains the values of the endomorphism
parameters in the gravity sector and substitutes
\ba
g_0^\star & \to &  g_0^\star - \frac {N_D}{2} \, , \nonumber  \\
g_1^\star & \to &  g_1^\star - (\alpha_D^M + \frac 1 { 6}) N_D \, ,
\label{RulesForFermions}\\
g_2^\star & \to &  g_2^\star + \frac 1 {2r_{min}}  (\alpha_D^M
+ \frac 1 { 6}) N_D \,\,\, ,
\nonumber
\ea
respectively,
\ba
g_0^\star & \to &  g_0^\star - \frac {N_D}{2} \, , \nonumber  \\
g_1^\star & \to & 
g_1^\star + \begin{cases}
 \frac 1 {12} N_D & {\mathrm {type \, II~\, reg.}} \\
- \frac 1 6 N_D  & {\mathrm {type \, I~\, reg.}} \\
\end{cases} \,\,\, , \label{RulesForFermionsT}\\
g_2^\star & \to &  
g_2^\star - \frac 1 {2r_{min}} \begin{cases}
 \frac 1 {12} N_D  & {\mathrm {type \, II~\, reg.}} \\
- \frac 1 6 N_D  & {\mathrm {type \, I~\, reg.}} \\
\end{cases} \,\,\, . \nonumber
\ea
This proves to be always possible independent of whether the coefficient of
the linear term is negative as, {\it e.g.}, for the type I regulator, or positive
as, {\it e.g.}, for the type II regulator. 

If one uses now the type II regulator for the fermions there will be a critical value of $N_D$ where
$g_1^\star$ becomes positive. For the solution (\ref{Q1}) this value is 
$N_D^{crit}=14.1$ whereas for the solution
(\ref{Q5}) it is $N_D^{crit}= 12.9$. 
If these values are exceeded the minimum turns to a maximum (but stays at
the same location) and the values of $g_1^\star$ and $g_2^\star$ change sign. As then $g_2^\star$
is negative $\phi(r) \to - \infty$ for $r\to \infty$, and thus the action becomes unbounded from
below.
Therefore, if a type II regulator is used for the fermions one can add only a finite number of
them and keep a physically meaningful solution in agreement with the results obtained
already in the previous section.

Adding now scalar and/or vector fields the degrees of the polynomials in \eqref{PolyRat}
increase. It turns out then that one cannot fix
the parameters $\alpha_S^M=\alpha_{V2}^M=0$ and $\alpha_{V1}^M=-1/4$, {\it i.e.}, 
to their respective type II values.
Although then no new singularities arise in the matter sector one can easily convince oneself that
from the fact that the expressions $\cT^{\rm scalar}$ and $\cT^{\rm vector}  $
in \eqref{matterflow} are of quadratic order one obtains for the numerator a  polynomial of degree six, and thus seven equations for six variables. A similar
situation arises, namely eight equations for seven variables etc., if one fixes only one or two
of the three parameters to the respective type II value.  Basically the same remark applies for fixing to type I values.

Exploring the possibility of adjusting the parameters $\alpha_S^M$ and $\alpha_{V1,2}^M$
to keep a global quadratic solution
one notes first that adding fermions is always straightforward by applying the rule
(\ref{RulesForFermions}). It proves to be easier to add scalar fields whereas when adding
vector fields one might loose quite fast track of the solution. To obtain a solution with the SM field content the following strategy has been applied: 
first, I added 45/2 Dirac fields (according to SM matter content) with
type I regulator by applying (\ref{RulesForFermions}) to the solution
(\ref{Q1}) and verified this numerically.
Second, on the top of this four scalar fields are added and the corresponding parameter
$\alpha_S^M$ was determined. From there on I increased $N_V$ in small steps until
the SM value 12 was reached. The results for pure gravity, gravity plus
fermions, gravity plus fermions and scalars as well as for gravity plus SM
matter content are  displayed in tables
\ref{tabQ1} and \ref{tabQ5}. In all cases one has 
$\alpha_S^M=\alpha_{V2}^M$.\footnote{As these solutions were obtained by performing very
small steps in $N_V$ this should probably not be taken as evidence that no solution
with $\alpha_S^M\not=\alpha_{V2}^M$ exists.} The stability coefficients $\theta_{0,1,2}$
have been calculated by diagonalising the stability matrix in the same way as in the 
previous subsection. As all $\theta_2$ are negative only the values of $\theta_0$ and 
$\theta_1$ are displayed in tables
\ref{tabQ1} and \ref{tabQ5}.

\begin{table*}[ht]
        \renewcommand{\arraystretch}{1.4}
\scalebox{0.825}{
\begin{tabular}{||c||c|c|c|c|c|c||c|c|c|c||c||}
\hline
$(N_S,N_D,N_V)$&$\alpha_S^G$ &$\alpha_V^G$ &$\alpha_T$ &$\alpha_S^M$ & 
$\alpha_D^M$  &$\alpha_{V1}^M$ &
$g_0^\star$ & $g_1^\star$ & $g_2^\star$ & $r_{min}$  & $\theta_0, \theta_1$ \\
\hline
(0,0,0)      & .0389 & .3189 & -.0978 & -           & -    & -        & .6800 & -1.179  & 0.4505 & 1.308 & 4,  2.02 \\
(0,45/2,0) & .0389 & .3189 & -.0978 & -           &  0  & -         & -10.57 & -4.929  & 1.884  & 1.308 & 4,  1.98 \\
(4,45/2,0) & -.0819 & .0389 & -.3778 & -.2111 &  0  & -         & -9.970 & -5.078  & 1.382 &  1.837 & 4,  2.35 \\
(4,45/2,12)&-.0190 & .1603 & -.2563 & -.0897 &  0  & -.3397 & -6.702 & -8.630  & 1.825 & 2.364 & 4,  2.36 \\
\hline
\end{tabular}
}
\caption{ Quadratic solutions for the fixed function with different matter content derived from the pure gravity solution (\ref{Q1}).
\label{tabQ1}}
\end{table*}
\begin{table*}[ht]
        \renewcommand{\arraystretch}{1.4}
\scalebox{0.825}{
\begin{tabular}{||c||c|c|c|c|c|c||c|c|c|c||c||}
\hline
$(N_S,N_D,N_V)$&$\alpha_S^G$ &$\alpha_V^G$ &$\alpha_T$ &$\alpha_S^M$ &$\alpha_D$  &$\alpha_{V1}^M$ &
$g_0^\star$ & $g_1^\star$ & $g_2^\star$ & $r_{min}$  & $\theta_0, \theta_1$ \\
\hline
(0,0,0)      & -.0638  & -.1472 & -.5638 & -          & -    & -         & 1.236  & -1.074 & 0.2134 & 2.518 & $\approx$ 16, 4 \\
(0,45/2,0) & -.0638  & -.1472 & -.5638 & -          &  0  & -          & -10.01 & -4.824 & 0.9581 & 2.518 &   4, 2.98 \\
(4,45/2,0) & -.0554  & -.1388 & -.5550 & -.3855 &  0  & -          & -9.415 & -4.637 & 0.9014 & 2.572 &   4, 3.05 \\
(4,45/2,12)& -.0308 & -.1140 & -.5308 & -.3644 &  0  & -.6143 & -5.813 & -9.368 & 1.705   & 2.746 &   4, 2.8  \\
\hline
\end{tabular}
}
\caption{ Quadratic solution for the fixed function with different matter content derived from the pure gravity solution (\ref{Q5}).
\label{tabQ5}}
\end{table*}

It has to be emphasised that a solution with a positive value for Newton's constant could be found
because the type I value $\alpha_D^M=0$ was used for the fermionic term.
The stabilising effect of the type I regulated fermions is very much needed.
{\it E.g.}, the solution with no fermions at all but four scalars and twelve vectors    possesses a negative value for Newton's constant.
As for the solution (\ref{Q5}) one obtains an interesting effect of the fermions for the critical exponents:
the pure gravity solution has a large critical exponent
which we estimate to be around 16. Adding now type I regulated fermions brings this one down to three
(which also restores the order such that the critical exponent 4 related to the cosmological constant is
the largest one). Adding scalars on top of gravity and fermions slightly increases the critical exponent
but has overall not much effect. The same can be said about the gauge fields.

Summarising this section, two solutions have been presented
with endomorphism parameters adjusted such that
a global quadratic solution exist for matter up to the SM matter content. This worked
because a type I regulator for the fermions has been used. At least for the type of 
solutions discussed
here, type II regulated fermions lead to a negative fixed point value of Newton's constant
and result thus in physically acceptable solutions.

\subsection{Asymptotic Behaviour for Large Curvature}

Studying the asymptotic behaviour for large curvature $r$ serves within this investigation
two purposes. On the one hand, this knowledge will be employed when numerically solving
for a fixed function. On the other hand, it will allow to identify a destabilising influence
of matter fields without actually searching for a numerical solution.

As shown below the possible  leading asymptotic behaviour for $r\gg 1$ is
either $\simeq r^2$ or $\simeq r^2 \ln r$ depending on the values of the endomorphism
parameters.\footnote{Note
that (4.9) in \cite{Ohta:2015fcu} were only correct if their
parameter $\alpha$ (our $\alpha_T^G$) would deviate slightly from
the type II value $\alpha=-1/6$. Allowing for a small change of this parameter, I have verified
this expression.}
The left hand side of \eqref{pdf4d} at the NGFP
reduces to a constant  plus linear term if $\varphi^\star (r)$ is a quadratic function due to a
cancelation (see  footnote \ref{FootCanc}), and it becomes a quadratic polynomial
if a term proportional to $r^2 \ln r$ is added.

Quite obviously cancellations in differences between terms play a significant role.
Therefore the most straightforward way to proceed is to infer the large curvature behaviour
in that terms  of the flow equation~(\ref{pdf4d}) which do not depend on the function
$\varphi (r)$ and its derivatives. This is especially useful as some of these terms 
belong to the leading order
terms. A counterexample is $\cT^{\rm Dirac}$
\eqref{Tdirac}. As it is a linear function in $r$, a leading quadratic behaviour 
of the type $\varphi^\star = Ar^2 +\cO(r)$ will
only be changed but not made impossible (as it is also evident from the discussion
of the global quadratic solution), and in the presence of a non-vanishing quadratic term
on the left hand side of the flow equation it is subleading.

As the scalar matter term $\cT^{\rm scalar}$ and the contribution from the gauge
ghost behave identical they can be discussed together. One clearly sees a qualitative
difference for $\alpha_S^M\not =0$, resp., $\alpha_{V2}^M\not= 0$ for which the
asymptotic behaviour of the corresponding terms is linear and for vanishing
parameter (which includes type I and type II coarse-graining) for which the
asymptotic behaviour is quadratic. In the first case these two terms provide
singularities at $r=-1/\alpha_S^M$ and $r=-1/\alpha_{V2}^M$, respectively.
In the latter case one has, of course, no singularities but the term is 
a leading one for large $r$.

As for the transverse vector matter fields one has to distinguish between the
type II case $\alpha_{V1}^M =-1/4$ for which there is no singularity but a
quadratic contribution (and thus the term contributes to the leading 
behaviour),  and all other cases with a singularity at
$r=-1/(\alpha_{V1}^M +1/4)$ and a sub-leading linear asymptotic behaviour.
A completely analogous discussion applies to the gravitational ghost term
$\cT^{\rm ghost}$ with the only difference that the type II corresponds
to $\alpha_{V}^G =+1/4$, and in a similar way to the first line in
\eqref{gravTT} (type II corresponds to $\alpha_{T}^G =-1/6$).

For the scalars and the gauge ghosts the type I and type II endomorphism
parameters coincide, $\alpha_S^M=\alpha_{V2}^M=0$. Therefore,
the ``dangers'' of type II apply for the related  two terms also for
type I coarse-graining. At this point, it is
interesting to note that, had one employed an interpolation scheme
based on the Euler-MacLaurin formula,  the scalar term had simplified very much
like the fermionic one does in the here used averaging interpolation:
\be
\cT^{\rm scalar} =  \frac {N_S} 2
 \frac 1 {(4\pi)^2}  (1+(\alpha_S^M+\tfrac 1 3 ) r ) \, .
\ee
With this behaviour the contributions of the scalars would be as
easily and semi-analytically taken into account as the ones
for the fermions here.\footnote{Although I did not find any interpolation
scheme which makes the transverse vector fields behave alike
one may entertain and further investigate the idea that matter fields in such
a setting might really contribute only constants and linear terms in the
curvature. If the sign of the linear term is the ``correct'' one, then
one could add an arbitrary amount of matter fields to a given pure
gravity solution and only change numbers, {\it i.e.},  without changing any
qualitative features of the solution.}

Last but not least, in order to obtain a global solution the moving singularities
in the second line of \eqref{gravTT}  and in both expressions in  \eqref{gravsinv}
need to cancel against the numerators, {\it cf.\ } the description of the
Frobenius method in, {\it e.g.}, \cite{Teschl}.
Even if this is arranged then these
three terms have leading quadratic behaviour.\footnote{However, there is one way
to avoid this: if $\varphi^\star \simeq r^2$ then
$\varphi^{\star \, \prime\prime\prime}  \to 0$ for  $r\to \infty$, and
the remaining two terms can be tuned to cancel. As a matter of fact, this 
is the mechanism how the global quadratic solution can be assumed.}

To summarise this discussion, especially with respect to the impact of
matter on the asymptotic behaviour, one notes that for type II coarse-graining
the generic leading behaviour on the right hand side of the flow equation
is quadratic. Given the form of the left hand side this implies that the
leading behaviour is then $\varphi^\star \simeq r^2\ln r $ for $r\gg 1$.
Note that this is different from the leading asymptotic behaviour found 
in \cite{Dietz:2012ic}. In this investigation, however, the linear split of the
metric and another gauge has been used.
As an advantage of using type II coarse-graining
one has that matter does not introduce any new
singularities, {\it i.e.}, the same counting of conditions with respect to the
solubility and the number of solutions for this non-linear differential
equation applies. For generic endomorphism parameters the leading asymptotic
behaviour of the matter contributions is linear, and thus will not qualitatively
change the leading asymptotic behaviour of the solution in the pure gravity
case. On the other hand, one introduces (even if one sets
$\alpha_S^M=\alpha_{V2}^M$ right away) one or two new singularities
which will make without specific choices ({\it e.g.}, to push them to values of
$r$ in which one is not interested, foremost to negative values) the differential
equation unsolvable. Note that for the transverse vectors and for the
Dirac fermions type I endomorphism parameters behave for this
purpose like general values. Therefore, if type I coarse-graining is used
the scalar fields and the gauge ghosts provide the leading quadratic terms. 
This results then in a leading behaviour $\simeq r^2 \ln r$ for the fixed function. 
It is then straightforward to verify that this is then selfconsistent, no terms 
growing faster than quadratic will be generated in the equation.
Analysing the flow equation for large $r$ one can infer the behaviour
\be
\label{AsympPhi}
\varphi^\star (r)  \simeq (2 \ln r - 1) \, r^2 + \cO \left( \frac {r^2}{\ln r} ,
\, \, r\, (\ln r)^2 \right)
\, .
\ee
It is interesting to note that potentially generated terms  of order 
$r^2 (\ln r)^3$ and of order $r^2 (\ln r)^2$ are canceled simultaneously.

\subsection{Numerical Solutions for Global Fixed Functions}

\begin{figure*}[ht!]
\includegraphics[width=0.49\textwidth]{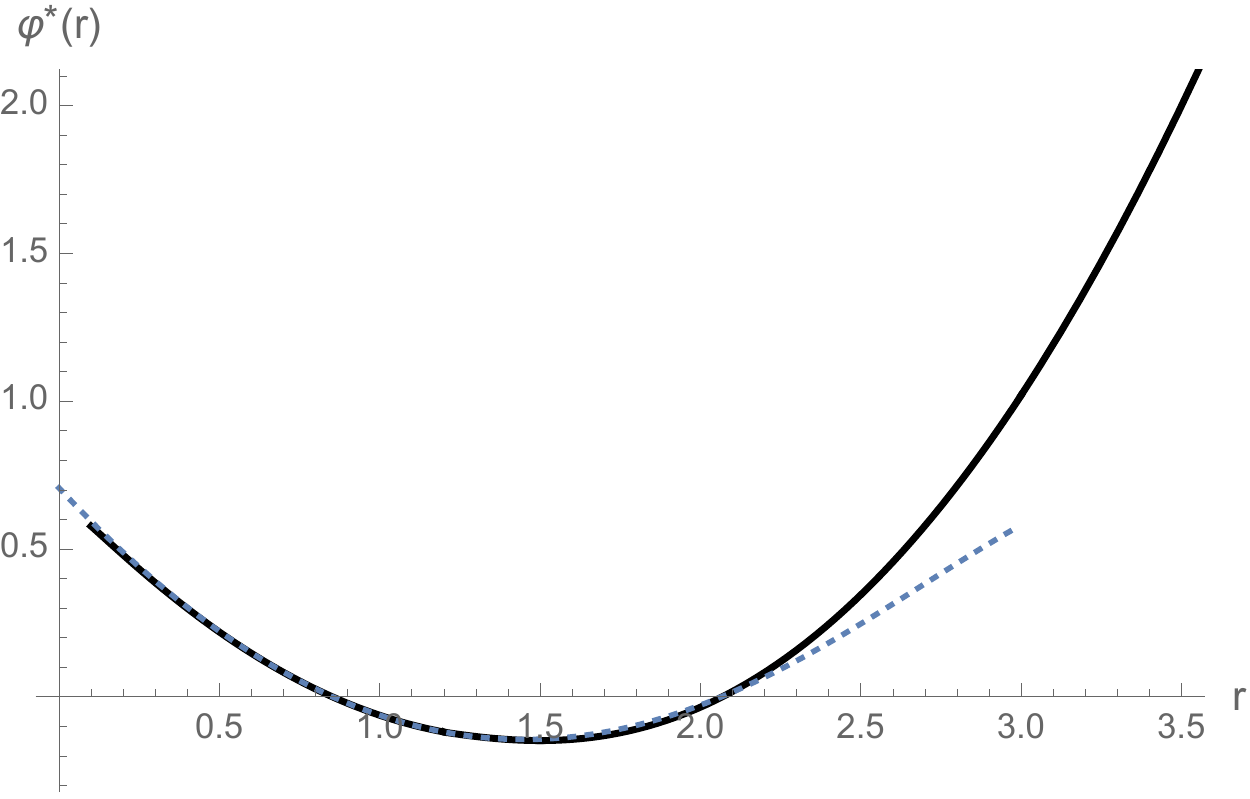}
\includegraphics[width=0.49\textwidth]{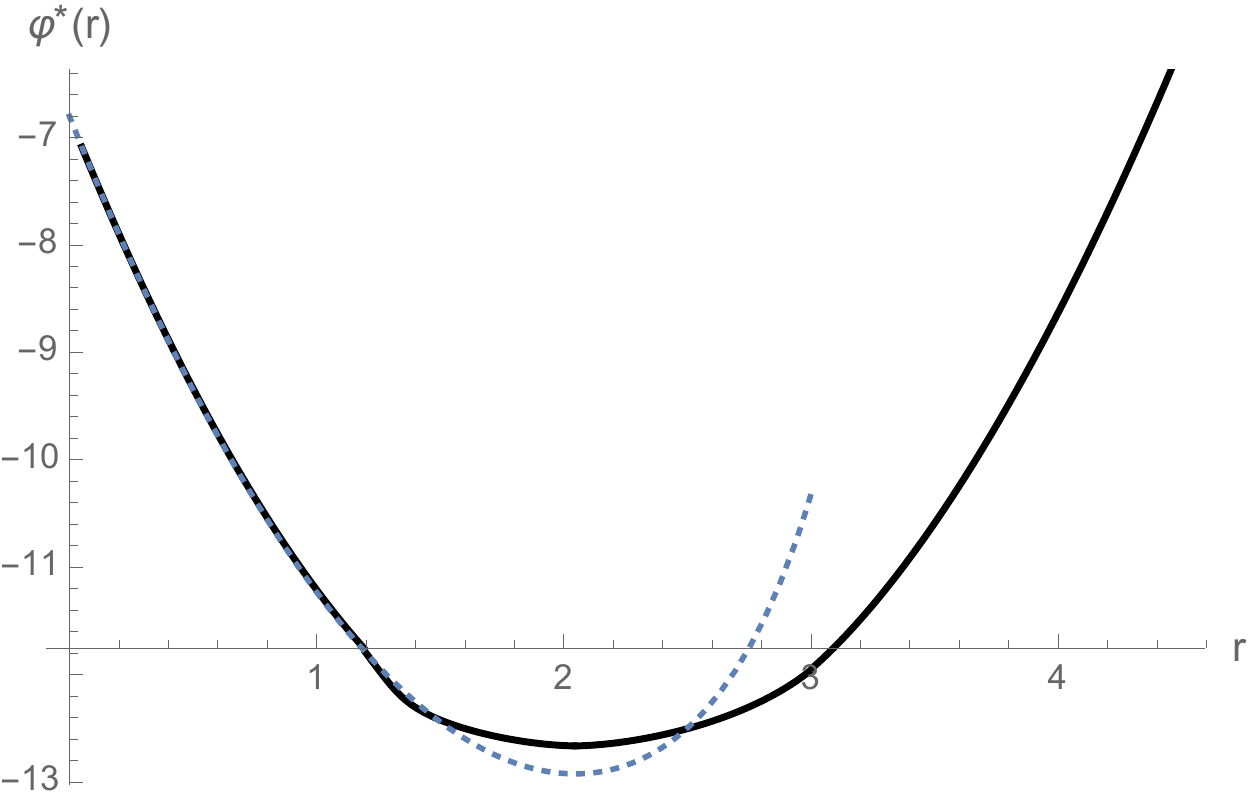}
\caption[Displayed  are the fixed functions for
the case of pure gravity and with SM matter
content. The respective  polynomial approximations of
order 14  using $r=0$ as
an expansion point are shown for comparison.]
{\label{FigNumSol}
Displayed  are the fixed functions (black lines) for
the case of pure gravity (left panel) and with SM matter
content (right panel). The respective  polynomial approximations of
order 14  using $r=0$ as
an expansion point are shown  as dashed lines for comparison.}
\end{figure*}

In this section two examples for a numerical solution will be presented,
one for pure gravity and another for SM content.
Given the fact that type II coarse-graining with SM matter content
is not going to work one may want to employ type I. However, already in
the pure gravity case the flow equation will not possess a solution for all
positive curvatures $r$.

The flow equation is a third order equation, and it is only then not over-constrained
if there are at most three singularities.
Therefore, if the solution had no extremum, and there were no moving singularity
one could allow for positive $r$ three fixed singularities.
However, the physical condition of a positive Newton constant
and thus a negative $g_1$ implies that $\varphi(r)$
decreases at small values of the curvature.
On the other hand, at large curvatures the function $\varphi (r)$ should assume
a positive value to make the functional integral well-defined which is achieved by
$\varphi(r)\to + \infty$ for $r\to \infty$. Consequently, $\varphi^\star (r)$
must possess at least one minimum, and one can allow for at most two fixed
singularities. However, for type I one has four additional fixed singularities at
$r^{\rm sing}_2 =0$,  $r^{\rm sing}_3 =6/5$, $r^{\rm sing}_4 =3$ and
$r^{\rm sing}_{\rm ghost} =4$,
where the last one originates from the ghost term $\cT^{\rm ghost}$.
Searching for solutions for strictly positive curvature one does not 
need to require
a condition at $r^{\rm sing}_2 =0$. The ghost singularity is  moved to negative
values of $r$ by choosing $\alpha_V^G=1/2$ which is well within the allowed
range of parameters, see above. This is then
the least modification of the flow equation as compared to the one
in type I coarse-graining which allows for a numerical solution.

To obtain numerical solutions  a multi-shooting method will be employed,
{\it cf.} \cite{Ohta:2015fcu,Demmel:2015oqa}. As for the pure gravity case:
to this end one enforces a minimum at the zero of the second summand in
$\cT^{\rm TT}$ \eqref{gravTT} at $r=3/2$. Shooting to the left
one constructs the solution left from the singularity by matching
the solution at $r^{\rm sing}_3\pm 10^{-4} = 6/5 \pm 10^{-4}$ such that
a singularity of the third derivative is avoided. This fixes one of the
two parameters. The other parameter is fixed by a similar matching 
at $r^{\rm sing}_4=3$. From there on the equation is straightforwardly 
integrated.\footnote{
Unfortunately, the asymptotic behaviour \eqref{AsympPhi}
only becomes reasonably precise at
very large values of $r$ and is thus only of limited use in the numerics.}
The result is displayed in the left panel of figure \ref{FigNumSol}.

When adding the SM matter content ($N_S=4, N_D=22.5$
and $N_V=12$ and type I coarse-graining)
one may plug in a polynomial approximation into the differential
equation for the fixed function to determine at which position the minimum
of $\varphi^\star(r)$ has to be located. From there the
analogous procedure as in the gravity case is followed. The result is displayed
in the right panel of figure \ref{FigNumSol}.

The most amazing observation hereby is the absence of any structure,
the global fixed functions are surprisingly close to parabolas, {\it i.e.},
all their features can be captured by a quadratic expression with only
three coefficients.

\section{Summary}
\label{sec:conclusions}

In this chapter I presented a study of the properties of NGFPs 
arising within $f(R)$-gravity minimally coupled to an 
arbitrary number of scalar, Dirac, and vector fields. The construction closely follows 
earlier work by Ohta, Percacci, and Vacca \cite{Ohta:2015efa,Ohta:2015fcu} covering the 
case of pure gravity: metric fluctuations are parameterised by the exponential split, the 
computation is carried out in ``physical'' gauge, and all operator traces are evaluated as 
averaged sums over eigenvalues. The result is the partial differential equation 
\eqref{pdf4d} which governs the scale-dependence of the dimensionless function 
$\varphi_k(r) \equiv f_k(R/k^2) k^{-4}$. The equation keeps track of a 7-parameter family 
of coarse-graining operators parameterising relative shifts of the eigenvalues associated 
with the fluctuations which are integrated out at the RG scale $k$. A direct 
consequence of the construction is that the gravitational sector of our partial 
differential equation agrees with \cite{Ohta:2015efa,Ohta:2015fcu}. 

Based on the partial differential equation \eqref{pdf4d},  a 
comprehensive picture detailing the existence and stability of interacting 
RG FPs in gravity-matter systems taking higher-order 
curvature terms into account has been developed.  
The main findings are summarised in table 
\ref{Tab.mainresults}. In the case where all coarse-graining operators are taken as the 
corresponding Laplacian operators, most of the matter sectors of phenomenological 
interest, including the SM of particle physics, admit a NGFP which is stable 
under the inclusion of higher-order curvature terms. In addition, they come 
with a low number of 
relevant directions. The fact that these gravity-matter FPs share many of the 
properties found in the case of pure gravity suggests to call this family of  
classes ``gravity-type'' NGFPs. 

In contrast to this success, the most commonly used set of non-trivial endomorphism 
parameters, given by the type II coarse-graining operators constructed from 
\eqref{endtypeII}, commonly leads to gravity-matter FPs which are unstable under 
the addition of higher order scalar curvature terms. While the instability of 
phenomenologically interesting gravity-matter FPs in the presence of a type II 
coarse-graining operator has already been observed several times,
see, {\it e.g.}, \cite{Dona:2012am,Dona:2013qba}, the present setup offers a logical
explanation: the inclusion of the $r^2$-terms reveals that the ``gravity-type'' NGFPs and 
the NGFPs found in the type II case {are not connected} by a continuous deformation 
of the coarse-graining operator, see figure \ref{Fig.typeIIannihilation}. The observation 
that the matter contributions destroy the typical behaviour found in the case of pure 
gravity suggests to refer to this family of FPs as ``matter-dominated'' NGFPs.

At this stage it is interesting to compare the classification of NGFPs in the Einstein-
Hilbert action obtained in this work (see figure \ref{fig.3}) with the one reported in 
\cite{Biemans:2017zca}.\footnote{Applying the approach taken in \cite{Biemans:2017zca} to 
the covariant setting leads to the same existence criteria for NGFPs in the 
$d_g$-$d_\lambda$-plane \cite{RSunpublished}.} The comparison reveals a qualitative 
difference in the FP structure for $d_g > 0, d_\lambda > 0$. Focusing on the 
type I case, the presented work shows no suitable FPs beyond the line $d_g = 179/12$ while 
\cite{Biemans:2017zca} identifies
suitable NGFPs in this region provided that $d_\lambda$ is sufficiently positive. This 
difference is related to the occurrence of the cosmological constant on the right hand 
side of the flow equation, manifesting itself in terms of denominators of the form 
$(1-c \lambda)$ with $c$ being a positive number. 
The resulting poles have been linked to a mechanism of gravitational catalysis 
\cite{Wetterich:2017ixo}.  
The comparison between this work and \cite{Biemans:2017zca} then reveals that these terms 
also play a crucial role in stabilising the NGFPs appearing in the upper-right corner of 
the $d_g$-$d_\lambda$ plane. Notably the NGFPs found in the case of pure gravity or 
gravity coupled to SM matter are not located in this region so that the 
stabilisation mechanism is not required to work in these cases.

For some well-chosen sets of coarse-graining
parameters global quadratic solution exists.  Although these
choices can hardly be motivated by physics
they explain an otherwise very astonishing behaviour of the
numerically obtained global fixed functions: for all studied
cases the deviations from a global quadratic form are
minor.\footnote{Respectively, the deviations from quadratic expressions augmented with a $r^2 \ln r$ term is even
smaller.} Given this situation one might even speculate that differences
to a quite simple form of the global fixed functions are only truncation
artefacts. This is in agreement with the basic assumption of the
asymptotic safety scenario of having only finitely many relevant
directions. In the present investigation we found only two
relevant directions, one of them is directly related to the constant term,
{\it i.e.}, the cosmological constant. The other one is with only very
small contributions from higher-order terms a linear
combination of a linear and a quadratic term.

\newpage
\thispagestyle{empty}


\chapter{{\color{MYBLUE}{Spectral Dimensions from the Spectral Action}}}\label{SpecDimSpecAct}
~\vspace{-10mm}

\centerline{This chapter is based on the following publication 
\cite{Alkofer:2014raa}:}
\vspace{3mm}
\centerline{N.~Alkofer, F.~Saueressig and O.~Zanusso.}
\centerline{Spectral dimensions from the spectral action.}
\centerline{Phys.\ Rev.\ D {91} (2015)  025025, arXiv:1410.7999 
[hep-th].}
\minitoc

\section{Motivation and Objective}
\label{sec:intro}

As briefly reviewed in section \ref{sec:SpecAct} almost-commutative geometries
lead to a spectral action as given in (\ref{SpAct1}). Its main ingredients
are a positive function $\chi$, a Dirac operator $\cD$ on an almost-commutative geometry,
and a physically motivated UV cutoff $\Lambda$.

The main objective of the study presented here is to calculate the generalised spectral dimension
$D_S(T)$ \cite{Ambjorn:2005db,Sotiriou:2011mu,Reuter:2011ah} from a spectral action
of the type (\ref{SpAct1}), and then  compare the obtained results to the corresponding 
ones of other approaches to quantum gravity. To this end, the work of  
\cite{Kurkov:2013kfa,Iochum:2011yq}, containing studies of the spectral action's properties 
in the far UV and thus beyond the framework of effective field theory, is followed, 
and the concept of  the generalised spectral dimension $D_S(T)$ is employed to provide a
quantitative measure for some of the more qualitative conclusions of 
\cite{Kurkov:2013kfa,Iochum:2011yq}. 

In appendix \ref{MathSpecDim} the spectral dimension and its relation to other types
of dimensions as well as some related basic formulas are summarised. In addition,
the generalisation to a ``time-dependent'' quantity $D_S(T)$ is defined and elucidated.
Over the last years this latter quantity which provides a measure for growth of a spectrum
has been calculated in many approaches to quantum gravity as well as  quantum gravity 
inspired models. The employed methods together with the key references for each of them
are also denoted in appendix \ref{MathSpecDim}.
A surprising feature common to many of the results of these investigations 
is a ``dynamical dimensional reduction'' when decreasing the ``time'' scale $T$: 
starting $D_S(T) = 4$ at macroscopic scales (as expected in a classical spacetime)  
one arrives at $D_S(T) = 2$ at the smallest length scales, for a synopsis of such 
observations for several different approaches to quantum gravity see, {\it e.g.},
 \cite{Carlip:2009kf,Carlip:2012md}. 
Clearly, in such a situation where seemingly different models and approaches provide
the same pattern it is highly desirable to obtain a detailed understanding
which features are actually encoded in $D_S(T)$ and how this relates to the quantum\
nature of the models' ingredients. Consequently, one objective of the study 
presented in this chapter is  to provide another example. In addition, 
a further and more important goal is to contribute to a better understanding of the
observed ``dynamical dimensional reduction'' seen in so many different approaches. 
As will become clear in the next two sections a fact which makes a model based on
a spectral action especially worthwhile to study is the unusual form of its
(inverse) propagators: they are non-analytic functions of the momentum. As we will
see this results in a quite astonishing behaviour of the spectral dimensions
presented below.

\section{The Spectral Action and its Bosonic Propagators}
\label{sec:SpecAct1} 

For the ``classical'' spectral action as given in (\ref{SpAct1}), 
$S_{\chi,\Lambda}\equiv \Tr \bigl( \chi(\cD^2/\Lambda^2) \bigr),$
with a given  function $\chi$ and a cutoff $\Lambda$,
a non-trivial spectral dimension can only arise if the propagators (which are as usual 
determined from the second variations of the action) acquire a  non-canonical momentum 
dependence. 
 For calculating this momentum dependence one may use
heat-kernel methods, see, {\it e.g.}, \cite{vandenDungen:2012ky,Kurkov:2013kfa}.

For constructing the spectral action one needs an almost-commutative product 
manifold $M \times F$, see section \ref{sec:SpecAct}. Hereby, the first factor  
$M$ is a Riemannian spin manifold  which takes the role of the Euclidean spacetime. 
The second factor $F$ is a finite, generally 
non-commutative, space which will be related to the internal degrees of freedom. 
The geometry of the manifold $M$ has an operator-algebraic description in terms of the 
so-called canonical triple
$(C^\infty(M), L^2(M,S), \cD) $
where, as usual,  $C^\infty(M)$ is the set of smooth functions on $M$, $L^2(M,S)$ is the 
Hilbert space
of square-integrable spinors on $M$, and $\cD$ is the (Euclidean) Dirac operator
acting on this Hilbert space. In an analogous way, the geometry of $F$ can be captured by 
a triple
$\left( \cA_F, \cH_F, D_F \right) \, , $
where $\cH_F$ is a finite-dimensional Hilbert space of complex dimension $N$, $\cA_F$ is 
an algebra
of $N \times N$ matrices acting on $\cH_F$, and $D_F$ is a hermitian $N\times N$ matrix.

For the purpose of determining the spectral dimension in this approach one can choose
the simplest internal space, taking $F$ as a single point. In this case the triple 
becomes $F = ({\mathbb C}, {\mathbb C}, 0)$, and correspondingly the Dirac operator 
 reduces to the one on $M$, {\it i.e.},
\be
\cD = \slashed{D} + \gamma_5 \, \phi \, ,
\ee
where the covariant derivative
\be
\slashed{D} = i \gamma^\mu \, \left( \nabla_\mu ^{LC} + i A_\mu \right) \, ,
\ee
contains the Levi-Civita spin connection \eqref{SpinConn} and the gauge potential $A_\mu$.
The resulting spectral action comprises a spin-2 field, the graviton, 
a massless $U(1)$ gauge field $A_\mu$ with field strength $F_{\mu\nu} = \p_\mu A_\nu - \p_\nu A_\mu$, 
and a scalar $\phi$. In a next step one projects on physical degrees of freedom, {\it i.e.},
for the graviton one takes into account  the transverse traceless fluctuations 
$h_{\mu\nu}$ with $\p^\mu h_{\mu\nu} = 0$, $\delta^{\mu\nu} h_{\mu\nu} = 0$ only, 
and the Landau gauge for the spin-1 field $\p^\mu A_\mu = 0$ is imposed.

The operator $\cD^2$ appearing in the spectral action (\ref{SpAct1}) can then be cast
into the standard form of a Laplace-type operator
\be
\cD^2 = -(\nabla^2 +E)
\ee
with the endomorphism $E$ given by
\be
E= -i \gamma^\mu \gamma_5 \nabla_\mu \phi - \phi^2 - \frac 1 4 R + \frac i 4
[\gamma^\mu, \gamma^\nu] F_{\mu\nu} .
\ee
The curvature related to $\nabla_\mu=\nabla_\mu^{LC}+iA_\mu$ is given by
\be
\Omega_{\mu\nu}  := [\nabla_\mu , \nabla_\nu ] = - \frac 1 4 \gamma^\rho \gamma^\sigma
R_{\rho\sigma\mu\nu} +i F_{\mu\nu} .
\ee 

A favourable choice for the function $\chi$ is $\chi(z) = e^{-z}$ \cite{Kurkov:2013kfa}. 
Then the spectral action (\ref{SpAct1}) coincides with the heat trace
\be
S_{\chi,\Lambda} = \Tr \left( e^{-{ t} \cD^2}\right) 
\quad \mathrm{with} \quad { t} := \Lambda^{-2} \, ,
\label{SpAct}
\ee
which is a well-studied object, see, e.g.,
\cite{Vassilevich:2003xt,Barvinsky:1987uw,Barvinsky:1990up,Iochum:2011yq,Codello:2012kq}.

\begin{figure}[h]
\centering
\includegraphics[width=0.63\textwidth]{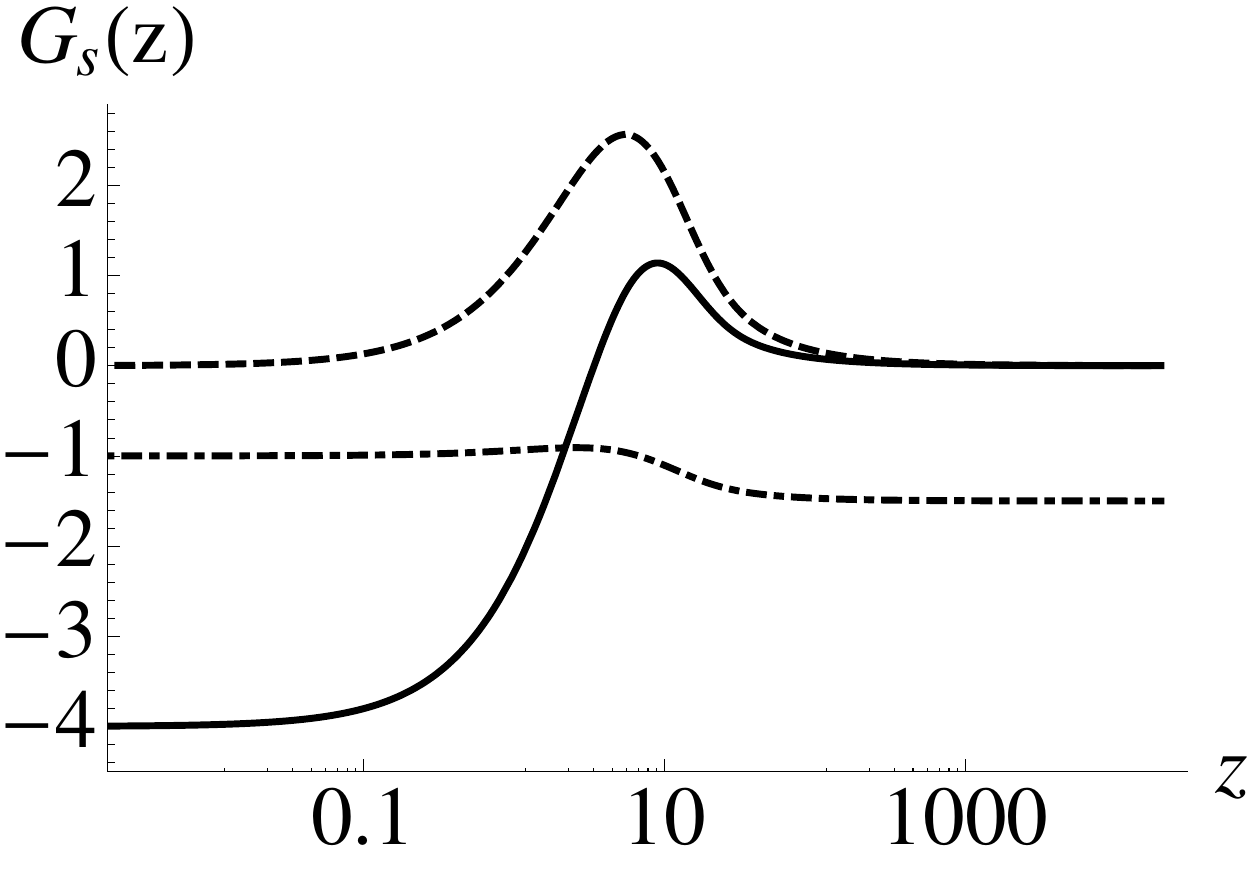}
\caption[Illustration of the momentum dependence of the structure functions ${ G}_{s}(z)$ for spins 0, 1 and 2.]
{\label{figsd_F012_F} Illustration of the momentum dependence of the structure functions ${ G}_{s}(z)$ (\ref{Gsfct}): spin 0 - solid thick line, spin 1- dashed thick line,  and spin 2 - dash-dotted thick line.}
\end{figure}

The propagators are extracted  by expanding \eqref{SpAct} up to second order
in the fields  $\phi$, $A_\mu$ and $h_{\mu\nu}$ where the latter is the 
fluctuating part of $g_{\mu\nu}$ around flat (Euclidean) space
\be\label{expmet}
g_{\mu\nu}= \delta_{\mu\nu} + \, \Lambda^{-1} \, h_{\mu\nu} \,   .
\ee
The inclusion of $\Lambda$ ensures that $h_{\mu\nu}$ has the same mass-dimension
as the matter fields and gives rise to the canonical form of the graviton propagator.
Comparing \eqref{expmet} with the standard expansion of $g_{\mu\nu}$ used in perturbation theory,
$g_{\mu\nu}= \delta_{\mu\nu} + \sqrt{16 \pi G_{\rm N}} \,  h_{\mu\nu} $ with $G_{\rm N}$ being
Newton's constant,
identifies the natural scale for $\Lambda$ as the Planck mass $m_{\rm Pl} = (8 \pi G_{\rm N})^{-1/2}$
(also see \cite{Devastato:2013wza} for a related discussion).
A straightforward although somewhat lengthy calculation
\cite{Kurkov:2013kfa,Barvinsky:1987uw,Barvinsky:1990up,Iochum:2011yq,Codello:2012kq}
yields the expression for the inverse propagators of the physical fields 
\ba
\label{Kmod}
K^{(2)}(\cD^2,{ t})  =   \int d^4x \Biggl[
 \phi F_{0} (- { t} \partial^2) \phi 
\; +  \,  A_\mu F_1(- { t} \partial^2)  A_\mu
+ \Lambda^{-2} \, h_{\mu\nu} F_{2} (- { t} \partial^2) h_{\mu\nu}
\Biggr] .
\label{K2}
\ea
Here the superscript on $K^{(2)}$ indicates that only the second order of the fields is retained.
Note that the factor $\Lambda^{-2}$  originates from the split of the metric \eqref{expmet}. The
structure functions $F_{s}$ describe the momentum dependence of the (inverse) propagators.
They coincide with the standard heat kernel result for spin-$s$ fields: 
\ba
F_{0} (z) &=& \frac { { t}^{-1}} {(4\pi)^2} \left( - 4 +  2 z h(z) \right),
\label{F0} \\
F_{1} (z) &=&  \frac { { t}^{-1}} {(4\pi)^2} \left(  -4 + 4 h(z) + 2 z h(z) \right),
\label{F1} \\
F_{2} (z) &=& \frac { { t}^{-2}} {(4\pi)^2} \left(  - 2  + h(z) +  \frac 1 4 zh(z)\right),
\label{F2}
\ea
where the function $h(z)$ can be written as the integral
\be
h(z)= \int_0^1 d\alpha \,e^{-\alpha (1-\alpha) z} .
\ee
It is important to note that the function $h(z)$ and thus the $F_{s}(z)$  are non-analytic in
\be
z= \frac{p^2}{\Lambda^2} = { t} \, p^2.
\ee
The inverses of these structure functions provide the classical propagators of the theory.
For illustration, suitably normalised versions of the structure functions 
$F_s(z)$ (\ref{F0}) - (\ref{F2}), 
\be\label{Gsfct}
G_s(z) = (4 \pi)^2 \, { t}^{\alpha_s} \, F_s(z) \, ,
\ee
with $\alpha_s = (1,1,2)$  are shown in figure \ref{figsd_F012_F}.

The structure functions $F_s(z)$  (\ref{F0}) - (\ref{F2}) possess an early-time expansion for
small momenta $p^2 \ll \Lambda^2$, i.e., $z < 1$. Using that
\be
h(z) = 1- \frac z 6+ \frac{z^2}{60}-\frac{z^3}{840} + \cO(z^4)
\ee
one finds
\be
\label{earlytime}
\begin{split}
(4\pi)^2 \, { t} \, F_{0} (z) =& - 4 + 2 z  - \frac{z^2}{3}  +
\frac{z^3}{30} + \ldots \, ,\\
(4\pi)^2 \, { t} \,  F_{1} (z) =& \frac 4 3 z -  \frac {4 }{15} z^2 +
\frac {1}{35} z^3  + \ldots \,  ,\\
(4\pi { t} )^2  \,  F_{2} (z) =& - 1 + \frac z{12} -  \frac {z^2}{40} +
\frac {z^3}{336} 
+ \ldots   \, .
\end{split}
\ee

\begin{figure}[h]
\centering
\includegraphics[width=0.63\textwidth]{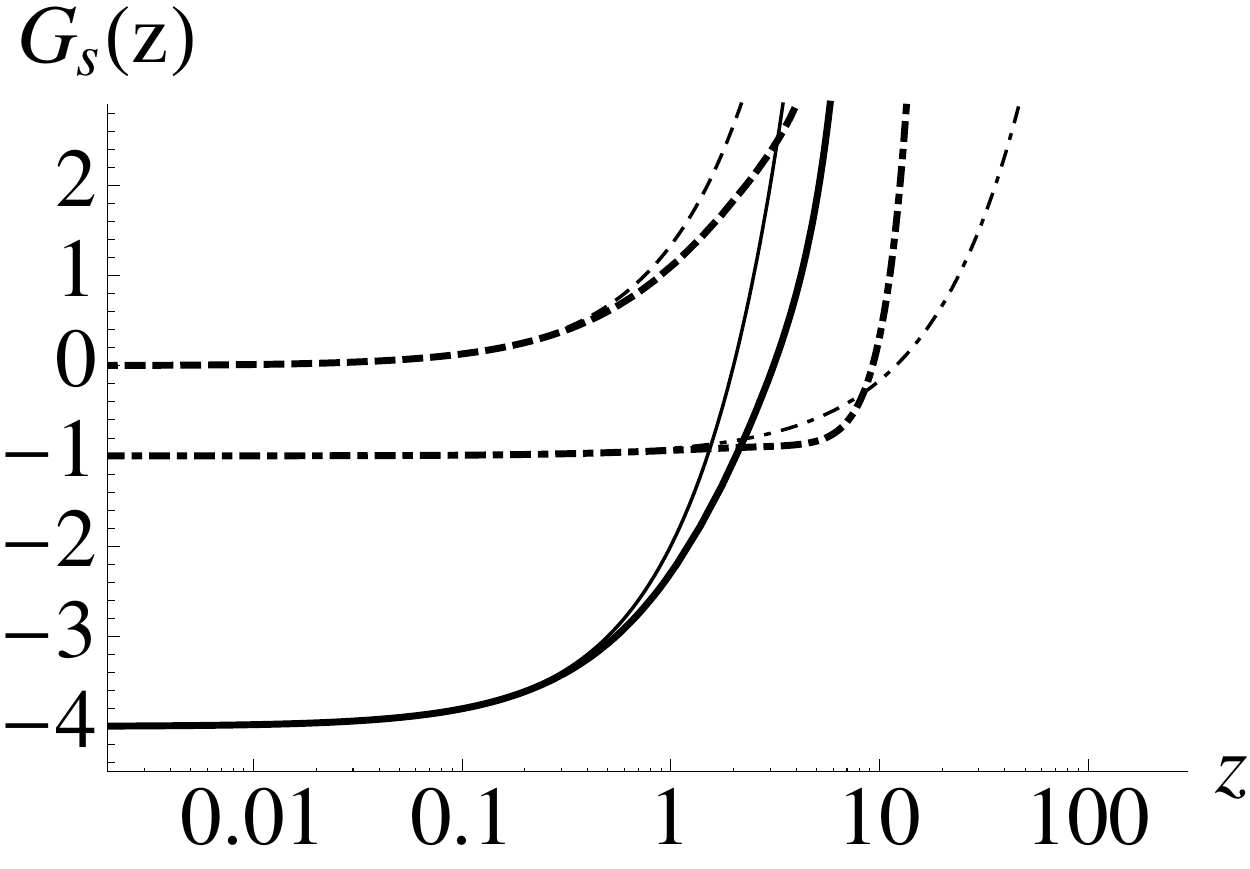}
\caption[Propagators obtained from truncating $G_s(z)$ at $z^3$,
 following the spirit of effective field theory.
The expansions up to linear order in $z$ are also shown.]
{\label{proptrunc} Propagators obtained from truncating $G_s(z)$ at $z^3$,
 following the spirit of effective field theory (spin 0 - solid thick line, spin 1- dashed thick line,  and spin 2 - dash-dotted). The expansions up to linear order in $z$ are shown as corresponding thin lines.}
\end{figure}

The early time expansion of the structure functions truncated at $\cO(z^3)$ and $\cO(z)$, respectively, 
is shown in figure \ref{proptrunc}. Comparing the result to the functions $F_s(z)$ including the full 
$z$-dependence it is easily seen that the truncation drastically modifies the behaviour of the (inverse)
propagators for large momenta. While the truncated $F_s(z)$ diverge the full structure functions remain finite. As it will be seen in section \ref{sec:IV}, this feature will have a drastic effect on the spectral dimension of the theory.

The constant term appearing
in the expansion of $F_{0}$ plays the role of a mass term for $\phi$. The sign thereby indicates
that the squared mass is negative. This is a remnant of the fact that the scalar $\phi$ acquires a
non-trivial vacuum expectation value via the Higgs mechanism
\cite{Chamseddine:1996rw,Chamseddine:2012sw,Estrada:2012te}. Since $K^{(2)}$, by construction,
contains the terms quadratic in $\phi$ only
the scalar potential is not included in \eqref{K2} so that the stabilisation of $\phi$
cannot be demonstrated in this approximation. The constant
term in $F_2$ signals the presence of a positive cosmological constant, acting like a
mass-term for the graviton, while the structure of $F_1$ reflects that the gauge field is massless.

To understand the behaviour of the theory at high energies, it is also useful to
carry out the late-time expansion of the structure functions, capturing the behaviour
for $z \gg 1$:
\be
h(z) = \frac 2 z+ \frac 4{z^2}+ \frac{24}{z^3}+ \frac{240}{z^4}+ \ldots
\ee
which yields
\be
\label{latetime}
\begin{split}
(4\pi)^2 \, { t} \,  F_{0} (z) =&  \frac 8 z + \frac {48}{z^2} + \frac{480}{z^3}+\ldots \, ,\\
(4\pi)^2 \, { t} \,  F_{1} (z) =& \frac {16} {z} + \frac{64}{z^2} + \frac {576}{z^3} +\ldots \, ,\\
(4\pi { t} )^2  \,  F_{2} (z) =&  - \frac 3 2 + \frac 3 z +  \frac {10}{z^2} + \frac {84}{z^3}  +
\ldots \, . \\
\end{split}
\ee

A typical viewpoint adopted in the spectral action approach to particle physics consider
the actions generated by \eqref{SpAct1} as effective actions which should be truncated
at a certain power of the cutoff $\Lambda^{-2}$. In this case the early-time expansion
\eqref{earlytime} allows to construct the effective action resulting from
an arbitrary function $\chi$. This uses the fact that the Trace \eqref{SpAct1} can
be related to the heat kernel \eqref{SpAct} using
\be
\label{mellin1}
S_{\chi, \Lambda} =  \, \Tr \bigl[ \chi (\cD^2/\Lambda^2 ) \bigr] 
= \int_0^\infty dy \, \tilde{\chi}(y) \, \Tr \bigl[ e^{-y \cD^2/\Lambda^2} \bigr] \, ,
\ee
where $\tilde{\chi}(y)$ is the inverse Laplace-transform of $\chi(z)$.
Evaluating the operator trace in \eqref{mellin1} based on the early-time expansion then yields
the systematic expansion of $S_{\chi, \Lambda}$ in (inverse) powers of the cutoff. 
The $\chi$-dependent coefficients
in this expansion are given by
\be
\label{Qdef}
\begin{split}
Q_n[\chi] \equiv & \, \int_0^\infty dy \, y^{-n} \, \tilde{\chi}(y)
\end{split}
\ee
and can be computed by standard Mellin-transform techniques \cite{Codello:2008vh}.
Thus the $Q_n \equiv Q_n[\chi]$ are real numbers which are normalised such that
 $Q_n=1$ for $\chi=\exp(- { t} z)$.

The part of the spectral action containing the terms quadratic in the fields
is then given by
\be
S^{(2)}_{\chi,\Lambda} = \frac{\Lambda^2}{(4\pi)^2} \int d^4x \Bigl[
  \phi \, \cF_{0,\chi}\left(-\partial^2/\Lambda^2\right) \phi 
  +  A_{\mu} \cF_{1,\chi}\left(-\partial^2/\Lambda^2\right)  A_{\mu} \label{Kmod2} 
  + h_{\mu\nu} \cF_{2,\chi}\left(-\partial^2/\Lambda^2\right) h_{\mu\nu}
\Bigr] 
\ee
with
\be
\label{earlytime2}
\begin{split}
\cF_{0,\chi} (z) =& \, - 4 Q_1 +2 {Q_0} z - \frac{Q_{-1}}{3}  z^2  + \frac{Q_{-2}}{30} z^3 + \dots \, ,\\
\cF_{1,\chi} (z) =& \, \frac {4Q_0} 3 z -  \frac {4 Q_{-1}}{15} z^2 + \frac {Q_{-2}}{35} z^3  + \dots \,  ,\\
\cF_{2,\chi} (z) =& \, - Q_2 + \frac {Q_1}{12}z -  \frac {Q_{0}}{40}z^2 + \frac {Q_{-1}}{336}z^3 + \dots \, .\\
\end{split}
\ee
The momenta \eqref{Qdef} can be adjusted by choosing a suitable function $\chi$.
Note, however, that the $Q_n$'s appearing in the matter and gravitational sector of
\eqref{Kmod2} \emph{cannot be adjusted} independently, however, since they are
generated by the same function $\chi$.
Possible truncations of the expansion \eqref{Kmod} include:

\noindent
{\it Truncating the moments $Q_n$.}
From the mathematical viewpoint it is tempting to choose a generating function $\chi$ whose moments
$Q_n[\chi]$ vanish for all values $n \ge n_{\rm max}$.
This leads to the rather peculiar property that the highest powers of  $-\p^2$ appearing in the matter and gravitational sector \emph{come with opposite signs}. In other words adjusting the $Q_n$ in such a way that the propagators in the matter sector are stable at high momenta implies an instability
of the gravitational propagator and vice versa. Thus ``truncating'' the theory by adjusting the momenta $Q_n$ gives rise to a dynamical instability of the
theory.

\noindent
{\it The effective field theory viewpoint.}
A similar (though not equivalent) strategy interprets the expansion \eqref{Kmod}
as an effective field theory, which should be truncated at a given power of the cutoff $\Lambda$.
Retaining the relevant and marginal operators then provides a good description
of the physics as long as $-p^2/\Lambda^2 \ll 1$. While it is possible
to systematically compute quantum corrections to an effective action, this expansion breaks down if the momenta are of the order of the Planck scale. A detailed analysis then reveals that the $Q_n$'s can be adjusted in such a way that all propagators of the theory are stable. Thus this case will be focused on the sequel.

\section{The Generalised Spectral Dimension}
\label{sec:specdim}

As explained in section \ref{SpecDimUV} 
a test particle diffusing on a given fixed background
feels certain features of this background as, e.g., its dimension. 
For a spin-less test-particle performing a Brownian random
walk on a Riemannian manifold with metric $g_{\mu\nu}$, the
diffusion process is described by the heat kernel $K_g(x,x^\prime;T)$
which gives the probability density for a particle diffusing from the
point $x$ to $x^\prime$ in the diffusion time $T$. The heat kernel
satisfies the heat equation
\be
\label{diffeq}
 \left( \p_T + \Delta_g \right) K_g(x,x^\prime;T) = 0 \, , 
\ee
where $\Delta_g \equiv -D^2$ is the Laplace-Beltrami operator,
and the initial condition is $K_g(x,x^\prime;0) = \delta(x-x^\prime)$. 
In flat space, the solution of this equation is ({\it cf.} section \ref{SpecDimUV})
\be
\label{flatspace1}
K(x,x^\prime;T) = \int \frac{d^dp}{(2\pi)^d} \, e^{ip \cdot(x-x^\prime)} \, e^{-p^2 T} \, .
\ee
In general $K_g(x,x^\prime;T)$ is the matrix element of the operator $\exp(-T \Delta_g)$.
For the diffusion process, its trace per volume gives the averaged return probability
\be
{\cal P}_g(T) =  \, V^{-1} \, \int d^dx \sqrt{g(x)} K_g(x,x;T) 
=  \, V^{-1} \, {\rm Tr} \, \exp(-T \Delta_g) \, ,
\ee
measuring the probability that the particle returns to its origin after a diffusion time $T$.
Here $V \equiv \int d^dx \sqrt{g(x)}$  denotes the total volume. For the flat-space solution
\eqref{flatspace1}
\be
{\cal P}(T) = ( 4 \pi \, T)^{-d/2} \, .
\ee

The (standard) spectral dimension $d_S$ is defined as the $T$-independent
logarithmic derivative
\be
\label{spec1}
d_S \equiv -2 \lim _{T\to 0} \, \frac{ \partial \ln {\cal P}(T)}{\partial \ln T} \, .
\ee
On smooth manifolds $d_S$ agrees with the topological dimension of the manifold $d$.
In order to also capture the case of diffusion processes exhibiting multiple scaling
regimes, it is natural to generalise the definition \eqref{spec1} to the $T$-dependent
spectral dimension
\be\label{spec2}
D_S(T) \equiv -2 \, \frac{ \partial \ln {\cal P}(T)}{\partial \ln T} \, .
\ee

In the classical spectral action \eqref{SpAct1} the propagation of the test
particles on a flat Euclidean background is modified by the higher-derivative
terms entering into the (inverse) propagators of the fields. In \eqref{diffeq}
this effect can readily be incorporated by replacing the Laplace-Beltrami operator
by the inverse propagators
\be
\label{diffeq2}
 \left( \p_T + F(-\p^2) \right) K_g(x,x^\prime;T) = 0 \, .
\ee
The solution of this equation can again be given in terms of its Fourier transform
\be
\label{flatspace2}
K(x,x^\prime;T) = \int \frac{d^dp}{(2\pi)^d} \, e^{ip \cdot(x-x^\prime)} \, e^{- F(p^2) T} \, .
\ee
For a generic function $F(p^2)$ there is no guarantee that the resulting
heat-kernel is positive semi-definite thereby admitting an interpretation as probability
density. This ``negative probability problem'' has been discussed in detail \cite{Calcagni:2013vsa,Calcagni:2014wba},
concluding that the spectral dimension remains meaningful.
The probability ${\cal P}(T)$ resulting from \eqref{flatspace2} is given by
\be
\label{returnprop}
{\cal P}(T) = \int \frac{d^dp}{(2\pi)^d} \, e^{- F(p^2) T} \, ,
\ee
and may still admit the interpretation of a (positive-semidefinite) return probability
even in the case where a probability interpretation of $K(x,x^\prime;T)$ fails. The generalised
spectral dimension may then be obtained by substituting the inverse propagators from \eqref{Kmod} and evaluating \eqref{spec2} for the corresponding return probabilities.

Following the ideas of \cite{Amelino-Camelia:2013gna,Amelino-Camelia:2013cfa} the spectral dimension
arising from \eqref{returnprop} permits an interpretation as the Hausdorff-dimension of the
theory's momentum space. Provided that the change of coordinates $k^2 = F(p^2)$ is bijective,
the inverse propagator in the exponential can be traded for a non-trivial measure on momentum space
\be
P(T) = \frac{{\rm Vol}_{S^d}}{(2\pi)^d} \, \int k dk \, \frac{\left(F^{-1}(k^2)\right)^{(d-2)/2}}{F^\prime(p^2)} \, e^{-T k^2} \, .
\ee
(NB: Here $F^\prime(p^2)$ is understood as the derivative of $F(z)$ with respect to its argument,
evaluated at $p^2 = F^{-1}(k^2)$.) Therefore \eqref{returnprop} is equivalent to 
the corresponding quantity for the case of a particle with canonical inverse propagator, 
$F(p^2) \propto p^2$ in a momentum space with non-trivial measure. 
This picture also provides a meaningful interpretation of $D_S(T)$ even in the case where 
the model is purely classical so that the non-trivial spectral dimension
cannot originate from properties of an effective quantum spacetime.

Due to the following argument it is helpful for the purpose of this study to split off the mass
term $m^2 = F(p^2)|_{p^2 = 0}$ in  the inverse momentum-space propagator $F(p^2)$:
\be
\label{masssplit}
F(p^2) = F^{(0)}(p^2) + m^2 \, .
\ee
Based on $F^{(0)}(k^2)$ one can then introduce the return probability
\be
\label{P0exp}
P^{(0)}(T) \propto \int \frac{d^4p}{(2\pi)^4 } \, e^{-TF^{(0)}(p^2)}
\ee
together with the spectral dimension seen by the massless field
\be
\label{Ds0}
D_S^{(0)}(T) \equiv -2  T \frac \partial {\partial T} \ln {\cal P}^{(0)}(T) \, .
\ee
Substituting \eqref{masssplit} into the return probability \eqref{returnprop}
 and extracting the mass-term from the integral
it is straightforward to establish
\be
\label{d0def}
D_S(T) = 2 \, m^2 \, T + D_S^{(0)}(T) \, .
\ee
Thus a mass-term just leads to a linear contribution in $D_S(T)$ and does not encode non-trivial information on the propagation of the particle. Therefore the quantity $D_S^{(0)}(T)$ will
be studied in the following.

\section{The Spectral Dimension from the Spectral Action}
\label{sec:IV}

Based on the discussion of the last two sections, it is now straightforward to compute the spectral dimensions from the spin-dependent propagators provided by the spectral action. I will start by investigating the truncated propagators based on \eqref{Kmod} and \eqref{Kmod2} in subsection \ref{sec:SpecAct2}
before including the full momentum dependence in subsection \ref{sect.IVb}.

\subsection{Effective Field Theory Framework}
\label{sec:SpecAct2}

In the effective field theory interpretation of \eqref{Kmod2} the functions
$\cF_{s,\chi}$ are truncated at a fixed power of the cutoff $\Lambda$. The resulting
massless parts of the bosonic propagators then become polynomials in
the particles' momentum,
\be
\label{fexp}
\cF_s^{(0)}(p^2) = \sum_{n=1}^{N_{\rm max}} \, a_n^s \, (p^2)^n \, ,
\ee
with obvious relations among the polynomial coefficients $a_n$ and the numbers $Q_n$
\eqref{earlytime2}. Note that limiting the expansion to the marginal and relevant operators,
coming with powers of the cutoff $\Lambda^{2n}$, $n \le 2$, fixes $N_{\rm max} = 1$ and
all propagators retain their standard $p^2$-form. Taking into account, however, power-counting
irrelevant terms containing inverse powers of the cutoff adds further
powers to the polynomial \eqref{fexp}. Thus the propagators include higher powers of momentum
in this case.

A positive semi-definite spectral dimension $D_S^{(0)}$ requires a positive function
$\cF_s^{(0)}$. This requirement puts constraints on the signs of the momenta $Q_n$ appearing
in \eqref{earlytime2}. In particular $a_1^s > 0$ is required for obtaining classical propagators
at low energies while $a_{N_{\rm max}}^s > 0$ is needed for stability at high energies.

\begin{figure}[h]
\centering
\includegraphics[width=0.63\textwidth]{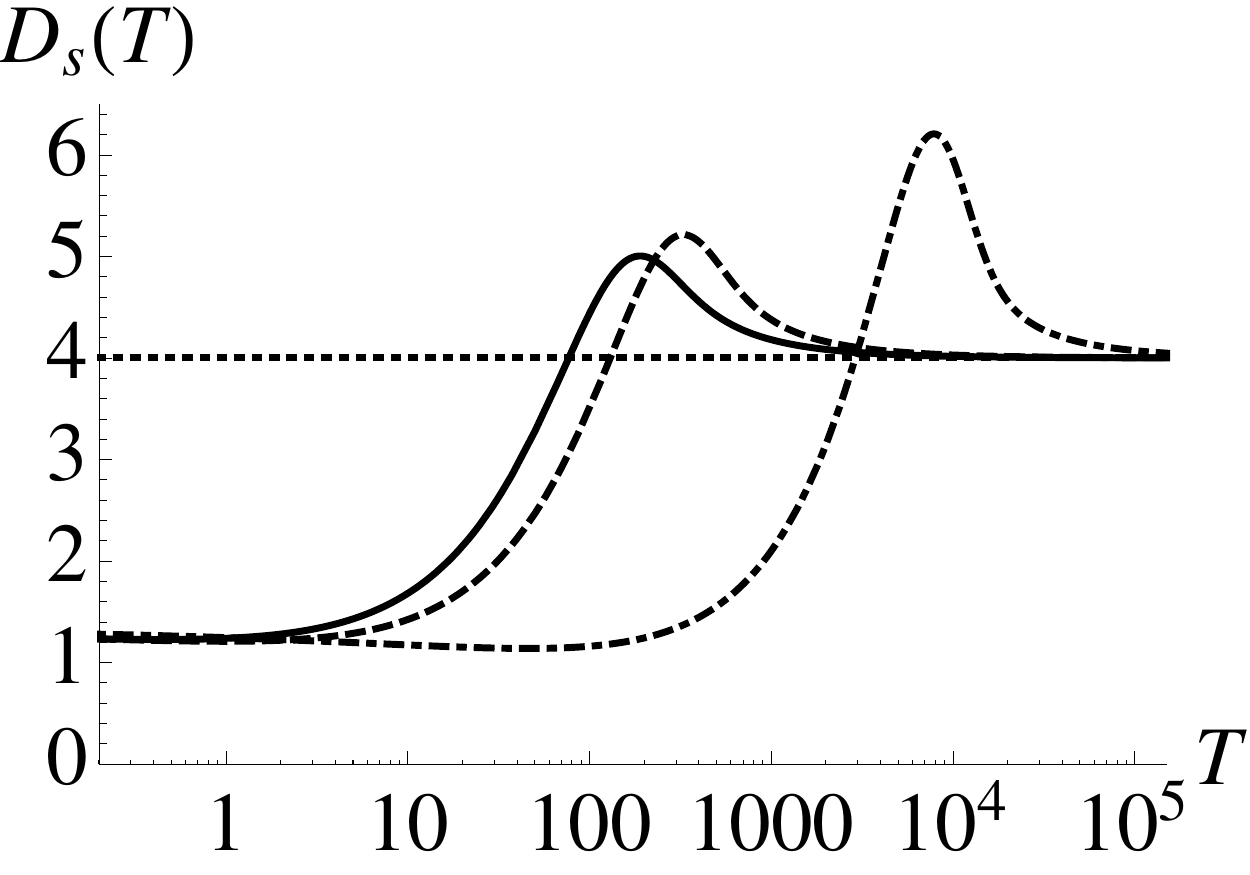}
\caption[The spin-dependent spectral dimension $D_S^{(0)}(T)$
with $N_{\rm max} = 1$ and $N_{\rm max} = 3$.]
{\label{figsd_F012_D} The spin-dependent spectral dimension $D_S^{(0)}(T)$
obtained from \eqref{fexp} with $N_{\rm max} = 1$ (dotted line) and $N_{\rm max} = 3$,
$Q_n = 1$, $n=1,0,-1,-2$ for spin 0 - solid line, spin 1- dashed line,  and spin 2 - dashed-dotted
line. The cross-over scale is set by ${ t} = \Lambda^{-2}$ and has been normalised to $ { t} =1$.
}
\end{figure}

The asymptotic behaviour of $D_S^{(0)}$ for short (long) diffusion time $T$
is governed by the highest (lowest) power of $p^2$ contained in \eqref{fexp}. Evaluating
\eqref{P0exp} and \eqref{Ds0} for the special cases $\cF^{(0)}_s(p^2) \propto p^2$ and $F(p^2) \propto (p^2)^{N_{\rm max}}$, a simple
scaling argument establishes
\be
\label{Dslimits}
 \lim_{T \rightarrow \infty} \, D_S^{(0)}(T) = 4  \quad {\mathrm {for}} \quad a_1 > 0 \, ,
\qquad {\mathrm {and}} \quad
 \lim_{T \rightarrow 0} \, D_S^{(0)}(T) = \frac{4}{N_{\rm max}} \, .
\ee
Thus $a_1 > 0$ ensures that the spectral dimension seen by particles for long
diffusion times coincides with the topological dimension of spacetime. Including
higher powers of momenta decreases $D_S^{(0)}(T)$ for short diffusion times. The
generalised spectral dimension then interpolates smoothly between these limits. This
feature is illustrated in figure \ref{figsd_F012_D}.

The case $N_{\rm max} = 1$, $a_1 > 0$ leads to a spectral dimension
which is independent of the diffusion time (dashed line). The spectral dimension
obtained for $N_{\rm max} = 3$ and momenta $Q_n = 1$, $n=1,0,-1,-2$. In this case,
$D_S^{(0)}(T)$ interpolates between $4$ at large $T$ and $4/3$ for small $T$ respectively.
The crossover occurs for $T/ { t} \approx (4\pi)^2$ (spin 0 and 1), resp.,
$T/ { t}^2 \approx 10 \times (4\pi)^2$ (spin 2). Notably, it is only the shape
of this crossover, which depends on the spin of the particle, while the asymptotic limits are universal for all spins.

\subsection{Propagators with Full Momentum Dependence}
\label{sect.IVb}

Let me now take a step beyond effective field theory and investigate
the spectral dimension arising from the inverse propagators \eqref{F2}
including the full momentum dependence. Comparing figures \ref{figsd_F012_F} and \ref{proptrunc} the structural difference is immediate: in the effective field theory framework $F_s(z)$ diverges as $z \rightarrow \infty$ while the inclusion
of the full momentum dependence renders $\lim_{z \rightarrow \infty} F_s(z)$ finite with the limit given by the leading term in \eqref{latetime}. As a consequence of the modified asymptotics, the integral \eqref{P0exp} diverges at the upper boundary, since the contribution of large momenta is no longer exponentially suppressed once the full momentum dependent propagators 
are considered.

In order to still be able to analyse the spectral dimension arising
in this framework, one regulates  \eqref{P0exp} by introducing
an UV cutoff $\Lambda_{\rm UV}$,
\be
\label{P0reg}
P^{(0)}(T; \Lambda_{\rm UV}) = \int^{\Lambda_{\rm UV}} \frac{d^4p}{(2\pi)^4} \, e^{-T F^{(0)}(p^2)} \, .
\ee
In the spirit of the discussion leading to \eqref{d0def}, I consider
the ``massless'' structure functions $F_s^{(0)}(z)$
where the constant terms appearing in the late-time expansion \eqref{latetime}
have been removed. The return probability \eqref{P0reg} then allows to
construct the spectral dimension as a function of $\Lambda_{\rm UV}$
\be
D^{(0)}_S(T; \Lambda_{\rm UV}) = - 2 \, T \, \p_T \, \ln(P^{(0)}(T; \Lambda_{\rm UV})) \, . 
\ee
 The scale $\Lambda$ appearing in the spectral action \eqref{SpAct1} is thereby held fixed and sets the transition scale between the UV and IR regime. A detailed analytical and numerical analysis based on
the expansion \eqref{latetime} then establishes
\be
\begin{split}
\lim_{\Lambda_{\rm UV} \rightarrow \infty} D^{(0)}_S(T; \Lambda_{\rm UV}) = 
  \lim_{\Lambda_{\rm UV} \rightarrow \infty} \, 4 T \, F_s^{(0)}(\Lambda_{\rm UV}) \, . 
\end{split}
\ee
Based on the late-time expansion \eqref{latetime}, one can conclude that including the full momentum dependence
in the structure function leads to a spectral dimension which \emph{vanishes for all diffusion times $T$}:
\be
D^{(0)}_S(T) = \lim_{\Lambda_{\rm UV} \rightarrow \infty} D^{(0)}_S(T; \Lambda_{\rm UV}) = 0 \, .
\ee
This result entails in particular that there is no scaling regime for large diffusion time where the spectral dimension matches the topological dimension. Thus including the full momentum dependence, one does not recover a ``classical regime'' were the spectral dimension would indicate the onset of classical low-energy physics.

\section{Summary}
\label{sec:Concl}
  
In this chapter a generalised spectral dimension $D_S(T)$ describing the propagation 
of (massless) scalars, vectors and gravitons based on the \emph{classical} spectral action \eqref{SpAct1}
has been calculated. These obtained results can be summarised as follows:
\begin{itemize}
\item 
If the spectral action is interpreted as an effective field theory restricted to the power-counting 
relevant and marginal terms, the generalised spectral dimension is independent of the diffusion time $T$ 
and matches the topological dimension of spacetime, {\it i.e.},  $D_S(T) = 4$.
\item 
If the effective field theory framework is extended to also include power-counting irrelevant terms, the 
generalised spectral action interpolates between $D_S(T) = 4$ for long diffusion time and 
$D_S(T) = 4/N_{\rm max}$ for short diffusion times. $N_{\rm max}$ is determined by the highest 
power of momentum contained in the propagator, $(p^2)^{N_{\rm max}}$. The crossover between these two 
asymptotic regimes is set by the cutoff $\Lambda$, and its shape explicitly depends on the spin of the 
propagating particle.
\item If the full momentum-dependence of the propagators is taken into account, the generalised spectral dimension becomes independent of the spin and vanishes identically $D_S(T) = 0$.
\end{itemize}
The last feature can be traced back to the fact that the full propagators approach a constant for momenta 
much larger than the characteristic cutoff scale $\Lambda$. In \cite{Kurkov:2013kfa} the peculiar 
behaviour was summarised by the pictorial phrase ``high-energy bosons do not propagate''. Indeed the
vanishing of the spectral dimension suggests that the momentum space of the 
theory resembles the one of a zero-dimensional field theory, or phrased otherwise,
to a picture where points in spacetime do not communicate.

It is interesting to note that 
the vanishing of the spectral dimension, $D_S(T) = 0$, is in agreement with the previous computations
obtained for other non-commutative spacetimes \cite{Alesci:2011cg}.
This is a welcome and surprising result, since the non-commutative nature of our spectral triple
construction differs substantially from that of \cite{Alesci:2011cg}, see also the discussion below in
chapter \ref{Chap:Concl}. 

Let me stress, however, that all computations carried out in this chapter are at the classical level. 
In particular the spacetime is given by classical Euclidean space. All effects are due to the 
change of measure in momentum space reflected by non-canonical form of the classical propagators, 
and thus do not capture properties of an underlying quantum spacetime. Further potential investigations
related to this point will be discussed in chapter \ref{Chap:Concl}. 
Nevertheless, I would like to end this chapter by expressing the expectation that   
the spectral action, comprising the three scenarios discussed above, provides a valuable 
test of the generalised spectral dimension as a novel candidate for a quantum gravity observable. 



\chapter{{\color{MYBLUE}{Quantum Gravity Signatures in the Unruh Effect}}}
\label{Chap6}
~\vspace{-10mm}

\centerline{This chapter is based on the following publication 
\cite{Alkofer:2016utc}:}
\vspace{3mm}
\centerline{N.~Alkofer, G.~D'Odorico, F.~Saueressig and F.~Versteegen.}
\centerline{Quantum Gravity signatures in the Unruh effect.}
\centerline{Phys.\ Rev.\ D {94} (2016)  104055, arXiv:1605.08015 [gr-qc].}
\minitoc

\section{Objective}
\label{intro}

In the last chapter the concept of a spectral dimension has been used to study the dimensional flow related 
to the classical (tree-level) propagators of the spectral action. 
However, as already noted, the spectral action provides only one out of many examples for
dimensional flows. They seem to be a property which can be found in virtually all approaches to quantum 
gravity and  quantum gravity inspired models if one is looking for it \cite{Carlip:2009kf,Carlip:2012md}. 
Being a rather theoretical concept the question arises whether any observable phenomenon is related 
to such a dimensional flow. The objective of the study presented here is to obtain an answer to 
the question: is there any experiment, or at least any ``Gedankenexperiment'', which makes
such a dimensional flow detectable? 
It will be seen in the course of this chapter that actually it is possible to exploit the Unruh 
effect  \cite{Fulling:1972md,Davies:1974th,Unruh:1976db} (also see 
\cite{Crispino:2007eb,Birrell:1982ix,Takagi:1986kn} for reviews)  
to make dimensional flows, at least in principle, observable. 

The Unruh effect is a quite intriguing phenomenon. In principle, it occurs within QFT 
in flat Minkowski spacetime, however, as it is related to an accelerated observer it is advantageous
to change the perspective and use the so-called Rindler space (see, {\it e.g.}, 
\cite{Mukhanov:2007zz} for a definition) as basis for the calculation. 
To summarise the Unruh effect in a brief manner: it predicts that to an accelerated observer 
(Rindler observer) the Minkowski vacuum appears as a thermal state whose temperature is proportional 
to the eigen-acceleration of the observer.
This acceleration radiation can leave imprints in a variety of phenomenological contexts:
for instance in the transverse polarisation of electrons and positrons in particle storage rings 
(Sokulov-Ternov effect) \cite{Akhmedov:2006nd,Akhmedov:2007xu},
at the onset of quark gluon plasma formation due to heavy ions collisions \cite{Kharzeev:2005iz},
on the dynamics of electrons in Penning traps, of ultra-intense lasers, and atoms in microwave cavities
(see \cite{Crispino:2007eb} and references therein), or in the Berry phase acquired by the accelerated 
detector \cite{MartinMartinez:2010sg}.

The standard way to derive the Unruh effect goes via defining creation and annihilation operators 
with respect to the positive and negative frequency modes associated with the Minkowski and Rindler space,
respectively, and relating them through a Bogoliubov transform, see, {\it e.g.},  
\cite{Mukhanov:2007zz} for a didactical treatment. Below we will follow, however, the detector approach
\cite{Agullo:2010iq}. Then it is evident that the origin of the thermal spectrum is, as will 
become clear in the following on the basis propagator properties \cite{Wipf:1998ss}, 
purely geometrical: it depends only on the presence of an horizon in the 
Rindler frame. As a geometric effect, the Unruh temperature is insensitive to the specific form of the 
action under consideration, and thermality of the spectrum is ensured by Lorentz invariance and the 
generic form of Lorentz transformations \cite{Unruh:1983ac}. As will be demonstrated within this chapter
this stays true for a broad class of quantum gravity corrections, at least the ones considered in 
the following.\footnote{For similar studies in the context of anisotropic dispersion relations and a 
minimal length scale, see \cite{Rinaldi:2008qt,Nicolini:2009dr,Agullo:2010iq,Husain:2015tna}. }
The main observation is now the following: 
while not affecting the thermal nature of the Unruh radiation, quantum gravity induced modifications 
of two-point functions change the profile functions multiplying the thermal distribution,
sometimes even substantially. And this provides then the example for observability of a dimensional 
flow.

It will be shown that different types of dimensional flows leave distinct  signatures in the detector rates.
In particular, in the case of dimensional reduction at high energies, one finds a suppression of the rates, 
whereas for a dimensional \emph{enhancement} at high energies, as in Kaluza-Klein models, the rate increases. 
Since the transition probability of the Unruh detector is clearly a signature which is observable at least in 
principle, it can be used to make phenomenological predictions from quantum gravity, at least in a 
Gedankenexperiment, allowing a  direct comparison between various approaches.

\section{Rates from Correlators}
\label{sect.2}

In order to make the connection between dimensional flows and modifications in the Unruh effect as close as 
possible, the detector approach \cite{Agullo:2010iq} (see also 
\cite{Birrell:1982ix,Unruh:1983ms,Hawking:1979ig}  will be followed. 
The main advantage of the detector approach is that it predicts observables, more precisely the emission and 
absorption rates of an accelerated detector.
The central idea is to consider a detector made from a two-level system with an upper, excited 
state $2$ and a lower state $1$ being separated by the energy $\Delta E \equiv E_2 - E_1 > 0$ coupled to a scalar 
field. The transition probabilities induced by the scalar can be expressed in terms of the positive-frequency 
Wightman function of the Minkowski vacuum state.  The emission rates of the detector
can be computed by evaluating a Fourier transform  of the two-point function along the worldline of an accelerated 
observer. For a standard massless scalar field, it is then rather straightforward to show that 
the Green's function evaluated on the worldline satisfies a Kubo-Martin-Schwinger (KMS) condition where the 
periodicity  in Euclidean time depends on the properties of the worldline only. The resulting Unruh temperature is
proportional to the acceleration $a$. This setup also makes clear that  corrections to the two-point functions, 
{\it e.g.}, induced by quantum fluctuations at small scales, may leave their fingerprints in the transition rate 
of the Unruh detector. Both, a dynamical dimensional flow and corrections to the transition rate, can be traced 
back to the same source: a non-trivial momentum dependence of the two-point function. 

\subsection{Particle Detectors and Two-Point Functions}
\label{sec21}

The simplest model of a particle detector \cite{Unruh:1983ms,Hawking:1979ig,Agullo:2010iq} is a quantum mechanical
system with two internal energy states $|E_2\rangle$ and $|E_1\rangle$, 
with energies $E_2>E_1$. 
The detector moves along a worldline $x(\tau)$ parameterised by the detector's proper time $\tau$ and interacts with a scalar field $\Phi(x)$ by absorbing or emitting its quanta. 
The coupling of $\Phi$ to the detector   is  modelled by a monopole moment operator $m(\tau)$ 
acting on the internal detector eigenstates through the Lagrangian
\be 
\label{detectorPhi} 
{L}_I=  g \ m(\tau)\Phi(x(\tau)) \, .
\ee

It is instructive to compare the two cases: a detector moving inertially in Minkowski space,
and one moving along a uniformly accelerated trajectory, which defines the Rindler space (see appendix
\ref{App.UnruhCoord}).
The Minkowski vacuum will be denoted by $|0_M\rangle$, the Rindler vacuum by $|0_R\rangle$, 
and  the one-particle state of the field $\Phi$ with spatial momentum $\vec k$ by $|\vec{k}\rangle $. 
There are three possible processes giving a non-zero rate.
First, the inertial detector can be in the excited state with energy $E_2$.
This leads to a spontaneous emission process and corresponds to the transition
$|E_2\rangle |0_M\rangle \  \to  |E_1\rangle |\vec{k}\rangle $ for an observer comoving with the detector. 
Second, the accelerating detector can be in the excited state with energy $E_2$.
This is related to an induced emission process and instead corresponds to the transition 
$|E_2\rangle |0_R\rangle \  \to  |E_1\rangle |\vec{k}\rangle $
for an inertial observer in Minkowski space (or equivalently $|E_2\rangle |0_M\rangle \  \to  |E_1\rangle |\vec{k}\rangle $ for an \emph{accelerating} one). 
Third, an accelerating detector in the ground state $E=E_1$ leads to
absorption, or the transition $ |E_1\rangle | 0_M \rangle \to |E_2\rangle |\vec{k} \rangle $.
Note that the term absorption here is meant purely as an analogy with two state systems,
since the one-particle state $|\vec{k}\rangle$ still appears as a final state.

In first order in time-dependent perturbation theory, the amplitude for the detector-field interaction  
factorises into a detector matrix element and a term containing the two-point function of the field:
\be
{\cal A}(\vec{k}) = ig \langle E_f|m(0)| E_i\rangle \int d\tau e^{i(E_f-E_i) \tau} \langle \vec{k}|\Phi(x(\tau))|0_M\rangle \ . 
\ee
The transition probability is the square of the amplitude, integrated over all possible final states
\be
P_{i \to f} = \int d^3k | {\cal A}(\vec{k})| ^2 \, .
\ee
For $E_f = E_1$ and $E_i = E_2$ this gives 
 the total, spontaneous plus induced, emission probability.

For the mode expansion of  the field $\Phi$ there are now two practical choices based on either
the annihilation (creation) operators in Minkowski space $a_{\vec{k}}$ ($a_{\vec{k}}^\dagger$)
such that $a_{\vec{k}} |0_M\rangle =0$,
or on those in Rindler space with $b_{\vec{k}} |0_R\rangle =0$:
\be
\Phi(x) = \int d^3k \left(u_{\vec{k}} a_{\vec{k}} + u^{*}_{\vec{k}} a^{\dagger}_{\vec{k}}\right) 
= \int d^3k \left(v_{\omega\vec{k}_{\bot}} b_{\omega\vec{k}_{\bot}} + v^{*}_{\omega\vec{k}_{\bot}} b^{\dagger}_{\omega\vec{k}_{\bot}}\right) \, , 
\ee
where $\vec{k}_{\bot}=(k_y,k_z)$ are the momenta left untransformed by changing to  Rindler coordinates.
In Minkowski one has as usual plane waves:
\be
u_{\vec{k}}= \frac{1}{\sqrt{2\omega (2\pi)^3}} e^{-i(wt-\vec{k}\vec{x})} \, ,
\ee
with $\omega = \sqrt{\vec k^2 + m^2}$.
In the Rindler coordinates $(\tau, \xi, \vec{x}_{\bot})$ the related solutions of the Klein-Gordon equation
are given in terms of a modified Bessel function $K_\nu(x)$ but with analytic continuation of the index to 
purely imaginary values \cite{Crispino:2007eb},
\be
\label{vRind}
v_{\omega\vec{k}_{\bot}} = \left( \frac{\sinh(\pi\omega/a)}{4\pi^2 a} \right)^{1/2} 
K_{i\omega/a}\left( \frac{\sqrt{\vec{k}_{\bot}^2 + m^2}}{a} e^{a\xi} \right)
e^{-i(\omega \tau-\vec{k}_{\bot}\cdot\vec{x}_{\bot})} \, .
\ee

As explained in \cite{Agullo:2010iq} one can then straightforwardly  calculate  
transition probabilities
\be
\label{Pif} 
P_{i \to f} = g^2 |\langle E_f |m(0)|E_i\rangle|^2 \, F(E_f-E_i) \ , 
\ee
where
$F(\Delta E)$ is the response function
\be
\label{FE1}
F(E_f-E_i)= 
\int_{-\infty}^{\infty}d\tau_1 \int_{-\infty}^{\infty}d\tau_2 \, 
e^{-i ( E_f - E_i) \Delta \tau}  G_M(\Delta \tau -i\epsilon) \, , \quad   \Delta \tau = \tau_1 -\tau_2 \, ,
\ee
with the usual identification $\epsilon\to 0^+$. 
Note that the response function can be recast as an integral over the Fourier transform of the 
Wightman two-point function $G_M(\Delta \tau -i\epsilon)$ evaluated on the detector's trajectory. 

For the purpose of this study it is sufficient to consider only emission. Therefore, 
$E_i=E_2$ and $E_f=E_1$ and $\Delta E \equiv E_2 - E_1$ is  positive.
For the accelerating detector the total transition probability \eqref{Pif} contains contributions 
from spontaneous and induced emission. Subtracting the spontaneous emission probability
one arrives at the induced emission response function given by
\be
\label{parkereq}
F_I(\Delta E)=
\int_{-\infty}^{\infty}d\tau_1d\tau_2 \, 
e^{i \Delta E \Delta \tau}  
 \left( G_M(\Delta \tau -i\epsilon) - G_{R}(\Delta \tau -i\epsilon) \right) \, . 
\ee
In this formula $G_{M}$ is the Wightman two-point function for an observer
on the ``accelerated'' trajectory of the detector in the Minkowski vacuum, 
\be
G_{M}\left(x,x^{\prime}\right)=\left\langle 0_{M}\right|\Phi\left(x\right)\Phi\left(x^{\prime}\right)\left|0_{M}\right\rangle \, ,  
\ee
whereas $G_{R}$ is the two-point function of an accelerating
observer in the Rindler vacuum,
\be
G_{R}\left(x,x^{\prime}\right)=\left\langle 0_{R}\right|\Phi\left(x\right)\Phi\left(x^{\prime}\right)\left|0_{R}\right\rangle \, . 
\ee
It is convenient to change to the induced transition rate per unit time given by
\be 
\dot P_{i \to f} = g^2 \, |\langle E_f |m(0)|E_i\rangle|^2 \, \dot F_I(\Delta E) \ , 
\ee
where 
\be
\label{parkereq2}
\dot F_I(\Delta E)  = 
\int_{-\infty}^{+\infty}d\Delta \tau \, 
e^{i \Delta E \Delta \tau} \, 
\left( G_M(\Delta \tau -i\epsilon) - G_{R}(\Delta \tau -i\epsilon) \right) \, . 
\ee
This expression of the physical rate in terms of two-point functions is the main result of this 
subsection and the basis for all further calculations.

Furthermore, the Wightman function for a massive scalar field with mass $m$ in Minkowski space 
is given by (see, e.g., \cite{Birrell:1982ix})
\be
\label{massiveWightman}
G_M (x, x^\prime) = - \frac{i m}{4\pi^2} 
\frac{K_1 \left( im \sqrt{\left( t - t^{\prime} -i\epsilon \right)^2 - \left( \vec{x} - \vec{x}^{\prime} \right)^2 } \right)}{\sqrt{\left( t - t^{\prime} -i\epsilon \right)^2 - \left( \vec{x} - \vec{x}^{\prime} \right)^2 }} \, . 
\ee
Here $K_1$ is the modified Bessel function of the second kind.  In the massless limit 
\eqref{massiveWightman} reduces to the Wightman function of a massless scalar field in position space 
\be
\label{masslessWightman}
G_M (x, x^\prime) = - \frac{1}{4\pi^2} \, 
\frac{1}{\left( t - t^{\prime} -i\epsilon \right)^2 - \left( \vec{x} - \vec{x}^{\prime} \right)^2 } \, . 
\ee

The Wightman function in Rindler space is just the same evaluated on the
worldline of the uniformly accelerated detector
\be
\label{rindlertrajectory}
t = a^{-1} \, \sinh(a \tau)  , \quad
 x = a^{-1} \, \cosh(a \tau)  , \quad
 y = 0  , \quad z = 0  . 
\ee

\subsection{Emergence of Thermal Spectrum}

The use of Rindler coordinates (see appendix \ref{App.UnruhCoord}) elucidates the emergence of the Unruh thermal spectrum
as a geometrical effect for every Lorentz-invariant QFT. 
Any generic Poincar\'e invariant Green's function $G_M(x,x^{\prime})=G_M(x-x^{\prime})$ for an interacting theory 
in Minkowski space, when evaluated on the worldline \eqref{rindlertrajectory} of a uniformly accelerating 
observer, must be a function of the Rindler coordinates 
$(\vec{x}_\bot, \tau)$ and $(\vec{x}^{\prime}_\bot, \tau^{\prime})$.
On the other hand, $G_M$ can only depend on $(x-x^{\prime})^2=(t-t^{\prime})^2 - (\vec x- \vec x^{\prime})^2 $.
Both statements together imply that $G_M$  is a function of 
\be
 a^{-2} \left[ ( \sinh a\tau - \sinh a\tau ^{\prime})^2 - 
 (\cosh a\tau - \cosh a\tau ^{\prime})^2 \right]   
 = \, 2 a^{-2} \left(  \cosh (a\Delta\tau) -1 \right)  \,,
\ee
with $\Delta\tau=\tau - \tau^{\prime}$. Consequently, 
the Rindler Green's function has a $\tau$ dependence of the form $G_R(\cosh a\Delta\tau)$.
Focusing for simplicity on $\tau^{\prime}=0$, a Wick rotation $t=it_E$ will induce, 
through $t=a^{-1} \sinh a\tau$, a corresponding Wick rotation
in Rindler time, $\tau = i \tau_E$. But this then means that a general Rindler two-point
function will be periodic in Rindler time, since $G_R(\cosh a\tau) \to G_R^{(E)}(\cos a\tau_E)=G_R^{(E)}(\cos (a\tau_E+2\pi))$.
We thus see\footnote{
There is a subtlety in the Wick rotation when working with Wightman functions.
Due to the different domains of analyticity of $G_+$ and $G_-$ in the complex $\tau$-plane,
one actually identifies $G_E(\tau_E)=G_+(i \tau_E)$ for $-2\pi<\tau_E<0$
and $G_E(\tau_E)=G_-(i \tau_E)$ for $0<\tau_E<2\pi$.
This is responsible for the change of sign of $\tau$ in the KMS condition.}
that the periodicity $\beta=2\pi/a$ implies a temperature $T=a/2\pi$.

Rotating back provides then the Kubo-Martin-Schwinger (KMS) condition 
\be
G_R(\tau) = G_R(-\tau-i\beta) \, . 
\ee
Since the detector rate is related to the Fourier transform of the Wightman function this yields 
\cite{Takagi:1986kn}
under the assumption that  $G_R(\tau)$ is analytic in the strip $-\beta<{\rm Im}\tau<0$
\ba
\dot{F}(E) &=& \int_{-\infty}^{+\infty} d\tau e^{-iE\tau} G_R(\tau-i\epsilon) 
= \int_{-\infty}^{+\infty} d\tau e^{-iE(\tau-i\beta+2i\epsilon)} G_R(\tau-i\beta+i\epsilon) \nonumber \\
&=& e^{-(\beta-2\epsilon)E} \int_{-\infty}^{+\infty} d\tau e^{iE\tau} G_R(\tau-i\epsilon) \, . 
\ea
Here in the second line the integration variable has been changed to $-\tau$.
Taking $\epsilon$ to zero, the KMS condition becomes\footnote{A general proof of the KMS condition for 
an interacting field theory in any dimension was given in \cite{Sewell:1982zz}. It can
also be derived   directly in the free massive case from
the parity properties of the integrands appearing in the rates \cite{Louko}.}
\be
\label{KMS}
\dot{F}(E) = e^{-\beta E} \dot{F}(-E) \, . 
\ee

It is important to note that 
the Unruh temperature $T=a/2\pi$ is therefore only determined by the Euclidean periodicity,
and it is protected against corrections as long as the Lorentz invariance of $G_M$ is preserved.

\subsection{Detector Response for Scalars}
\label{sect.23}

In the massless case the rate integral can be computed directly.  The integration contour is then closed   
by a large semicircle in the upper complex-$\tau$ half-plane leading to a Matsubara-type sum over the 
infinitely many poles of the integrand located along the imaginary axis which then yields 
the Planckian thermal factor. Alternatively one may employ the KMS condition.
With $\dot{F}_A$ being the absorption rate and $\dot{F}_E$ the emission rate,
the formulas for the detector rates in section \ref{sec21} imply  $\dot{F}_A(-E) = \dot{F}_E(E)$.
Note that the emission rate is the sum of spontaneous and induced emission rates,
 $\dot{F}_E = \dot{F}_S + \dot{F}_I$.
Using the KMS condition (\ref{KMS}) one obtains
\be
\dot{F}_A(E) =  e^{-\beta E}\dot{F}_A(-E) = e^{-\beta E}\dot{F}_E(E) =
  e^{-\beta E} [ \dot{F}_I(E) + \dot{F}_S(E) ] \, ,
\ee
and if the induced emission and absorption rates coincide
\be
\label{equal}
\dot{F}_A(E) = \dot{F}_I(E)
\ee
it follows that
\be
\label{induced}
\dot{F}_I(E) = \frac{\dot{F}_S(E)}{e^{\beta E}-1} \, . 
\ee
Thus one only needs to compute the spontaneous rate to obtain that for induced emission.
In the massless case this is easily computed to give $\dot{F}_S(E)=E/2\pi$.

Condition (\ref{equal}) unfortunately does not hold for a (free) massive scalar field.
An explicit calculation in this case \cite{Takagi:1986kn} gives for the total rate
\be
\label{4.1.11}
\dot{F} (E) = \int_{-\infty}^{+\infty}d\tau e^{-iE\tau} G_R(\tau-i\epsilon)
 = 2\pi \int d^2 k_{\bot} \left| v_{\omega\vec{k}_{\bot}} \right|^2 
\left( \theta(E) N(\frac E a ) + \theta(-E) (1+ N(\frac {|E|}  a) ) \right) \, , 
\ee
where $N$ is the Bose distribution
\be
N(x)= \frac{1}{e^{2\pi x}-1} \, . 
\ee
An explicit calculation of the second (``spontaneous'') term in  (\ref{4.1.11}) 
following \cite{Agullo:2010iq} (see (3.11) in that reference)  yields
\be
\label{3.11m}
\dot{F}_S(E) = 2\pi \int d^2 k_{\bot} d\omega \left| K_{i\omega/a}\left( \frac{\sqrt{\vec{k}_{\bot}^2 + m^2}}{a} \right) \right|^2 \frac{\sinh(\pi\omega/a)}{4\pi^4 a}\delta(\omega-E) \, .
\ee
Unfortunately, this does not in general coincide with the true spontaneous rate,
(defined as the rate of a detector at rest in Minkowski space)
due to the absence of a mass gap in Rindler space.

\begin{figure}[h]
\centering
\includegraphics[width=0.8\textwidth]{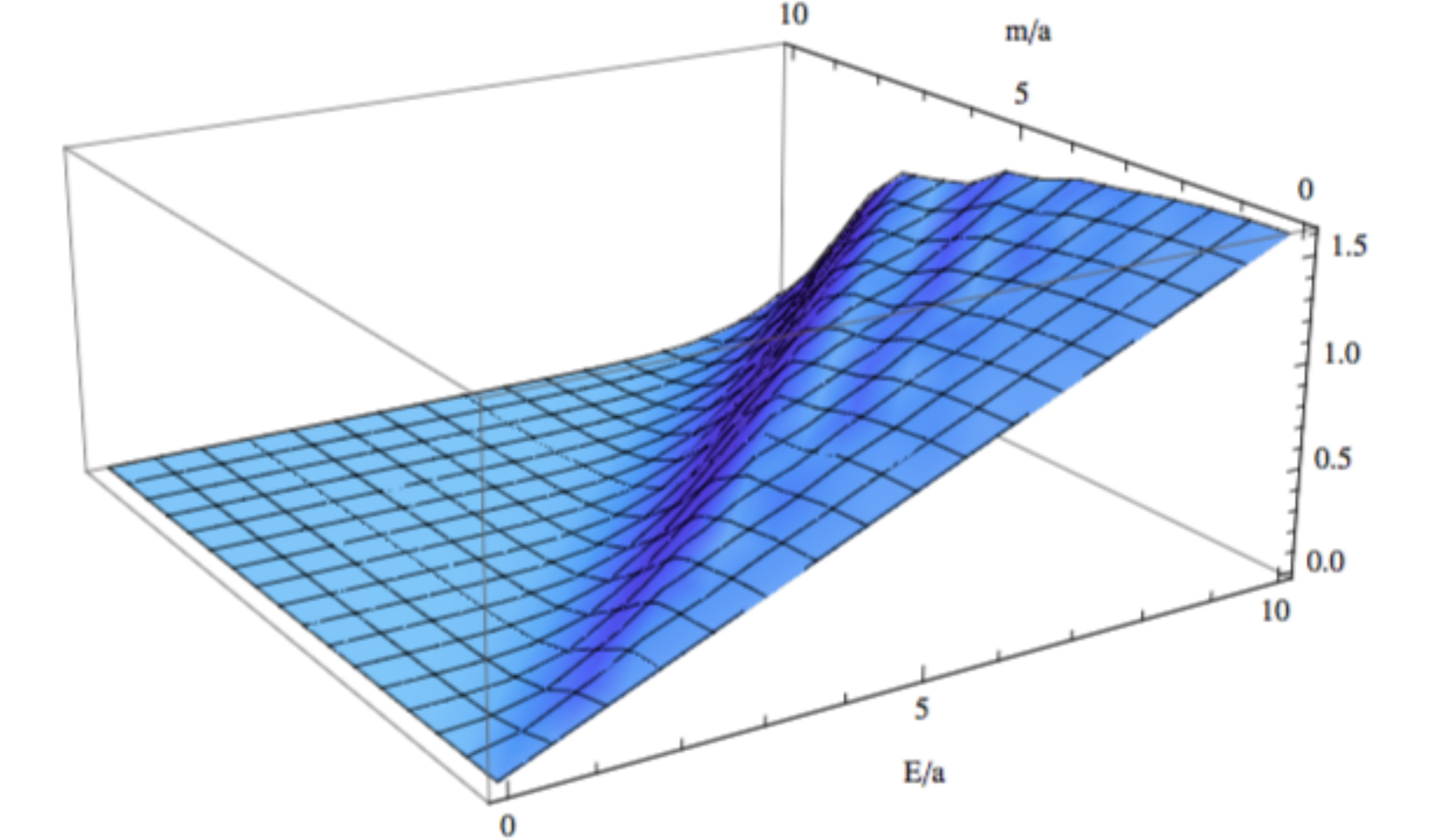}
\caption[Numerical integration of \eqref{3.11m} 
as a function of dimensionless ratios.]
{\label{gap} Numerical integration of \eqref{3.11m}, as a function of the dimensionless ratios $E/a, m/a$.}
\end{figure}

A direct calculation for  detector at rest in the Minkowski vacuum, in general dimension $d$, 
provides \cite{Birrell:1982ix}
\ba
\label{spont}
\dot{F}_S(E) &=&
\int \frac{d^{d-1} k}{(2\pi)^{d-1}} \frac{1}{2 \sqrt{k^2+m^2}} 
\int_{-\infty}^{+\infty} d\tau e^{- i (\sqrt{k^2+m^2} - E) \tau}
\nonumber \\
&=& \frac{\pi^{\frac{d-1}{2}}}{\Gamma(\frac{d-1}{2})(2\pi)^{d-2}} 
\left(E^2-m^2 \right)^{\frac{d-3}{2}}\theta(E-m) \,.
\ea
The relations \eqref{3.11m} and \eqref{spont} coincide in the limit where $E\gg m$.
A crucial difference between the two results is that \eqref{spont} exhibits
a mass gap which is absent in \eqref{3.11m}.
The numerical integration of \eqref{3.11m}, displayed in figure \ref{gap},
shows that this expression well approximates \eqref{spont} when $E<m$,
and therefore this one will be used in the following.

Employing (\ref{induced}), 
the induced rate function per unit time of the accelerating detector in $d=4$ is determined to be 
\be
\label{Fdotfinal}
\dot F = \frac{1}{2 \pi} \, \sqrt{E^2 - m^2} \, \theta(E-m) \, \frac{1}{e^{\frac{2\pi E}{a}} -1} \, .
\ee
This rate function constitutes the main result of this subsection. Taking the limit $m \rightarrow 0$, 
it agrees with the derivation for the massless case given in \cite{Birrell:1982ix,Agullo:2010iq}. 

The structure of \eqref{Fdotfinal} then motivates the definition of a profile function $\mathcal{F}(E)$
via
\be
\label{defresf}
\dot F = \frac{1}{2 \pi} \, \mathcal{F}(E) \, \frac{1}{e^{\frac{2\pi E}{a}} -1} \, .
\ee
For a massless and massive scalar field obeying the Klein-Gordon equation one then has
\be
\mathcal{F}^{\rm massless}(E) = E \, , \quad 
\mathcal{F}^{\rm massive}(E) = \sqrt{E^2 - m^2} \, \theta(E-m) \, . 
\ee
For general dimension the profile function is
\be
\label{rategen}
{\cal F}(E) = \frac{\pi^{\frac{d-1}{2}}}{\Gamma(\frac{d-1}{2})(2\pi)^{d-3}} 
\left(E^2-m^2 \right)^{\frac{d-3}{2}}\theta(E-m) \, . 
\ee
As will be demonstrated below, it is this profile function 
that actually carries information about quantum gravity corrections to
the Unruh rate.\footnote{As emphasised previously,
 the Planckian thermal factor is independent of the details of the field considered.
The fact that the mass dependence enters through the prefactor tells us that
the signatures of the fields involved will only be present in physical rates,
and not in number densities $\left\langle n \right\rangle$. }

\section{Master Formulas for Modified Detector Rates}
\label{sect.3}

In the presence of a dimensional flow, $\tilde G(p^2)$ 
acquires a non-trivial momentum dependence.\footnote{As noted before, this does not necessarily entail the breaking of Lorentz symmetry since $\tilde G(p^2)$ may still be a Lorentz invariant function depending on the square of the momentum four-vector only.} 
It is useful to distinguish the two cases where $[\tilde G(p^2)]^{-1}$ is a polynomial in $p^2$ or given by a more general function with a finite number (typically one) of zeros in the complex $p^0$-plane. These two cases will be discussed in sections \ref{sect.31} and \ref{sect.32}, respectively.

\subsection{Detector Rates from the Ostrogradski Decomposition}
\label{sect.31}

First, the case in which the inverse propagator $\widetilde{G}^{-1}(p^2) \equiv {\cal P}_n (p^2)$ 
is an inhomogeneous polynomial of order $n$ will be considered.
This covers the class of theories with a general quadratic effective Lagrangian ${\cal L} = \tfrac{1}{2} \phi \, {\cal P}_n(-\partial^2) \, \phi$ where ${\cal P}_n$ is a local function of the flat space d'Alem\-bertian operator that admits a Taylor expansion around zero momentum. This comprises all local theories in which higher order corrections come in definite powers of momenta.
The limiting case $n\to\infty$ can also be considered. In this case the profile function ${\cal F}(E)$, \eqref{defresf}, can be constructed from the Ostrogradski decomposition for a higher-derivative field theory.

The polynomial ${\cal P}_n (z)$ has $n$ roots, $\mu_i, i = 1,\ldots,n$ in the complex $z$-plane. It can then be factorised according to
\be
\label{poly1}
{\cal P}_n (z) = c \, \prod_{i=1}^{n} \left(z - \mu_i \right)
\ee
where $c$ is a normalisation constant. In order to connect to the case of a massive scalar field, the momentum space propagator is decomposed according to
\be
\label{poly2}
\left[ {\cal P}_n (z) \right]^{-1} = \frac{1}{c} \, \sum_{i=1}^n \frac{A_i}{\left(z - \mu_i \right)}
\ee
where the coefficients $A_i$ are functions of the roots $\mu_i$:
\be
\label{Aisol}
A_i = \left( {\prod_{j \not = i} (\mu_i - \mu_j)} \right)^{-1} \, . 
\ee
For future reference, it is convenient to give the coefficients $A_i$ entering   the decomposition \eqref{poly2} for the cases $n=2$ and $n=3$ explicitly. For $n=2$,
\be
A_1 = \frac{1}{\mu_1 - \mu_2} \, , \quad 
A_2 = \frac{1}{\mu_2 - \mu_1}
\, ,  
\ee
while for $n=3$ one has 
\be
A_1 = \frac{1}{(\mu_1 - \mu_2)(\mu_1 - \mu_3)} \, , \; 
A_2 = \frac{1}{( \mu_2- \mu_1)(\mu_2 - \mu_3)} \, , \; 
A_3 = \frac{1}{(\mu_3 - \mu_1)(\mu_3 - \mu_2)} \, . 
\ee

At this stage the following remark is in order. On mathematical grounds the decomposition \eqref{poly2} works as long as all roots of the polynomial have order one. On physical grounds there are extra conditions on the roots:  
$\mu_i = m^2$ should be identified with the square of the particle mass. This implies that roots located at the 
negative real axis correspond to modes with a negative mass squared. In this case the isolated poles at $p_0 = \pm 
\sqrt{\vec{p}^2 + \mu_i}$ are turned into branch cuts and we will not consider this tachyonic case in the 
following. Moreover, complex roots always come in pairs $\mu, \bar \mu$. This implies that the positive frequency 
Wightman function contains unstable modes which grow exponentially in the far past and far future (also see
\cite{Aslanbeigi:2014zva} for a detailed discussion of this feature).
On this basis, we restrict ourselves to polynomials $P_n(p^2)$ whose roots
are located at the positive real axis.

Since the rate function \eqref{parkereq2} is linear in the Wightman function, it is rather straightforward to obtain the detector response function for the case \eqref{poly2}. Following the steps of section \ref{sect.23}, we can compute the profile function ${\cal F}(E)$ determining the rate  \eqref{defresf}. Substituting the explicit form of the $A_i$ from \eqref{Aisol} the result reads
\be
\label{moddetrate}
{\cal F}  (E) = \frac{1}{c} \, \sum_{i=1}^n \; \left( {\prod_{j \not = i} (\mu_i - \mu_j)} \right)^{-1} \, 
\sqrt{E^2 - \mu_i}\; \theta( E -\sqrt{\mu_i}) \, . 
\ee
The rate function is completely determined by the roots of the polynomial ${\cal P}_n(p^2)$. It receives new contributions once new channels become available, i.e., if the energy gap $E$ crosses a threshold $\mu_i$ where new degrees of freedom enter. Ordering the roots $\mu_i$ by their magnitude, i.e., $\mu_j > \mu_i$ for $j > i$, one sees that the sector with $\mu_j$, $j > i$ does not affect the ``low-energy'' part of the rate function with $E < \mu_i$: the energy gap $E$ of the detector is not large enough to absorb a particle of mass $\sqrt{\mu_j}$, $j > i$. This, in particular, implies that if the polynomial \eqref{poly1} arises from an effective field theory description of a system, there are no corrections to the massless Unruh effect below the first threshold $\mu_2 > 0$, provided that the polynomial ${\cal P}_n$ is properly normalised. The master formula \eqref{moddetrate} then constitutes the main result of this section.

\subsection{Detector Rates from the K{\"a}llen-Lehmann Representation}
\label{sect.32}

Not all two-point functions proposed in the context of quantum gravity fall in the class where the 
Ostrogradski-type decomposition is admissible. In these cases it is still possible to obtain an explicit formula 
for the profile function ${\cal F}(E)$ based on the K{\"a}llen-Lehmann representation of the two-point function.

The K{\"a}llen-Lehmann representation of the positive frequency Wightman function in position space is given by
\be
G_+(t,\vec{x}) = \int_0^{\infty} dm^2 \, \rho(m^2) \, G_+^{(0)}(t,\vec{x}; m) \, . 
\ee
Here $\rho(m^2)$ denotes a spectral density and $G_+^{(0)}(t,\vec{x}; m)$ is the positive-frequency Wightman function given
in \eqref{massiveWightman}. Substituting the K{\"a}llen-Lehmann representation into \eqref{parkereq2} and exchanging the order of integration, the computation of the rate function reduces to the one for the massive scalar field carried out in section \ref{sect.23}. The resulting profile function ${\cal F}(E)$,  \eqref{defresf}, is given by  
\be
\label{spectralF}
\begin{split}
{\cal F}  (E)  
= & \,  \int_0^{E^2} dm^2 \, \rho(m^2) \, \sqrt{E^2 - m^2} \, . 
\end{split}
\ee
Hence the profile function obtained from the K{\"a}llen-Lehmann representation is given by the superposition of contributions with mass $m$ weighted by the spectral density $\rho(m^2)$. Only excitations with mass 
below the energy gap of the detector contribute to the rate function, which is consistent with the expectation that contributions with $m^2>E^2$ will not excite the detector. The result from the Ostrogradski decomposition,  \eqref{moddetrate}, can then be understood as a special case where $\rho(m^2)$ is given by a sum of $\delta$-distributions located at $m^2 = \mu_i$. 

At this stage, the following remark is in order. The dimensional reduction discussed in this work is not necessarily in conflict with the unitarity of the underlying model. In this context, it is important to stress that the construction of the spectral dimension relies on effective propagators dressed by quantum corrections and \emph{not} on the two-point functions appearing in the fundamental action. On a manifold with spectral dimension $d_s$, the asymptotic form of the \emph{effective} two-point function is
\be
G(p^2) \sim (p^2)^{d/d_s} \, . 
\ee
Expressing a general two-point function through the K{\"a}llen-Lehmann representation as in the previous section, we see that, as soon as $d_s<d$, its fall-off properties can only be consistent with the $p^{-2}$ behaviour of the spectral representation if we relax the positivity properties of the spectral function $\rho(m^2)$. At the level of the fundamental action, the non-positivity of $\rho(m^2)$ would signal the presence of negative-normed states and thus a departure from unitarity, {\it cf.\/} the discussion in \cite{Oehme:1979ai}. At the level of the effective propagator this feature is acceptable though and does not signal an intrinsic sickness of the theory. A prototypical example for this behaviour is given by Yang-Mills theory with gauge group $SU(N)$: in this case the propagator appearing in the fundamental action falls off as $p^{-2}$ compatible with unitarity while the K{\"a}llen-Lehmann representation of the fully dressed gluon propagator in Landau gauge gives rise to a spectral density $\rho(m^2)$ which is not positive definite \cite{Alkofer:2003jj,Cucchieri:2004mf,Strauss:2012dg,Dudal:2013yva}.

\section{Scaling Dimensions}
\label{sect.3b}

As discussed in detail in the last chapter the two-point function 
$\widetilde{G}(p^2)$ provides the essential input
for computing the spectral dimension $D_s$ seen by a scalar field propagating on the spacetime.
In addition,  it is also determining the rate function of the Unruh detector. 
This indicates that there is a relation between the rate function of the Unruh detector and the 
spectral dimension. In this section the definitions needed to make this relation precise will be 
introduced.

Analysing the scaling behaviour in \eqref{returnprop} one finds that for the case where $F(p^2) \propto p^{2 + \eta}$ the spectral dimension is given by \cite{Reuter:2011ah}
\be
\label{spectraldimension}
D_s = \frac{2d}{2+\eta} \, . 
\ee
The case of a massless scalar field with $\widetilde{G}(p^2) = p^{-2}$ corresponds to $\eta = 0$ and the spectral dimension agrees with the topological dimension $d$ of the spacetime. In case of a multiscale geometry
the scaling law $F(p^2) \propto p^{2 + \eta}$ is obeyed for a certain interval of momenta only. In this case the spectral dimension will depend on the diffusion time $T$. If the scaling regime extends over a sufficiently large order of magnitudes, $D_s(T)$ will be approximately constant in this regime, realising a plateau structure. Typically, such plateaus where $D_s(T)$ is approximately constant are connected by short transition regions where $D_s$ changes rather rapidly, see figure \ref{Fig.dimflow1} for an explicit example illustrating this type of crossover. 

In a similar spirit, one can define the effective dimension of spacetime seen by the Unruh detector. \eqref{rategen} indicates that the profile function for a massless scalar field obeying the Klein-Gordon equation in a $d$-dimensional spacetime scales as
\be
{\cal F}(E) \propto E^{d-3} \, . 
\ee
This motivates defining the effective dimension  seen by the Unruh rate, the Unruh dimension $D_U$, according to
\be
\label{UnruhDimension}
D_U(E) \equiv \frac{d \ln {\cal F}(E)}{d \ln E} + 3 \, . 
\ee
For a massless scalar field with $\widetilde{G}(p^2) = p^{-2}$ or a massive
scalar field with energy $E^2 \gg m^2$, $D_U$ is independent of $E$ and coincides with the classical dimension $d$ of the underlying spacetime.
Paralleling the discussion of the spectral dimension, this feature changes, however, if $\widetilde{G}(p^2)$ has a non-trivial momentum profile. The examples presented in section \ref{sect.4} indicate that $D_U$ may agree with the spectral dimension in certain cases, but in general the two are different quantities. The Unruh dimension may yield a characterisation of quantum spacetimes which is accessible by experiment, at least in principle. 
Note that the dimensions are only well-defined in plateau regions of sufficient extent and have to be taken with caution during crossovers \cite{Reuter:2011ah}.

A direct comparison between $D_U$ and $D_s$ requires an identification of $E$ and the diffusion time $T$. The matching of dimensions in the 
classical case suggests using
\be
\label{scalesetting}
T = E^{-2n} \, , 
\ee
where $2n$ is the mass-dimension of $\widetilde{G}(p^2)$. 

The emission/absorption rates can be related to the density of states of the system
interacting with the detector.
The density of states as a function of momentum can be defined as $\rho(k) = d\Omega(k)/dk$,
where $\Omega(k)$ is the volume of momentum space.
Since the spectral dimension $D_s$ is the Hausdorff dimension of momentum space,
we can assume that $\Omega$ will scale as $\Omega(k) \sim c k^{D_s}$.
Then we see that $\rho(k) \propto k^{D_s-1}$, and a smaller value of $D_s$ entails
a suppression of the density of states.
This in turn will imply a suppression of the various transition rates.
Due to the relation between this density of states and the transition rates,
we expect a relation between the spectral and Unruh dimensions, $D_s$ and $D_U$.
This relation will indeed be made more precise in the next sections.

\section{Unruh Rates and Dimensional Flows}
\label{sect.4}

\subsection{Dynamical Dimensional Reduction}
\label{sect.41}

In this subsection modifications of the Unruh rate arising from a particular class of quantum-gravity 
inspired two-point functions $\widetilde{G}(p^2)$ typically encountered when discussing the flows 
of the spectral dimension will be investigated. 

\subsubsection*{Two-Scale Models}

The simplest way to obtain a system exhibiting dynamical dimensional reduction is based on a polynomial, \eqref{poly1} with $n=2$, containing a single mass scale $m$:
\be
\label{ansatz1}
\cP_2(p^2) = - \frac{1}{m^2} \, p^2 \, \left( p^2 - m^2 \right) \, . 
\ee
Here the normalisation $c$ has been chosen such that the model gives rise to a canonically normalised two-point function at low energy.
The scaling of this ansatz is given by
\be
\cP_2(p^2) \propto
\left\{	
\begin{array}{ll}
p^2 \, , \qquad & p^2 \ll m^2 \\[1.1ex]
p^4 \, , \qquad & p^2 \gg m^2 \, , 
\end{array}
\right.
\ee
with the crossover occurring at $m^2$. Evaluating \eqref{spectraldimension}, the spectral dimension based on this model interpolates between a classical regime with $D_s = 4$ for long diffusion times and $D_s = 2$ for short diffusion times.

The Ostrogradski decomposition \eqref{poly2} of \eqref{ansatz1} yields
\be
\label{2point}
\widetilde{G}(p^2)  = \frac{1}{p^2} - \frac{1}{p^2 - m^2} \, . 
\ee
The master formula \eqref{moddetrate}  gives the following expression for the profile function
\be
\label{2scalemodel}
{\cal F}  (E) = E - \sqrt{ E^2 - m^2} \, \theta(E - m) \, . 
\ee
Expanding $\cF$ for small and large $E$ leads to the scaling behaviour
\be
\label{2scaleasym}
\begin{array}{llcl}
	E < m: \qquad & {\cal F}(E) = E    & \quad \Longleftrightarrow \quad & D_U = 4 \, ,  \\[1.2ex]
	E \gg m: \qquad & {\cal F}(E) = \frac{1}{2E} + \cO(E^{-2})  & \quad \Longleftrightarrow \quad & D_U = 2 \, . 
\end{array}
\ee
This expansion implies that a kinetic term including higher-derivative contributions leads to detector rates which are suppressed at high energies.
In particular, whereas for a massless (free or interacting) scalar field with
a standard kinetic term the prefactor of the rate grows linearly with energy,
the profile function vanishes proportional to  $E^{-1}$ at high energies.
This also entails that the Unruh dimension $D_U$ interpolates between the classical dimension $D_U = 4$ for small energy and $D_U=2$ for $E \gg m$.

\begin{figure}[h]
\centering
\includegraphics[width =0.49\textwidth]{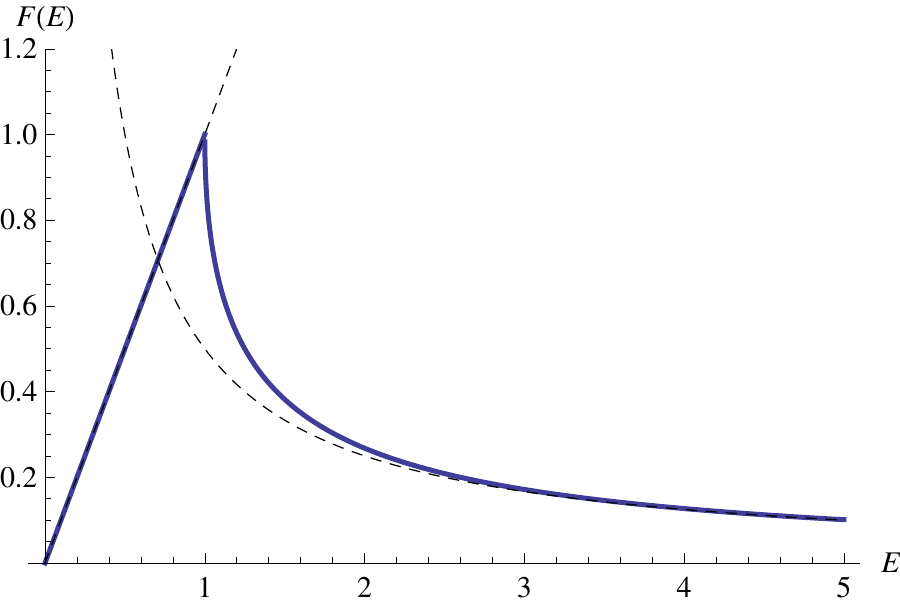} 
\includegraphics[width =0.49\textwidth]{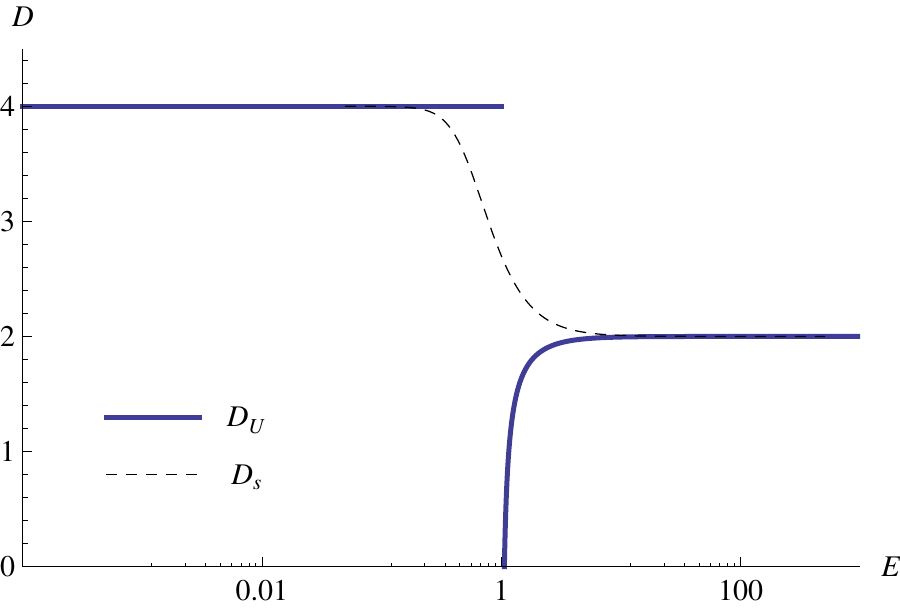}
\caption[Profile function $\cF(E)$ for $m=1$ (left panel). The 
asymptotics are illustrated by the dashed lines. The right 
panel shows the dimensions $D_S$ and $D_U$ resulting from the  
two-point function.]
{\label{Fig.dimflow1} Profile function $\cF(E)$, \eqref{2scalemodel}, for $m=1$ (left panel). The 
asymptotics given in \eqref{2scaleasym} are illustrated by the dashed lines. The right 
panel shows the dimensions $D_S$ (dashed line) and $D_U$ (solid line) resulting from the 
two-point function \eqref{2point}.}
\end{figure}

For $m=1$ this profile function is shown in the left panel of figure  \ref{Fig.dimflow1}. Despite the inclusion of modes with a wrong sign kinetic term (``ghosts'') in \eqref{2point} the Unruh rate is positive definite, indicating that the model is stable in this respect. The right panel of figure \ref{Fig.dimflow1} shows the spectral dimension (dashed line) and effective dimension seen by the Unruh effect (solid line) where the construction of the spectral dimension is based on the identification \eqref{scalesetting}. Both dimensions interpolate between $D = 4$ for $E < m$ and $D = 2$ for $E \gg m$. $D_U$ displays a discontinuity at $E^2 = m^2$ which can be tracked back to the derivative of the square-root becoming singular at this point.

\subsubsection*{Multi-Scale Models}

Before one comes to conclusions based upon the two-scale model
it is instructive to consider a multiscale model which may exhibit more than two scaling regions. The simplest model of this form is build from a third order polynomial $\cP_3(p^2)$ with vanishing mass $m_1 = 0$
\be
\label{ansatz2}
\cP_3(p^2) = \frac{1}{m_2^2 \, m_3^2} \, p^2 \, (p^2-m_2^2) \, (p^2 - m_3^2) \, , \qquad m_3 > m_2 \, . 
\ee
Provided that $m_3 \gg m_2$ this ansatz exhibits three scaling regimes 
\be
\label{3scalemod}
\cP_3(p^2) \propto
\left\{	
\begin{array}{lll}
	p^2 \, , \qquad & p^2 \ll m^2_2  \, , & D_s = 4\\[1.1ex]
	p^4 \, , \qquad & m_2^2 \ll p^2 \ll m^2_3  \, , \qquad & D_s = 2 \\[1.2ex]
	p^6 \, , \qquad & m_2^3 \gg p^2  \, , & D_s = \frac{4}{3} \, ,  \\[1.1ex]
\end{array}
\right.
\ee
where the spectral dimension has been determined by evaluating \eqref{spectraldimension}. 

Performing the Ostrogradski decomposition for $\cP_3(p^2)$ gives
\be
\widetilde{G}(p^2) = \frac{1}{p^2} - \frac{m_3^2}{m_3^2 - m_2^2} \, \frac{1}{p^2 - m_2^2} + \frac{m_2^2}{m_3^2 - m_2^2} \, \frac{1}{p^2 - m_3^2} \, . 
\ee
The resulting profile function then reads
\be
\cF(E) = E - \frac{m_3^2}{m_3^2 - m_2^2} \, \sqrt{E^2 -m_2^2} \, \, \theta(E-m_2) + \frac{m_2^2}{m_3^2 - m_2^2} \, \, \sqrt{E^2 -m_3^2} \, \, \theta(E-m_3) \, . 
\ee
Expanding $\cF$ for small and large $E$ leads to the scaling behaviour
\be
\label{3scaleasym}
\begin{array}{llcl}
	E < m_2: \qquad & {\cal F}(E) = E    & \quad \Longleftrightarrow \quad & D_U = 4 \, ,  \\[1.2ex]
	E \gg m_3: \qquad & {\cal F}(E) = - \frac{m_2^2 \, m_3^2}{8E^3} + \cO(E^{-4})  & \quad \Longleftrightarrow \quad & D_U = 0 \, . 
\end{array}
\ee
 
\begin{figure}
\centering
\includegraphics[width =0.49\textwidth]{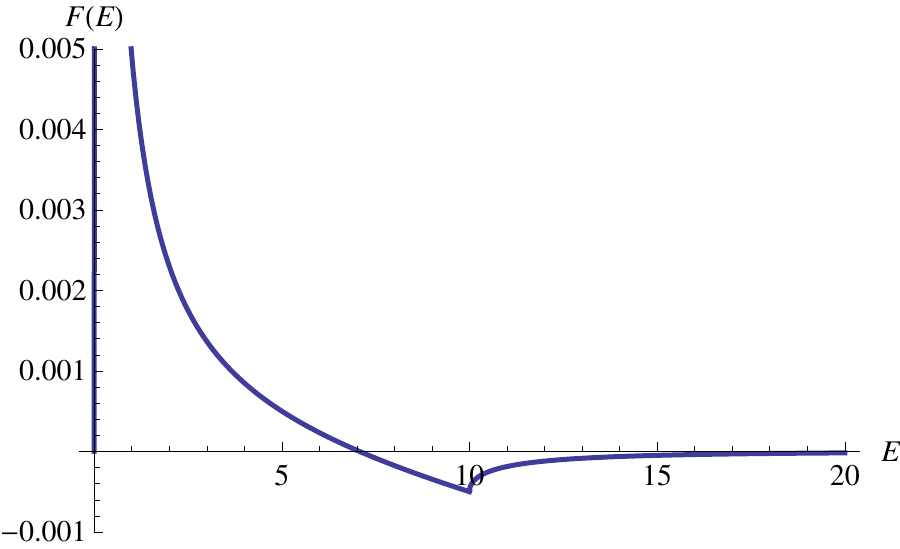} 
\includegraphics[width =0.49\textwidth]{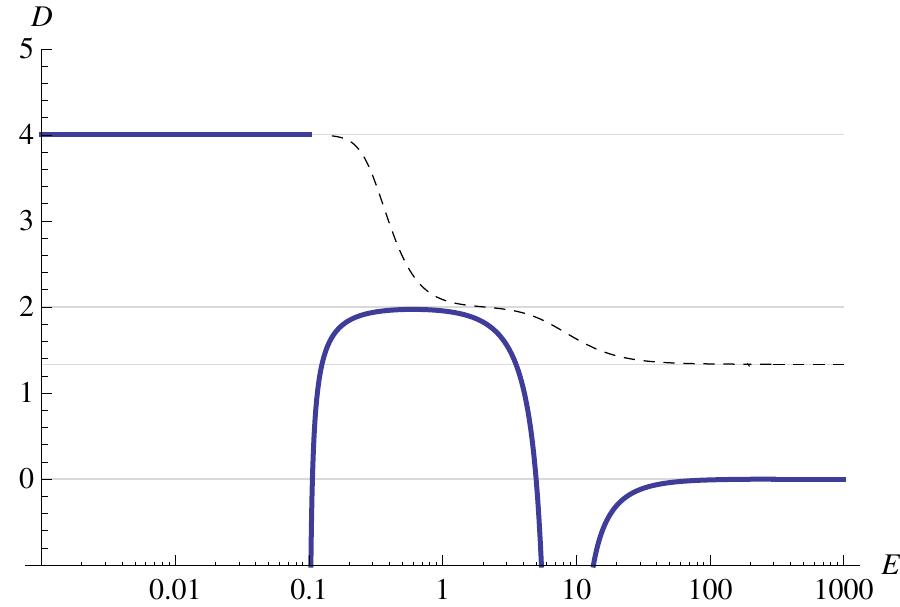}
\caption{\label{Fig.dimflow2} Illustration of the Unruh effect in a $n=3$ multiscale model with $m_1 = 0$, $m_2 = 0.1$ and $m_3 = 10$. The resulting profile function $\cF(E)$ is shown in the left panel while $D_U$ and $D_s$ are displayed in the right panel. The horizontal grey lines indicate the plateau values of the dimensions at $4,2,4/3$ and $0$. Notably, $D_U$ and $D_s$ exhibit different asymptotics for $E \gg m_3$.}
\end{figure}

Two remarks are in order. In contrast to the two-scale model, the $n = 3$ case exhibits regions where the profile function $\cF(E)$ actually becomes negative. This is illustrated in the example shown in figure \ref{Fig.dimflow2}. The form where $\lim_{E \rightarrow \infty}F(E) \rightarrow 0$ from below then indicates that this feature holds for all values $m_2$ and $m_3$. Thus the Unruh rate exhibits an instability for a generic $n=3$ model.

Furthermore, the spectral and Unruh dimensions shown in the right panel of figure  \ref{Fig.dimflow2} show that, contrary to the two-scale model, the asymptotics for $D_U$ and $D_s$ do not agree for $E \gg m_3^2$. In the general case, this may be understood as follows. Considering the general expression \eqref{moddetrate} for $m_1 = 0$, $D_U$ is given by the classical dimension as long as $E < m_2$. Each additional term in the sum creates a new scaling region where $D_U$ decreases by two compared to its previous value. In contrast the pattern for the spectral dimension follows from \eqref{spectraldimension}. Combining these relations allows to express the effective dimension seen by the Unruh effect in terms of the spectral dimension
\be
\label{dimrel}
D_U = 6 - \frac{8}{D_s} \, . 
\ee
Thus, while there is a clear relation between $D_U$ and $D_s$, the effective dimensions seen by a random walk and the Unruh effect generically do not coincide within the class of multiscale models studied here.

\subsubsection*{Logarithmic Correlation Functions}

An interesting model which does not fall into the class of multiscale models where the Ostrogradski decomposition can be applied arises from
\be
\label{p4model}
\widetilde{G}(p^2) = p^{-4} \, . 
\ee
This is the typical fall-off behaviour of correlation functions in quantum gravity models which lead to $D_s = 2$ in the UV. 
In this case a short calculation shows that the positive-frequency Wightman function takes the form
\be
G_+(\vec{x},t) = \frac{1}{8\pi^2} \left( \log \left(\sqrt{\left( t - t^{\prime} -i\epsilon \right)^2 - \left( \vec{x} - \vec{x}^{\prime} \right)^2 } \right)  + \mbox{const} \right) \, . 
\ee
Substituting the Wightman function into the formula for the Unruh rate,  \eqref{parkereq2}, yields
\be
\dot{F} (E) = \frac{1}{8\pi^2} \int_{-\infty}^{\infty} d\tau e^{i E \tau} \,
\left[ \log \left( \frac{2\sinh(\frac{a\tau}{2})}{a\tau} \right) + \mbox{const} \right] \, . 
\ee
The constant terms give rise to terms proportional to $\delta(E)$, indicating an IR instability of the setup. Since the propagator \eqref{p4model} is thought of describing the asymptotic behaviour of the system at high energies these terms will be ignored. Since the argument of the logarithm is an even function in $\tau$ the integral can be expressed as a (regularised) Fourier cosine transform
\be
\dot{F} (E)  = \lim_{\epsilon\to 0^+}\frac{1}{2a\pi^2} \int_{0}^{\infty} dx e^{-\epsilon x} \log \left( \frac{\sinh(x)}{x} \right) \, 
\cos(\omega x)  
\ee
written in terms of the new variables $x=a\tau/2$ and $\omega=2E/a$.
This integral can be performed, the resulting detector rate is given by
\be
\dot{F} (E) = \frac{1}{4\pi E} \, \frac{1}{1-e^{\frac{2\pi E}{a}}} \, ,
\ee
implying that the profile function resulting from a $p^{-4}$ propagator
is given by
\be
{\cal F}(E)= \frac{1}{2E} \qquad \, \Longleftrightarrow \qquad  D_U = 2 \, .   
\ee
This is precisely the asymptotic behaviour \eqref{2scaleasym} found in the two-scale model in the limit $E \gg m$. Thus the direct computation of the detector rate in the $p^4$-case confirms the drop of the Unruh rate at high energies and constitutes an independent verification of the rate function found in the two-scale case.

\subsection{Kaluza-Klein Theories}
\label{sect.43}

A scenario where the number of dimensions increases towards the UV is provided by 
Kaluza-Klein theories.\footnote{A related discussion of the Unruh detector in Kaluza-Klein theories 
can be found in  \cite{Chiou:2016exd}.}
In this case the (classical) spacetime is assumed to possess four non-compact and a number of compact spatial dimensions whose typical extension is given by the compactification scale $R$. At length scales $l \gg R$ the effect of the extra-dimensions is invisible and physics is effectively four-dimensional. Now, as the number of effective dimensions increases when going to high energies the detector rates for energies above the inverse compactification scale are actually \emph{enhanced} as compared to the four-dimensional rate as will be seen
shortly.

\begin{figure}[h]
\centering
\includegraphics[width=0.49\textwidth]{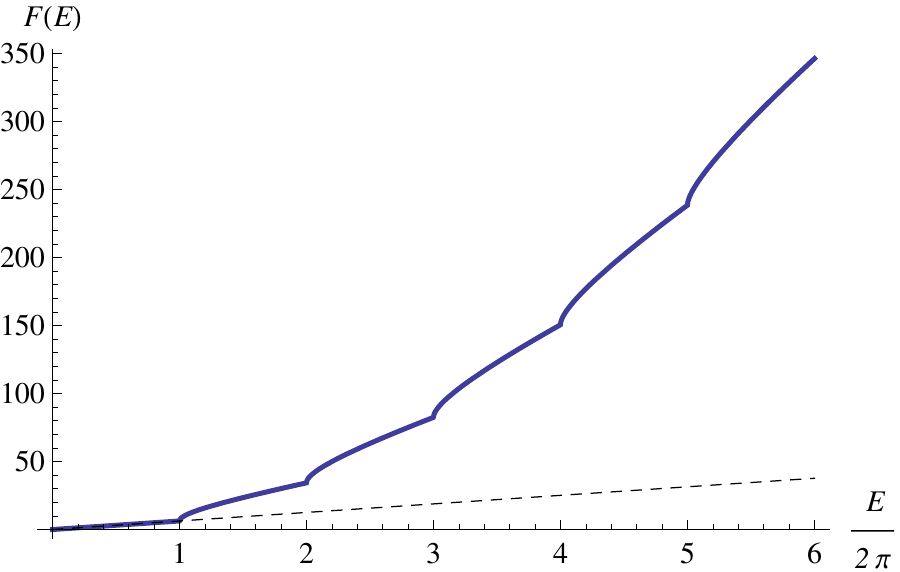} \, 
\includegraphics[width=0.49\textwidth]{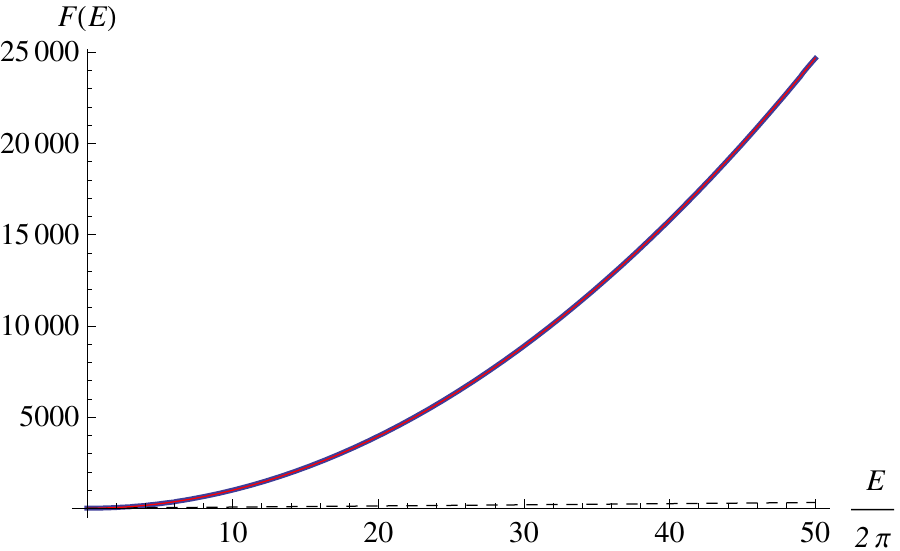}
\caption[Profile  
function ${\cal F}(E)$ for a $5$-dimensional Kaluza-Klein theory.
For $E < R^{-1}$ the profile function is linear in $E$, while for $E \gg R^{-1}$ it increases proportional to  $E^2$.]
{\label{FigKKprofile} Profile function ${\cal F}(E)$ for a $5$-dimensional Kaluza-Klein theory \eqref{kkprofile} with $R = 1/(2\pi)$ (blue, solid line). For guidance the lines ${\cal F}(E) = E$ (black, dashed line)  and ${\cal F}(E) = E^2/4$ (red line, right diagram) have been included.
			For $E < R^{-1}$ the profile function is linear in $E$, while for $E \gg R^{-1}$ it increases proportional to  $E^2$.
		}
\end{figure}

For concreteness the focus will be on the case of a five-dimensional spacetime ${\mathbb R}^4 \times S^1_R$ where the extra dimension is given by a compact circle of radius $R$. A scalar field $\phi$ living on this spacetime has a Fourier-expansion in the circle coordinate $x_5$
\be
\phi(x,x_5) = \sum_{n=-\infty}^{+\infty} \phi_n(x) \, e^{i\frac{n}{R}x_5} \, , \qquad x_5 \in [0, 2 \pi R[ \, . 
\ee
The Fourier coefficients $\phi_n(x)$ depend on the coordinates on ${\mathbb R}^4$ and are called Kaluza-Klein modes. For a real scalar field $\phi$ they obey the reality condition $\phi_{-n} = \phi_n^*$. Substituting this mode expansion into the action of a free scalar field in five dimensions yields
\be
\int d^5x \, \tfrac{1}{2} \, \left[ (\partial_{\mu}\phi)^2 - (\partial_5 \phi)^2 \right] =
 \, 2 \pi R \int d^4x \sum_{n=-\infty}^{+\infty} \tfrac{1}{2} \, \left[ |\partial_{\mu}\phi_n |^2 - \frac{n^2}{R^2} |\phi_n |^2 \right] \, . 
\ee
Each Kaluza-Klein mode $\phi_n$ has a two-point function of a scalar field with mass $m_n = n/R$. Taking into account the entire tower of modes, the resulting function $\widetilde{G}(p^2)$ is given by
\be
\widetilde{G}(p^2) = \frac{1}{2 \pi R} \, \sum_{n=-\infty}^{\infty} \, \left( p^2 - \frac{n^2}{R^2}\right)^{-1} \, . 
\ee
Applying the master formula \eqref{moddetrate} to this case then yields the profile function
\be
\label{kkprofile}
{\cal F}  (E) = \frac{1}{2\pi R} \, \left( E + 2 \sum_{n=1}^\infty \sqrt{ E^2 - (n/R)^2} \; \theta( E - n/R) \right) \, . 
\ee
The shape of this profile function is illustrated in figure \ref{FigKKprofile}.

In contrast to the case of a dynamical dimensional reduction at high energies, all Kaluza-Klein modes contribute to the profile function with the same sign. This leads to an effective enhancement of the profile function for $E > R^{-1}$.
Explicitly,
\be
\begin{array}{llcl}
E < 1/R: \qquad & {\cal F}(E) \propto E    & \quad \Longleftrightarrow \quad & D_U = 4 \, ,  \\[1.2ex]
E \gg 1/R: \qquad & {\cal F}(E) \propto E^2  & \quad \Longleftrightarrow \quad & D_U = 5 \, . 
\end{array}
\ee
 The profile function \eqref{kkprofile} interpolates between these two behaviours. Thus also the presence of extra dimensions leaves its imprint on the Unruh rate, adapting the scaling law of the profile function once the energy $E$ exceeds the inverse compactification scale.

\subsection{Spectral Actions}
\label{sect.44}

As discussed in the last chapter, a framework which naturally gives rise to two-point functions
$\widetilde{G}(p^2)$ with the properties discussed above are spectral
actions (\ref{SpAct1}). For the purpose of this subsection it is sufficient to restrict to the case 
where $\cD^2$ is given in terms of the Laplace operator on flat Euclidean space 
supplemented by an endomorphism including a real scalar field $\phi$:
\be
\label{lapop}
\cD^2 = - \left( \nabla^2 + E \right) \, , \quad E = - i \gamma^\mu \gamma_5 \partial_\mu \, \phi - \phi^2 \, .  
\ee

\subsubsection*{Non-local Analytic Models} 

First, as in the last chapter,  $\chi (z) = e^{-z}$ is chosen, and the fact that then the
spectral action \eqref{SpAct1} coincides with the heat-trace of the Laplace-type operator \eqref{lapop}
is again exploited. In particular the two-point function of the model is 
\be
\label{spectral2}
S^{(2,\phi)}_{\chi,\Lambda}  =  \frac  {\Lambda^2}{(4\pi)^2} \int d^4x \left[
\, \phi \, F_{0} (-\partial^2_E / \Lambda ^2) \, \phi \, 
\right] \, , 
\ee
where the structure function $F_{0}$ is then given by (\ref{F0}):
\be
F_{0} \left (z\right) =     
2 \, z \, h\left(z\right) - 4  \, ,
\quad {\mathrm{with}} \quad
h(z)= \int_0^1 d\alpha \,e^{-\alpha (1-\alpha) z} .
\ee
The function $h(z)$ is an entire analytic function which is nowhere vanishing in the complex plane. The momentum-dependent two-point function for this model is then obtained by analytically continuing \eqref{spectral2} to Lorentzian signature
\be
\label{ss1}
\widetilde{G}(p^2) = - \frac{8 \pi^2}{\Lambda^2} \, \frac{1}{F_0(-p^2/\Lambda^2)} \, , 
\ee
where $p^2$ is the Lorentzian momentum four-vector.

A careful study of the two-point function \eqref{ss1} reveals several remarkable features. First, the model naturally gives rise to a Higgs mechanism for $\phi$. The propagator exhibits a pole at $p^2 \simeq -3.41 \Lambda^2$ indicating that the expansion of $\phi$ around vanishing field value corresponds to expanding at an unstable point in the potential. Restoring the $\phi^4$ term\footnote{For a discussion of the Higgs mechanism in almost-commutative geometry see section 11.3.2 of \cite{vanSuijlekom:2015iaa}.}
leads to a scalar potential
\be
V(\phi) = -\mu_H^2 \phi^2 + \lambda \phi^4 + \ldots  \, , 
\ee
with $\mu_H^2 = 2 \Lambda^2$. Neglecting the higher-order terms, the potential gives a non-vanishing vacuum expectation value $\langle \phi \rangle = \pm \frac{\mu_H}{\sqrt{2\lambda}}$. Expanding the field around this minimum leads to a potential for the fluctuation field $\tilde \phi$
\be
V(\tilde \phi) = 2 \, \mu_H^2 \, \tilde \phi^2 + \ldots \, . 
\ee
Thus, when expanded around the minimum of the scalar potential, the structure function entering into \eqref{ss1} should be given by
\be
\label{Fhiggsed}
F_H(z) = 2 \, z \, h(z) + 8 \, . 
\ee
$F_H(z)$ has a single real root located at $p^2 \simeq 2.56 \Lambda^2$. This root corresponds to a positive mass pole in \eqref{ss1}. In addition there are complex roots located, e.g., at
\be
p^2 = - \left( 1.32 \pm 21.98 i \right) \Lambda^2 \, . 
\ee
These roots can be traced back to the mass-term contribution in $F_0$ or $F_H$ and are absent if one considers the $z h(z)$ part only. The presence of complex roots signals that the Wightman function contains modes which increase exponentially for large times. These modes introduce an instability in the Unruh effect, which we will not investigate further. It would be very interesting to see if there are functions $\chi$ which give rise to a non-local theory avoiding this instability.

\subsubsection*{Ostrogradski-Type Models}

By making a suitable choice for the function $\chi$ one can also generate spectral actions which are local in the sense that the (inverse) two-point function is given by a finite polynomial in $p^2$.\footnote{This is closely related to the zeta-function spectral action proposed in \cite{Kurkov:2014twa}.}
The simplest choice, leading to a two-scale model, uses
\be
\label{chi1}
\chi(z) = (a + z) \, \theta (1-z) \, \, , \quad a > 0 \, .
\ee
Replacing the polynomial multiplying the step function by a polynomial of order $n$ leads to a multiscale model whose inverse propagator is given by a polynomial of order $n$ in $p^2$.

\begin{figure}[h]
\centering
\includegraphics[width=0.5\textwidth]{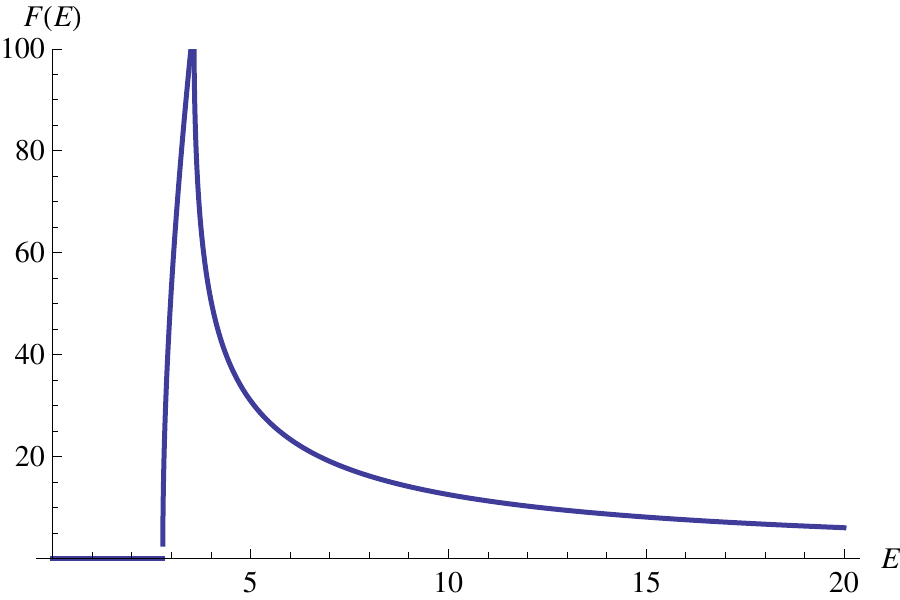}
\caption{\label{spectralrate} Profile function \eqref{spectralprofile} for $a=3.2$.
		}
\end{figure}

The spectral action for these cases can be found explicitly by combining the early-time expansion of the heat-kernel in $s \equiv \Lambda^{-2}$  
\be
F_H =  \, \frac{1}{(4\pi)^2} \frac{1}{s} \, \sum_{m=0}^\infty \, a_m \, (p_E^2 \, s)^m 
=  \frac{1}{(4\pi)^2} \frac{1}{s} \left( 8 + 2 \, s \, p_E^2  - \frac{1}{3} \left( s \, p_E^2 \right)^2 + \ldots \right)
\ee
with standard Mellin transform techniques \cite{Codello:2008vh}
\be
\label{spectralas}
S^{(2,\phi)}_{\chi,\Lambda}  =  \frac  {1}{(4\pi)^2} \int \frac{d^4p}{(2\pi)^4} 
\, \phi \, \left[ \sum_{m=0} \, Q_{m+1}[\chi] \, a_m \,   (p_E^2)^m \,\right] \, \phi \,  .
\ee
The moments $Q_n$ depend on the function $\chi$ and, for $n \in {\mathbb Z}$ are given by
\be
\begin{array}{ll}
Q_n[\chi] =  \frac{1}{\Gamma(n)} \int^\infty_0 dz \, z^{n-1} \, \chi(z) \, , \qquad & n > 0 \, , \\[1.3ex]
Q_{-n}[\chi] =  (-1)^n \, \chi^{(n)}(0) \, , \qquad & n \ge 0 \, .
\end{array}
\ee
For the ansatz \eqref{chi1} the moments are
\be
Q_1[\chi]  = a + \frac{1}{2} , \quad Q_0[\chi] = a, \quad Q_{-1}[\chi] = -1 , \quad
Q_{-2}  = Q_{-3} = \ldots =0.
\label{Qchi1}
\ee
Converting to Lorentzian signature, the inverse two-point function based on the expansion of $F_H$, \eqref{Fhiggsed} is
\be
\label{p3poly}
\cP_2(p^2) = - \frac{1}{8\pi^2} \left( 8a+4 - 2 a p^2 + \frac{1}{3} p^4 \right) \, . 
\ee
The two roots of the system are located at
\be
\mu_{1,2} = 3 a \mp \sqrt{ 9 a^2 - 24 a - 12 } \, . 
\ee
Provided that $2(2 + \sqrt{7})/3 < a < (3 + \sqrt{15})/2$, both roots are on the positive real axis. Thus the model falls into the class discussed in section  \ref{sect.41}. The profile function is readily obtained by applying the Ostrogradski decomposition to \eqref{p3poly}
\be
\label{spectralprofile}
{\cal F}  (E) = \frac{24 \pi^2}{\mu_2 - \mu_1} \left( \sqrt{E^2 -\mu_1} \; \theta( E -\sqrt{\mu_1}) - \sqrt{E^2 - \mu_2} \; \theta( E - \sqrt{\mu_2}) \right) \, . 
\ee
The behaviour of this profile function is illustrated in figure \ref{spectralrate}.

For $E^2 < \mu_1$ the profile function vanishes, indicating that the energy gap is too small for the detector to interact with the two massive fields. For $7.77 < E^2 < 12.77$ the profile corresponds to the standard Unruh rate for a field with mass $m^2 = 7.77$. Once $E^2$ crosses the threshold at $12.77$ the profile function decreases and falls off asymptotically as $E^{-1}$ for high energies. Thus spectral actions may give rise to similar profile functions as the multiscale models discussed at the beginning of this section.

\subsection{Causal Set Inspired Theories}
\label{sect.45}

A second framework which naturally gives rise to corrections to the Unruh effect are the non-local two-point functions emerging in the context of causal set theory. In this case the two-point functions extrapolate between a classical massless or massive propagator at energy scales well below the discretisation scale and a discrete d'Alembertian naturally associated with the causal set at high energies 
\cite{Aslanbeigi:2014zva,Belenchia:2014fda}. 

\begin{figure}
\centering
\includegraphics[width=0.49\textwidth]{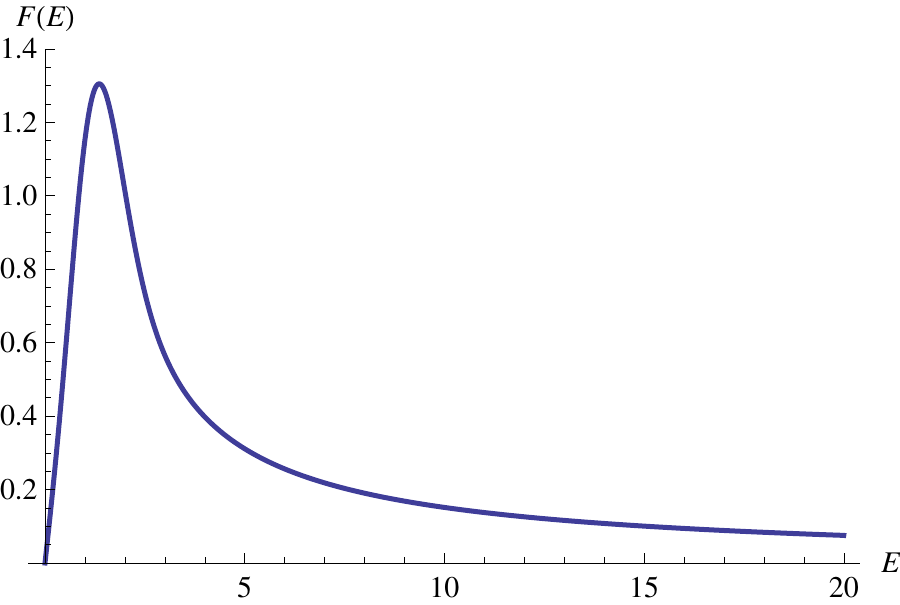} \, 
\includegraphics[width=0.49\textwidth]{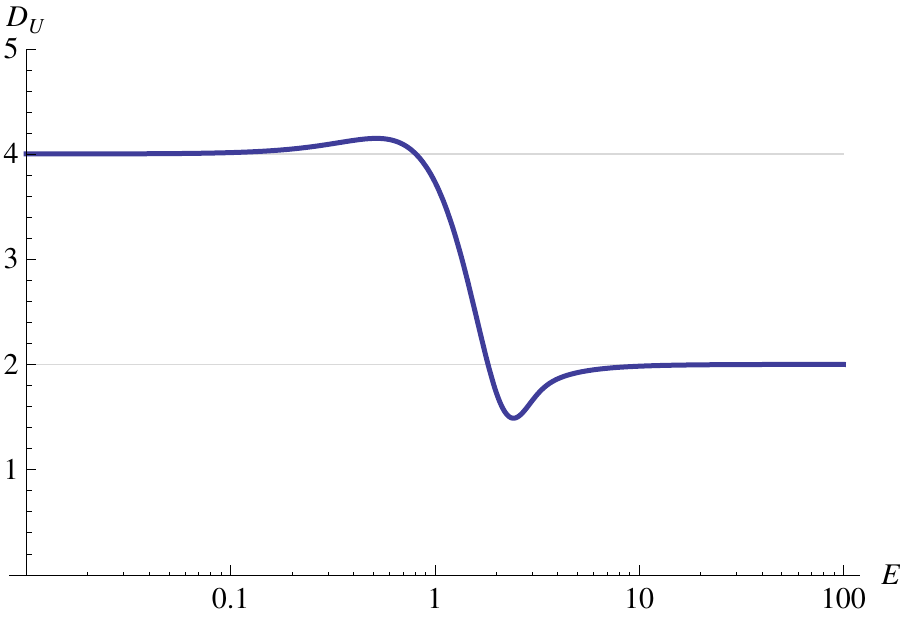}
\caption[Profile function ${\cal F}(E)$ and  Unruh dimension $D_U$ arising from a model to causal set theory.]{\label{Fig.causal} Profile function ${\cal F}(E)$ and  Unruh dimension $D_U$ arising from a model to causal set theory, for details see \cite{Versteegen:2016aar}.}
\end{figure}

The resulting Unruh signature arising from this setting as well as from causal set inspired toy models
can be found in the master thesis \cite{Versteegen:2016aar}. 
For completeness of the presentation here, the main result of a  model, constructed such that its
Unruh rate is expected to match the one of  causal set theory, is shown in figure  \ref{Fig.causal}. 
Both the profile function and the Unruh dimension undergo a transition when the energy scale 
meets the discretisation scale.

\section{Summary}
\label{sect.5}

In this chapter I reported on an investigation of 
the Unruh effect in quantum gravity inspired models exhibiting dynamical dimensional flows.
Both the detector approach to the Unruh effect 
and dimensional flows originate from a non-trivial momentum dependence of the two-point correlation functions
which leads to the anticipation of a natural connection between the two. This was not only verified
but also quantified. 
Explicitly, results on two-point functions arising within the context of phenomenologically motivated models 
for dynamical dimensional reduction, multiscale models, Kaluza-Klein theories, spectral actions, and causal set theory were obtained and used for the investigation of the question whether the related dimensional
flows are observable via the Unruh effect.
From the viewpoint of two-point functions, these models come in two distinguished classes. 
In the first case the inverse two-point function has a polynomial expansion in momentum space. 
This case is realised within dynamical dimensional reduction, multiscale models, Kaluza-Klein theories, 
and certain classes of spectral actions.
The models forming the second class possess two-point functions which are quasi-local in the sense that their inverse consists of a first order polynomial multiplying a function which is analytic in the complex plane. This setup is realised by causal set theory. The here presented study of these models exhibits two universal features. 
First, despite incorporating quantum (gravity) corrections in the two-point function, the Unruh radiation remains thermal in all cases. Moreover, the low-energy spectrum is robust with respect to corrections of the two-point functions at high energies, {\it i.e.}, the response of an Unruh detector is not modified below the characteristic scale where the dimensional flow sets in.

The two-point functions occurring in the first class of models can be reduced to a sum of (massive) second order propagators through an Ostrogradski-type decomposition. In this case we derive a master formula which expresses the response function of the Unruh detector as a function of the mass poles. As a generic feature, one finds that dynamical dimensional reduction leads to a suppression of the Unruh effect at high energies while the opening up of extra dimensions leads to an enhancement above the compactification scale. In particular, models where the spectral dimension asymptotes to $D_s \rightarrow 2$ at high energies also exhibit a universal falloff in the rate function \eqref{defresf} of the Unruh effect ${\mathcal F}(E) \propto 1/E$.
We proposed here to quantify this non-trivial asymptotic behaviour of the profile function through a new
parameter, which we called the \emph{Unruh dimension} of the system.
This is defined through the scaling of the profile function, as in  (\ref{UnruhDimension}).
Differently from other proposed parameters characterising the high energy behaviour
induced by quantum gravity effects, this one is directly related to a physical quantity
that is accessible experimentally, at least in principle. Moreover, it is directly related to the spectral dimension via the relation \eqref{dimrel}.
These specific examples already indicate that different quantum gravity models come with a very distinguished signature in terms of their Unruh detector response function. This may serve as an interesting starting point towards identifying universal features among different approaches to quantum gravity. This requires the computation of positive-frequency Wightman functions within different quantum gravity programs. 
This and related issues will be discussed on the concluding chapter \ref{Chap:Concl}.

\newpage
\thispagestyle{empty}


\chapter{{\color{MYBLUE}
{Black holes in Asymptotically Safe Gravity}}} \label{Chap7}
~\vspace{-10mm}

\centerline{This chapter is based on the following publication
\cite{Saueressig:2015xua}:}
\vspace{3mm}
\centerline{F.~Saueressig, N.~Alkofer, G.~D'Odorico and F.~Vidotto.}
\centerline{Black holes in Asymptotically Safe Gravity.}
\centerline{PoS FFP {14} (2016) 174,  104055, arXiv:1503.06472 [hep-th].}


\bigskip

Black holes have now become objects routinely observed in astrophysics
\cite{Falcke:2013ola}, and in addition, their existence has been impressively 
verified by the detection of gravitational waves originating from black hole 
mergers, see \cite{Abbott:2016blz}.
GR describes very well their exterior, as well as
their horizon. The theory is expected to fail close to and at the central singularity.  
On physical grounds, we expect the physics of the deep central
region to be strongly affected by quantum effects, therefore using general
relativity all the way up to the singularity is pushing the theory outside
its domain of validity. A theory reconciling GR with
quantum mechanics is needed to describe this central region.
In this section we analyse the consequences of a NGFP as, {\it e.g.},
studied in chapter \ref{Chap4}. 

\begin{figure}[t]
\centering
\includegraphics[width=0.65\textwidth]{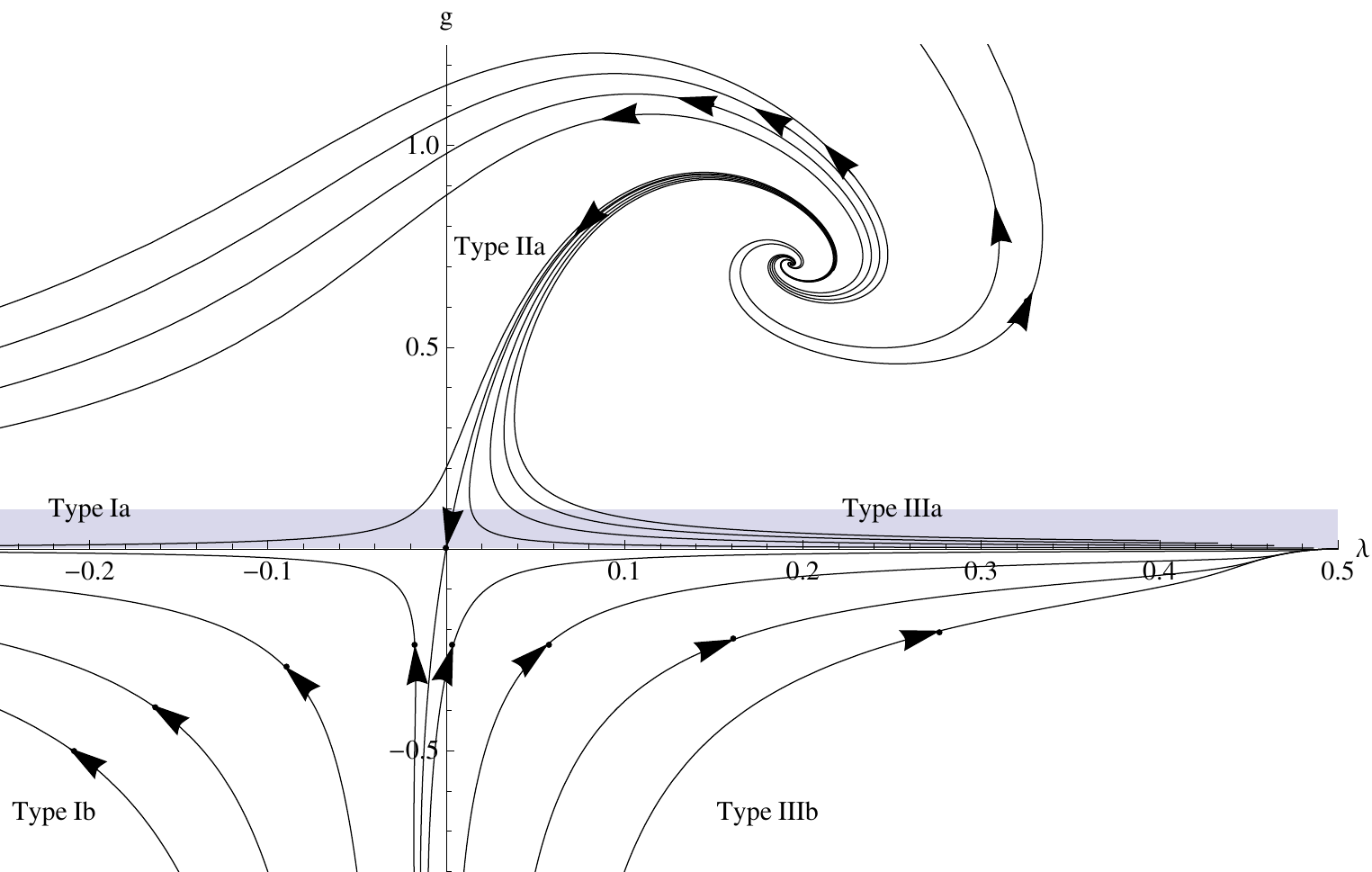}
\caption[Phase diagram showing the gravitational RG flow of the 
EH truncation in terms of dimensionless coupling constants .
The flow is governed by the interplay of the NGFP located at 
$g^\star > 0, \lambda^\star > 0$ and the GFP at the origin. ]
{\label{phasedia}
Phase diagram showing the gravitational RG flow of the EH truncation 
in terms of the dimensionless coupling constants $g_k \equiv G_k k^{2}$ and 
$\lambda_k \equiv \Lambda_k k^{-2}$. The flow is governed by the interplay of the NGFP 
located at $g^\star > 0, \lambda^\star > 0$ and the GFP at the origin. 
The arrows point towards the IR, i.e., in the direction of lower 
$k$-values. Adapted from \cite{Reuter:2001ag}.}
\end{figure}

For simplicity the following arguments will be based on 
the RG flow obtained by approximating $\Gamma_k$ by the EH action
\be
\Gamma_k^{\rm grav} = \frac{1}{16 \pi G_k} \int d^4x \sqrt{|g|} 
\left( 2 \Lambda_k - R \right) 
\ee
with scale-dependent (dimensionless) Newton's constant $g_k \equiv k^2 \, G_k$ 
and cosmological constant $\lambda_k \equiv \Lambda_k/k^2$,
{\it cf.} the discussion on the properties of the NGFP  in section \ref{AsympSafe}.
The flow diagram underlying the results presented in this chapter 
is shown in figure \ref{phasedia}. 
Depending on whether the RG trajectory ends at the GFP or flows to its left (right) a 
zero or negative (positive) IR value of the cosmological constant is recovered. The 
scaling of the coupling constants at the NGFP is easily deduced from the dimensionless 
couplings becoming constant,
\be
\label{NGFPscaling}
\mbox{NGFP:} \qquad G_k = g^\star \, k^{-2} \, , \quad 
\Lambda_k = \lambda^\star \, k^2 \, , 
\ee
while in the IR close to the GFP \cite{Reuter:1996cp}
\be
\label{lowenergy}
G_k = G_0 \left( 1 - \omega \, G_0 \, k^2 + \cO(G_0^2 \, k^4) \right) \, ,
\ee
with $\omega > 0$ a fixed number dependent on the particular choice of 
regularisation scheme.
For the regulator used in figure \ref{phasedia}, $\omega = 11/6\pi$.


Classical Schwarzschild black holes are exact vacuum solutions of Einstein's
field equations. The geometry is characterised by the line element
\be
\label{ssmetric}
ds^2 = f(r) \, dt^2 - f(r)^{-1} \, dr^2 - r^2 d\Omega_2^2
\ee
with $d\Omega_2^2$ denoting the line-element of the two-sphere and the radial function
\be
\label{frfct}
f(r) = 1 - \frac{2 \, G \, m}{r} \, . 
\ee 

Following \cite{Bonanno:1998ye,Bonanno:2000ep} quantum gravity corrections to the 
classical black hole geometry may be incorporated by RG improving the classical 
solution.\footnote{For an investigation of the RG improved black holes from the 
perspective 
of black hole thermodynamics see \cite{Falls:2012nd}. A recent review with
further references can be found in \cite{Koch:2014cqa}.}
The basic idea underlying the RG improvement is to promote the constant $G$ to depend on 
the RG scale $k$, replacing 
$G \mapsto G_k$. {Such a procedure is common in many areas of physics. 
Two prominent examples are the RG based derivation of the 
Uehling correction to the Coulomb potential in massless QED \cite{Dittrich:1985yb}
or the calculation of the factor provided by the instanton's scaling zero mode
by evaluating the running strong coupling at the inverse of the instanton radius
\cite{Coleman:1978ae}.} 
Subsequently, the RG scale is identified with a physical scale of the (classical) 
geometry.
For the spherically symmetric Schwarzschild solution, it is natural to relate $k$ to 
 the absolute value of the radial proper distance $d_r(r)$ between a point $P(r)$ in the 
spacetime and the centre of the black hole
\be
d_r(r) = \int \sqrt{|ds^2|} \, .
\ee
Close to the origin and at asymptotic infinity, $d_r(r)$ has the expansions
\be
\label{asympkvonP}
d_{r}(r)|_{r\ll 2 \, G_0 \, m} \simeq \frac{2}{3}\frac{1}{\sqrt{2 G_0 m}} \, 
r^{3/2} + {\mathcal{O}}(r^{5/2}) 
\, , \qquad
d_{r}(r)|_{r\gg 2 \, G_0 \, m} \simeq r+ {\mathcal{O}}(r^{0}) \, .  
\ee
The cutoff identification then relates the momentum scale $k$ to this distance 
according to
\be
\label{kvonP}
k(r)=\frac{\xi}{d_r(r)}\, , 
\ee
with $\xi$ being a free parameter. There is no predetermined recipe for determining the 
identification of $k$ with a physical scale of the system. An equally valid 
choice relates $k$ to the proper time measured by a freely falling observer 
starting at $P(r)$ to reach the black hole singularity. Notably, these alternative 
choices lead to similar results as the ones reported below. Applying this RG improvement procedure to
 the classical radial function \eqref{frfct} yields the RG improved geometry where $f(r)$ 
is given by
\be
\label{frimp}
f(r) = 1 - \frac{2 \, G(k(r)) \, m}{r} \, . 
\ee
The RG improvement changes the classical Schwarzschild metric
to a Hayward-type effective geometry \cite{Hayward:2005gi} 
with $f(r) = 1 - 2 M(r)/r$. Hereby, the function
$M(r)$ is determined from an RG trajectory constructed within the 
fundamental theory and the RG improvement  \eqref{kvonP}.

\begin{figure}[t]
\centering
\includegraphics[width=0.49\textwidth]{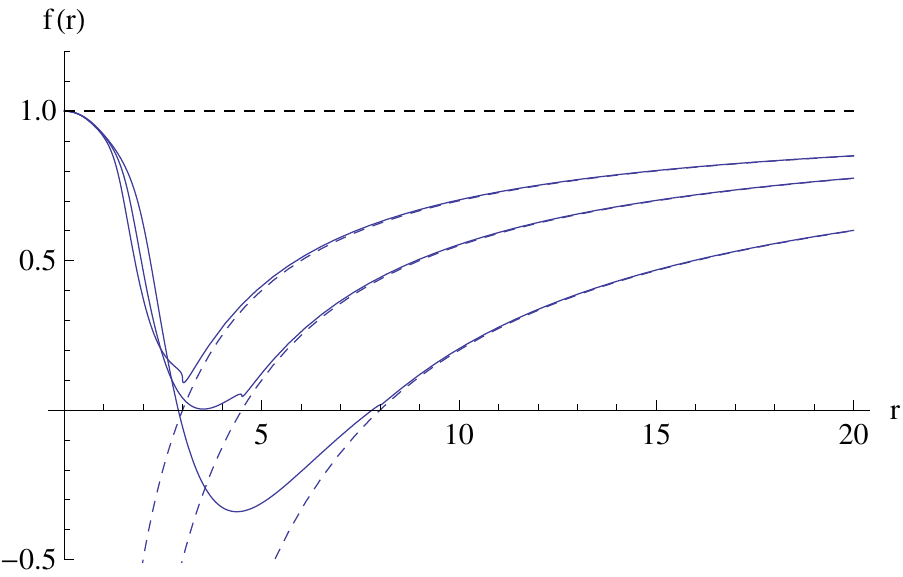} 
\includegraphics[width=0.49\textwidth]{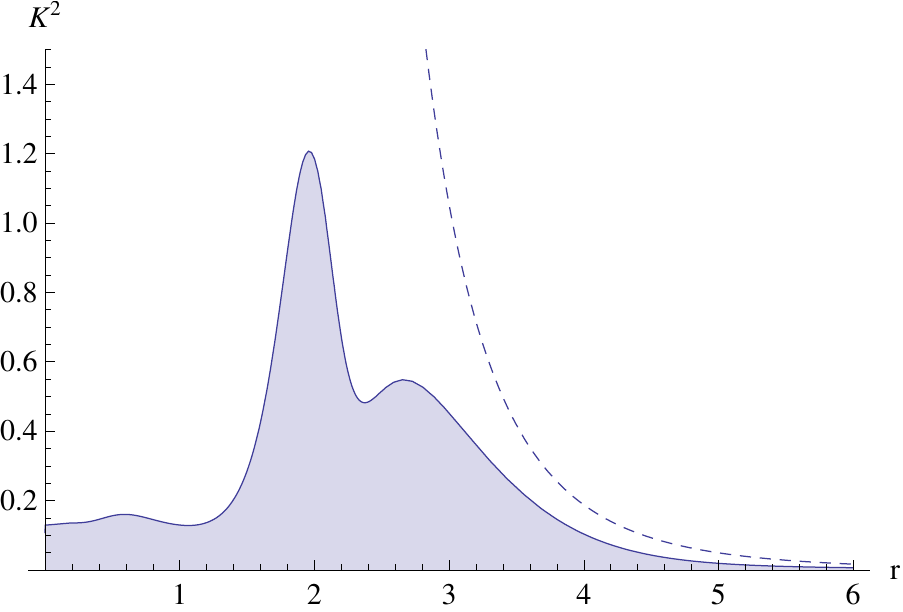}
\caption[The left diagram illustrates the horizon structure for the 
RG improved Schwarz\-schild black holes. The Kretschmann scalar curvature
$K^2 = R_{\mu\nu\rho\sigma} R^{\mu\nu\rho\sigma}$ for the case 
$m = 4 > m_{\rm crit}$ is shown in the right diagram.]
{\label{p.horizon}
The left diagram illustrates the horizon structure for the RG improved Schwarz\-schild 
black holes with $m = 1.5$ (top curve), $m = m_{\rm crit} \approx 2.25$ (middle curve) 
and $m = 4$ (bottom curve), while the Kretschmann scalar curvature 
$K^2 = R_{\mu\nu\rho\sigma} R^{\mu\nu\rho\sigma}$ for the case 
$m = 4 > m_{\rm crit}$ is shown in the right diagram. All quantities are 
measured in Planck units. The classical result is visualised by the dashed 
curves for comparison.}
\end{figure}
In the asymptotic regimes of the black hole spacetime,
the effect of the RG improvement can be traced analytically. Substituting
the low energy expansion \eqref{lowenergy} into \eqref{frimp} and evaluating the cutoff
identification \eqref{kvonP} for small $k$, yielding $k^2 = \xi^2/r^2 + \cO(r^{-3})$,
results in
\be
\left. f(r) \right|_{r\gg 2 \, G_0 \, m} \simeq 1 - \frac{2 G_0 m}{r} 
\left( 1 -  \frac{\tilde{\omega} \, G_0}{r^2} \right) \, , 
\ee 
with $\tilde \omega = \omega \xi^2$. The improved line-element naturally incorporates 
the 1-loop corrections found in effective field theory 
\cite{Donoghue:1993eb} and can be matched
by adjusting the free parameter in the cutoff identification to be 
$\xi^2 = \tilde \omega/\omega$.
Using $\omega = 11/6\pi$ and $\tilde \omega = 118/15\pi$ we obtain 
$\xi^{\rm 1-loop} \approx 2.07$,
which we will use in numeric evaluations below. Close to the black hole singularity, the 
RG improvement is based on the FP scaling \eqref{NGFPscaling}.
Substituting the asymptotic cutoff identification based on \eqref{asympkvonP} then yields
\be
\left. f(r) \right|_{r\ll 2 \, G_0 \, m} \simeq 1 - \tfrac{1}{3} \, 
\Lambda_{\rm eff} \, r^2 \, , \quad \mbox{with} \qquad \Lambda_{\rm eff}
 = \frac{4}{3} \frac{g_*}{G_0 \xi^2} \, . 
\ee 
Thus the RG improvement correctly incorporates the one-loop corrections
determined in effective field theory (fixing the only free parameter in the procedure)
and resolves the black hole singularity by giving rise to a de Sitter type behaviour 
close to the centre.

The complete RG improved radial function can be constructed numerically. For concreteness 
we choose the underlying RG trajectory to be the type IIa trajectory 
(see figure \ref{phasedia}) connecting the GFP with the NGFP, setting 
$\Lambda_0 = 0, G_0 = 1$. The resulting improved $f(r)$ depends on the asymptotic 
mass of the black hole $m$ only and is shown in the left diagram of 
figure \ref{p.horizon}. For $m > m_{\rm crit}$ the improved geometry has an 
outer and inner horizon. For $m = m_{\rm crit}$, with $m_{\rm crit}$ being 
of the order of the Planck mass, the two horizons coincide while for 
$m < m_{\rm crit}$ there is no horizon. The Kretschmann scalar curvature 
of the improved geometry (right diagram) peaks below the inner horizon 
and its maximum value is (approximately) given by the Planck scale. 

\begin{figure}[t]
\centering
\includegraphics[width=0.49\textwidth]{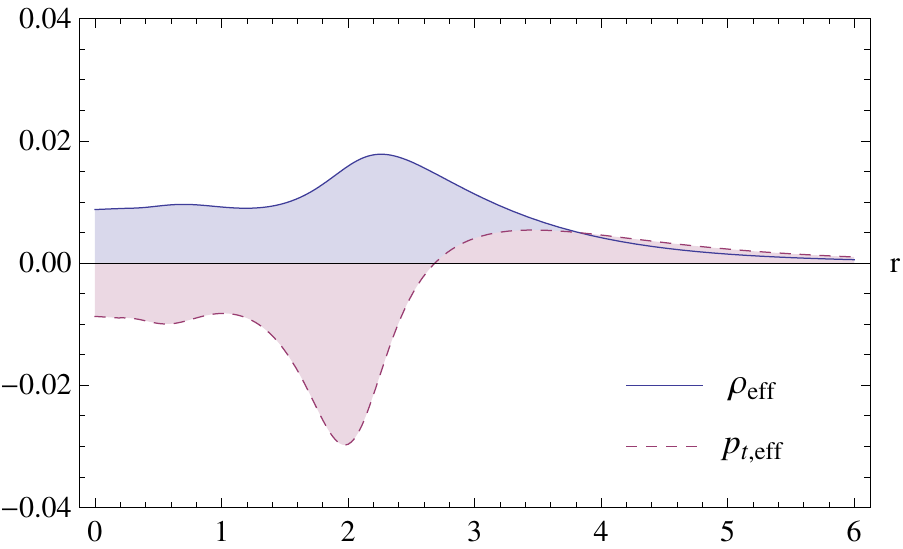} 
\includegraphics[width=0.49\textwidth]{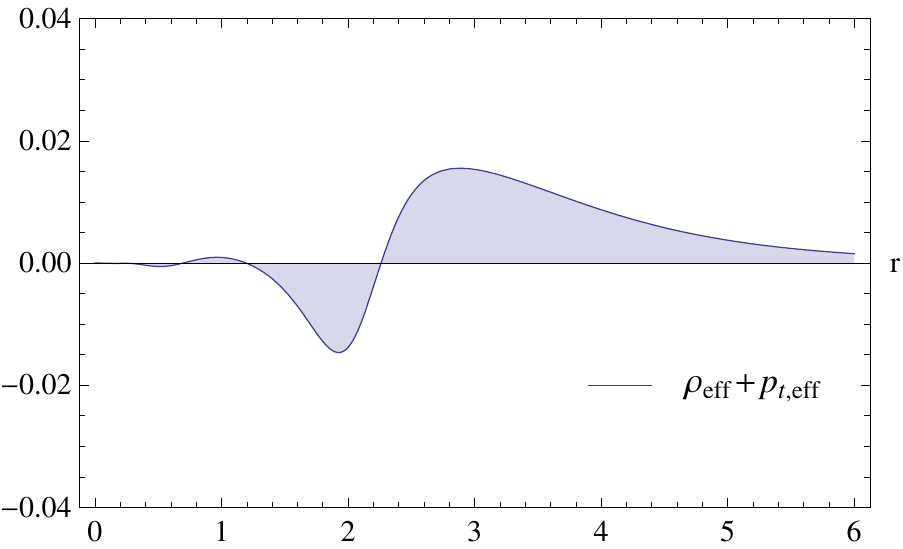}
\caption{\label{p.wec}
Effective energy density and pressure profiles for the RG improved Schwarz\-schild black 
hole with $m=4, G_0 = 1$. The radial component of the weak energy condition is zero 
everywhere, while the transversal contribution, shown in the right diagram, violates the 
condition on scales below the inner horizon.
}
\end{figure}
Substituting the RG improved geometry into the {\it classical} Einstein 
equations allows to interpret the resulting modifications in the classical 
black hole geometry as a quantum contribution to the energy momentum tensor. 
The resulting effective energy density $\rho_{\rm eff}$ and transverse 
pressure $p_{t, {\rm eff}}$ are shown in the left diagram of figure \ref{p.wec}. 
The radial pressure $p_{r, {\rm eff}} = - \rho_{\rm eff}$, so that the RG 
improvement acts like a cosmological constant in the radial direction. The 
right diagram of figure \ref{p.wec} displays the weak energy condition 
$\rho_{\rm eff} + p_{\rm eff}$ for the effective energy momentum tensor. 
Notably, 
it is violated at subhorizon scales due to strong transversal pressure.

Recently, \cite{Koch:2013owa} generalised the study of the RG 
improved Schwarzschild black hole to Schwarzschild-(Anti-) de Sitter
black holes, with the radial function \eqref{frfct} also including a non-zero 
cosmological constant
\be
\label{frcosmo}
f(r) = 1 - \frac{2 \, G \, m}{r} - \frac{1}{3} \, \Lambda \, r^2 \, . 
\ee
This extension is motivated by the observation that, even for the case where 
$\Lambda_0 = 0$
a non-zero cosmological constant will be generated along the RG flow 
({\it cf.}\ figure \ref{phasedia}).
Moreover, in the vicinity of the NGFP the scaling \eqref{NGFPscaling} 
implies that the {\it dimensionful} Newton's constant goes to zero while 
the dimensionful cosmological constant
actually diverges when $k \rightarrow \infty$. Thus, contradicting the 
intuition that the cosmological 
constant is important at large distances only, its inclusion may also 
influence the structure
of microscopic black holes. Indeed, applying the RG improvement procedure for the
Schwarzschild case to the radial function \eqref{frcosmo} and evaluating the result 
for the FP scaling \eqref{NGFPscaling}  
the RG improved line-element \emph{valid at the NGFP} is again of the 
form \eqref{frcosmo}
\be
f^\star(r) = 1  - \frac{2 \, G_0 \, m}{r} \left( \frac{3}{4} 
\lambda^\star \xi^2  \right) - \frac{1}{3} \, \left( \frac{4 g^\star}
{3 G_0 \xi^2} \right) \, r^2 \, .
\ee
Thus the RG improved Schwarzschild-de Sitter black holes become self-similar 
in the sense that 
the line-element takes the same form in the IR and UV.
Notably, the inclusion of the scale-dependent cosmological constant has also reintroduced 
a singularity which, for the case of the Schwarzschild black hole, has been removed by 
the RG improvement process. It is worth stressing that the physical nature of this new 
singularity is actually quite different to the one found in the classical black hole 
solution: applying the RG improvement procedure to flat Lorentzian spacetime also 
introduces a singular behaviour of the RG improved line-element even in the absence of 
any matter. This nurtures the speculation that the ``quantum'' singularity introduced by 
the cosmological constant may actually reflect a feature of quantum spacetime which is 
actually unrelated to the study of black holes \cite{Koch:2013owa}.

In summary, the RG improved Schwarzschild black holes found within the
asymptotic safety scenario \cite{Bonanno:1998ye,Bonanno:2000ep} naturally fall into the 
class of 
Hayward metrics \cite{Hayward:2005gi} which have been proposed as effective models for 
non-singular black holes.

\newpage
\thispagestyle{empty}


\chapter{{\color{MYBLUE}{Conclusions and Outlook}}}
\label{Chap:Concl}
~\vspace{-10mm}

\rightline{\footnotesize{\it{``Habe Mut, dich deines eigenen Verstandes zu bedienen."}}}
\rightline{\footnotesize{\it{(I. Kant)}}}
\minitoc

\bigskip



One of the main achievements presented in this thesis are the results for
the fixed functions arising within $f(R)$-gravity minimally coupled to an 
arbitrary number of scalar, Dirac, and vector fields. On the one hand, they
allowed a comprehensive picture detailing the existence and stability of 
interacting RG FPs in gravity-matter systems 
taking higher-order curvature terms into account. Furthermore, some conflicts 
related to   previously achieved  results in the literature (see, {\it e.g.},
\cite{Daum:2010bc,
  Folkerts:2011jz,Eichhorn:2011pc,
  Eichhorn:2012va,Dona:2012am,Henz:2013oxa,Dona:2013qba,
  Labus:2015ska,Oda:2015sma,Meibohm:2015twa,Dona:2015tnf,
  Meibohm:2016mkp,Eichhorn:2016esv,Henz:2016aoh,Eichhorn:2016vvy,
  Christiansen:2017gtg,Eichhorn:2017eht,Christiansen:2017qca,
  Eichhorn:2017ylw,Eichhorn:2017egq,Christiansen:2017cxa}) could be resolved 
by demonstrating that the obtained NGFPs for two different coarse-graining schemes belong to different classes, and only one of these classes is stable by including higher-order operators in the action. This provides evidence that the use of these two different schemes leads to different quantisation prescriptions for the same
``classical'' theory. Such a question deserves certainly further investigations.

While working on this project the same and related topics have been studied by 
others, and hereby the remaining differences to two publications which appeared
very recently call special attention. Fixed functions with a sphere as background have been
calculated, however, using a linear split of the metric and a vertex expansion 
\cite{Christiansen:2017bsy}. Although these authors have a very different way 
of presenting this function it is obvious on first sight that their results 
are qualitatively different. A further investigation has two possible starting
points to resolve this tension: first, as suggested by comparing, {\it e.g.}, to
\cite{Ohta:2015fcu}, using either the linear or the exponential split is the 
cause for the differences. If this is the case this calls then for arguments
why at least one of the two splittings should not be used. Second, the calculated
functions are not really the same quantity. Nevertheless, they should be related
and the task would be to elucidate the relation between them.

As by others before, see, {\it e.g.}, \cite{Dona:2013qba} as a key reference, 
a dependence of the existence of a NGFP on the coarse-graining operators
is seen in this work.
In other words, ``matter matters'' for the asymptotic safety scenario. In more
recent publications \cite{Meibohm:2015twa,Christiansen:2017cxa} arguments were
presented that  asymptotic safety for gravity-matter systems follows always from 
asymptotic safety of pure gravity independent of how much matter is added,
{\it i.e.}, ``gravity rules''. These studies are based on a vertex expansion 
but around flat backgrounds. On the one hand, the vertex expansion is certainly 
a more sophisticated truncation scheme than the single-metric background
approach used here. On the other hand, my results make it evident that 
some important features can only be detected with curvature also present in 
the background. And therefore, my approach might contain some 
of the essential clues to obtain a final resolution on this important question.


Two further projects are concerned with the supposedly non-manifold like structure
of spacetime at very small distances. To this end, the spectral dimension
of the spectral action has been studied.
The vanishing of this spectral dimension is in agreement with the 
previous computations obtained for other non-commutative spacetimes 
\cite{Alesci:2011cg}.
Aside for the limiting case of vanishing spectral dimension, both results 
display the same qualitative features:
the spectral dimension interpolates between the topological dimension and zero,
and has a local maximum situated close to a transition scale.
Additionally, the onset of the transition is, in both cases, controlled by the 
parameter which is introduced by the spectral action.

In agreement with the conjecture of ``asymptotic silence'' pushed forward in 
\cite{Carlip:2009kf},
it is a general feature of our computation that the spectral dimension 
decreases in the UV. However,  these results are in a striking contrast to 
the generalised spectral dimension typically obtained within 
quantum gravity approaches where it 
interpolates between  four (classical, macroscopic phase) and 
a non-vanishing number smaller than four (typically at or around two) at 
microscopic scales \cite{Carlip:2009kf}.

At this stage, it is tempting to speculate  that the vanishing spectral 
dimension is an artefact of extrapolating the classical spectral action 
into the trans-Planckian regime without taking quantisation effects into account.
In other words, vacuum fluctuations seem to be fundamental to obtain 
the value for the spectral dimension of spacetime at the smallest distances.

An UV completion of the spectral action, taking quantum fluctuations
into account, could be achieved through  the asymptotic safety mechanism, which 
seems a natural choice given the field content and symmetries of the model. 
This might serve then as a starting point for a further investigation based 
on the results presented in this thesis.


A similar remark applies to the project about the Unruh effect. For a 
further detailed investigation of the ``Unruh dimension" proposed by 
my coauthors and me, it would be quite natural to apply the formalism 
described in this thesis to the gravitational asymptotic safety program. 
In this context, the momentum dependence of two-point functions has 
recently been studied in \cite{Dona:2013qba,Christiansen:2015rva,Meibohm:2015twa}. 
It is clear that an investigation of the Unruh effect should be based on the renormalised 
propagators where all quantum (gravity) fluctuations have been incorporated. 
The corresponding expression for the positive-frequency Wightman function is 
unfortunately currently not available.
Nevertheless, much progress has been made in recent years towards the construction 
of renormalised fully-dressed two-point functions
\cite{Codello:2013fpa,Christiansen:2012rx,Christiansen:2014raa,Christiansen:2015rva,
Meibohm:2015twa,Eichhorn:2016esv}.
This will make it feasible to compute the signatures of asymptotic safety 
in the Unruh effect which will likely also be relevant for understanding the fate of black 
holes within the asymptotic safety scenario
\cite{Bonanno:1998ye,Bonanno:2000ep,Bonanno:2006eu,Reuter:2006rg,Reuter:2010xb,
Becker:2012js,Falls:2012nd,Becker:2012jx,Koch:2013owa,Koch:2014cqa,Saueressig:2015xua,
Litim:2013gga}.

Finally, we have not analysed the class of models displaying a minimal length.
These models are important for quantum gravity phenomenology, since this effect
is believed to appear quite generically \cite{Garay:1994en}.
It would be interesting to see if a connection to the here presented results can be made.

Another natural extension is the application to Hawking radiation. Here it was argued that the 
low-energy Hawking spectrum is actually insensitive to Planck scale effects \cite{Agullo:2009wt}. 
The situation is quite similar to the one encountered in the present work, where the Unruh 
spectrum at energy scales below the scale where the dimensional flow sets in is actually unaltered. 
At the same time there are indications that quantum gravity effects could stop the black hole 
evaporation process and leave a cold remnant. In particular, it was argued in \cite{Carlip:2011uc} 
that the black hole evaporation could come to an end once the spectral dimension drops to three.
This would be relevant for the information problem as well \cite{Chen:2014jwq}.
Applying the techniques based on two-point correlation functions used in the present work may 
actually allow one to develop these ideas based on a first-principle calculation.


The project described in the previous chapter is about RG improved Schwarzschild 
black holes found within the asymptotic safety scenario 
\cite{Bonanno:1998ye,Bonanno:2000ep}. It is shown that they naturally fall into the 
class of  Hayward metrics \cite{Hayward:2005gi} which have been proposed as 
effective models for non-singular black holes.
The disappearance of the central singularity has also been observed
in other approaches to quantum gravity, as, {\it e.g.}, in loop quantum
gravity.   This physical scenario has been recently investigated in 
\cite{Rovelli:2014cta,DeLorenzo:2014pta,Haggard:2014rza}, where the 
potential non-singular central core of the black hole is called ``Planck star''. 
Interestingly, this opens a new 
window for quantum gravity phenomenology \cite{Barrau:2014hda,Barrau:2014yka} 
as the resulting ``bounce'' of the ``Planck star'' should give a characteristic 
astrophysical signal. 
This substantiates the hope that this possible link between fundamental theories of gravity 
 will be useful towards improving our understanding of  black hole evolution.


In this thesis I have explored different routes to quantum gravity. The obtained 
results provide some insight about the meaning of the quote of 
John Archibald Wheeler 
``No question about quantum gravity is more difficult than the question
``what is the question?''.''
Although some of the starting questions which triggered the research 
presented here have now been answered, the most 
useful output has been quite obviously to generate more questions.
In this sense I hope that I could contribute a little if one day in the 
future somebody will find the decisive key question and maybe even 
an answer to it.


\newpage
\thispagestyle{empty}


\begin{appendix} 

\chapter{\color{MYBLUE}{Laplace Operators}}
\label{App.Laplace}

To make this thesis reasonably self-contained the properties of some widely-used
different types of Laplace operators are here briefly reviewed, see, {\it e.g.}
section 5.3 of \cite{Percacci:2017fkn} on which the presentation here is based. 

On a Riemannian manifold one may simply use the covariant derivative containing 
the Levi-Civita connection to define the Laplacian $\Delta = - D^2$.  However, it 
does not have typically required properties as, {\it e.g.}, respecting symmetries
of tensors it acts on.

To arrive at a more suitable definition let us recall the equivalence of differential forms
and antisymmetric covariant tensors. On such forms one defines the differential $d$ 
as a map from the space of $p$-forms into the $p+1$-forms
\be
(d\omega)_{\mu_1 \ldots \mu_{p+1}} = (p+1) \partial _{[\mu_1} 
\omega _{\mu_2 \ldots \mu_{p+1}]} \, 
\ee
where the parentheses $[\ldots ]$ indicate antisymmetrisation.  
On a Riemannian manifold with metric $g$ one can define the usual inner product of $p$-forms
which then allows one to define the co-differential $\delta$ via the relation
\be
(d\omega , \rho ) = (\omega , \delta \rho) \, 
\ee
which makes evident that $\delta \rho$ is a $p$-form. Within a given coordinate
 system one has
\be
(\delta \omega )^{\mu_1 \ldots \mu_{p-1}} = 
- \frac 1 {\sqrt{g}} \partial_\lambda \left( \sqrt{g} \,  
\omega ^{\lambda \mu_1 \ldots \mu_{p-1}} \right) 
= - D_\lambda \omega ^{\lambda \mu_1 \ldots \mu_{p-1}} \, .
\ee
The Laplacian on $p$-forms, also called Laplace-Beltrami operator, is then given by
\be
\label{DeltaLB}
\Delta_{LB} = d\delta + \delta d \, .
\ee
Its action on a scalar is equal to the action $\Delta = - D^2$ but on one-forms it differs already,
$\Delta_{LB} \omega_\mu = -D^2 \omega_\mu + R_\mu^{~\nu} \omega_\nu $. Its action 
on two-forms involves then several terms, also including the Riemann tensor.

The nilpotency of the differential and the co-differential, $d^2=0$ and $\delta ^2=0$, imply
that these operators commute with the  Laplace-Beltrami operator\footnote{Note, however,
the different order of forms the respective left and right hand sides act upon.}
\be
d\Delta_{LB} =\Delta_{LB} d \, , \quad \delta \Delta_{LB} =\Delta_{LB}  \delta \, ,
\ee
which is a very welcome property.

The generalisation of this construction is called Lichnerowicz Laplacians. Its action on a 
general covariant $p$-tensor is given by
\be
(\Delta_{Lp} T)_{\mu_1\ldots \mu_p} = - D^2 T _{\mu_1\ldots \mu_p}
+ \sum _k R_{\mu_k}^{~\rho}  T _{\mu_1\ldots \rho \ldots \mu_p} - 
\sum _{k\not = l} R _{\mu_k}^{~\rho}{}_{\mu_l}^{~\sigma } 
T _{\mu_1\ldots \rho \ldots \sigma  \ldots \mu_p} \, .
\ee
Hereby, the indices $\rho$ and $\sigma$ are in the positions $k$ and $l$, respectively. 
The metric compatibility condition $D_\rho g_{\mu\nu} =0$ allows to freely raise and lower 
indices. 

The Lichnerowicz Laplacian preserves the type and the symmetries of the tensor it acts on, 
it is self-adjoint, and it commutes with contractions. It coincides with the Laplace-Beltrami 
operator \eqref{DeltaLB} when acting on totally antisymmetric tensors.
On manifolds of the Einstein type this Laplacian commutes with the covariant derivative.
With $\phi$, resp.,  $\xi^\mu$, being an arbitrary scalar, resp.,  vector, field one has on these manifolds
\ba
\Delta_{L1}D_\mu \phi &=& D_\mu \Delta_{L0} \phi \, ,\\
D_\mu \Delta_{L1} \xi^\mu &=& \Delta_{L0} D_\mu \xi^\mu  \, , \\
\Delta_{L2} (D_\mu D_\nu \phi) &=& D_\mu D_\nu  \Delta_{L0} \phi \, ,\\
\Delta_{L2} D_{\{ \mu } \xi_{\nu \}} &=& D_{\{ \mu }  \Delta_{L1}  \xi_{\nu \}} \, , \\
\Delta_{L2}  g_{\mu\nu} \phi &=& g_{\mu\nu} \Delta_{L0} \phi \, ,
\ea
where, of course, $\Delta_{L0} = \Delta = -D^2$. 

Examples for the use of these Lichnerowicz Laplacians in the context of gravity can be found 
in the book \cite{Percacci:2017fkn} or in \cite{Benedetti:2009rx}.


\chapter{\color{MYBLUE}{Dirac Operator on 
Spheres $S^d$: Eigenvalues and Degeneracies}}
\label{App.Dirac}

For the calculation of the eigenvalues of the Dirac operator on $S^d$ I will follow \cite{Camporesi:1995fb}.
This is done first for the unit sphere, and the dependence on the scalar curvature will 
be concluded from dimensional arguments.
As usual for the consideration of the Dirac operator even and odd number of dimensions 
behave quite differently and will be treated separately.

First, the construction of Dirac matrices is needed, i.e., one searches for $d$ matrices 
$\Gamma^\mu$ fulfilling the Clifford algebra
\begin{equation}
\{\Gamma^k, \Gamma^j \} = 2 \delta^{kj} {\mathbf 1} .
\end{equation}
These can be constructed inductively, and the dimension of the matrices is then 
$2^{\lfloor d/2\rfloor }$ where $\lfloor d/2\rfloor  = d/2$ for even $d$ and
$(d-1)/2$ for odd $d$. For $d=2$ one chooses the first two Pauli matrices, 
$\Gamma^1=\sigma^1$ and $\Gamma^2=\sigma^2$,
for $d=3$ one adds $\Gamma^3=\sigma^3$.

For $d=4$:
\begin{equation}
\Gamma^4 = \begin{pmatrix} 0 & {\mathbf 1}_2 \\ {\mathbf 1}_2 & 0 \end{pmatrix} \,  , 
\quad
\Gamma^j = \begin{pmatrix} 0 & i \Gamma^j_{d=3} \\  - i \Gamma^j_{d=3} & 0 \end{pmatrix} 
\,  , j=1,2,3.
\end{equation}

For $d=5$ we add to these four matrices $\Gamma^5 = (-i)^2 \Gamma^1 \Gamma^2 \Gamma^3 
\Gamma^4 =\begin{pmatrix}  {\mathbf 1}_2 & 0 \\ 0& - {\mathbf 1}_2  \end{pmatrix}$.

From here on the general pattern should be obvious. A discussion of the properties of 
these matrices as well as their connection to the
fundamental representations of the groups $Spin(d)$ (which for $d>2$ is the universal 
covering of $SO(d)$) is given in section 2 of \cite{Camporesi:1995fb}. In the following the $\frac 1 2 d(d-1)$ matrices (=generators of 
a representation of $Spin(d)$)
\begin{equation}
\Sigma^{kl} = \frac 1 4 [\Gamma^k, \Gamma^l]
\end{equation}
are needed.

The metric on $S^d$ can be constructed from the one on $S^{d-1}$:
\begin{equation}
ds_d^2 = d\theta^2 + \sin^2\theta ds_{d-1}^2 = d\theta^2 + 
\sin^2\theta  \tilde g_{ij} d\omega^i \otimes d\omega^j \, .
\end{equation}
Correspondingly one can construct from the vielbeins on 
$S^{d-1}$ ($\tilde{ \mathbf e} _i$)
the ones on $S^d$ ($ \mathbf e _\mu$), and for the Levi-Civita connections one obtains
(the tilde always refers to the lower dimensional case)
\begin{equation}
\label{eqB5}
\omega_{ijk} = \frac 1 {\sin \theta} \tilde  \omega_{ijk} \, , \quad
\omega_{idk} = - \omega_{ikd} = \cot \theta \delta_{ik} \, , \quad i,j,k=1, \ldots d-1.
\end{equation}
The covariant derivative in spinor representation is then given by
 $D_i = {\mathbf e}_i^{~j}\partial_j- \frac 1 2 \omega_{ijk}\Sigma^{jk}$,
and the Dirac operator is then $\Dslash = \Gamma^jD_j$.

\bigskip

For even $d$ one has
\begin{equation}
\Dslash = (\partial _\theta + \frac 1 2 (d-1) \cot \theta ) \Gamma^d 
+ \frac 1 {\sin \theta}
\begin{pmatrix} 0 & i \tilde \Dslash \\  - i \tilde \Dslash & 0 \end{pmatrix}
\,\, .
\end{equation}
The decisive step is to assume that the solution of the eigenvalue equation on $S^{d-1}$ 
is of the form
\begin{equation}
\tilde \Dslash \,  \chi^\pm_{lm} (\Omega) = \pm i \bigl( l + \frac1 2 (d-1)\bigr)  \chi^
\pm_{lm} (\Omega), \quad l=0,1,2,\ldots
\end{equation}
and to verify this assumption a posteriori.  Hereby, the $\chi^\pm_{lm} (\Omega)$ are
suitable spinors which depend on the angles $\Omega$ on $S^{d-1}$, the index $m$ compromises
all other indices except the index $l$.
The resulting differential equation for the 
variable $\theta$ can be solved in terms of Jacobi polynomials explicitly  \cite{Camporesi:1995fb}.
As the solutions are explicitly known the degeneracies can be calculated directly:
\begin{equation}
D_d(n)= 2^{d/2} {n+d-1 \choose n}.
\end{equation}
As anticipated, $D_d(n)$ is equal to the dimension of the spinor representation of 
$Spin(d+1)$.

On $S^4$ one has $\lambda_n = \pm i (n+2) $  and $D_4(n)=\frac 4 3 (n+1)(n+2)(n+3)$.

\bigskip

For odd $d$ the starting point is
\begin{equation}
\Dslash = (\partial _\theta + \frac 1 2 (d-1) \cot \theta ) \Gamma^d + 
\frac 1 {\sin \theta} \tilde \Dslash \, .
\end{equation}
By explicit calculation one can then show that the eigenvalues are also then 
$\pm i(n+d/2)$ and
\begin{equation}
D_d(n)= 2^{(d-1)/2} {n+d-1 \choose n}.
\end{equation}

\bigskip
\bigskip

As stated in the beginning of the section this calculation has been done for the unit 
sphere. As the eigenvalue has the dimension of an
inverse length, for a sphere of general radius $a$ the eigenvalue should be multiplied by 
$1/a = \sqrt{R/d(d-1)}$. The final result is therefore:
\begin{equation}
\lambda_n = \pm i \sqrt{\frac R {d(d-1)}} \left( n+ \frac d 2 \right) , \quad D_d(n)= 2^
{[d/2]} {n+d-1 \choose n} , \quad n=0,1,2, \ldots .
\end{equation}
This agrees with the expression used in \cite{Dona:2012am} (and which was given 
there without reference).


\chapter{\color{MYBLUE}{Derivation of the Functional Renormalisation Group Equation}}
\label{App.FRGderiv}

Here, I provide a brief presentation of the derivation of the Wetterich equation  \cite{Wetterich:1992yh}.
The starting point is the generating functional $Z[J]$ where $J(x)$ is a source coupled to the
field(s) which will be generically denoted by $\Phi$.  
Taking the logarithm provides the generating functional of the connected Green
functions $W[J]$. A Legendre transform leads to the effective action $\Gamma [\bar \Phi]$
 which is the  generating functional of the one-particle irreducible Green functions. 
Hereby, $\bar \Phi$ is the expectation value of the fields $\Phi$.   In the
Wilsonian RG approach  one also performs these three steps, however, with a  $k$-dependent
regulator function included, {\it i.e.}, one starts with\footnote{Only
Euclidean QFTs are considered.  
In addition, I will use the DeWitt notation $J\cdot \Phi = \int d^Dx J(x) \Phi (x)$.}
\be
Z_k[J] = \int {\cal D} \Phi \exp \left( -S[\Phi] -\Delta S_k[\Phi] +J\cdot \Phi \right) \,\,.
\ee
In this functional integral the term $\Delta S_k[\Phi]$ is included to provide a smooth
momentum cutoff such that the UV modes are unchanged and the IR modes are suppressed.
The standard choice is a quadratic cutoff which, in momentum space, reads
\be
\Delta S_k[\Phi] =  \frac 1 2 \int \dq \, \Phi(-q) \, R_k(q^2) \, \Phi(q) \,\,.
\ee
For the UV modes one thus requires
\be
\lim_{q^2/k^2\to \infty}  R_k(q^2) = 0
\ee
and for the IR modes
\be
R_k(q^2)  
> 0 \quad {\mathrm {for}} \quad q^2/k^2\to 0 \, , 
\ee
{\it cf.} also the discussion in \cite{Litim:2008tt}.

The flow for the generating functional $W_k$ is then straightforwardly determined to be
\be
\partial_t W_k[J] = \partial _t \ln Z_k[J] = - \langle \partial_t \Delta S_k \rangle
= - \frac 1 2 \int \dq \, \partial_t R_k(q^2) \, \langle \Phi(q) \Phi(-q) \rangle \,\,.
\ee
Next one introduces the Legendre transformed functional
\be
\Gamma_k[\bar \Phi] = \sup_J \left( J\cdot \bar \Phi - \ln Z_k [J] - \Delta S_k[\bar \Phi] \right)
\ee
where $\bar \Phi$ is the expectation value of $\Phi$ for a fixed external current $J$,
$\bar \Phi = \langle \Phi \rangle _J$. Returning for clarity for the time being to a real-space
representation
one employs then the definition of the connected two-point function
\be
G(x,y) := \frac{\delta^2\ln Z_k [J]}{\delta J(x) \, \delta J(y) }
\ee
and the fact that the quantum equation of motion
\be
J(x) = \frac{\delta \Gamma_k [\bar \Phi]}{\delta \bar \Phi (x)} +
(R_k\cdot \bar \Phi ) (x)
\ee
leads to
\be
G(x,y) = \Bigg(
\frac{\delta^2 \Gamma_k [\bar \Phi]}{\delta \bar \Phi (x)\,\delta \bar \Phi (y) }
+ R_k (x-y) \Bigg)^{-1}
\ee
to show that in momentum space  the  equation
\be
\partial _t \Gamma_k [\bar \Phi] = \frac 1 2 \int \dq \Bigg(
\frac{\delta^2 \Gamma_k [\bar \Phi]}{\delta \bar \Phi (q)\,\delta \bar \Phi (-q) }
+ R_k (q^2) \Bigg)^{-1} \partial_t R_k (q^2) \,\,
\ee
is fulfilled.
In case discrete indices are present they have to be summed over. Denoting by
$\mathrm{Tr}$ these sums as well as the integrals leads then to
the compact notation for the Wetterich equation \cite{Wetterich:1992yh}
\be
\partial _t \Gamma_k = \frac 1 2 {\mathrm{Tr}}
{\Bigg(\left( \Gamma_k^{(2)}
+ R_k  \right)^{-1} \partial_t R_k \Bigg)  } 
\label{eq:B.23}
\ee
with
\be
\Gamma_k^{(2)}
:= \frac{\delta^2 \Gamma_k [\bar \Phi]}{\delta \bar \Phi (q)\,\delta \bar \Phi (-q) }
\,\,.
\ee

In this thesis the Wetterich equation has been employed for  gravity and gravity-matter systems. 
To this end the gauge has been 
fixed and correspondingly ghosts have been added. In addition, the York decomposition
\eqref{york} has been applied. The Hessian $ \Gamma_k^{(2)}$ becomes
then effectively a matrix, and the regulator has to be chosen accordingly also as a matrix.


\chapter{{\color{MYBLUE}{Topological, 
Hausdorff and Spectral Dimensions}}}
\label{MathSpecDim}

Fractals are geometrical structures which cannot be described with ordinary Riemannian 
(or pseudo-Riemannian) geometry. Typically  a part of a fractal is similar to the
whole fractal such that it is invariant under certain scale transformations. 
Phrased otherwise, a fractal is self-similar under scale transformations. 

A line like the coast of England might show over and over again the same structures
when zooming into the map. If this goes on forever the length of such a line diverges
but it also does not cover an area. Attributing a dimension other than the topological 
dimension of the line (which is clearly one) to it the related number 
should be larger than one and smaller than two. Such non-integer numbers are 
provided by the Hausdorff dimension. 

For its definition one introduces first the Hausdorff measure \cite{Falconer:2014} of a 
fractal $F$. Given a finite and countable collection of non-empty sets 
$U_i  \in \mathbb R^d$ with a finite largest distance each, $|U_i|< \delta$ 
which covers $F$ the $\{ U_i \}$ are called a $\delta$-cover of $F$.

For any positive power $s$ and any $\delta>0$ one defines the quantity 
\be
\cH_\delta^s(F) = {\mathrm{ inf}} \left( \sum_{i}\, |U_i|^s \, : \,
\{ U_i \}  \,\, {\mathrm{ is \, a}} \,\, \delta{\mathrm {-cover \,\, of}} \, \, F \, \right)
\, ,
\ee
and takes the limit 
\be
\cH^s (F) = \lim_{\delta \to 0} \cH_\delta^s(F) \, .
\ee
The latter quantity is called $s$-dimensional Hausdorff measure of $F$. 
Increasing $s$ for a certain set $F$ leads to a jump of the value of
$\cH^s(F)$ from $\infty$ to $0$ at a critical value $d_H$ . This value is 
called the Hausdorff dimension of $F$.

A visualisation of the Hausdorff dimension  $d_H$ is given by the 
number $N$ of balls of radius $a$ to cover the set $F$. In this case 
$|U_i|=2a$ for all $i$ and the number of balls will scale
as $N(a) \propto 1/a^{d_H}$. 

The Hausdorff dimension coincides with the usual topological dimension 
for regular structures, in particular for manifolds, and yields in general 
non-integer values for fractals. 

The basic idea underlying the concept of the spectral dimension $d_S$ 
is that a test particle diffusing on a given background probes also some 
kind of dimension of this background, see, {\it e.g.}, \cite{Avraham:2005}
for a related textbook. In addition, let us recall that 
random walks are good models for diffusion processes. In a $d$-dimensional
euclidean space its mean square distance after time $t$ from its starting point scales
 with $t$, $\langle r^2(t)\rangle \propto t $. On fractals one will obtain another
power law
\be
\langle r^2(t)\rangle \propto t^{2/d_W}
\ee
which defines the walk dimension. Those anomalous diffusion processes 
can be divided into two cases. First, there are recurrent diffusion processes, 
also called compact diffusion processes, in which the walker returns to its origin 
with unit probability. In this case one has $d_W > d_H$. Second, we have 
nonrecurrent random walks with $d_W < d_H,$ where the return probability 
$\cP(T)$ vanishes with increasing time 
\be
\label{Ppower}
\cP (T) \propto T^{-d_H/d_W} = T^{-d_S/2} \,  .
\ee
This defines then the spectral dimension $d_S$. If $d_S < 2$ the walk is compact 
whereas for $d_S > 2$ the random walk is non-recurrent. 

From \eqref{Ppower} one can then also extract the relation
\be
d_S =  -2 \lim _{T\to 0} \, \frac{ \partial \ln {\cal P}(T)}{\partial \ln T} \,  ,
\ee
and in addition define the generalised spectral dimension $D_S(T)$ as an effectively
scale-dependent measure of the dimension of the underlying space,
\be
D_S(T)  =  -2 \, \frac{ \partial \ln {\cal P}(T)}{\partial \ln T} \,  ,
\ee
see section \ref{sec:specdim} for the derivation of this relation.


\chapter{{\color{MYBLUE}{Interpolations of Staircase-Type Results for the Traces}}}
\label{App.fRA}

The evaluation of the flow equation in section \ref{sec:spectrum} is based on
averaging over the upper staircase and lower staircase interpolations. This has been
called averaging approximation. In this appendix, 
the explicit expressions derived from it are given in appendix \ref{App.A0}.
In addition,  two alternative interpolation schemes
which avoid non-analytic expressions are presented. 
The ``middle-of-the-staircase'' interpolation to be discussed in appendix \ref{App.A1} 
performs the sums at the averaged eigenvalues, setting $p^{(s)} = \tfrac 1 2$. 
The Euler-MacLaurin interpolation which will be introduced in appendix \ref{App.A2} 
replaces  the sum by a continuous integral and neglects the discrete correction terms.

In a general number of dimensions $d\ge 3$ the traces 
are in the three different interpolations given by
\begin{subequations}\label{gravflowD}
	\begin{align}
	(4\pi)^{d/2} \cT^{\rm TT} = & \, 
\frac{(d+1)(d-2)}{4\Gamma(d/2+1)} \Biggl( 
\tfrac{\left( \dot{\varphi}^\prime + (d-2) \varphi^\prime - 2 r \varphi^{\prime\prime}
\right) (1+ \alpha^G_T r ) + 2 \varphi^\prime }
{\varphi^\prime\left( 1+ \left(\alpha^G_T + \tfrac{2}{d(d-1)}\right)r  \right)}
\left( 1 + \frac d 2 \alpha^G_T r + \frac {C_{T}^I}{12} \, r + \cO(r^2) \right)
\nonumber \\
& \qquad \qquad - \tfrac{\dot{\varphi}^\prime + (d-2) \varphi^\prime - 2 r \varphi^{\prime\prime}}
{\varphi^\prime\left( 1+ \left(\alpha^G_T + \tfrac{2}{d(d-1)}\right)r  \right)}
\left( \frac d{d+2} + \frac d 2 \alpha^G_T r + \frac {\tilde C_{T}^I}{12} \, r + 
\cO(r^2) \right) \Biggr) \, ,
 \\ 
\label{gravsinvD}
	(4\pi)^{d/2} \cT^{\rm sinv}  = & \, 
\frac{1}{2\Gamma(d/2+1)} \Biggl( 
\tfrac{\left( \dot{\varphi}^{\prime\prime} + (d-4) \varphi^{\prime\prime}  
- 2 r \varphi^{\prime\prime\prime}
\right) (1+ \alpha^G_S r ) + 2 \varphi^{\prime\prime} }
{\varphi^{\prime\prime}\left( 1+ \left(\alpha^G_S - \tfrac{1}{d-1}\right)r \right)
+\tfrac{d-2}{2(d-1)} \varphi^\prime }
\left( 1 + \frac d 2 \alpha^G_S r + \frac {C_{S}^I}{12} \, r + \cO(r^2) \right)
\nonumber \\
& \quad \!\!\! - \tfrac{ \dot{\varphi}^{\prime\prime} + (d-4) \varphi^{\prime\prime}  
- 2 r \varphi^{\prime\prime\prime}}
{\varphi^{\prime\prime}\left( 1+ \left(\alpha^G_S - \tfrac{1}{d-1}\right)r \right)
+\tfrac{d-2}{2(d-1)} \varphi^\prime }
\left( \frac d{d+2} + \frac d 2 \alpha^G_S r + 
\frac d{d+2} \frac {\tilde C_{S}^I}{12} \, r 
+ \cO(r^2) \right) \Biggr) \, ,
	 \\
	(4\pi)^{d/2} \cT^{\rm ghost} = &  \frac 1 2 
\frac{1+ \frac d 2 \left( \alpha^G_{V} + \tfrac{1}{d}\right) r}
{1 + \left( \alpha^G_{V} - \tfrac{1}{d} \right) r} 
\left( (d-1)  + \frac{C_{V}^I}{12} \,  r  + \cO (r^2)  \right) \, 
	\end{align}
\end{subequations}
and
\begin{subequations}\label{matterflowD}
	\begin{align}
(4\pi)^{d/2} \cT^{\rm scalar} = & \, \frac{N_S}{2 } \, 
\frac{1+ \tfrac d 2 \alpha_S^M r}{1 + \alpha_S^M r} \,  
\left(1 + \frac {C_S^I}{12} \, r  + \cO (r^2) \right)\, , 
 \\ 
\label{TdiracD}
(4\pi)^{d/2} \cT^{\rm Dirac} = & \, - N_D 2^{\lfloor d/2 -1 \rfloor } 
\frac{1+ \tfrac d 2 \alpha_D^M r}{1 + \alpha_D^M r} \,  
\left(1 + \frac {C_D^I}{24} \,  r  + \cO (r^2) \right)\, , 
\\
(4\pi)^{d/2} \cT^{\rm vector} = & \,  \frac{N_V}{2} \, 
\frac{1+ \frac d 2 \left( \alpha^M_{V_1} + \tfrac{1}{d} \right) r}
{1 + \left( \alpha^M_{V_1} + \tfrac{1}{d} \right) r} 
\left( (d-1)  + \frac {C_{V}^I}{12} \, r  + \cO (r^2)  \right) \\
 & \qquad \qquad \nonumber
- \frac{1+ \frac d 2 \alpha^M_{V_2} r }{1 + \alpha^M_{V_2} r} 
\left( 1  + \frac {C_{S}^I}{12} \,  r  + \cO (r^2)  \right)  
\bigg)
\, ,  
	\end{align}
\end{subequations}
where in the constants $C_A^I$ and $\tilde C_A^I$ the superscript 
$I\in \{A, M, E  \}$ labels the 
averaging, middle-of-the-staircase and Euler-McLaurin interpolation and the 
subscript $A\in \{T,S,V,D \}$ the type of tensor structures. The $C_A^I$ and $\tilde C_A^I$ 
are rational functions in $d$. The explicit
expressions are given in the respective sections \ref{App.A0}, \ref{App.A1} 
and \ref{App.A2} below. Note that $d=2$ is for most of these expressions 
special, and one has to return to the original definitions of the sums 
to derive them correctly. As, however, for $d=2$ they degenerate to linear,
constant or even vanishing terms all these calculations are straightforward.

\section{The Averaging Interpolation}
\label{App.A0}

The constants $C_A^A$ and $\tilde C_A^A$ are for the averaging approximation given by
\begin{align}
&C^A_{T} = \frac{d^3-2d^2-13d+2}{(d-1)(d-2)} \, ,  \quad
\tilde C^A_{T}  =  \frac{d^2-2d-9}{d-1}  \, , \quad 
C^A_{S} =  {d+1} \, , \nonumber  \\ 
&\tilde C^A_{S} = \frac{d^2-2d+3}{d-1}  \, ,   \qquad \quad \quad
C^A_{V} = {d^2-6d-1} \, , \quad 
C^A_{D} = {d-2} \, .
\end{align}

The dimensionless counterparts of the traces for $d=4$ are (with all endomorphism
parameters kept) given by
\begin{subequations}\label{gravflow}
	\begin{align}\label{gravTT}
	(4\pi)^2 \cT^{\rm TT} = & \, 	\frac{5}{2} \, 
\frac{1}{1 + \left(\alpha^G_T + \tfrac{1}{6}\right)r} 
\left(1 + \left(\alpha^G_T - \tfrac{1}{6}\right)r\right)
\left(1 + \left(\alpha^G_T - \tfrac{1}{12}\right)r\right) \\ \nonumber
		+& \frac{5}{12 } \, 
\tfrac{\dot{\varphi}^\prime + 2 \varphi^\prime - 2 r \varphi^{\prime\prime}}{\varphi^\prime}
 \left(1 + \left(\alpha^G_T - \tfrac{2}{3}\right)r\right)
\left(1 + \left(\alpha^G_T - \tfrac{1}{6}\right)r\right) \, , \\ \label{gravsinv}
	(4\pi)^2 \cT^{\rm sinv}  = & \, 
	\frac{1}{2} 
	\tfrac{ \varphi^{\prime\prime}}
{\left(1+ \left(\alpha^G_S - \tfrac{1}{3}\right)r\right) \varphi^{\prime\prime} 
+ \tfrac{1}{3} \varphi^\prime} 
	\left(1 + \left(\alpha^G_S - \tfrac{1}{2}\right)r\right)
	\left(1 + \left(\alpha^G_s + \tfrac{11}{12}\right)r\right)
	 \\ \nonumber 
	 +& \frac{1}{12} 
	 \tfrac{\dot{\varphi}^{\prime\prime} - 2 r \varphi^{\prime\prime\prime}}
{\left(1+ \left(\alpha^G_S - \tfrac{1}{3}\right)r\right) \varphi^{\prime\prime} 
+ \tfrac{1}{3} \varphi^\prime} 
	 \left(1 + \left(\alpha^G_S + \tfrac{3}{2}\right)r\right)
	 \left(1 + \left(\alpha^G_s - \tfrac{1}{3}\right)r\right)
	 \left(1 + \left(\alpha^G_S - \tfrac{5}{6}\right)r\right) \, , 
	 \\
	(4\pi)^2 \cT^{\rm ghost} = & \, - \frac{1}{48} \, 
\frac{1}{1 + (\alpha^G_V - \tfrac{1}{4})r} \, 
\left( 72 + 18 r (1 + 8 \alpha^G_V) - r^2 (19 - 18 \alpha^G_V - 72 (\alpha^G_V)^2) \right) \, ,  \label{gravghost}
	\end{align}
\end{subequations}
together with the matter results
\begin{subequations}\label{matterflow}
	\begin{align}
(4\pi)^2 \cT^{\rm scalar} = & \, \frac{N_S}{2 } \, \frac{1}{1 + \alpha_S^M r} \,  
\left( 1 + \left( \alpha_S^M + \tfrac{1}{4} \right) r \right) 
\left(1 + \left( \alpha_S^M +  \tfrac{1}{6} \right) r\right) \, , \\ \label{Tdirac}
(4\pi)^2 \cT^{\rm Dirac} = & \, - {2 N_D} \, 
\left(1 + \left(\alpha^M_D + \tfrac{1}{6}\right)r\right) \, , \\
(4\pi)^2 \cT^{\rm vector} = & \, \frac{N_V}{2 } \, \bigg(
\frac{3}{1 + \left( \alpha^M_{V_1} + \tfrac{1}{4} \right) r} 
\left(1 + \left(\alpha^M_{V_1} + \tfrac{1}{6} \right) r \right) 
\left(1 + \left(\alpha^M_{V_1} + \tfrac{1}{12} \right)r\right) \\ & \qquad \qquad \nonumber
- \frac{1}{1 + \alpha^M_{V_2} r} \left( 1 + (\alpha^M_{V_2} + \tfrac{1}{2}) r \right) 
\left(1 + ( \alpha^M_{V_2} - \tfrac{1}{12}) r\right) 
\bigg) \, .  
	\end{align}
\end{subequations}
Of course, if all $\alpha$ are put to zero these expressions reduce to equations
\eqref{gravflowTypeI} and \eqref{matterflowI}.

\bigskip 

\section{The Middle-of-the-Staircase Interpolation}
\label{App.A1}

By definition, the middle-of-the-staircase interpolation evaluates the spectral sums 
\eqref{mattersums} and \eqref{gravitonsum} on the average of the eigenvalues bounding 
a plateau of the staircase. The resulting values $N^{(s)}_{\rm max}$ are given by 
\eqref{Ncutoff} evaluated for
\be
\label{intfct}
p^{(s)} = \tfrac 1 2 \, , \qquad q^{(s)} = 0 \, . 
\ee
The analogue of (\ref{matterevaluated}) 
and (\ref{gravevaluated}) for this interpolation scheme
is obtained from the replacements
\be
\label{reprules}
T_d^{(s)}(N) \to S_d^{(s)}(N^{(s)}_{\rm max}) \, , \qquad
\widetilde{T}_d^{(s)}(N) \to \widetilde{S}_d^{(s)}(N^{(s)}_{\rm max}) \, , 
\ee 
with $N^{(s)}_{\rm max}$ defined in \eqref{Ncutoff} and evaluated at \eqref{intfct}. 
Notably, the middle-of-the-staircase scheme also removes all non-analytic terms in $r$.

Comparing the spectral sums resulting from this interpolation scheme to the early-time 
expansion of the heat-kernel one (again) finds a deviation in the linear term. The 
two terms can be brought into agreement by setting $q^{(s)} = \tfrac{1}{3} (d-1)$.

The signs of these parameters are opposite to the corresponding 
 ones obtained for the averaging approximation, (\ref{qopt}). Phrased differently, 
the linear terms found in the averaging and the  middle-of-the-staircase interpolations 
differ from the corresponding heat-kernel results in opposite directions.  
 
For completeness we present here the results for the traces 
obtained within the middle-of-the-staircase 
 interpolation. 
The constants $C_A^M$ and $\tilde C^M_A$ for this interpolation are given by
\begin{align}
&C^M_{T} = \frac{d^3- \tfrac 7 2 d^2- \tfrac {17}2 d-1}{(d-1)(d-2)} \, ,  \quad
\tilde C^M_{T}  =  \frac{d^2-\tfrac 7 2 d- \tfrac {21} 2 }{d-1}  \, , \quad 
C^M_{S} =  {d- \tfrac 1 2} \, , \nonumber  \\ 
&\tilde C^M_{S} = \frac{d^2-\tfrac 9 2 d+ \tfrac 1 2}{d-1}  \, ,   \qquad \quad \quad
C^M_{V} = {d^2- \tfrac {15}2 d + \tfrac 1 2 } \, , \quad 
C^M_{D} = {d+1} \, .
\end{align}

The traces for $d=4$  result in
\begin{subequations}\label{gravflowI}
 	\begin{align}
 	(4\pi)^2 \cT^{\rm TT} = & \,
 	\frac{5}{2} \, \frac{1}{1 + \left(\alpha^G_T + \tfrac{1}{6}\right)r} 
\left(1 + \left(\alpha^G_T - \tfrac{19}{48}\right)r\right)
\left(1 + \left(\alpha^G_T + \tfrac{1}{48}\right)r\right) \\ \nonumber & \; \;
 	+ \frac{5}{12} \, 
\tfrac{\dot{\varphi}^\prime + 2 \varphi^\prime - 2 r \varphi^{\prime\prime}}
{\varphi^\prime \left( 1 + \left(\alpha^G_T + \tfrac{1}{6}\right)r \right) } 
 	\left(1 + \left(\alpha^G_T - \tfrac{19}{48}\right)r\right)
 	\left(1 + \left(\alpha^G_T - \tfrac{1}{24}\right)r\right) 
 	\left(1 + \left(\alpha^G_T + \tfrac{1}{48}\right)r\right) \, , \\
 	(4\pi)^2 \cT^{\rm sinv} = & \,
 	\frac{1}{2}
 	\tfrac{ \varphi^{\prime\prime}}
{\left(1+ \left(\alpha^G_S - \tfrac{1}{3}\right)r\right) \varphi^{\prime\prime}
 + \tfrac{1}{3} \varphi^\prime}
 	\left(1 + \left(\alpha^G_S - \tfrac{9}{16}\right)r\right)
 	\left(1 + \left(\alpha^G_S + \tfrac{41}{48}\right)r\right)
 	\\ \nonumber & \; \;
 	+ \frac{1}{12}
 	\tfrac{\dot{\varphi}^{\prime\prime} - 2 r \varphi^{\prime\prime\prime}}
{\left(1+ \left(\alpha^G_S - \tfrac{1}{3}\right)r\right) \varphi^{\prime\prime} 
+ \tfrac{1}{3} \varphi^\prime}
 	\left(1 + \left(\alpha^G_S - \tfrac{9}{16}\right)r\right)  \times \\ 
 	& ~ \hspace{50mm} 
 	\left(1 + \left(2\alpha^G_S + \tfrac {55}{48} \right)r +  
 	\left((\alpha^G_S)^2  + \tfrac {55}{48}  \alpha^G_S - \tfrac{865}{1152} \right) r^2 
 	\right) \, , \nonumber
 	\\
 	(4\pi)^2 \cT^{\rm ghost} = & \, - \frac{3}{2} \, 
\frac{1}{1 + (\alpha^G_V - \tfrac{1}{4})r} \, 
\left( 1 + \left( \alpha^G_V - \tfrac {23}{48} \right) r  \right) 
 	\left( 1 + \left( \alpha^G_V + \tfrac {29}{48} \right) r  \right) \, ,
 	\end{align}
 \end{subequations}
for the gravity part and
 \begin{subequations}\label{matterflowM}
 	\begin{align}
 	(4\pi)^2 \cT^{\rm scalar} = & \, \frac{N_S}{2 } \, \frac{1}{1 + \alpha_S^M r} \, 
 \left( 1 + \left( \alpha_S^M + \tfrac{5}{48} \right) r \right) 
  \left(1 + \left( \alpha_S^M +  \tfrac{3}{16} \right) r\right) \, , \\ \label{Tdirac1}
 	(4\pi)^2 \cT^{\rm Dirac} = & \, - {2 N_D} \, 
\frac 1 {1 + \left( \alpha_D^M + \tfrac 1 4 \right) r}
 	\left(1 + \left(\alpha^M_D + \tfrac{1}{16}\right)r\right) 
 	\left(1 + \left(\alpha^M_D + \tfrac{11}{48}\right)r\right) \, , \\
 	(4\pi)^2 \cT^{\rm vector} = & \, \frac{N_V}{2} \, \bigg(
 	\frac{3}{1 + \left( \alpha^M_{V_1} + \tfrac{1}{4} \right) r}
 	\left(1 + \left(\alpha^M_{V_1} - \tfrac{1}{16} \right) r \right)
 	\left(1 + \left(\alpha^M_{V_1} + \tfrac{3}{16} \right)r\right) \\ & \qquad \qquad \nonumber
 	- \frac{1}{1 + \alpha^M_{V_2} r} \left( 1 + (\alpha^M_{V_2} + \tfrac{7}{16}) r \right) 
 	\left(1 + ( \alpha^M_{V_2} - \tfrac{7}{48}) r\right)
 	\bigg) \, 
 	\end{align}
\end{subequations}
for the matter part. Note that, in contrast to (\ref{Tdirac}), 
the middle-of-the-staircase interpolation does not lead to a cancellation between 
the numerator and denominator in the fermion sector. Comparing the expressions 
obtained from the two interpolation schemes clearly shows that the procedures of 
summing and averaging do not commute: the trace contributions obtained from summing 
first and averaging afterwards (averaging interpolation) differ from averaging first 
and summing afterwards (middle-of-the-staircase interpolation).

\goodbreak
 
\section{The Euler-MacLaurin Interpolation}
\label{App.A2}

A third interpolation scheme which avoids non-analytic terms in the spectral sums 
is provided by the Euler-MacLaurin interpolation see, {\it e.g.},  \cite{Dona:2012am}. 
In this case the finite sums are approximated through the Euler-MacLaurin formula,
\be
\sum_{l=n}^m \, f(l) = \int_n^m dl \, f(l) + \ldots \, , 
\ee
and neglecting the discrete terms. Applying this strategy to the spectral sums 
\eqref{mattersums} and \eqref{gravitonsum}, identifying $N_{\rm max}^{(s)}$ with 
\eqref{Ncutoff} based on the values
\be
\label{intfct2}
p^{(s)} = 0 \, , \qquad q^{(s)} = 0 \, , 
\ee
leads to replacement rules similar to \eqref{reprules}. By construction, the terms 
appearing at zeroth and first order in the scalar curvature agree with the 
early-time expansion of the heat-kernel.

For completeness, we also give the explicit expression for the operator traces 
entering into \eqref{pdf4d} based on the Euler-MacLaurin interpolation.

The constants $C_A^E$ and $\tilde C_A^E$ are for the averaging approximation given by
\begin{align}
&C^E_{T} = \frac{(d+2)d(d-5)}{(d-1)(d-2)} \, ,  \quad
\tilde C^E_{T}  =  \frac{(d+2)(d-5)}{d-1}  \, , \quad 
C^E_{S} =  {d} \, , \nonumber  \\ 
&\tilde C^E_{S} = \frac{(d+2)(d-2)}{d}  \, ,   \qquad 
C^E_{V} = {d(d-7)} \, , \quad 
C^E_{D} = {d} \, .
\end{align}

For $d=4$ one obtains
\begin{subequations}\label{gravflowM}
        \begin{align}
        (4\pi)^2 \cT^{\rm TT} = & \,
        \frac{5}{2} \, \frac{1}{1 + \left(\alpha^G_T + \tfrac{1}{6}\right)r} 
\left(1 + \left(\alpha^G_T - \tfrac{2}{3}\right)r\right)
\left(1 + \left(\alpha^G_T + \tfrac{1}{3}\right)r\right) \\ \nonumber & \; \;
                + \frac{5}{12 } \, 
\tfrac{\dot{\varphi}^\prime + 2 \varphi^\prime - 2 r \varphi^{\prime\prime}}
{\varphi^\prime \left( 1 + \left(\alpha^G_T + \tfrac{1}{6}\right)r \right) } 
\left(1 + \left(\alpha^G_T - \tfrac{2}{3}\right)r\right)^2 
\left(1 + \left(\alpha^G_T + \tfrac{5}{6}\right)r\right) \, , \\
        (4\pi)^2 \cT^{\rm sinv} = & \,
        \frac{1}{2}
        \tfrac{ \varphi^{\prime\prime}}
{\left(1+ \left(\alpha^G_S - \tfrac{1}{3}\right)r\right) \varphi^{\prime\prime} 
+ \tfrac{1}{3} \varphi^\prime}
        \left(1 + \left(\alpha^G_S + \tfrac{7}{6}\right)r\right)
        \left(1 + \left(\alpha^G_S - \tfrac{5}{6}\right)r\right)
         \\ \nonumber & \; \;
         + \frac{1}{12}
         \tfrac{\dot{\varphi}^{\prime\prime} - 2 r \varphi^{\prime\prime\prime}}
{\left(1+ \left(\alpha^G_S - \tfrac{1}{3}\right)r\right) \varphi^{\prime\prime} 
+ \tfrac{1}{3} \varphi^\prime}
         \left(1 + \left(\alpha^G_S - \tfrac{5}{6}\right)r\right)^2 
         \left(1 + \left(\alpha^G_S + \tfrac{13}{6}\right)r\right),  \\ 
        %
        (4\pi)^2 \cT^{\rm ghost} = & \, - \frac{3}{2} \, 
\frac{1}{1 + (\alpha^G_V - \tfrac{1}{4})r} \, 
\left( 1 + \left( \alpha^G_V - \tfrac {3}{4} \right) r  \right) 
\left( 1 + \left( \alpha^G_V + \tfrac {11}{12} \right) r  \right) \, ,
        \end{align}
\end{subequations}
and
\begin{subequations}
        \begin{align}
(4\pi)^2 \cT^{\rm scalar} = & \, \frac{N_S}{2 } \,  
\left( 1 + \left( \alpha_S^M + \tfrac{1}{3} \right) r \right)  \, , \\ 
\label{Tdirac2}
(4\pi)^2 \cT^{\rm Dirac} = & \, - {2 N_D} \, 
\frac 1 {1 + \left( \alpha_D^M + \tfrac 1 4 \right) r}
\left(1 + \left(\alpha^M_D - \tfrac{1}{12}\right)r\right) 
\left(1 + \left(\alpha^M_D + \tfrac{5}{12}\right)r\right) \, , \\
(4\pi)^2 \cT^{\rm vector} = & \, \frac{N_V}{2} \, \bigg(
\frac{3}{1 + \left( \alpha^M_{V_1} + \tfrac{1}{4} \right) r}
\left(1 + \left(\alpha^M_{V_1} - \tfrac{1}{4} \right) r \right)
\left(1 + \left(\alpha^M_{V_1} + \tfrac{5}{12} \right)r\right) \\ & \qquad \qquad \nonumber
- \frac{1}{1 + \alpha^M_{V_2} r} \left( 1 + (\alpha^M_{V_2} + \tfrac{2}{3}) r \right) 
\left(1 + ( \alpha^M_{V_2} - \tfrac{1}{3}) r\right)
\bigg) \, .
        \end{align}
\end{subequations}
Note that in this case a cancellation of numerator and denominator takes place 
for the scalar matter field.

\newpage
\thispagestyle{empty}


\chapter{{\color{MYBLUE}{Fixed Point Structure of Selected Gravity-Matter Systems}}}
\label{App.fRB}

In this appendix we collect the FP data for the convergent 
NGFP solutions passing the $f(R)$-stability test in table \ref{Tab.mainresults}. The results for pure gravity are given in tables \ref{Tab.grav.I} (type I coarse-graining operator) and \ref{Tab.grav.II} (type II coarse-graining operator). In this case the critical exponents with positive real part coincide with the ones obtained in \cite{Ohta:2015efa,Ohta:2015fcu}. The FP data for the gravity-matter models featuring matter sectors based on frequently studied models for BSM physics are displayed in tables \ref{Tab.FPSMdm} - \ref{Tab.FPMSSM}, respectively. Throughout the presentation, we give results up to $N=8$, and all gravity-matter FPs show a rapid convergence in the FPs' position and stability coefficients. Extended computations along the lines of table \ref{Tab.FPstandardmodel}, covering the critical exponents up to $N=9$ and the polynomial coefficients of the FP solution up to $N=14$, confirm this picture.

Following the discussion related to figure \ref{Fig.typeIIannihilation}, the stable gravity-matter FPs for a type II coarse-graining scheme can be understood as a deformation of their type I counterparts. For the matter sectors listed in table \ref{Tab.3}, these  deformations do not extend to a coarse-graining operator of type II. Hence our lists of stable gravity-matter FPs comprise results for the type I coarse-graining operator only.    

\medskip
\newpage
\begin{landscape}

\begin{table}
	\centering
	\begin{tabular}{c|ccccccccc}
		$N$ & \hspace*{5mm}  $g_0^*$ \hspace*{4mm}  & \hspace*{4mm} $g_1^*$ \hspace*{4mm}  & \hspace*{4mm} $g_2^*$ \hspace*{2mm} & \;  $g_3^* \times 10^{-3}$ & \hspace*{1mm} $g_4^* \times 10^{-4}$ & \hspace*{1mm}  $g_5^* \times 10^{-4}$  & \hspace*{1mm}  $g_6^* \times 10^{-5}$  &  \hspace*{2mm} $g_7^*$ \hspace*{4mm}  & \hspace*{5mm} $g_8^*$ \hspace*{5mm} \\ \hline \hline
		$1$ & $0.46$ & $-1.24$  \\
		$2$ & $0.70$ & $-0.75$ & $0.27$ \\
		$3$ & $0.69$ & $-0.74$ & $0.26$ & $-2.30$   \\
		$4$ & $0.70$ & $-0.75$ & $0.26$ & $-1.27$ & $-6.33$  \\
		$5$ & $0.70$ & $-0.74$ & $0.26$ & $-1.83$ & $-6.38$ & $-1.04$ \\
		$6$ & $0.70$ & $-0.74$ & $0.26$ & $-1.76$ & $-6.87$& $-1.04$ & $-1.99$ \\ 
		$7$ & $0.70$ & $-0.74$ & $0.26$ & $-1.81$ &  $-6.90$ & $-1.13$ & $-2.08$ & $\approx 0$ \\
		$8$ & $0.70$ & $-0.74$ & $0.26$ & $-1.80$ & $-6.93$ & $-1.14$ & $-2.23$ & $\approx 0$ &  $\approx 0$  \\ \hline \hline
		\multicolumn{10}{c}{}\\
		$N$ & $\theta_0$ & $\theta_1$ & $\theta_2$ & $\theta_3$ & $\theta_4$ & $\theta_5$ & $\theta_6$ & $\theta_7$ & $\theta_8$  \\ \hline \hline
		$1$ & $4$ & $2.78$ \\
		$2$ & $4$ & $2.29$ & $-1.50$  \\
		$3$ & $4$ & $2.00$ & $-1.50$ & $-4.01$ \\
		$4$ & $4$ & $2.17$ & $-1.80$ & $-3.99$ &  $-6.23$ 
		\\
		$5$ & $4$ & $2.10$ & $-1.79$ & $-4.37$ & 
		$-6.26$ & $-8.39$
		\\
		$6$ & $4$ & $2.13$ & $-1.87$ & $-4.41$ &
		$-6.62$ & $-8.39$ & $-10.50$ 
		\\ 
		$7$ & $4$ & $2.11$ & $-1.88$ & $-4.50$ & 
		$-6.70$ & $-8.72$ & $-10.50$ & $-12.57$ \\
		$8$ & $4$ & $2.11$ & $-1.90$ & $-4.52$  & $-6.79$ & $-8.80$ & $-10.79$
		& $-12.57$ & $-14.63$ 
		\\ \hline \hline
	\end{tabular}
	\caption[FP structure of $f(R)$-gravity without matter field and a coarse-graining operator of type I.]{\label{Tab.grav.I} FP structure of $f(R)$-gravity without matter fields $N_S = N_D = N_V = 0$ and a coarse-graining operator of type I. The value of couplings smaller than $10^{-5}$ is indicated by $\approx 0$.}
\end{table}

\begin{table}
	\centering
	\begin{tabular}{c|ccccccccc}
		$N$ & \hspace*{5mm}  $g_0^*$ \hspace*{4mm}  & \hspace*{4mm} $g_1^*$ \hspace*{4mm}  & \hspace*{4mm} $g_2^*$ \hspace*{2mm} & \;  $g_3^* \times 10^{-2}$ & \hspace*{1mm} $g_4^* \times 10^{-4}$ & \hspace*{1mm}  $g_5^* \times 10^{-4}$  & \hspace*{1mm}  $g_6^* \times 10^{-4}$  &  \hspace*{2mm} $g_7^*$ \hspace*{4mm}  & \hspace*{5mm} $g_8^*$ \hspace*{5mm} \\ \hline \hline
		$1$ & $0.46$ & $-1.78$  \\
		$2$ & $0.67$ & $-1.21$ & $0.51$ \\
		$3$ & $0.65$ & $-1.07$ & $0.53$ & $-4.29$   \\
		$4$ & $0.64$ & $-1.06$ & $0.54$ & $-4.42$ & $8.17$  \\
		$5$ & $0.64$ & $-1.06$ & $0.54$ & $-4.66$ & $6.88$ & $-6.26$ \\
		$6$ & $0.64$ & $-1.06$ & $0.54$ & $-4.63$ & $3.72$& $-6.53$ & $-1.57$ \\ 
		$7$ & $0.64$ & $-1.06$ & $0.54$ & $-4.67$ &  $2.65$ & $-8.04$ & $-1.92$ & $\approx 0$ \\
    	$8$ & $0.64$ & $-1.06$ & $0.53$ & $-4.68$ & $1.45$ & $-8.51$ & $-2.46$ & $\approx 0$ &  $\approx 0$  \\ \hline \hline
		\multicolumn{10}{c}{}\\
		$N$ & $\theta_0$ & $\theta_1$ & $\theta_2$ & $\theta_3$ & $\theta_4$ & $\theta_5$ & $\theta_6$ & $\theta_7$ & $\theta_8$  \\ \hline \hline
		$1$ & $4$ & $2.75$ \\
		$2$ & $4$ & $1.98$ & $-1.22$  \\
		$3$ & $4$ & $1.74$ & $-1.11$ & $-3.96$ \\
		$4$ & $4$ & $1.83$ & $-1.46$ & $-4.02$ &  $-6.68$ 
		\\
		$5$ & $4$ & $1.75$ & $-1.40$ & $-4.42$ & 
		$-6.67$ & $-9.33$
		\\
		$6$ & $4$ & $1.78$ & $-1.46$ & $-4.40$ &
		$-7.11$ & $-9.15$ & $-12.45$ 
		\\ 
		$7$ & $4$ & $1.76$ & $-1.46$ & $-4.46$ & 
		$-7.10$ & $-9.56$ & $-11.52$ & $-16.72$ \\
		$8$ & $4$ & $1.77$ & $-1.48$ & $-4.47$  & $-7.13$ & $-9.54$ & $-11.84$
		& $-13.74$ & $-23.33$ 
		\\ \hline \hline
	\end{tabular}
	\caption[FP structure of $f(R)$-gravity without matter fields and a coarse-graining operator of type II.]{\label{Tab.grav.II} FP structure of $f(R)$-gravity without matter fields $N_S = N_D = N_V = 0$ and a coarse-graining operator of type II. The value of couplings smaller than $10^{-4}$ is indicated by $\approx 0$.}
\end{table}

\begin{table}
	\centering
	\begin{tabular}{c|ccccccccc}
		$N$ & \hspace*{5mm}  $g_0^*$ \hspace*{4mm}  & \hspace*{4mm} $g_1^*$ \hspace*{4mm}  & \hspace*{4mm} $g_2^*$ \hspace*{2mm} & \;  $g_3^* \times 10^{-3}$ & \hspace*{1mm} $g_4^* \times 10^{-4}$ & \hspace*{1mm}  $g_5^* \times 10^{-5}$  & \hspace*{1mm}  $g_6^* \times 10^{-5}$  &  \hspace*{2mm} $g_7^*$ \hspace*{4mm}  & \hspace*{5mm} $g_8^*$ \hspace*{5mm} \\ \hline \hline
		$1$ & $-7.17$ & $-5.72$  \\
		$2$ & $-6.64$ & $-5.10$ & $1.11$ \\
		$3$ & $-6.64$ & $-5.15$ & $1.13$ & $4.74$   \\
		$4$ & $-6.64$ & $-5.15$ & $1.13$ & $4.61$ & $-2.58$  \\
		$5$ & $-6.64$ & $-5.15$ & $1.12$ & $4.79$ & $-2.39$ & $2.96$ \\
		$6$ & $-6.64$ & $-5.15$ & $1.12$ & $4.87$ & $-1.87$& $3.83$ & $1.26$ \\ 
		$7$ & $-6.64$ & $-5.15$ & $1.13$ & $5.03$ &  $-1.51$ & $6.48$ & $1.79$ & $\approx 0$ \\
		$8$ & $-6.64$ & $-5.15$ & $1.13$ & $5.13$ & $-1.08$ & $7.79$ & $2.77$ & $\approx 0$ &  $\approx 0$  \\ \hline \hline
		\multicolumn{10}{c}{}\\
		$N$ & $\theta_0$ & $\theta_1$ & $\theta_2$ & $\theta_3$ & $\theta_4$ & $\theta_5$ & $\theta_6$ & $\theta_7$ & $\theta_8$  \\ \hline \hline
		$1$ & $4$ & $2.13$ \\
		$2$ & $4$ & $2.36$ & $-1.77$  \\
		$3$ & $4$ & $2.29$ & $-1.83$ & $-6.20$ \\
		$4$ & $4$ & $2.29$ & $-1.92$ & $-6.08$ &  $-9.52$ 
		\\
		$5$ & $4$ & $2.29$ & $-1.92$ & $-6.12$ & 
		$-9.52$ & $-12.17$
		\\
		$6$ & $4$ & $2.29$ & $-1.91$ & $-6.10$ &
		$-9.50$ & $-12.34$ & $-14.50$ 
		\\ 
		$7$ & $4$ & $2.29$ & $-1.90$ & $-6.08$ & 
		$-9.49$ & $-12.27$ & $-14.82$ & $-16.69$ \\
		$8$ & $4$ & $2.29$ & $-1.90$ & $-6.06$  & $-9.46$ & $-12.27$ & $-14.72$
		& $-17.14$ & $-18.81$ 
		\\ \hline \hline
	\end{tabular}
	\caption[FP structure of $f(R)$-gravity coupled to the matter content of the SM supplemented by one additional DM scalar.]{\label{Tab.FPSMdm} FP structure of $f(R)$-gravity coupled to the matter content of the SM supplemented by one additional DM scalar where $N_S=5, N_D=\tfrac{45}{2}, N_V = 12$. The value of couplings smaller than $10^{-5}$ is indicated by $\approx 0$.}
\end{table}

\begin{table}
	\centering
	\begin{tabular}{c|ccccccccc}
		$N$ & \hspace*{5mm}  $g_0^*$ \hspace*{4mm}  & \hspace*{4mm} $g_1^*$ \hspace*{4mm}  & \hspace*{4mm} $g_2^*$ \hspace*{2mm} & \;  $g_3^* \times 10^{-3}$ & \hspace*{1mm} $g_4^* \times 10^{-4}$ & \hspace*{1mm}  $g_5^* \times 10^{-5}$  & \hspace*{1mm}  $g_6^* \times 10^{-5}$  &  \hspace*{1mm} $g_7^* \times 10^{-5}$  & \hspace*{1mm} $g_8^*$ \hspace*{1mm} \\ \hline \hline
		$1$ & $-8.04$ & $-6.08$  \\
		$2$ & $-7.52$ & $-5.46$ & $1.20$ \\
		$3$ & $-7.53$ & $-5.51$ & $1.22$ & $5.25$   \\
		$4$ & $-7.52$ & $-5.51$ & $1.21$ & $5.11$ & $-2.85$  \\
		$5$ & $-7.53$ & $-5.51$ & $1.21$ & $5.31$ & $-2.64$ & $3.45$ \\
		$6$ & $-7.53$ & $-5.51$ & $1.21$ & $5.40$ & $-2.05$& $4.42$ & $1.48$ \\ 
		$7$ & $-7.53$ & $-5.51$ & $1.21$ & $5.58$ &  $-1.66$ & $7.39$ & $2.06$ & $\approx 0$ \\
		$8$ & $-7.53$ & $-5.51$ & $1.21$ & $5.68$ & $-1.18$ & $8.83$ & $3.16$ & $1.12$ &  $\approx 0$  \\ \hline \hline
		\multicolumn{10}{c}{}\\
		$N$ & $\theta_0$ & $\theta_1$ & $\theta_2$ & $\theta_3$ & $\theta_4$ & $\theta_5$ & $\theta_6$ & $\theta_7$ & $\theta_8$  \\ \hline \hline
		$1$ & $4$ & $2.12$ \\
		$2$ & $4$ & $2.33$ & $-1.62$  \\
		$3$ & $4$ & $2.26$ & $-1.67$ & $-5.93$ \\
		$4$ & $4$ & $2.27$ & $-1.75$ & $-5.83$ &  $-9.24$ 
		\\
		$5$ & $4$ & $2.27$ & $-1.75$ & $-5.85$ & 
		$-9.29$ & $-11.91$
		\\
		$6$ & $4$ & $2.27$ & $-1.74$ & $-5.83$ &
		$-9.23$ & $-12.10$ & $-14.26$ 
		\\ 
		$7$ & $4$ & $2.27$ & $-1.73$ & $-5.80$ & 
		$-9.21$ & $-12.02$ & $-14.60$ & $-16.46$ \\
		$8$ & $4$ & $2.27$ & $-1.72$ & $-5.79$  & $-9.18$ & $-12.01$ & $-14.48$
		& $-16.92$ & $-18.59$ 
		\\ \hline \hline
	\end{tabular}
	\caption[FP structure of $f(R)$-gravity coupled to the matter content of the SM supplemented by three right-handed neutrinos.]{\label{Tab.FPSM3nu} FP structure of $f(R)$-gravity coupled to the matter content of the SM supplemented by three right-handed neutrinos where $N_S=4, N_D=24, N_V = 12$. The value of couplings smaller than $10^{-5}$ is indicated by $\approx 0$.}
\end{table}

\begin{table}
	\centering
	\begin{tabular}{c|ccccccccc}
		$N$ & \hspace*{5mm}  $g_0^*$ \hspace*{4mm}  & \hspace*{4mm} $g_1^*$ \hspace*{4mm}  & \hspace*{4mm} $g_2^*$ \hspace*{4mm} & \;  $g_3^* \times 10^{-3}$ \; & \; $g_4^* \times 10^{-4}$ \; &   $g_5^* \times 10^{-5}$  &  $g_6^* \times 10^{-5}$  & \hspace*{4mm} $g_7^*$ \hspace*{4mm} & \hspace*{4mm} $g_8^*$ \hspace*{4mm} \\ \hline \hline
		$1$ & $-7.79$ & $-5.87$  \\
		$2$ & $-7.25$ & $-5.24$ & $1.14$ \\
		$3$ & $-7.25$ & $-5.29$ & $1.15$ & $4.65$   \\
		$4$ & $-7.25$ & $-5.29$ & $1.15$ & $4.51$ & $-2.61$  \\
		$5$ & $-7.25$ & $-5.29$ & $1.15$ & $4.68$ & $-2.43$ & $2.76$ \\
		$6$ & $-7.25$ & $-5.29$ & $1.15$ & $4.75$ & $-1.95$& $3.56$ & $1.16$ \\ 
		$7$ & $-7.25$ & $-5.29$ & $1.15$ & $4.90$ &  $-1.60$ & $6.07$ & $1.66$ & $\approx 0$ \\
		$8$ & $-7.25$ & $-5.29$ & $1.15$ & $5.00$ &  $-1.18$ & $7.33$ & $2.58$ & $\approx 0$ & $\approx 0$  \\ \hline \hline
		\multicolumn{10}{c}{}\\
		$N$ & $\theta_0$ & $\theta_1$ & $\theta_2$ & $\theta_3$ & $\theta_4$ & $\theta_5$ & $\theta_6$ & $\theta_7$ & $\theta_8$  \\ \hline \hline
		$1$ & $4$ & $2.13$ \\
		$2$ & $4$ & $2.36$ & $-1.81$  \\
		$3$ & $4$ & $2.29$ & $-1.88$ & $-6.31$ \\
		$4$ & $4$ & $2.29$ & $-1.97$ & $-6.18$ &  $-9.66$ 
		\\
		$5$ & $4$ & $2.29$ & $-1.97$ & $-6.22$ & 
		$-9.66$ & $-12.33$
		\\
		$6$ & $4$ & $2.29$ & $-1.96$ & $-6.21$ &
		$-9.64$ & $-12.49$ & $-14.67$ 
		\\ 
		$7$ & $4$ & $2.29$ & $-1.95$ & $-6.18$ & 
		$-9.63$ & $-12.43$ & $-14.99$ & $-16.86$ \\
		$8$ & $4$ & $2.29$ & $-1.95$ & $-6.16$  & $-9.59$ & $-12.42$ & $-14.88$
		& $-17.31$ & $-18.99$ 
		\\ \hline \hline
	\end{tabular}
	\caption[FP structure of $f(R)$-gravity coupled to the matter content of the SM supplemented by right-handed neutrinos and two additional scalars.]{\label{Tab.FPSMext3} FP structure of $f(R)$-gravity coupled to the matter content of the SM supplemented by right-handed neutrinos and two additional scalars where $N_S=6, N_D=24, N_V = 12$. The value of couplings smaller than $10^{-5}$ is indicated by $\approx 0$.}
\end{table}

\begin{table}
	\centering
	\begin{tabular}{c|ccccccccc}
		$N$ & \hspace*{5mm}  $g_0^*$ \hspace*{4mm}  & \hspace*{4mm} $g_1^*$ \hspace*{4mm}  & \hspace*{4mm} $g_2^*$ \hspace*{4mm} & \;  $g_3^* \times 10^{-4}$ \; & \; $g_4^* \times 10^{-5}$ \; & \hspace*{3mm}  $g_5^*$ \hspace*{4mm} & \hspace*{4mm} $g_6^*$ \hspace*{4mm} & \hspace*{4mm} $g_7^*$ \hspace*{4mm} & \hspace*{4mm} $g_8^*$ \hspace*{4mm} \\ \hline \hline
		$1$ & $-5.67$ & $-2.47$  \\
		$2$ & $-4.64$ & $-1.46$ & $0.28$ \\
		$3$ & $-4.65$ & $-1.52$ & $0.29$ & $3.50$   \\
		$4$ & $-4.64$ & $-1.51$ & $0.29$ & $3.24$ & $-3.32$  \\
		$5$ & $-4.64$ & $-1.51$ & $0.29$ & $3.22$ & $-3.35$ & $\approx 0$ \\
		$6$ & $-4.64$ & $-1.51$ & $0.29$ & $3.18$ & $-3.56$& $\approx 0$ & $\approx 0$ \\ 
		$7$ & $-4.64$ & $-1.51$ & $0.29$ & $3.23$ &  $-3.45$ & $\approx 0$ & $\approx 0$ & $\approx 0$ \\
		$8$ & $-4.64$ & $-1.51$ & $0.29$ & $3.27$ &  $-3.30$ & $\approx 0$ & $\approx 0$ & $\approx 0$ & $\approx 0$  \\ \hline \hline
		\multicolumn{10}{c}{}\\
		$N$ & $\theta_0$ & $\theta_1$ & $\theta_2$ & $\theta_3$ & $\theta_4$ & $\theta_5$ & $\theta_6$ & $\theta_7$ & $\theta_8$  \\ \hline \hline
		$1$ & $4$ & $2.33$ \\
		$2$ & $4$ & $5.18$ & $-4.98$  \\
		$3$ & $4$ & $3.21$ & \multicolumn{2}{c}{$-11.03 \pm3.75i$} \\
		$4$ & $4$ & $3.01$ & \multicolumn{2}{c}{$-11.57 \pm6.40i$} &  $-17.60$ 
		\\
		$5$ & $4$ & $2.93$ & \multicolumn{2}{c}{$-12.44 \pm 7.65i$} & 
		$-16.50$ & $-20.79$
		\\
		$6$ & $4$ & $2.90$ & \multicolumn{2}{c}{$-12.70 \pm 8.36i$} &
		$-17.53$ & $-19.66$ & $-23.06$ 
		\\ 
		$7$ & $4$ & $2.88$ & \multicolumn{2}{c}{$-12.89 \pm 8.76i$} & 
		$-18.02$ & $-19.95$ & $-22.41$ & $-25.20$ \\
		$8$ & $4$ & $2.88$ & \multicolumn{2}{c}{$-12.96 \pm 
			8.96i$} & $-18.11$ & \multicolumn{2}{c}{$-21.38 \pm 0.71i$}
		& $-25.05$ & $-27.31$ 
		\\ \hline \hline
	\end{tabular}
	\caption[FP structure of $f(R)$-gravity coupled to the matter content of the MSSM.]{\label{Tab.FPMSSM} FP structure of $f(R)$-gravity coupled to the matter content of the MSSM where $N_S=49, N_D=61/2, N_V = 12$. The value of couplings smaller than $10^{-6}$ is indicated by $\approx 0$.}
\end{table}
\end{landscape}

\newpage
\thispagestyle{empty}


\chapter{{\color{MYBLUE}{Uniformly Accelerated Frames}}}
\label{App.UnruhCoord}

Throughout chapter \ref{Chap6} different coordinates of the worldline of an 
accelerated observer are used. 
A uniformly accelerated observer in special relativity is an observer having constant 
acceleration in the frame in which its instantaneous velocity is zero.
The coordinate transformation to the uniformly accelerated frame defines the so-called 
\emph{Rindler frame}.

\begin{figure}[h]
\centering
\includegraphics[width=90mm]{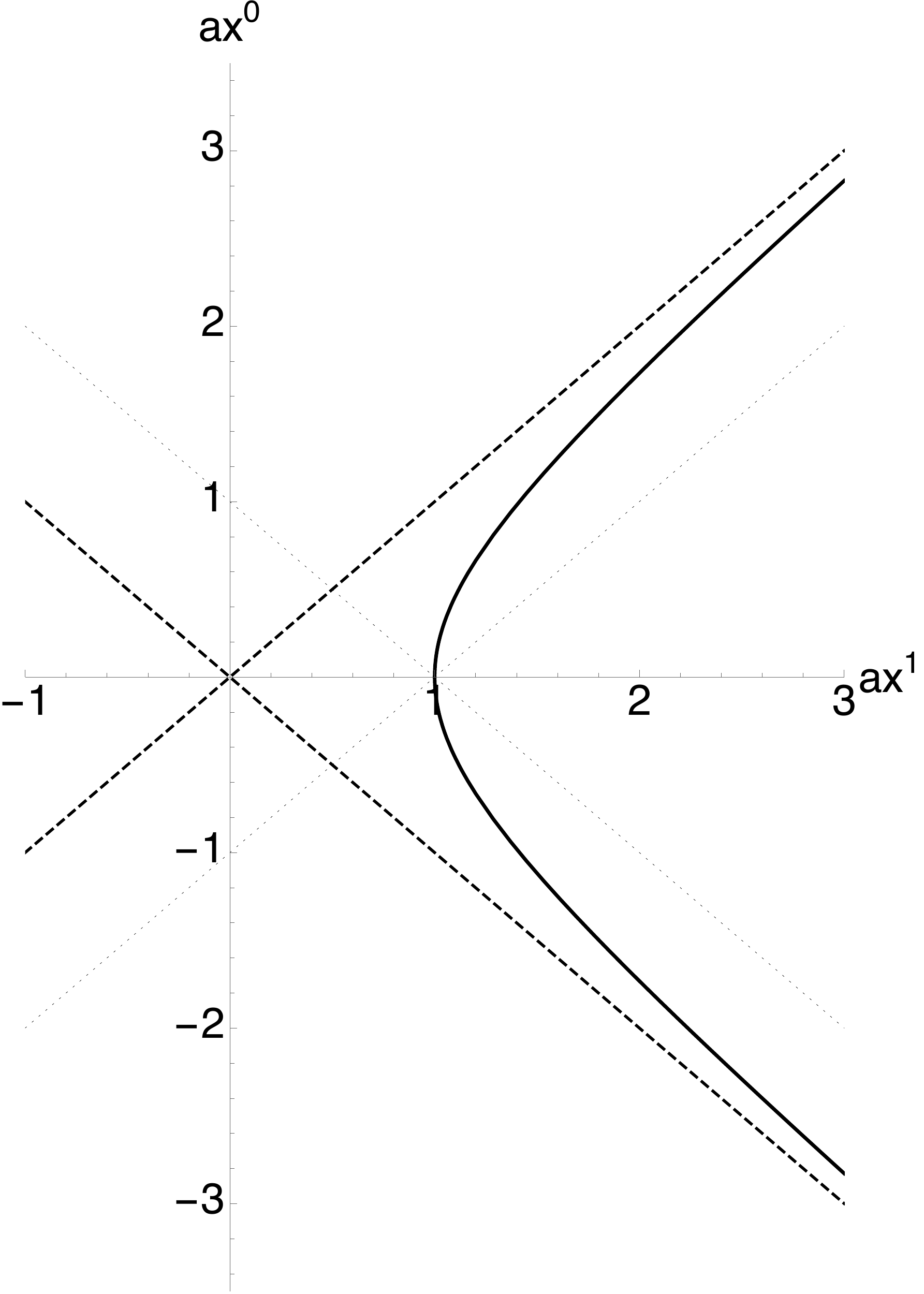}
\caption[The worldline of an observer with constant proper
acceleration $a$. The corresponding horizons (boundary of the right Rindler
wedge) are displayed as dashed lines.]
{\label{UnruhHyperbola} The worldline of an observer with constant proper
acceleration $a$ (full line). The coordinates $(x^0=t,x^1=x,x^2=y,x^3=z)$ are
chosen such that the hyperbola lies in the $(x^0,x^1)$ plan. The sign convention
is such that a positive $ds^2=(dx^0)^2 - (dx^1)^2 - (dx^2)^2 - (dx^3)^2$ 
is a timelike distance.
The corresponding horizons (boundary of the right Rindler
wedge) are displayed as dashed lines. (Dotted lines indicate the
lightcones with origin at $(x^0=0,x^1=1/a)$.)}
\end{figure}

Following chapter 8 of \cite{Mukhanov:2007zz} the proper time $\tau$ is used 
to parameterise the observer's trajectory $x^\alpha (\tau) = (t(\tau), x(\tau), y ,z)$
where already the fact has been used that for a motion along the $x=x^1$-axis the coordinates
$y=x^2$ and $z=x^3$ do not change. These two ``inert'' coordinates will not be 
displayed explicitly in the following. Denoting with a dot the partial derivative 
with respect to $\tau$ it is easy to verify that the ``two''-velocity $(\dot t (\tau), 
\dot x(\tau))$ is normalised to one, and that with the metric convention $ds^2=dt^2-dx^2$
one has in any inertial frame $\ddot x_\alpha(\tau) \ddot x^\alpha(\tau) = -a^2$, where $a$
is the proper constant acceleration. Using the inertial lightcone coordinates 
$u\equiv t-x$ and $v\equiv t+x$ one immediately deduces 
\be
\dot u(\tau) \dot v(\tau) = 1 \, , \quad \ddot u(\tau) \ddot v(\tau) = -a^2 \, ,
\ee
which can be solved straightforwardly,
\be
u(\tau) = - \frac 1 a e^{-a\tau} \, , \quad v(\tau) =  \frac 1 a e^{a\tau} \, . 
\ee
Note that the additive integration constant has been put here to zero, the 
multiplicative one to one. This leads to  
\be
x(\tau) = \frac 1 2 (v-u) = a^{-1} \cosh(a\tau),\;\;\; t(\tau)= \frac 1 2 
(v+u) a^{-1} \sinh(a\tau) \, , 
\ee
which obviously fulfils $x(\tau)^2 - t(\tau)^2 = a^{-2}$.
The worldline of an observer with constant proper acceleration $a$ is therefore
given by a hyperbola, {\it cf.} figure \ref{UnruhHyperbola}.

Again, following chapter 8 of \cite{Mukhanov:2007zz} conformally flat comoving
coordinates are introduced:
\begin{eqnarray}
x^0&=&\frac 1 a e^{a\xi^1}\sinh(a\xi^0) \, , \nonumber \\
x^1&=&\frac 1 a e^{a\xi^1}\cosh(a\xi^0) \, , \label{ComovCoord}
\end{eqnarray}
and as $x^2=\xi^2$, $x^3=\xi^3$  these two coordinates will also not be 
displayed explicitly in the following.
A short calculation shows that
\begin{equation}
ds^2=(dx^0)^2 - (dx^1)^2  = e^{2a\xi^1} \left( (d\xi^0)^2 - (d\xi^1)^2  \right) .
\end{equation}
For the hyperbola depicted in figure \ref{UnruhHyperbola} one shows straightforwardly
that $\xi^1=0$, {\it i.e.}, $\xi^1$ is not only a constant along the trajectory of the
observer with constant proper acceleration $a$ but it vanishes for the whole
trajectory.

The coordinate $\xi^0$ is the proper time along the trajectory. This can be most
straightforwardly seen by the fact that the hyperbola is tangential to the Killing field
generated by
\begin{equation}
\partial_{\xi^0} = a (x^1 \partial_0 + x^0 \partial_1) \, .
\end{equation}

For the calculation presented in chapter \ref{Chap6} 
one needs an expression for the Lorentz invariant
distance for two events along the hyperbola at two different proper times
$\xi^0$ and $\xi^{0\prime}$:
\begin{eqnarray}
(x-x^\prime)^2 &=& \left( (x^0-x^{0\prime})^2 - (x^1-x^{1\prime})^2\right)_{\xi^1=0}
\nonumber \\
&=& \frac 1 {a^2}  \left( -2 +2 (\cosh(a\xi^0) \cosh(a\xi^{0\prime}) -
\sinh(a\xi^0) \sinh(a\xi^{0\prime}) )\right)
\nonumber \\
&=& \frac 1 {a^2}  \left( -2 + 2 \cosh ( a (\xi^0 - \xi^{0\prime})) \right)
= \frac 4 {a^2}  \sinh^2 \left( \frac a 2 (\xi^0 - \xi^{0\prime}) \right) \, ,
\end{eqnarray}
respectively,
\begin{equation}
\pm \sqrt{(x-x^\prime)^2} = \pm
\frac 2 a \sinh \left( \frac a 2 (\xi^0 - \xi^{0\prime}) \right)
\, .
\end{equation}

\newpage
\thispagestyle{empty}

\end{appendix} 


\cleardoublepage
\phantomsection
\markboth{\MakeUppercase{\color{MYBLUE}Bibliography}}
{\MakeUppercase{\color{MYBLUE}Bibliography}}
\addcontentsline{toc}{chapter}{{\color{MYBLUE}Bibliography}}
\bibliographystyle{utphys}
\bibliography{phd_thesis}

\newpage
\thispagestyle{empty}


\chapter*{{\color{MYBLUE}Acknowledgements}}
\addcontentsline{toc}{chapter}{{\color{MYBLUE}Acknowledgements}}

In the last four years many persons have contributed to the success of this thesis. Let me start by
thanking Renate Loll for giving me the possibility to work in her group and my adviser, 
Frank Saueressig, for suggesting the many interesting topics which became part of this exciting journey,
and his valuable input. In 
addition I am grateful to both of them for a critical reading of my thesis and their comments.

While working on the different projects I had the pleasure to discuss with many different outstanding 
physicists, and I would like to take the opportunity to explicitly thank some of them here: Daniel Becker, Tu\u{g}ba B\"{u}y\"{u}kbe\c{s}e, Holger Gies, Benjamin Knorr, Daniel F. 
Litim, Jan M. Pawlowski, Roberto Percacci, Chris Ripken, H\`{e}lios Sanchis Alepuz, Fleur Versteegen and Omar Zanusso.

Hereby I would like to especially thank Jan M. Pawlowski for all the helpful and insightful discussions. I 
really enjoyed all the thrilling discussions with Jan, being about physics or not, they were always very 
enjoyable, specially the ones at late hours!

My thanks also go to my colleagues in the quantum gravity group in Nijmegen, where in particular I want to 
mention my office mate, Chris, with whom I had a really nice time together, and Reiko Toriumi for the fun 
times at the Cultuurcaf\'{e}.

As one can easily be swamped by a lot of work, I am really thankful to my (non-physicists) friends, Daphne 
Broeks and Leonie Lautz, who brought me back to Earth and all the fun we had together.

Furthermore, I thank cordially the manuscript commission for the critical reading and comments to my 
thesis.

And of course, I am very grateful to the Netherlands Organization for Scientific Research (NWO) within the 
Foundation for Fundamental Research on Matter (FOM) grant 13PR3137 for all the financial support.

Last but not least, I am in deep gratitude to my family and to my husband for all their support and providing  mental strength.

\newpage
\thispagestyle{empty}


\chapter*{{\color{MYBLUE}Curriculum Vitae}}
\addcontentsline{toc}{chapter}{{\color{MYBLUE}Curriculum Vitae}}

\section*{Personal Information}
\begin{tabular}{L!{\VRule}R}
Date of Birth & May 6th, 1987\\[5pt]
Place of Birth & Niter\'oi, RJ, Brazil\\
\end{tabular}

\section*{Education}
\begin{tabular}{L!{\VRule}R}
2002--2004 & Col\'egio Rede MV1 (High School), Brazil\\[5pt]
2005--2009 & Bachelor of Science in Physics at the Universidade do Estado do Rio de Janeiro and Karl-Franzens-Universit\"at Graz; Bachelor Thesis: ``Applying the Worm Algorithm to the 2D Ising Model''; adviser: Christof Gattringer\\[5pt]
2010--2013 & Master of Science in Physics at the Karl-Franzens-Universit\"at Graz; Master Thesis: ``Renormalisation Group for Gravity and Dimensional Reduction''; adviser: Daniel F. Litim (University of Sussex, UK) and co-adviser: Bernd-Jochen Schaefer (Karl-Franzens-Universit\"at Graz)\\[5pt]
2014--2018 & Doctoral candidate in Physics at the Radboud Universiteit Nijmegen; PhD Thesis: ``Quantum Gravity from Fundamental Questions to Phenomenological Applications''; adviser: Frank S. Saueressig
\end{tabular}

\end{document}